\documentclass[twoside,openright,a4paper,11pt]{book}
\usepackage{amssymb,amsfonts,amsmath}   
\usepackage{graphicx,float}   
\usepackage{physics}
\usepackage{dsfont}   	         		
\usepackage{mathrsfs}   	       		
\usepackage{tcolorbox}      	        
\usepackage[normalem]{ulem}         	
\usepackage{todonotes}					
\usepackage{xcolor}
\usepackage{esvect}
\usepackage{mathtools}
\usepackage[toc,page]{appendix}
\usepackage[square,sort,comma,numbers,sort&compress]{natbib}
\usepackage{epigraph}

\usepackage[Glenn]{fncychap}
\ChNameAsIs
\ChTitleAsIs
\ChTitleVar{\Huge \bf}

\usepackage{bold-extra}

\definecolor{myblueviolet}{rgb}{0.4, 0.0, 0.9}
\definecolor{antiquefuchsia}{rgb}{0.57, 0.36, 0.51}
\definecolor{amethyst}{rgb}{0.6, 0.4, 0.8}
\definecolor{darkblue}{rgb}{0.0078, 0.0275, 0.3647}	
\definecolor{darkred}{rgb}{0.55, 0.0, 0.0}
\definecolor{darkviolet}{rgb}{0.58, 0.0, 0.83}	
\definecolor{midnightblue}{rgb}{0.1, 0.1, 0.44}	
\definecolor{oxfordblue}{rgb}{0.0, 0.13, 0.28}
\definecolor{prussianblue}{rgb}{0.0, 0.19, 0.33}
\definecolor{goldenrod}{rgb}{0.85, 0.65, 0.13}
\definecolor{darkgreen}{rgb}{0.0, 0.5, 0.0}
\definecolor{Ultramarine}{rgb}{18,10,143}
\definecolor{blue-violet}{rgb}{0.54, 0.17, 0.89}
\definecolor{ceruleanblue}{rgb}{0.16, 0.32, 0.75}
\definecolor{darkslateblue}{rgb}{0.28, 0.24, 0.55}
\definecolor{hanpurple}{rgb}{0.32, 0.09, 0.98}
\definecolor{indigo(web)}{rgb}{0.29, 0.0, 0.51}

\usepackage[pagebackref=false,pdftex,pdfborderstyle={/S/U/W 1},hyperfootnotes=true,colorlinks,
linkcolor={myblueviolet},
citecolor={darkred},
urlcolor={darkred}]{hyperref}

\usepackage{setspace}
\usepackage{wrapfig,sidecap,,fullpage,morefloats}
\usepackage{grffile,float,xspace}
\usepackage[sort&compress]{natbib}     
\usepackage{slashed}
\usepackage{makecell}
\usepackage{multirow,array,bm,bbm,esint}

\usepackage[numbib]{tocbibind}  
\usepackage{fancyhdr} 

\newcounter{daggerfootnote}

\usepackage{caption}                     
\newcommand{\bec}{\begin{center}}
	\newcommand{\eec}{\end{center}}
\newcommand{\beq}{\begin{equation}}
	\newcommand{\eeq}{\end{equation}}
\newcommand{\bea}{\begin{eqnarray}}
	\newcommand{\eea}{\end{eqnarray}}
\newcommand{\nn}{\nonumber}

\newcommand{\hf}{\frac{1}{2}}
\newcommand{\qtr}{\frac{1}{4}}
\newcommand{\psib}{{\overline{\psi}}}
\newcommand{\B}{{\mathcal{B}}}
\newcommand{\Q}{{\mathcal{Q}}}
\newcommand{\Qb}{{\overline{\mathcal{Q}}}}
\newcommand{\mcS}{\mathcal{S}}

\newcommand{\cD}{{\cal D}}

\newcommand{\cN}{{\cal N}}
\newcommand{\cO}{{\cal O}}

\newcommand{\vsphf}{\vspace{0.5cm}}

\newcommand{\thesistitle}{Non-Perturbative Simulations of Quantum Field Theories using Complex Langevin Dynamics}

\newcommand{\degree}{Doctor of Philosophy}
\newcommand{\candidate}{Arpith Kumar}

\newcommand{\department}{\href{https://web.iisermohali.ac.in/dept/physics/}{Department of Physical Sciences}}
\newcommand{\university}{\href{https://www.iisermohali.ac.in/}{Indian Institute of Science Education and Research  Mohali\\Knowledge City, Sector 81, SAS Nagar, Manauli PO, Mohali 140306 Punjab, India}}

\usepackage[titles]{tocloft}

\rmfamily
\begin{document}
	
	\pagenumbering{roman}
	\begin{titlepage}
	
	\vfill
	\begin{center}
	\vfill
	{\huge { {\textsc{\thesistitle}}} \par}\vspace{0.01cm}
	\vfill
	{\LARGE \sc \candidate \vskip 0.3cm} 
	
	\vfill
	\textit{\large A thesis submitted for the partial fulfillment of }\\[0.1cm] 
	\textit{\large the degree of \degree}\\[0.4cm]
	\vfill
	
	\centering
	\includegraphics[width=3.5cm]{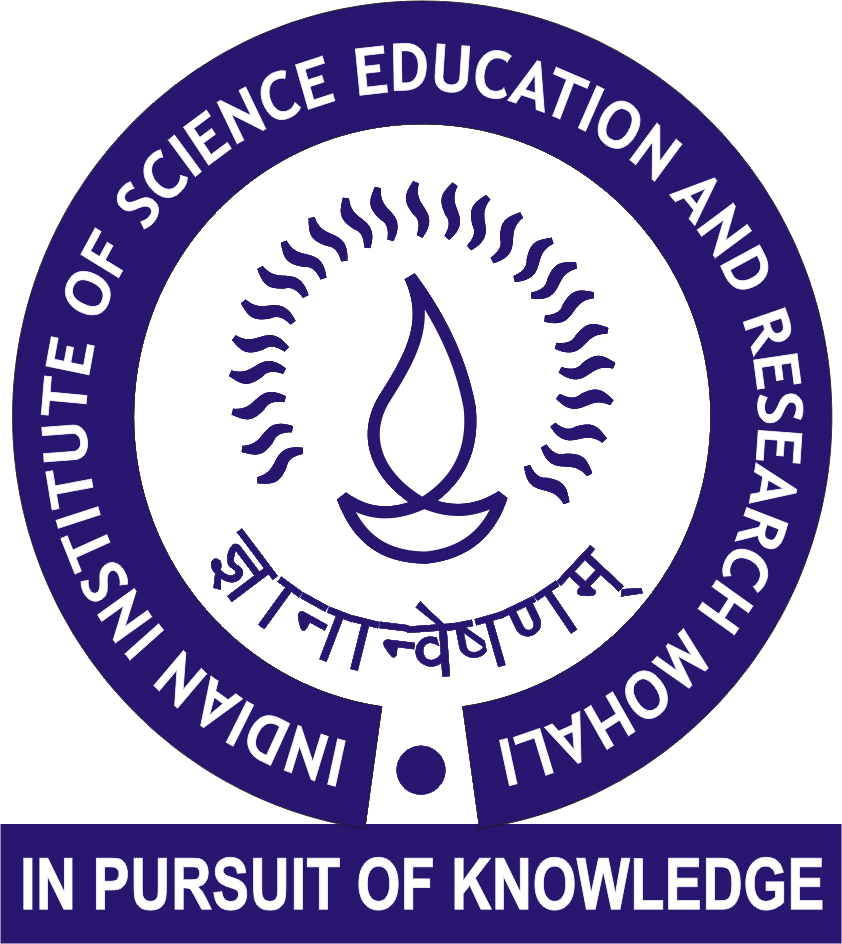}
	\vfill
	{\sc \small \department}
	\\
	\vspace{0.cm}
	{\sc \small \university}
	\vfill
	{\sc March 2023}
	\end{center}
	\vfill
	
\end{titlepage}

	\clearpage\mbox{}\clearpage
	\setlength\oddsidemargin{\dimexpr(\paperwidth-\textwidth)/2 - 0.8in\relax}
	\setlength\evensidemargin{\dimexpr(\paperwidth-\textwidth)/2 - 1.2in\relax}
	
	
	\vspace*{\fill}
	\begin{center}
		\Large \emph{dedicated to my parents and my brother  \dots}
	\end{center}
	\vspace*{\fill}
	\setstretch{1.4}		
	\cleardoublepage
\phantomsection
\addcontentsline{toc}{chapter}{Abstract}
\begin{center}
	\section*{Abstract}
\end{center}
Non-perturbative formulations of field theories are essential to capture numerous intriguing physical phenomena, including confinement in quantum chromodynamics, spontaneous supersymmetry (SUSY) breaking, and dynamical compactification of extra dimensions in superstring theories. Regularizing field theories on a spacetime lattice provides a robust framework for studying their non-perturbative features. The underlying theory can be quantized on a spacetime lattice using Euclidean path integrals. Conventionally, these path integrals are evaluated using numerical methods based on Monte Carlo importance sampling, where generating field configurations requires the Boltzmann factor to be interpreted as a probability weight. However, various interesting physical systems have complex actions, rendering the Boltzmann factor complex, and thus, path integral Monte Carlo encounters the sign problem. The complex Langevin method (CLM) based on stochastic quantization aims to overcome the sign problem by analyzing the associated Langevin dynamics to evaluate complex integrals. This thesis employs the CLM to investigate various non-perturbative aspects of field-theoretic systems with complex actions.

Physicists have long sought a unified description of all fundamental interactions of nature, and SUSY is now widely accepted as a necessary ingredient for such unifying approaches. However, since experimental evidence suggests that low-energy physics is manifestly non-supersymmetric, SUSY must be spontaneously broken at some energy scale. This thesis probes the possibility of spontaneous SUSY breaking in the simplest realizations of supersymmetric field theories. These systems generally have complex actions arising from a complex determinant of the fermion operator, and the phase of the determinant plays a critical role in determining the correct vacuum. We studied various interesting classes of, in general, complex superpotentials, including the ones exhibiting $\mathcal{PT}$-symmetry. Non-Hermitian $\mathcal{PT}$-symmetric theories are fascinating because they have real and below-bounded energy spectra. We first considered zero-dimensional supersymmetric systems with one bosonic and two fermionic variables. In the case of spontaneous SUSY breaking, the partition function (Witten index) vanishes, and the normalized expectation values encounter an indefinite form. We overcome this difficulty by using twisted boundary conditions on fermionic fields and then taking the vanishing limit of the twist parameter. Our CLM simulations reliably predicted the presence or absence of SUSY breaking for various superpotentials. We then considered $\mathcal{N}=2$ supersymmetric quantum mechanical models with appropriate lattice regularization. Here also, we overcame the indefinite form of normalized observables by using twisted boundary conditions. While applying the CLM, we noticed that some models suffered from the singular-drift problem. In such cases, we introduced appropriate deformation parameters such that the CLM correctness criteria are respected and then recovered the original theory by taking the vanishing limits of the deformation parameters. Our analysis demonstrated that the CLM could reliably probe dynamical SUSY breaking in various quantum mechanics models with real and complex actions. We then extend our zero- and one-dimensional analysis to two-dimensional field-theoretic systems. As a warm-up, we first laid out the lattice construction for bosonic field theories, including $\mathcal{PT}$-invariant potentials. We then introduced fermions and considered the $\mathcal{N}=1$ Wess-Zumino model, a two-dimensional model with minimal fields. We then applied the CLM for double-well superpotential and examined the relationship between parity symmetry and supersymmetry.

Another exciting aspect of non-perturbative physics we explore in this thesis is the dynamical compactification of extra dimensions in superstring theories. Superstrings are the most promising theories for unifying all interactions, including gravity. However, these theories are consistently defined in ten dimensions. The connection to the real world, where only four dimensions are macroscopic, is realized in the non-perturbative definition of superstrings via compactification of the six extra dimensions. Matrix models in the large-$N$ limit are conjectured as non-perturbative formulations of superstring theories. In this thesis, we study a constructive formulation of the type IIB superstring, the IKKT (type IIB) matrix model. A smooth spacetime manifold is expected to emerge from the eigenvalues of the ten bosonic matrices in this model. When this happens, the SO(10) symmetry in the Euclidean signature must be spontaneously broken. The Euclidean version has a severe sign problem due to the inherently complex nature of the Pfaffian. This thesis probes the possibility of spontaneous rotational symmetry breaking in the Euclidean version of the IKKT matrix model. We resolved the singular-drift problem associated with CLM by introducing supersymmetry-preserving deformations with a Myers term. The original IKKT model can be recovered at the vanishing deformation parameter limit. Our preliminary analysis indicates that the phase of the Pfaffian indeed induces spontaneous SO(10) symmetry breaking in the Euclidean IKKT model.

The investigations performed in this thesis suggest that the CLM can successfully simulate the non-perturbative aspects of quantum field theories by taming the associated sign problem.


	\tableofcontents
	\clearpage
	\mbox{}
	\clearpage
	\listoffigures
	\listoftables
	\clearpage\mbox{}\clearpage
	\clearpage\mbox{}\clearpage
	\pagenumbering{arabic}
	
	\setstretch{1.35}	
	\pagestyle{fancy}
	\fancyhead{}
	\fancyfoot{}
	\fancyhead[LE]{\footnotesize \textbf{\thepage} \ \ \ \   \ Chapter \thechapter \ \ \ \textit{\leftmark}}
	\fancyhead[RO]{\footnotesize  \thesection \ \ \ \textit{\rightmark}  \ \ \ \   \ \textbf{\thepage}}
	\renewcommand{\sectionmark}[1]{\markright{#1}}
	\renewcommand{\subsectionmark}[1]{}
	
	\renewcommand{\chaptermark}[1]{\markboth{#1}{}}
	\renewcommand{\headsep}{30pt}
	\setlength{\headheight}{12pt}

	\chapter{Introduction}
\label{chap:introduction}

Knowledge of the fundamental interactions among the constituents of matter is imperative to comprehend nature. Quantum field theory (QFT), a theoretical framework born of the inevitable necessity of combining the special theory of relativity and quantum mechanics, provides deep and profound insights into fundamental interactions in nature. In particular, the electromagnetic, weak, and strong interactions between
elementary particles can be understood through respective field-theoretic descriptions, videlicet quantum electrodynamics (QED), quantum flavourdynamics (QFD), and quantum chromodynamics (QCD). The standard model (SM) of particle physics is a consistent quantum field theoretical unification of three of the four known fundamental interactions, with gravity excluded. For decades, the SM has successfully explained, to a great extent, the plethora of particles experimentally discovered.

A perturbative approach to QFT has yielded impressive results for weakly interacting theories. The anomalous magnetic moment of the electron (first calculated by Julian Schwinger in 1948 \cite{Schwinger:1948iu}), computed by regularizing QED order by order in the coupling, is one of the best-understood physical quantities from the perturbative approaches. However, perturbation theory is only an asymptotic expansion, and the sum of all orders is divergent. Moreover, it is well known that in strongly coupled theories, such as QCD at low energies, the perturbative regularization is entirely ineffective. The non-Abelian gauge symmetry and the resulting asymptotic freedom (discovered in 1973 by David Gross and Frank Wilczek \cite{Gross:1973ju, Gross:1973id}, and independently by David Politzer \cite{Politzer:1973fx}) inherent in QCD predicts the existence of exchange particles referred to as gluons, whose interactions essentially confine quarks to bound hadrons like mesons and baryons. At low energies, the confinement of quarks and gluons within composite hadrons leads to the breakdown of conventional perturbation theory and makes elementary dynamics of SM fermions inaccessible. Therefore, to capture numerous intriguing and physical non-perturbative phenomena such as confinement, chiral symmetry breaking, and Higgs mechanism (in 1960, Yoichiro Nambu offered the conjecture leading to a series of astounding developments \cite{Nambu:1961tp, Goldstone:1961eq, Goldstone:1962es, Anderson:1963pc, Higgs:1964ia}), there is a need to define a formulation beyond perturbation theory. The lattice regularized path integrals provide a robust framework for studying the non-perturbative aspects of QFTs.

\section{Euclidean path integral}
Feynman path integrals are functional integrals over all quantum mechanically possible space of trajectories satisfying some boundary conditions \cite{Feyman:thesis, Feynman:1948ur}. A prominent way to extract non-perturbative physics is through the exact evaluation of these path integrals. Let us consider functional integral formalism for the theory of a real scalar field $\phi(\vv{x},t)$ in four-dimensional Minkowski spacetime. The action $S[\phi(\vv{x},t)]$ of the theory is a spacetime integral of the Lagrangian density $\mathcal{L}(\phi,\partial_{\mu}\phi)$, that is
\beq
S[\phi(\vv{x},t)] = \int_{0}^{T} dt \int_{\Omega} d^3x~ \mathcal{L}(\phi,\partial_{\mu}\phi) = \int_{0}^{T} dt \int_{\Omega} d^3x~ \left[	 \hf \partial_{\mu} \phi \partial^{\mu} \phi -V(\phi)  \right].
\eeq
where the time integration is over a fixed interval from $0$ to $T$, $\Omega$ represents a finite spatial volume, and $V(\phi)$ represents a potential term. In the path integral formulation.

In the path integral formulation, propagators of fields, or more generally, the position space Green's functions are the natural objects to derive physical information of the system. Then the propagator to go from field configuration $\phi_1(\vv{x})$ to $\phi_2(\vv{x})$ in say time $T$, that is $G(\phi_2(\vv{x}), \phi_1(\vv{x});T)$ is given by the probability amplitude
\bea
G(\phi_2(\vv{x}), \phi_1(\vv{x});T) = \langle \phi_2(\vv{x}) | e^{-iHt} | \phi_1(\vv{x})   \rangle
\equiv \underbrace{\int_{\phi(\vv{x},0)\equiv \phi_1(\vv{x})}^{\phi(\vv{x},T)\equiv \phi_2(\vv{x})} \mathcal{D} \phi}_{_{ \substack{\text{over the space of} \\ \text{continuous trajectories}}}}~ e^{iS[\phi(\vv{x},t)]},
\eea
where $H$ is the Hamiltonian of the system. The integrand in these path integrals has a highly oscillatory nature of form $\exp(iS[\phi])$, where $i$ is the imaginary unit. If the oscillations were suppressed, it might be possible to define a sensible measure on the set of paths. With this in mind, much of the rigorous work in path integral formalism is concerned with analytically continued {\it Euclidean time}. To understand this analytical continuation, let us study the analogy with statistical mechanics and consider the canonical system at inverse temperature $\beta$ with partition function
\beq
Z(\beta) = {\rm Tr} \left(  e^{-\beta H} \right) = \sum_{n}e^{-\beta E_n}. 
\eeq
From the basis independent property of trace operation, we have the partition function in position space given by
\beq
Z(\beta) = \int  \mathcal{D} \phi (\vv{x})~ \langle \phi(\vv{x}) | e^{-\beta H} | \phi(\vv{x})   \rangle. 
\eeq
The analogy between the partition function and the propagator is now reasonably noticeable. If we perform a {\it Wick rotation} to {Euclidean time}  \cite{Wick:1954eu, Baym:1961}, that is $t\to -i\tau $, we can write
\bea
\langle \phi(\vv{x}) | e^{-\beta H}  | \phi(\vv{x})   \rangle = {\int_{\phi(\vv{x},0)\equiv \phi(\vv{x})}^{\phi(\vv{x},\beta)\equiv \phi(\vv{x})} \mathcal{D} \phi}~ e^{-S_{E}[\phi(\vv{x},\tau)]} = G(\phi(\vv{x}), \phi(\vv{x});-i \beta),\\
S_{E}[\phi(\vv{x},\tau)] = \int_{0}^{\beta} d\tau \int_{\Omega} d^3x~  \mathcal{L}_{E}(\phi,\partial_{\mu}\phi) =\int_{0}^{\beta} d\tau \int_{\Omega} d^3x~ \left[	 \hf \partial_{\mu} \phi \partial^{\mu} \phi + V(\phi)  \right], 
\eea
where $S_{E}[\phi(\vv{x},\tau)] $ is the Euclidean action, $\mathcal{L}_{E}$ is the Euclidean Lagrangian density and the field configurations $\phi(x,\tau)$ describe the state of the system in Euclidean spacetime. Then the partition function in Euclidean signature can be written as 
\bea
Z(\beta) &=& \int  \mathcal{D} \phi (\vv{x})~ \langle \phi(\vv{x}) | e^{-\beta H} | \phi(\vv{x})   \rangle \\ 
&=&  \int  \mathcal{D} \phi (\vv{x})~ {\int_{\phi(\vv{x},0)\equiv \phi(\vv{x})}^{\phi(\vv{x},\beta)\equiv \phi(\vv{x})} \mathcal{D} \phi}~ e^{-S_{E}[\phi(\vv{x},\tau)]} \\
&=&  \oint_{PBC} \mathcal{D} \phi ~ e^{-S_{E}[\phi(\vv{x},\tau)]}, 
\eea
where we landed up with periodic boundary conditions (PBC), that is, $\phi(\vv{x},0) = \phi(\vv{x},\beta)$  for scalar fields. The corresponding observables in the Euclidean QFTs are then computed as follows,
\bea
{\langle \mathcal{O}(\phi) \rangle} = \frac{1}{Z(\beta)} \int \mathcal{D} \phi~ \mathcal{O}(\phi) ~e^{-S_{E}[\phi]},
\eea 
and then analytically continued back to real-time dynamics by inverse Wick rotation. Although the functional integral has much better mathematical behavior, the path integral is still infinite-dimensional, and the exact computation is intractable. This is where lattice regularized QFTs and simulation algorithms come into play. 

\section{Lattice regularization}
The underlying theory can be quantized on a spacetime lattice using Euclidean path integrals. The idea is to replace a field variable defined at every spacetime point with a variable defined at some location on a lattice. The concept of discrete spacetime lattice is probably older than the field variable itself. Lattice is, in fact, a physical quantity in some condensed matter systems, such as in the description of electrons in a crystal. However, in 1974, Kenneth G. Wilson was the first to specifically replace a continuum gauge field theory with a lattice gauge theory, from which this field grew. He showed confinement at the strong coupling limit in lattice QCD \cite{Wilson:1974sk}. The simplest lattice prescription for the naive discretization of a four-dimensional spacetime is as follows;
\bea
\int_{0}^{\beta}dt \int_{\Omega} d^3 x  &\to& a^4 \sum_{r} \\
\phi(\vv{x}, t) &\to& \phi_r \\
\partial_{\mu} \phi(\vv{x}, t) &\to& \sum_{\mu} \frac{\left( \phi_{r+e_{\mu}} - \phi_{r}  \right)}{a} 
\eea
where $r, e_{\mu}$ are four-dimensional lattice and unit vector, respectively, $\phi_{r}$ is the field at lattice point represented by vector $r$ and $a$ is the lattice spacing. The temporal and spatial extents in the continuum theory can be related to the number of lattice sites, $\beta = L_t a$ and $\Omega = (L_x a)^3$ (in general, we can have different lattice spacing in different directions), where $L_t$ and $L_x$ are the total numbers of sites in lattice temporal and spatial direction, respectively. The original infinite-dimensional continuous path integral is equivalent to the lattice regularized finite-dimensional form in the continuum limit, that is
\bea
{Z} = \int \mathcal{D} \phi~ e^{-S[\phi]} \equiv \lim_{\substack{a\to 0\\ L_t\to \infty \\ L_x\to \infty}} \int \left( \prod_{r} d\phi_r \right)~e^{-S_{ lat}[\phi_r]},
\eea
where $S_{ lat}[\phi_r]$ is the lattice action. We can use this lattice regularized expression to compute observables of the theory using numerical methods.

Conventionally, these path integrals in the Euclidean signature are evaluated using Monte Carlo based on importance sampling \cite{Metropolis:1953am}. Therefore, the computation of a Euclidean QFT turns into a simulation of a statistical system. For a real-valued action $S[\phi(\vv{x}, t) ]$, a set of field configurations, $\{\phi_r \}$  are generated by interpreting the normalized Boltzmann factor $Z^{-1}\exp(-S_{lat}[\phi_r ])$ as a probability weight. Then, with the help of the ergodicity property, the expectation values of observables are defined as averages over a large number of such sampled configurations ($N_{\{\phi_r \}}$), that is
\bea
{\langle \mathcal{O}(\phi) \rangle}_{E} &=& \frac{1}{Z} \int \mathcal{D} \phi~ \mathcal{O}(\phi)~ e^{-S[\phi]}\\
&\equiv& \lim_{\substack{a\to 0\\ L_t\to \infty \\ L_x\to \infty}} \int \left( \prod_{r} d\phi_r \right)~\mathcal{O}(\phi_r)e^{-S_{  lat}[\phi_r]} \\
&\approx& \frac{1}{N_{\{\phi_r \}}} \sum_{\{\phi_r \}} \mathcal{O}(\phi_r).
\eea
It is essential to question whether the above-described procedure works even when the Euclidean action is manifestly complex.

\section{Complex actions and sign problem}

Path integral Monte Carlo works reliably for real actions. However, many physical systems of theoretical and experimental interest have complex Euclidean actions. In fact, models with a complex Euclidean action are so widespread that they are arguably in the majority instead of being exceptions. Some important examples of such complex action systems are mentioned below: 
\begin{itemize}
	\item {\it Field Theories in Minkowski space}: Even though a Wick rotation usually produces a real Euclidean action, it is nevertheless intriguing to look for a technique that could address the primary issue and not the subsidiary one in the Euclidean signature \cite{Anzaki:2014hba, deAguiar:2010ue}.
	
	\item {\it Systems with chemical potential}: Exploring strongly interacting QCD systems for non-vanishing chemical potential at finite temperatures is of great physical interest. However, the fermion determinant becomes complex in SU$(N)$ gauge theories for $N\ge 2$, leading to a complex effective action \cite{Aarts:2010gr, Schmalzbauer:2016pbg, Nagata:2017pgc}.
	
	\item {\it Theories with external charges}: A gauge theory with external static charges has a non-trivial Euclidean action. The computation of the string tension and width between two static charges is physically significant, particularly because of correspondence to the confinement of quarks \cite{Ambjorn:1986fz}. 
	
	\item {\it System with fermions}: Fermions in the conventional path integral formalism do not have a real $c$-number representation. Simulating fermions directly is difficult (practically even impossible); therefore, we use indirect approaches resulting in fermion determinant/Pfaffian, where one encounters complex actions and negative probabilities \cite{Li_2019, Wang:2015vha, chen2004lattice}.
	
	\item {\it Effective actions in the presence of topological $\theta$-term or Chern-Simons theories}: In QFT, topological terms such as $\theta$-term, Chern-Simons, and WZW may have non-trivial effects on the low energy theory. Chern–Simons gauge theory, the first example of a topological QFT, generalizes compact Lie group to complex Lie group \cite{Witten:1989ip, Gukov:2003na}. A Wick rotation cannot make the Boltzmann factor real and positive for such actions. The $\mathcal{CP}$-symmetry preservation in QCD experiments (the strong $\mathcal{CP}$ problem) is potentially correlated with the presence of a topological $\theta$-term coupled with topological charge $Q$. Such terms in the action are purely imaginary, that is, $S_{\theta} =-i\theta Q$, and introduce a complex phase problem \cite{Bongiovanni:2014rna, Hirasawa:2020bnl, Matsumoto:2021zjf}.

	\item {\it Non-equilibrium physics of quantum many-body systems}: Quantum Monte Carlo methods are fruitful in understanding the physics in equilibrium and can be easily extended to non-equilibrium situations. However, this entails estimating integrals that contain combinations of oscillating factors, and the computational cost rises exponentially with simulation time \cite{Eisert:2014jea, PhysRevLett.115.266802}.
	
	\item {\it Condensed matter systems of strongly correlated electrons}: In the condensed matter context, the most striking example is the repulsive Hubbard model on a bipartite or triangular lattice, which faces a severe sign problem upon introducing a finite doping \cite{Loh:1990zz, Blanc:2015, Yamamoto:2015ura}. Other exciting candidates include the spin-polarized electron gas, frustrated magnetic systems, Shastry-Sutherland antiferromagnetic spin model, to name only a few \cite{Huffman:2013mla, Wessel_2018, D_Emidio_2020, Pan:2022fgf}.
\end{itemize}

Field theories with complex actions are challenging to address non-perturbatively since the Boltzmann factor is non-positive or more generally complex. Although the partition function is well defined, the Boltzmann factor cannot be interpreted as probability weight, and the path integral Monte Carlo encounters the notorious {\it sign problem}. The sign problem goes by many names in different areas of physics. Sometimes, it is also known as {\it numerical sign problem}, {\it complex phase problem}, {\it complex action problem}, or even {\it negative sign problem}. It ranks among the most important and infamous problems in modern computational physics and is a widespread numerical hurdle preventing equilibrium behavior analysis of diverse physical systems at the frontier of physics.

Sign problem inhibits the application of importance sampling without significant modifications; choosing a positive-definite probability distribution becomes difficult (or even impossible). The most straightforward and brute-force idea to circumvent the sign problem is the re-weighting procedure \cite{Ferrenberg:1988yz}, which incorporates the non-positive part (complex phase) of the Boltzmann weight into the observable. The complex Boltzmann weight can be rewritten as $e^{-S[\phi]} = |e^{-S[\phi]}|e^{i\omega}$, with $e^{i\omega}$ being the complex phase. Then, the field configurations are sampled using the magnitude of weight or phase-quenched weight, $|e^{-S[\phi]}|$, as a probability measure. The expectation values of observables have the form 
\bea
{\langle \mathcal{O}(\phi) \rangle} = \frac{1}{Z} \int \mathcal{D} \phi~ \mathcal{O}(\phi)~ e^{-S[\phi]} = \frac{\int \mathcal{D} \phi~ \mathcal{O}(\phi) ~|e^{-S[\phi]}|e^{i\omega}}{\int \mathcal{D} \phi ~|e^{-S[\phi]}|e^{i\omega}}. 
\eea
By multiplying both the numerator and denominator by the phase-quenched partition function, $Z_{\rm pq}= \int \mathcal{D} \phi~|e^{-S[\phi]}|$, we get
\bea
{\langle \mathcal{O}(\phi) \rangle} &=& \frac{ \int \mathcal{D} \phi~ \mathcal{O}(\phi) ~|e^{-S[\phi]}|e^{i\omega}}{\int \mathcal{D} \phi~|e^{-S[\phi]}|} \times \frac{\int \mathcal{D} \phi~ |e^{-S[\phi]}|}{\int \mathcal{D} \phi~|e^{-S[\phi]}|e^{i\omega}} \\
&=& \frac{{\langle \mathcal{O}(\phi) e^{i\omega} \rangle}_{Z_{\rm pq}}}{{\langle e^{i\omega} \rangle}_{Z_{\rm pq}}},
\eea
where ${\langle \cdot  \rangle}_{Z_{\rm pq}}$ denotes expectation values with respect to phase-quenched weight $|e^{-S[\phi]}|$. Thus, re-weighting has redefined the original problem of computing complex observables with respect to real and positive Boltzmann weight. Despite the elegance of the procedure, in practice, due to the highly oscillating nature of $e^{i\omega}$, both numerator and denominator have extremely small values and vanish exponentially as the physical extent of the spacetime lattice is increased. The severity of the sign problem is measured by the expectation value of the complex phase, that is
\bea
\langle e^{i\omega} \rangle_{Z_{\rm pq}} = \frac{Z}{Z_{\rm pq}} = e^{-\Omega \Delta f},
\eea
where $\Omega$ is the spacetime lattice volume, and $\Delta f = f-f_{\rm pq}$ is the difference in free energy densities for original and phase-quenched theory.\footnote{$Z_{\rm pq}$ corresponds to a bosonic ensemble containing sum over non-negative real numbers while $Z$ is a fermionic path integral in which the phase is taken into account. Then, $\Delta f$, the free energy density difference between bosonic and fermionic systems is necessarily positive. See Refs. \cite{Chandrasekharan:1999cm, Chandrasekharan:1999ys, Alford:2001ug, Berger:2019odf} for detailed discussions.} The minuscule nature of $\langle e^{i\omega} \rangle$ is apparent since $Z\leq Z_{\rm pq}$ and vanishes as $\Omega \to \infty$. What is even worse is that the statistical uncertainty $\sigma$ in Monte Carlo, which for $N_{\phi}$ samples decreases as ${N_{\phi}}^{-{1}/{2}}$, is overpowered by the exponentially decaying behavior of ${\langle e^{i\omega} \rangle}_{Z_{\rm pq}}$, that is 
\bea
\frac{\sigma}{{\langle e^{i\omega} \rangle}_{Z_{\rm pq}}} = \frac{e^{\Omega \Delta f}}{\sqrt{N_{\phi}}}.
\eea

The equation above demonstrates the enormity of taming the sign problem with straightforward approaches like re-weighting; the sign problem may be regarded as the reappearance of an exponential type of computational wall, which affects non-stochastic methods in the guise of memory requirements and statistical methods in the form of a signal-to-noise problem.

\section{Approaches to circumvent sign problem}

Several approaches have been proposed for solving systems with severe sign problem. Although some of them are recent (even still in development) and have their own merits and demerits, they offer significant potential for taming the sign problem. A few of these successful approaches are listed below: 

\begin{itemize}
	\item {\it Meron-cluster and fermion bag approach} \cite{Rossi:1984cv, Chandrasekharan:1999cm,  Wolff:2008km, deForcrand:2009dh, Chandrasekharan:2009wc,  Chandrasekharan:2010iy, Huffman:2017swn}
	\item {\it Majorana fermions algorithm}	\cite{Li:2014tla, Liu:2015mxb, Li:2017sbk, Li_2019, Wang:2015vha, Wei:2016sgb, Chen:2003vy} 
	\item {\it Density-of-states} \cite{Wang:2000fzi, Bazavov:2012ex, Langfeld:2012ah, Ejiri:2007ga, Fodor:2007vv, Springer:2021liy}
	\item {\it Complexification of space} 
	\begin{itemize}
		\item {\it Path optimization method} \cite{Ohnishi:2017zxh, Mori:2017pne, Mori:2019rql, Mori:2019tux, Ohnishi:2018jjw, Kashiwa:2018vxr, Kashiwa:2019lkv, Detmold:2020ncp, Bursa:2018ykf}
		\item {\it Imaginary asymmetry} \cite{Dagotto:1989fw, Alford:1998sd, deForcrand:2003vyj, DElia:2002tig, DElia:2004ani, DElia:2007bkz, Karbstein:2006er, Philipsen:2012nu, Cea:2012ev, Bonati:2014kpa, Wu:2018oed, Guenther:2017hnx, Borsanyi:2020fev}
		\item {\it Complex Langevin method}: In Chapter \ref{chap:complex-Langevin}, we briefly review this method. 
		\item {\it Lefschetz Thimble method} \cite{Cristoforetti:2012su, Fujii:2013sra, DiRenzo:2015foa, Tanizaki:2015rda, Fujii:2015vha, Alexandru:2015xva}
	\end{itemize}
\end{itemize}

Among these, the approaches based on the complexification of field configuration space have the appeal of general applicability. In this thesis, we have employed one of the most prominent methods in complexification approaches, the complex Langevin method, to investigate non-perturbative aspects of beyond SM field-theoretic systems with complex actions.

\section{Beyond the standard model: non-perturbative aspects}

Over the last half-century, our understanding of the fundamental particles and forces of nature has evolved beyond recognition. The SM has reproduced most known phenomena up to the energy of the order of electroweak scale ($\sim10^2$ GeV). However, there remain many fundamental problems the SM needs to address. First of all, gravity is not incorporated in the SM; there is no mention of one of the four fundamental forces of nature. Another is the famous hierarchy problem. Higgs boson mass is extremely sensitive to any new physics at higher energies (open to alterations by radiative corrections from every scale), and if the SM is valid up to that scale, then due to large quantum corrections, its natural value should be of the order of the Planck mass. Then, the question arises as to what makes the Higgs boson much lighter than the Planck mass (also called the naturalness problem). Also, we live at or below the electroweak scale, far below the Planck scale ($\sim10^{19}$ GeV). How can two such widely separated scales exist in a world quantum theory describes? Moreover, why is the electroweak scale what it is and not much larger? These issues form the hierarchy problem, and because the Planck scale is associated with strong gravitational interactions, the problem is related to constructing a quantum theory of gravity. In addition, the SM does not describe the dark matter or the dark energy of the universe. It still needs to explain why the charges of elementary particles are quantized or the observed neutrino masses and oscillations.

In the past few decades, it has become clear that the SM is a work in progress, and most research in theoretical high-energy physics is focused on finding the theory that will extend the SM to describe physics at higher energies. Arguably the most important issue in particle physics, the hierarchy problem, along with some other major problems discussed above, is partly but elegantly resolved by incorporating a spacetime symmetry, supersymmetry (SUSY).

SUSY relates two fundamental classes of particles, bosons and fermions. In the early 1970s, SUSY was independently discovered in the context of QFTs, unifying spacetime and internal symmetries, by Jean-Loup Gervais and Bunji Sakita \cite{Gervais:1971ji}, Yuri Golfand and Evgeny Likhtman \cite{Golfand:1971iw}, and Dmitrij Volkov and Vladimir Akulov \cite{Volkov:1973ix}. A supersymmetric transformation changes a bosonic state into a fermionic state, and vice versa, through an anti-commuting spinor operator $\mathcal{Q}$, that is 
\beq
\mathcal{Q} | {\rm Boson} \rangle = | {\rm Fermion} \rangle,~~~~ \mathcal{Q} | {\rm Fermion} \rangle = | {\rm Boson} \rangle.
\eeq
Spinors are intrinsically complex objects, and both $\mathcal{Q}$ and its Hermitian conjugate $\mathcal{Q}^{\dagger}$ are SUSY generators. These operators transform as spin-1/2 particles and change the spin of a particle and hence its spacetime properties. Thus, SUSY is not an internal symmetry but a symmetry of spacetime. However, possible forms of such supersymmetric transformations are highly restricted by Coleman-Mandula theorem \cite{Coleman:1967ad}. The theorem implies that the generators $\mathcal{Q}$ and $\mathcal{Q}^{\dagger}$ satisfy a graded Lie algebra, which closes under a combination of commutation and anti-commutation relations. The anti-commutation properties have the following schematic (spinor indices are suppressed) form
\bea
\{\mathcal{Q}, \mathcal{Q}^{\dagger} \} \sim P^{\mu}, \\
\{\mathcal{Q}, \mathcal{Q} \} =\{\mathcal{Q}^{\dagger}, \mathcal{Q}^{\dagger} \} =  0,
\eea
where $P^{\mu}$ is the four-momentum generator of the spacetime translations. These properties imply that a general coordinate transformation is equivalent to local SUSY, with spin-3/2 particle, the gravitino as gauge mediator. An important consequence is that local SUSY and general relativity are tied together. In a supersymmetric world, each one-particle state has a superpartner, and one has to deal with (super)multiplets of particle states instead of single-particle states. SUSY operators commute with spacetime translation but not with Lorentz generators $M_{\mu \nu}$, that is
\bea
\left[P^{\mu} , \mathcal{Q}\right] =  [ P^{\mu} ,\mathcal{Q}^{\dagger} ] = 0, \\
\left[M^{\mu\nu} , \mathcal{Q}\right] \neq 0 \neq \left[M^{\mu\nu} , \mathcal{Q}^{\dagger}\right]. 
\eea
These properties imply that particles belonging to the same supermultiplet have different spin but the same mass. SUSY has long been touted as a beautiful, elegant theory that resolves the naturalness problem through stupendous cancellations between SUSY partners in computing the Higgs mass and similar observables. Also, physicists have long sought a unified description of all fundamental interactions of nature, and SUSY is now widely accepted as a necessary ingredient for such unifying approaches.

As discussed earlier in this chapter, Lattice regularization offers a systematic tool to investigate the non-perturbative aspects of QFTs, and in the past few decades, a lot of effort has been put into formulating lattice regularized supersymmetric models. However, the discretization explicitly breaks the Lorentz invariance and, in general, the Poincar\'e invariance. The theory on the lattice respects only a finite discrete subgroup of the (Euclidean) Poincar\'e symmetry. One might expect that the lattice action must preserve all the symmetries of the target theory, but due to the emergence of {\it accidental symmetry}, this is not necessary. An accidental symmetry refers to a symmetry that emerges in the infrared continuum limit of the lattice theory, even though only a subgroup of the symmetry is respected by lattice action. 
Typically, the field theories only allow irrelevant operators that violate exact continuum symmetry, and these become unimportant in the continuum infrared limit, leading to the emergence of continuum symmetry. Accidental symmetry can automatically recover the (Euclidean) Poincare symmetry of the target theory in the infrared limit. Now, since SUSY algebra dictates that the anti-commutator of supercharges yields an infinitesimal translation, SUSY cannot be an exact symmetry on the lattice. The idea is to formulate lattice theories that preserve as many symmetries of the target theory as possible, thereby limiting the number of exact symmetry-breaking operators and then tuning their coefficients to yield the supersymmetric target theory in the infrared. As a consequence, due to accidental symmetry, SUSY can emerge from a lattice action with minimal or no fine-tuning. Ref. \cite{Catterall:2009it} provides an excellent overview in this context. 

Non-perturbative aspects are imperative for the phenomenological admissibility of the SM extensions incorporating SUSY. In this thesis, we probe two exciting non-perturbative aspects, namely spontaneous SUSY breaking and dynamical compactification of extra dimensions.

\subsection{Spontaneous supersymmetry breaking}

Despite the elegance and beauty of SUSY, the utter lack of experimental evidence for SUSY suggests that low-energy physics is manifestly non-supersymmetric. In the spectrum of elementary particles, at least at energies of order $10^2$ GeV or below, we do not observe any mass degeneracy. Then it follows that at some scale $M_{s}$, SUSY is broken so that at energies $E > M_s$, the theory behaves supersymmetrically, while at energies $E<M_s$, it does not. Generally, SUSY can either be broken spontaneously or explicitly:
\begin{itemize}
	\item {\it Spontaneous SUSY breaking}: The theory is supersymmetric, yet it contains scalar potentials that can admit sufficiently long-lived stable or meta-stable, supersymmetry breaking vacua.
	
	\item {\it Explicit SUSY breaking}: The Lagrangian itself contains terms that do not preserve SUSY. Although these terms should be irrelevant in the far UV, in such a scenario, SUSY is softly broken, and the SUSY breaking scale $M_s$ enters the Lagrangian explicitly.
\end{itemize}

However, non-renormalization theorems in four dimensions ensure that SUSY is preserved at any finite order of perturbation theory for tree-level supersymmetric theories \cite{Witten:1981nf}. Therefore, SUSY has to be spontaneously broken at some energy scale. In this thesis, we probe the possibility of spontaneous SUSY breaking in the simplest realizations of supersymmetric field theories.

\subsection{Dynamical compactification of extra dimensions}

Another exciting aspect of non-perturbative physics that we explore in this thesis is the dynamical compactification of extra dimensions in superstring theories. In particular, matrix models, non-perturbative definitions of superstrings, allow investigations of phenomenological admissibility and dynamical emergence of spacetime in superstrings via the compactification of six extra dimensions.

Even though QFT and general relativity have proven to be the most successful theories in their respective realms, a consistent unification of the two remains elusive. A reconciliation of this kind would address some of the most fundamental questions in theoretical physics, including gravitational singularities and black hole information paradox, where quantum aspects of spacetime are consequential. String theory was initially proposed in the late 1960s as a never-entirely successful theory of strong interactions. In 1974, Tamiaki Yoneya \cite{Yoneya:1975gh}, and independently John Schwarz and Jo\"el Scherk \cite{Scherk:1974ca} argued that the string theory is the theory of gravity and not hadrons due to the existence of a massless spin-two particle, graviton, in the spectrum. SUSY arises naturally in string theory, and superstrings became the most promising theories for unifying all interactions, including gravity. However, one of the salient features of these string theories is the requirement for extra dimensions since superstrings are consistently defined in ten dimensions. A realistic possibility that allows connection to the real world, where only four dimensions are macroscopic, is that these six extra dimensions are compact enough to escape any experimental detection. These notions stem from the ideas of Theodor Kaluza in 1919, who proposed an extra dimension to unify electromagnetism and general relativity \cite{Wuensch:2003xy}. Then, in 1926, Oskar Klein physically interpreted the unobservant extra dimension to be compact, wrapped into a small circle \cite{Klein:1926tv}. A quantum theory of gravity admits dynamical spacetime, emerging from non-gravitational degrees of freedom. Since we are accustomed to the idea of a spacetime manifold existing a priori, it is not immediately apparent what the fundamental degrees of freedom should be. Developments in the direction of matrix models from the 1990s have proposed matrix degrees of freedom as plausible candidates for fundamental degrees of freedom. In matrix models, spacetime does not exist a priori but is dynamically generated from the matrix degrees of freedom, where the matrices are analogous to the coordinates and the eigenvalues of the matrices to points in spacetime.

Matrix models in the large-$N$ limit are conjectured to be non-perturbative formulations of superstring theories. The connection of the matrix model to superstring theory can be made transparent by considering a matrix regularization of superstrings following Goldstone-Hoppe regularization \cite{Hoppe:1982}. The procedure, broadly, amounts to mapping functions on supermembranes to finite-sized matrices. The large-$N$ limit ensures an exact connection between the structure constants of membrane spherical harmonics and their matrix analogs. In 1996, BFSS matrix model \cite{Banks:1996vh}, named after its discoverers Tom Banks, Willy Fischler, Stephen Shenker, and Leonard Susskind, was proposed as a quantum mechanical conjecture of M-theory defined on eleven-dimensional spacetime. In the same year, the IKKT matrix model \cite{Ishibashi:1996xs}, discovered by Noboyuki Ishibashi, Hikaru Kawai, Yoshihisa Kitazawa, and Asato Tsuchiya was proposed as the first constructive nonperturbative formulation of the superstring theory, the type IIB superstring in ten-dimensional spacetime. The IKKT (type IIB) matrix model is linked to the type IIB superstrings by a Goldstone-Hoppe matrix regularization of the Green-Schwarz action \cite{Green:1983wt} in the Schild gauge \cite{Schild:1976vq}. In this thesis, we probe the possibility of spontaneous rotational symmetry breaking in the Euclidean version of the IKKT matrix model.

	\chapter{Review of Complex Langevin Method} \label{chap:complex-Langevin}

The complex Langevin method aims to overcome the {sign problem} by extending the idea of stochastic quantization for ordinary field theoretic systems with real actions to the cases with complex actions \cite{Klauder:1983nn, Klauder:1983zm, Klauder:1983sp, Parisi:1984cs}. In this chapter, we will review the basic concepts behind the complex Langevin dynamics and stochastic quantization. Canonical and path integral quantization are the two most encountered schemes to quantize field theories. In the 1980s, Parisi and Wu established a connection between Euclidean field theories and statistical systems coupled to a heat bath \cite{Parisi:1980ys} and proposed an alternative quantization scheme by analyzing associated stochastic differential equations, that is, Langevin equations. This method is known as stochastic quantization. In this approach, the Euclidean field theory is regarded as the equilibrium limit of a statistical system governed by a stochastic process. 
The expectation values of the observables for the original real variables with the complex weight can be obtained by measuring the observables for the complexified variables produced by the Langevin process and computing their expectation values at a sufficiently large simulation time.



\section{Basic concepts of a stochastic process}
A stochastic process represents the evolution in stochastic time $\theta$ of a random variable. Let us consider the phenomenon of Brownian motion, also known as the Wiener process (for simplicity, only one-dimensional), to understand the mathematical formalism behind these stochastic processes \cite{Langevin:1908}. The jittery motion of a particle freely suspended in a liquid is governed by the following stochastic differential equation,
\beq
\label{eqn:stoch-process}
m \dot{v}(\theta) = -\gamma {v}(\theta) + \eta(\theta),
\eeq 
where $m$ is the particle mass, and $\gamma$ is the coefficient of friction arising from the viscosity of the liquid. The derivative with respect to $\theta$ is denoted by a dot. The noise $\eta(\theta)$ pertains to the stochastic contributions, that is, the random forces due to the surrounding particles in the liquid. The above-mentioned stochastic differential equation is famously known as the Langevin equation of free Brownian motion (considering the absence of external potentials such as gravity, spring, etc). The Langevin equations describe the evolution of a random variable under the effect of a random force. After multiplying by an integrating factor $e^{\gamma \theta/m}$, the above differential equation can be integrated to give
\beq
{v}(\theta) = {v}(0)e^{-\gamma \theta/m} +  \frac{1}{m} \int_{0}^{\theta} d\theta' \eta(\theta')e^{-\gamma (\theta-\theta')/m},
\eeq
and computing the physical quantities, such as position, velocity, and their correlation functions, requires utilizing properties of the noise. We can consider the simplest case, that is, noise $\eta(\theta)$ obeys a Gaussian distribution in the time domain, satisfying the constraint
\beq
\langle \eta (\theta) \rangle = 0,~~
~	\langle \eta (\theta) \eta (\theta') \rangle =\alpha \sigma^2 \gamma^2  \delta(\theta-\theta'),
\eeq
where $\sigma^2$ is the variance and the coefficient $\alpha$ governs the strength of the correlations. One can show that 
\bea
\langle x(\theta) \rangle &=& x(0) +\frac{m}{\gamma} \dot{x}(0) \left(1-e^{-\gamma \theta/m} \right), \\
\langle {v}(\theta) \rangle &=& {v}(0)e^{-\gamma \theta/m}, \\
\langle {v}(\theta){v}(\theta') \rangle &=& \langle v(\theta)\rangle \langle v(\theta')\rangle + \frac{\alpha \sigma^2\gamma}{m}  \left(e^{-\gamma (\theta-\theta')/m} - e^{-\gamma (\theta+\theta')/m}  \right).
\eea
For very large time, $\theta = \theta' \to \infty, \langle{v}^2(\theta) \rangle$ goes to $\frac{\alpha\sigma^2\gamma}{2m} = \frac{K_B T}{m}$, which is in agreement with the equipartition theorem. However, to compute a higher-order correlation function, we would need higher moments of the probability distribution of $\eta(\theta)$. Instead of specifying all these moments, it is convenient to specify the probability distribution directly. The generalized functional probability distribution that gives the noise read
\beq
P[\eta(\theta)]\propto \exp \left( -\int_{-\infty}^{\infty}d\theta ~\frac{\eta^2(\theta)}{2\sigma^2 \alpha\gamma^2} \right).
\eeq
\section{The approach of Parisi and Wu}

Now to implement stochastic quantization, let us consider a real scalar field $\phi(x)$ in \\
$d$-dimensions with a real Euclidean action $S[\phi(x)]$. We are interested in computing the expectation values of physical observables $\mathcal{O}(\phi)$, given by the path integral
\beq
\langle \mathcal{O}(\phi)\rangle = \frac{1}{Z}{\int \mathcal{D} \phi(x)~ \mathcal{O}(\phi) e^{-S[\phi(x)]}};~~ Z = \int \mathcal{D} \phi(x)~ e^{-S[\phi(x)]}.
\eeq
In stochastic quantization, the expectation values of observables are obtained as equilibrium values of a stochastic process. To implement this, the system evolves according to Langevin dynamics in fictitious Langevin time $\theta$, subject to a Gaussian noise \cite{Parisi:1980ys}. At some Langevin time $\theta$, the Langevin evolution reads 
\beq
\label{eqn:real-Langevin}
\frac{d\phi(x;\theta)}{d\theta} = - \frac{\partial S[\phi(x;\theta)]}{\partial \phi(x;\theta)} + \eta(x;\theta),
\eeq 
where $\eta(x;\theta)$ is a Gaussian noise satisfying
\beq
\langle \eta (x;\theta) \rangle = 0,~~
\langle \eta (x;\theta) \eta (x;\theta') \rangle = 2\delta(x-x') \delta(\theta-\theta').
\eeq
Let us assume that the solution of the above Langevin equation, say $\phi_{\eta}(x;\theta)$, gives rise to the desired equilibrium configurations. Then, at very large Langevin time $\Theta$, the Langevin time average of the observable over this solution is supposed to provide the path integral expectation value, that is
\beq
\frac{1}{\Theta} \int_{\theta_0}^{\theta_0+\Theta}d\theta ~\mathcal{O}(\phi_{\eta}(x;\theta)) \xrightarrow[]{\Theta\to \infty} \langle \mathcal{O}(\phi(x))\rangle.
\eeq
The solution $\phi_{\eta}(x;\theta)$ depends on the noise $\eta(x;\theta)$ and different realization of the noise gives rise to a probability distribution  of field configurations, $P(\phi;\theta)$ at Langevin time $\theta$. The average of $\mathcal{O}(\phi_{\eta})$ with respect to noise can thus be written as
\beq
{\langle \mathcal{O}(\phi_{\eta}(x;\theta))\rangle}_{\eta}
= \int \mathcal{D} \phi(x)~ P(\phi;\theta) \mathcal{O}(\phi(x;\theta)), 
\eeq   
where the brackets on the left-hand side denote a noise-averaged expectation value. The dynamics of distribution $P(\phi;\theta)$ can be translated from the microscopic dynamics of the Langevin equation, 
such that $P(\phi;\theta)$ satisfies the famous Fokker-Planck equation
\bea
\frac{\partial P(\phi;\theta)}{\partial \theta} = -H_{\rm FP} P(\phi;\theta);~~ P(\phi;0) =  \delta(\phi-\phi_{0}), \\
H_{\rm FP} =  \int d^d x~\frac{\partial}{\partial \phi} \left( \frac{\partial}{\partial \phi} + \frac{\partial S}{\partial \phi} \right),
\eea
where $\phi_0$ is the initial field configuration and $H_{\rm FP}$ is the Fokker-Planck Hamiltonian. 
Now, with a similarity transformation $\widetilde{P}(\phi;\theta) = e^{S[\phi(x;\theta)]/2} P(\phi;\theta)$, the Fokker-Planck equation reads
\bea
\frac{\partial \widetilde{P}{(\phi;\theta)}}{\partial t} = - \widetilde{H}_{\rm FP} \widetilde{P}{(\phi;\theta)}, \\
\widetilde{H}_{\rm FP} = e^{S[\phi]/2} ~H_{\rm FP}~ e^{-S[\phi]/2} =  \int d^d x~\left( -\frac{\partial}{\partial \phi} + \hf\frac{\partial S}{\partial \phi} \right)\left( \frac{\partial}{\partial \phi} - \hf\frac{\partial S}{\partial \phi} \right).
\eea
When the action, $S[\phi]$, is real, $\widetilde{H}_{\rm FP}$ is a Hermitian semi-positive definite operator. Then, in the large Langevin time limit, the field distribution $\widetilde{P}{(\phi;\theta)}$ mimics the original Boltzmann weight $e^{-S[\phi]}$ in the path integral, that is
\beq
\lim_{\theta\to \infty}{P}{(\phi;\theta)} \propto e^{-S[\phi]},
\eeq
which ensures correct convergence of the real Langevin dynamics. 

\section{An extension to complex actions}
Not long after Parisi and Wu's work, it was realized that the concept of stochastic quantization  could be extended to the case of complex actions in a fairly straightforward manner. In 1983, Klauder and Parisi \cite{Klauder:1983nn, Klauder:1983zm, Klauder:1983sp, Parisi:1984cs} independently proposed an extension aimed to simulate the complex measure with entire holomorphic action on a real manifold $\mathcal{M}$. For this purpose, they set up a stochastic process on the complexification $\mathcal{M}_c$ of $\mathcal{M}$ in such a way that the expectation values of entire holomorphic observables $\mathcal{O}$ obtained in this stochastic process converge to the ones computed using the original complex measure.  For such complex actions, the solution to the Langevin equation also becomes complex. Therefore, we need to extend the domain of field variables into the complex plane (for simplicity, we consider zero dimension), that is $\phi(\theta) = \phi_r(\theta) +i\phi_i(\theta)$. The complex Langevin equation is similar to the Eq. \eqref{eqn:real-Langevin} for real Langevin
\beq
\frac{d\phi(\theta)}{d\theta} = K (\phi;\theta) + \eta(\theta); ~~~~ K (\phi;\theta)=- \frac{\partial S(\phi(\theta))}{\partial \phi(\theta)},
\eeq
where all quantities are complex, except for stochastic noise  $\eta(\theta)$, which is kept to be real. See Ref. \cite{Aarts:2013uza} for a detailed analysis with complex noise.

The relaxation dynamics of the above complex Langevin equation can be understood by defining a complex-valued density $\rho(\phi_r;\theta)$ on $\mathcal{M}$ parameterized over real variable $\phi_r$, which evolves according to
\bea
\frac{\partial }{\partial \theta} \rho(\phi_r;\theta) = L^T_0 \rho(\phi_r;\theta);~~ \rho(\phi_r;0) =  \delta(\phi_r-\phi_{r_0}), 
\eea 
where the complex Fokker-Planck operator, $L_0$ has the form
\beq
L^T_0 \equiv \nabla_r \left[\nabla_r - \nabla_r S(\phi_r)\right], 
\eeq
with $\nabla_r= \frac{\partial}{\partial \phi_r}$. A slight generalization for any constant $\phi_{i_0}\in \mathcal{M}$, gives us $L_{c0}$. The generalized complex Fokker-Planck operator for any $\phi_{i_0}$,
\beq
L^T_{c0} \equiv \nabla_r \left[\nabla_r -  \nabla_r S(\phi_r+i\phi_{i_0})\right]
\eeq
acts on complex valued density (measure) on $\mathcal{M}$, again parameterized over real variable $\phi_r$ \cite{Klauder:1983zm}. However, these operators restrict a probabilistic interpretation since they do not preserve positivity.

The trajectories of complex Langevin evolution obviously migrate into the complex directions, but it is crucial to emphasize that the process is still a real stochastic process; however, now on the complexification $\mathcal{M}_c$. This can be realized by writing the process in real and imaginary parts,
\bea
\frac{d\phi_r}{d\theta} = K_r + \eta(\theta),&& ~~\frac{d\phi_i}{d\theta} = K_i, \\
{\rm where~} K_r = - {\rm Re} \left[ \partial_r S(\phi_r+i\phi_i)\right],&&~~K_i = - {\rm Im} \left[ \partial_i S(\phi_r+i\phi_i)\right].
\eea
For this real stochastic process, we can define a real and positive definite probability density $P(\phi_r,\phi_i;\theta)$ on $\mathcal{M}_c$, which evolves according to the Fokker-Planck equation
\bea
\label{eqn:clm-holomorphic-L}
\frac{\partial }{\partial \theta} P(\phi_r,\phi_i;\theta) = L^T P(\phi_r,\phi_i;\theta); ~~ P(\phi_r,\phi_i;0) = \delta(\phi_r - \phi_{r_0})\delta(\phi_i), 
\eea
where the real Fokker-Planck operator, $L$, has the form
\beq
L^T \equiv \nabla_{r} \left[\nabla_r - K_r \right] -\nabla_i K_i,
\eeq
such that $\nabla_r= \frac{\partial}{\partial \phi_r}$ and $\nabla_i= \frac{\partial}{\partial \phi_i}$.



Then, the unavoidable question is whether the real and complex evolutions are consistent such that they lead to an identical evolution of expectation values of holomorphic observables $\mathcal{O}$; that is whether
\bea
\langle \mathcal{O}\rangle_{P(\theta)}
= \frac{\int d\phi_rd\phi_r~ \mathcal{O}(\phi_r+i\phi_i) P(\phi_r,\phi_i;\theta)}{\int d\phi_rd\phi_r~ P(\phi_r,\phi_i;\theta)} {\rm ~and~}
\langle \mathcal{O}\rangle_{\rho(\theta)}
= \frac{\int d\phi_r~ \mathcal{O}(\phi_r) \rho(\phi_r;\theta)}{\int d\phi_r~ \rho(\phi_r;\theta)}
\eea
remain equal if they agree at initialization, $\theta=0$. That is, the idea is to show, $\langle \mathcal{O}\rangle_{P(\theta)}
=\langle \mathcal{O}\rangle_{\rho(\theta)}
$ provided the initial condition
\beq
\label{eqn:clm-density-initial}
P(\phi_r,\phi_i;0) =  \rho(\phi_r;0) \delta(\phi_i-\phi_{i_0}).
\eeq 

The above question was formally addressed in Refs. \cite{Aarts:2009uq, Aarts:2011ax}. The authors established that the relation holds only for holomorphic observables, as long as the action and its gradient are holomorphic functions of complex field $\phi$.  It is crucial to emphasize that only holomorphic observables are considered so that we can extend the action of the operator $L^T_{c0}$ (extend to complex Langevin operator) to observables having analytic continuation to all of $\mathcal{M}_c$.  We have the complex Langevin operator $\widetilde{L}$,
\beq
\widetilde{L}^T \equiv \nabla_\phi \left[\nabla_\phi -  \nabla_\phi S(\phi)\right],
\eeq
such that its action on these analytic continuations agrees with that of real Fokker-Planck operator $L$, that is, difference $L-\widetilde{L}$ vanishes with the help of Cauchy-Riemann (CR) equations.

The introduction of three different Langevin/Fokker-Planck operators, $L, L_{c0}, \widetilde{L}$, may seem daunting. To clear up and summarize, the Langevin operator $\widetilde{L}$ is obtained as an analytic continuation to act on complexification $\mathcal{M}_c$, of the complex Fokker-Planck operator $L_{c0}$ acting on real manifold $\mathcal{M}$. The Langevin operator $\widetilde{L}$ and real Fokker-Planck operator $L$ act on functions on complexification $\mathcal{M}_c$, agreeing only for holomorphic functions.

Now, the idea is to consider the time evolution of holomorphic observables rather than densities and make use of CR equations. The operator $\widetilde{L}$ or $L$ can be interchangeably used to evolve holomorphic observables in the following manner
\bea 
\partial_{\theta} \mathcal{O}(\phi;\theta) =  {L} \mathcal{O}(\phi;\theta)
\eea
with some initial condition $\mathcal{O}(\phi;0)= \mathcal{O}(\phi)$. Solving the above equations formally yields 
\beq
\label{eqn:clm-LO-evolve}
\mathcal{O}(\phi;\theta) = \exp[\theta {L}]~ \mathcal{O}(\phi).
\eeq

Finally, for justification of the method, we need to establish equality, at large Langevin time $\Theta = \theta \to \infty$, between the expectation values of real probability density $\langle \mathcal{O}\rangle_{P(\Theta)}$
and complex density $\langle \mathcal{O}\rangle_{\rho(\Theta)}$. For this purpose we define a quantity $F(\Theta, \theta)$ for $0\le \theta \le \Theta$ such that
\beq
F(\Theta, \theta) \equiv \int d\phi_r d\phi_i~P(\phi_r,\phi_i;\Theta-\theta)~ \mathcal{O}(\phi_r+i\phi_i;\theta)
\eeq
admits the interpolation
\beq
F(\Theta, 0) =  \langle \mathcal{O}\rangle_{P(\Theta)}~ {\text{and }} 
F(\Theta, \Theta) = \langle \mathcal{O}\rangle_{\rho(\Theta)}.
\eeq
The first equality is trivial. The second can be understood as
\bea
F(\Theta, \Theta) &=& \int d\phi_r d\phi_i~P(\phi_r,\phi_i;0)~ \mathcal{O}(\phi_r+i\phi_i;\Theta) \\
&=& \int d\phi_r d\phi_i~P(\phi_r,\phi_i;0)~ \exp[\Theta L]\mathcal{O}(\phi_r+i\phi_i;0),
\eea
where we used the property in Eq. \eqref{eqn:clm-LO-evolve} to evolve the observable. Now using the initial condition in Eq. \eqref{eqn:clm-density-initial}, we get
\bea
F(\Theta, \Theta)
&=&\int d\phi_r d\phi_i ~\rho(\phi_r;0) \delta(\phi_i-\phi_{i_0})~\exp[\Theta L]\mathcal{O}(\phi_r+i\phi_i;0) \\
&=&\int d\phi_r ~\rho(\phi_r;0)~\exp[\Theta L_{c0}]\mathcal{O}(\phi_r+i\phi_{i_0};0) \\
&=&\int d\phi_r d\phi_i~ \mathcal{O}(\phi_r+i\phi_i;0)~\exp[\Theta L^T_{c0}]\rho(\phi_r;0)\\
&=&\langle \mathcal{O}\rangle_{\rho(\Theta)}, 
\eea
where we also used integration by parts in $\phi_r$ ignoring the boundary terms at $\infty$ and at the poles of drift. These boundary terms will vanish if we have $F(\Theta,\theta)$ independent of $\theta$. To visualize this, consider $F(\Theta,\theta)$ derivative
\bea
\frac{\partial}{\partial \theta}F(\Theta,\theta) &=&\int d\phi_r d\phi_i~  P(\phi_r,\phi_i;\Theta-\theta) \widetilde{L}\mathcal{O}(\phi_r+i\phi_i;\theta) \nn \\ 
&&-\int d\phi_r d\phi_i~ \left(L^T P(\phi_r,\phi_i;\Theta-\theta)  \right) \mathcal{O}(\phi_r+i\phi_i;\theta). 
\eea
Then once the equilibrium is reached, at large Langevin time, we obtain the condition for stationarity of observables over the stochastic process, 
\beq
\label{eqn:clm-cc}
\lim_{\theta \to \infty}\partial_\theta \langle \mathcal{O} \rangle = \langle \widetilde{L}\mathcal{O} \rangle \equiv \int d\phi_r d\phi_i~ P(\phi_r,\phi_i;\infty) \widetilde{L}\mathcal{O}(\phi_r+i\phi_i;0) = 0,
\eeq
which resembles the Schwinger-Dyson equations and is also known as the {\it consistency condition}. The consistency condition, along with some additional conditions, is sometimes sufficient to ensure the correctness of the equilibrium measure of the complex Langevin method \cite{Aarts:2011ax}.

\subsection{Justification and correctness criteria}
After a large Langevin time, the probability distribution of field configurations should mimic the Boltzmann factor, and we have observed that this is indeed the case for real actions. However, to date, there is no exact mathematical proof of convergence for complex actions. Because of this problem, for many years complex Langevin method eluded the interest of physicists. The above-discussed formal arguments raise a few major mathematical questions:
\begin{itemize}
	\item {\it Langevin/Fokker-Planck operators}: Whether the proliferation of Langevin/Fokker-Planck operators $L, L_{c0}, \widetilde{L}$ be exponentiated? More precisely, the existence of a unique stochastic process and the time evolutions generated by Langevin/Fokker-Planck operators are unknown. In numerical terms, this is never a problem as long as {\it adaptive step size} (discussed in coming sections) is taken into account \cite{Aarts:2011ax}.
	
	\item {\it Convergence to an equilibrium measure}: It is still not mathematically proven that positive density converges to the equilibrium measures and requires information about the spectrum of Langevin operators \cite{weingarten2002complex}. In 1985, Klauder and Peterson \cite{Klauder:1985kq, Klauder:1985ks} commented on the ``{\it conspicuous absence of general theorem}" for non-self-adjoint operators.
	
	\item {\it Boundary terms}: How justifiable are the various integration by parts \cite{Aarts:2008wh}, which control the shifting of time evolution from measure to observables and back again? Do we need to worry about the boundary terms?
\end{itemize}

Despite the lack of mathematical rigor, physicists were not deterred and, in fact, progressed pragmatically. The bright side is that an equilibrium measure exists numerically in all interesting physical cases. The complex Langevin method was used in many successful and inspiring research studies \cite{Gausterer:1985sm, Klauder:1985kq, Karsch:1985cb,  Horowitz:1986dt, Flower:1986hv, Ambjorn:1986fz, Haymaker:1987ey, Catterall:1990qn, Kieu:1994xg, Gausterer_1998, adami2001complex, Berges:2005yt, Berges:2007nr}. In the recent past, proofs of convergence of probability distribution $P(\phi_r,\phi_i;\theta)$ to the complex measure $\rho=e^{-S[\phi]}$, have been subjected to great interest, and certain correctness criteria have been proposed for the reliability of the method.
\begin{itemize}
	\item {\it Langevin-operator criterion}: In 2009, interest in the complex Langevin method was successfully revived with the help of correctness criteria by Gert Aarts, Erhard Seiler, and Ion-Olimpiu Stamatescu \cite{Aarts:2009uq}. Based on the consistency condition mentioned in Eq. \eqref{eqn:clm-cc}, the Langevin operator acting on the observables should vanish, that is 
	\beq
	\langle \widetilde{L} \mathcal{O} \rangle = 0.
	\eeq
	It can be treated as a reliability criterion for numerical simulations to verify the correct convergence of distribution $ P(\phi_r,\phi_i;\theta)$ to equilibrium measure.  In principle, the criterion for correctness needs to be satisfied for the entire set of observables $\cO[\phi]$, in a suitably chosen basis \cite{Aarts:2011ax}. It leads to an infinite tower of identities resembling the Schwinger-Dyson equations. 
	
	\item {\it Probability drift criterion}: A fairly recent criterion was introduced in 2016 by Keitaro Nagata, Jun Nishimura, and Shinji Shimasaki \cite{Nagata:2015uga}. The presence or absence of boundary terms depends on the growth of holomorphic function and hence the associated drift term. This can lead to {\it tails} in the probability distribution $P(\phi_r,\phi_i;\theta)$ even at very large Langevin time. The criterion defines a magnitude of drift,
	\beq
	u = \Big | \frac{\partial S[\phi]}{\partial \phi} \Big |
	\eeq
	and for reliable simulations, necessitates an exponential or faster decay of the probability of the drift term at larger magnitudes.
\end{itemize}

In the coming chapters, we will make use of these correctness criteria to verify the reliability of our complex Langevin simulations.



\section{Complex Langevin simulations}
Let us briefly discuss the complex Langevin method from a numerical simulation perspective. The complex Langevin equation (in zero dimension, for simplicity) reads
\beq
\frac{d\phi(\theta)}{d\theta} = K (\phi;\theta) + \eta(\theta); ~~ K (\phi;\theta)=- \frac{\partial S(\phi(\theta))}{\partial \phi(\theta)},
\eeq
where $\eta(\theta)$ is a real continuous noise function with variance 2. For simulations purpose, we need to discretize the continuous complex Langevin equation. A naive discretization with the Langevin time discretized as an integer multiple of the Langevin step size, that is, $\theta = n \epsilon$, would lead to the following discretized equation;
\beq
\phi_{\theta +\epsilon} = \phi_{\theta } - \frac{\partial S[\phi]}{\partial \phi} \Big |_{\phi_{\theta}}  { \epsilon}+ \eta^{d}_{\theta}  { \epsilon},
\eeq
where $\eta^{d}_{\theta}$ is the new discretized noise. Although straightforward at first glance, the discretization procedure for noise function is slightly non-trivial. To understand further, let us consider a continuous noise function $\eta(\theta)$ with variance $\sigma^2$, then the correspondence with discretized noise $\eta^d_{\theta}$ can be shown as
\bea
\langle \eta (\theta ) \eta (\theta') \rangle = \sigma^2 \delta(\theta-\theta') ~~~~~ 
&\xrightleftharpoons[\rm continuous]{\rm discretized} & ~~~~~~~\langle \eta^d_{\theta} \eta^d_{\theta'} \rangle = c \delta_{\theta \theta'}\\
\Big \downarrow {\substack{\text{integrate} \\ \text{over continuous } \theta}  }~~~~~~~~~~&&~~~~~~~~~~ \Big \downarrow {\substack{\text{sum} \\ \text{over discretized } \theta}  } \nn \\
\int d\theta ~\langle \eta (\theta ) \eta (\theta') \rangle = \sigma^2 ~~~~~~~~ &\xrightleftharpoons[\rm continuous]{\rm discretized}& ~~~~~\epsilon \sum_{n=1}^{\infty} \langle \eta^d_{\theta} \eta^d_{\theta'}  \rangle =\sigma^2. 
\eea
From the above discretization procedure, we can find the value of arbitrary constant $c$, that is
\beq
\epsilon \sum_{n=1}^{\infty} c\delta_{\theta \theta'} = \sigma^2 {\implies} c= \frac{\sigma^2}{\epsilon}.
\eeq 
Now, the discretized Gaussian noise $\eta^d_{\theta}$ satisfies $\langle \eta^d_{\theta} \rangle = 0,
\langle \eta^d_{\theta} \eta^d_{\theta'}\rangle = {\sigma^2}\delta_{\theta\theta'}/{\epsilon} 
$ and a simple rescaling of the noise $\eta^d_{\theta} \to {\eta_{\theta} }/{\sqrt{\epsilon}} $ can remove the dependence of noise correlation on Langevin step-size $\epsilon$.
Substituting the variance $\sigma^2 = 2$, we have the discretized complex Langevin equation 
\beq
\label{eqn:disc-clm}
\phi_{\theta +\epsilon} = \phi_{\theta } - \frac{\partial S[\phi]}{\partial \phi} \Big |_{\phi_{\theta}}  { \epsilon}+ \eta_{\theta}  \sqrt{ \epsilon},
\eeq
where the discretized Gaussian noise $\eta_{\theta}$ satisfies 
\beq
\langle \eta_{\theta} \rangle = 0,~~
\langle \eta_{\theta} \eta_{\theta'}\rangle = 2 \delta_{\theta\theta'}.
\eeq

\begin{figure}[htp]
	\begin{center}
		\includegraphics[width=.7\textwidth,origin=c,angle=0]{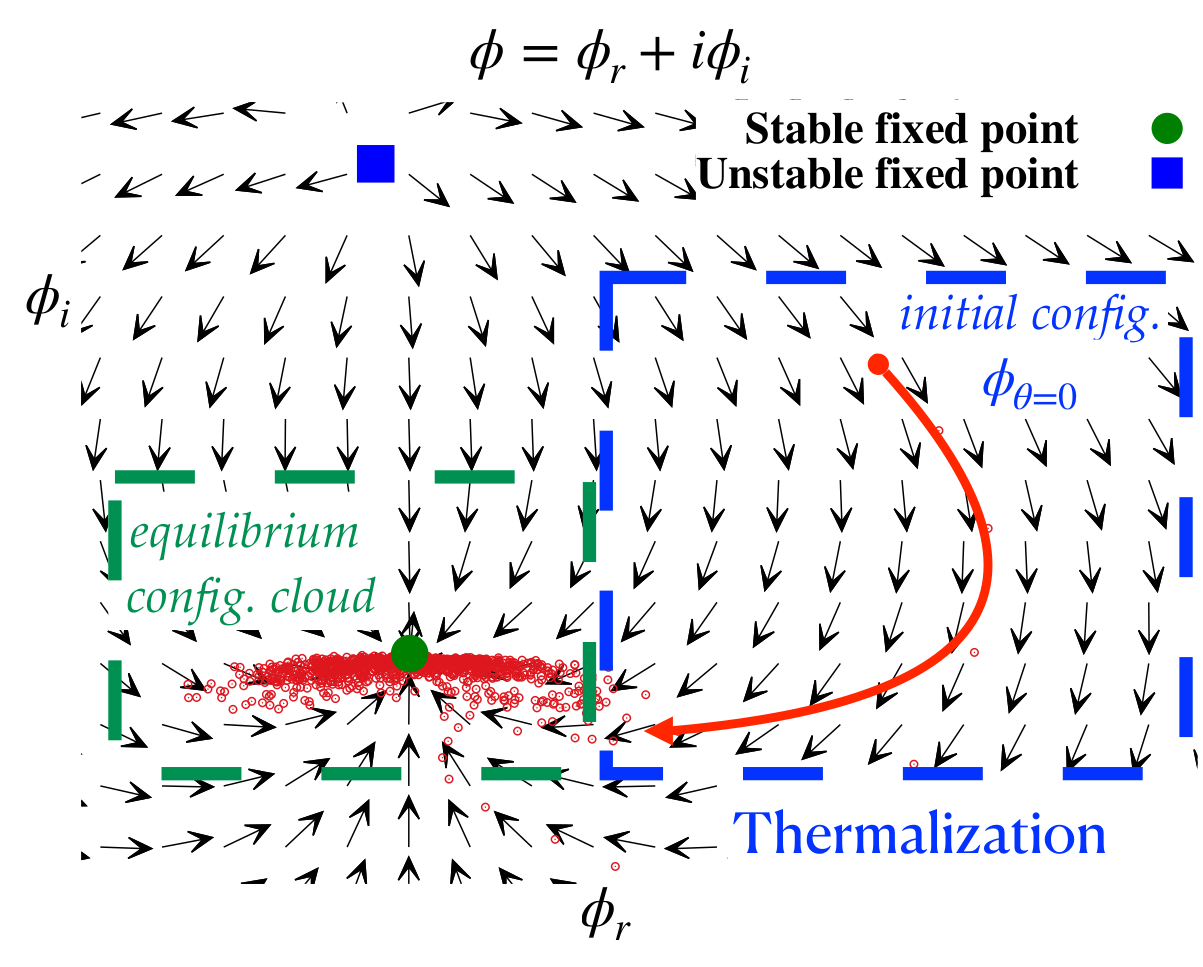}
	\end{center}
	\caption{Schematic scatter plot of the evolution of complex fields in the complex Langevin dynamics.}
	\label{fig:schematic-clm}	
\end{figure}
Finally, we use the discretized complex Langevin equation in Eq. \eqref{eqn:disc-clm} to evolve the fundamental degrees of freedom of the system. That is, the fields evolve in Langevin time, according to the Langevin dynamics, with the help of drift $-\frac{\partial S[\phi]}{\partial \phi} \big |_{\phi_{\theta}} $. The schematic scatter plot in Fig. \ref{fig:schematic-clm} explains the evolution of complex fields under Langevin dynamics for a complex system. Suppose we start from some initial field configuration, say $\phi_0$ at Langevin time $\theta = 0$. With the help of the drift, the arrows represent the direction of drift, and the field reaches the neighborhood of a stable fixed point. Upon reaching the vicinity of the stable fixed point, because of the noise, the field does not just collapse into the stable fixed point and, in fact, forms a cloud around it. After thermalization, at considerably large Langevin time $\theta$, the collected cloud of field configurations give rise to a real probability distribution $P(\phi_r, \phi_i; \theta)$, which is supposed to be the equilibrium solution of the Fokker-Planck equation. Then, once equilibrium distribution has been reached, using ergodicity, the noise expectation values of observables are obtained according to the probability distribution $P(\phi_r, \phi_i; \theta)$, that is 
\bea
\langle \mathcal{O} \rangle_{\eta} = \lim_{\theta \to \infty } \langle \mathcal{O (\phi(\theta))} \rangle_{\eta} &=& \lim_{\theta \to \infty }  \int d\phi_r d\phi_i~\mathcal{O}(\phi_r + i\phi_i) P(\phi_r, \phi_i; \theta) 
\\
&\approx& \frac{1}{N} \sum_{n=0}^{N} \mathcal{O}(\phi_r + i\phi_i).
\eea

\subsection{Numerical hurdles and stabilization techniques}

The complex Langevin method became very popular when it was first proposed in the 1980s. 
The method did not rely on a probability interpretation of the weight, so it can, in principle, be applied even where there is a severe sign problem. 
However, certain problems were encountered shortly after, and despite the initial flurry, the problems of numerical instability and incorrect convergence hindered early complex Langevin studies \cite{Ambjorn:1985iw, Ambjorn:1986fz, Gausterer:1992jz, Gausterer:1998jw}. This section briefly discusses these numerical problems and proposed stabilization techniques to resolve them. These issues are a numerical consequence of the convergence of probability distribution $P(\phi_r, \phi_i; \theta) $ to equilibrium measure. The first problem is runaways, where the field configurations would not converge even for a large Langevin time. The second is an even worse numerical problem, the convergence to a wrong limit. \\

Recent developments have resulted in the successful resurgence of the complex Langevin method, which produces correct results even when the sign problem is severe. In most cases, the classical flow will have unstable fixed points, but the introduction of the stochastic noise term has the effect of kicking off these trajectories and therefore keeping the dynamics stable. The complexification of the fields introduces new degrees of freedom, which are typically unbounded and can potentially follow divergent trajectories, which renders numerical simulations unstable. For instance, when configurations are in the vicinity of unstable directions. Thus, taking sufficient care in the numerical integration of the Langevin equations is necessary. To cure these unstable trajectory problems, some prominent stabilization techniques are \textit{adaptive step size}, \textit{gauge cooling}, and \textit{dynamical stabilization}. We briefly discuss them below.

\subsubsection{Adaptive step size}
A Langevin trajectory can make extensive excursions into imaginary directions, and as a naive solution, a small enough step size may suffice. But this only solves instabilities in some situations. Moreover, a smaller step size will slow the evolution, requiring many updates to explore the configuration space. An efficient algorithm was proposed to cure these runaway trajectories; it involves considering adaptive Langevin step size in the numerical integration of Langevin equations \cite{Aarts:2009dg}. At each Langevin sweep, the absolute value of maximum drift is computed, that is $K_{\rm max}(\theta)$ and the step size for the next evolution sweep is obtained as
\beq
\epsilon= \frac{\gamma}{K_{\rm max}(\theta)},
\eeq
where the parameter $\gamma$ can be appropriately selected depending on the model.

\subsubsection{Gauge cooling}
The inherently complex nature of the action can result in excursions of the fields (say matrix fields $X_{\mu}$) into anti-Hermitian or imaginary directions. These excursions, in turn, complexify and enlarge the group space, say from SU($N$) to SL($N, \mathbb{C}$). We encounter the excursion problem when field configurations wander too far from SU($N$). A proposed solution to this problem is gauge cooling \cite{Seiler:2012wz}. The method defines a {\it Hermiticity norm } \cite{Nagata:2016vkn}
\beq
\mathcal{N}_{\rm H} \equiv -\frac{1}{10N} \sum_{\mu}{\rm tr} \left( \left[ X_\mu - X_\mu^\dagger \right]^2 \right)
\eeq
to track the deviation of $X_\mu$ from Hermitian configurations. The matrix fields $X_\mu$ are invariant under the enlarged gauge symmetry,
\bea
X_\mu \rightarrow g X_\mu g^{-1},~ g \in {\rm SL}(N,\mathbb{C}),
\eea
where $g$ is chosen to be ${g = {\rm e}^{-\alpha \delta{\mathcal{N}}_{\rm H}}}$ for a real, positive tuning parameter $\alpha$, and 
\bea
\delta{\mathcal{N}}_{\rm H} = \frac{1}{N} \sum_\mu\left[ X_\mu, X_\mu^\dagger \right].
\eea
It is crucial to note that $\mathcal{N}_{\rm H}$ is not invariant under this gauge transformation. This property allows the above gauge transformation to be repeated successively to minimize the norm $\mathcal{N}_{\rm H}$ and reach closer to Hermitian directions. The gauge cooling procedure has been proven to respect complex Langevin correctness criteria \cite{Nagata:2016vkn}.

\subsubsection{Dynamical stabilization}
This method was introduced recently by Felipe Attanasio and Benjamin J\"ager. Here, the unitary norm is decreased by adding a hand-crafted drift term to the complex Langevin process, which vanishes at continuum limit \cite{Attanasio_2019}. The drift term $K_x$ at lattice site $x$ is modified in the following manner
\beq
K_x \to \widetilde{K}_x = K_x + i \alpha_{\rm DS} M_x,
\eeq
where $\alpha_{\rm DS}$ is the control parameter and $M_x$ is chosen to act only in the imaginary direction, that is, orthogonal to the SU($N$) manifold, and grows with the unitary norm. This modification of the drift term vanishes in the continuum limit, but because it is incorporated manually by hand, the method violates the complex Langevin correctness criteria. Because of this, it is difficult to claim that simulations with dynamical stabilization produce correct results.
\vsphf

\begin{figure}[t]
	\begin{center}
		\includegraphics[width=0.8\textwidth,origin=c,angle=0]{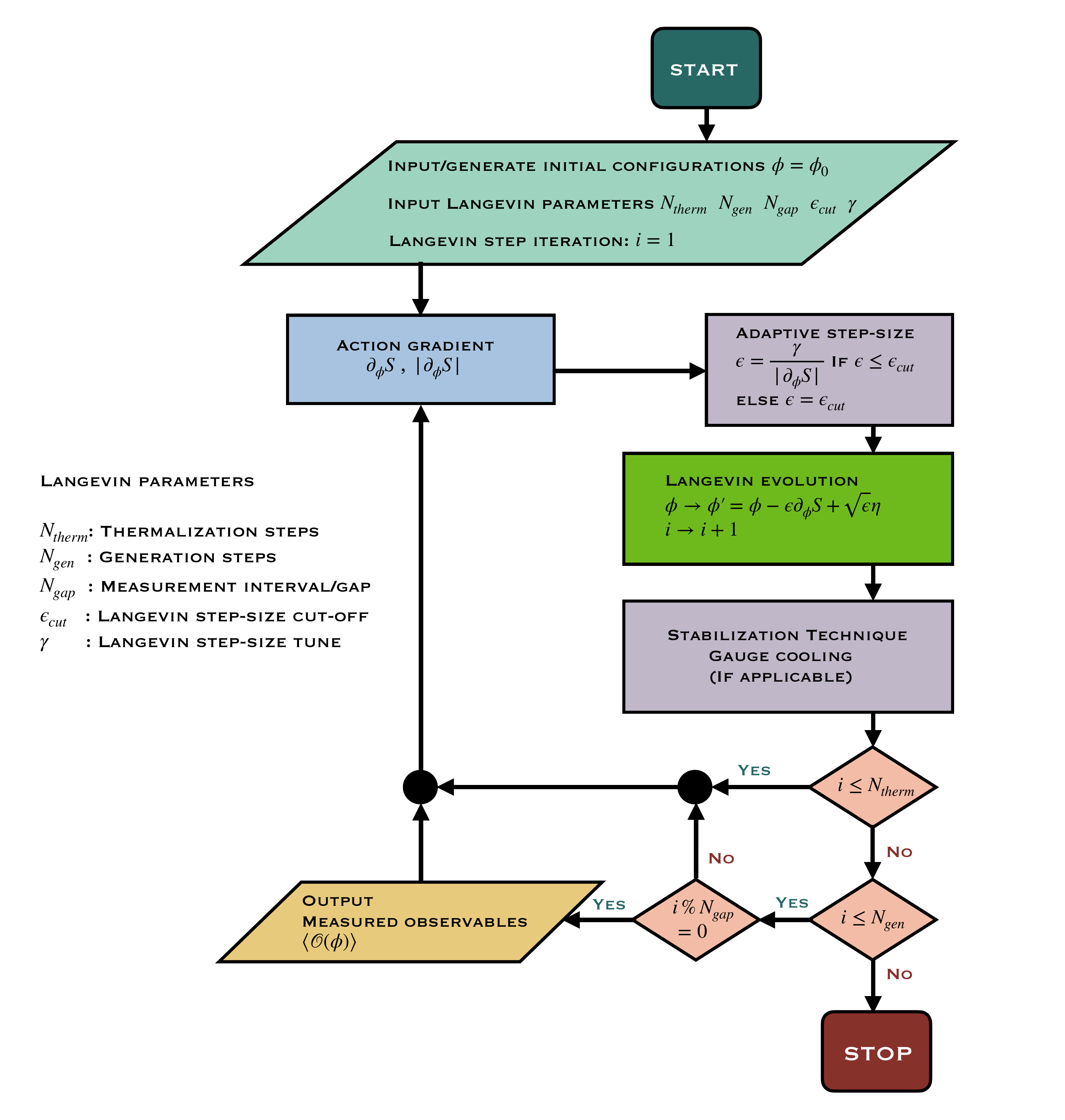}
	\end{center}
	\caption{Flowchart for the complex Langevin algorithm.}
	\label{fig:flowchart-clm}	
\end{figure}

The flowchart in Fig. \ref{fig:flowchart-clm}, provides an overview of the complex Langevin algorithm. The numerical calculations in the coming chapters are based on implementing and optimizing this algorithm in the C/C++ programming language. In the complex Langevin study of the two-dimensional QFTs, we utilized various parts of the object-oriented code introduced in Ref.  \cite{Catterall:2011cea}. The simulations of the IKKT matrix model are performed on the PARAM Smriti supercomputing system (part of India’s National Supercomputing Mission) using MPI-based parallel architecture.

\section{Recent studies employing complex Langevin method}

This section briefly mentions some recent successful applications of the complex Langevin method. In Refs. \cite{Damgaard:1987rr, Okano:1992hp}, a pedagogical review of the complex Langevin method is provided. See Ref. \cite{Berger:2019odf} for a recent review of the sign problem in quantum many-body physics. Ref. \cite{Seiler:2017wvd} provides a precise, coherent overview and status report on the complex Langevin method.

Complex Langevin studies have been essential to understanding non-equilibrium QFTs. Unlike well-established thermal equilibrium systems, these out-of-equilibrium systems are not amenable to a Euclidean formalism and must be formulated in real-time. See Ref. \cite{Berges:2005yt} for simulations of non-equilibrium quantum fields, Ref. \cite{Berges:2006xc} for scalar and non-Abelian gauge fields, and Ref. \cite{Berges:2007nr} for gauge theories in Minkowski spacetime with optimized updating. In Ref. \cite{deAguiar:2010ue}, the authors used the stochastic quantization method to study finite temperature field theory in real-time formalism.

Non-perturbative studies of QCD at a finite chemical potential are challenging because of the complex fermion determinant and the sign problem. The reliable applicability of the complex Langevin dynamics to finite-density lattice QCD was demonstrated in Refs. \cite{Aarts:2008rr, Aarts:2008yw, deForcrand:2009zkb} led to a resurgence of interest in the method. Refs. \cite{Aarts:2012yal, Felipe:2017wuf, Nagata:2018mkb, Scherzer:2019weu, Scherzer:2020kiu} explore the application of complex Langevin dynamics and other approaches to tackle the sign problem in lattice QCD at non-zero baryon density and its relationship to the {\it overlap} and {\it Silver Blaze problem}. An in-depth overview of the progress of the complex Langevin simulations of the QCD phase diagram is provided in Refs. \cite{Aarts:2014fsa, Sinclair:2017zhn, Bloch:2018sof,  Attanasio:2020spv}. The renewed interest in complex Langevin dynamics led to progress in optimizing and stabilizing lattice QCD simulations \cite{Aarts:2011zn,  Seiler:2012wz, Aarts:2017vrv, Attanasio_2019}. See Ref. \cite{Nagata:2021ugx} (translated from Japanese to English by Masanori Hanada and Etsuko Itou) by Keitaro Nagata for an excellent summary of progress and current status of problems in lattice QCD along with the possible solutions.

Numerical approaches have yet to produce detailed solutions in some regions of the QCD phase diagram. However, simpler models that partly contain the phenomenology of QCD have been studied. In Refs. \cite{Aarts:2008wh, Aarts:2009hn}, relativistic Bose gas at finite chemical potential was investigated, which has a Silver Blaze and sign problem, similar to lattice QCD. An extensive comparison of different algorithms to circumvent the sign problem for the O(3) non-linear sigma model in $1+1$ dimensions was carried out in Ref. \cite{Katz:2016azl}. The ability of adaptive step size to mitigate the unstable complex Langevin trajectories in lattice QCD was first demonstrated in the three-dimensional XY model, see Refs. \cite{ Aarts:2010aq, Aarts:2010vk, Aarts:2012ft, James:2012dna}. The random matrix theory shares several essential aspects of QCD, including the spontaneous breakdown of chiral symmetry, the finite density phase transition, and the complex fermion determinant. The authors of Refs. \cite{Mollgaard:2013qra, Mollgaard:2014mga, Nagata:2015ijn, Ichihara:2016uld, Nagata:2016alq, Bloch:2016jwt, Bloch:2017sex} used the random matrix model to investigate the properties of the complex Langevin method near the chiral limit in the cold and dense regimes of QCD. Detailed studies of spatially reduced low-dimensional QCD models have been conducted, particularly in $(0+1)$ dimensions in Refs. \cite{Aarts:2010gr, Pawlowski:2013pje} and $(1+1)$ dimensions in Refs. \cite{Bloch:2017sfg, Schmalzbauer:2016pbg, Bloch:2015coa}. These studies have been extremely helpful in gaining deep insights into the behavior of the complex Langevin method in lattice QCD.

The strong $\mathcal{CP}$ problem remains an unsolved QCD question in the SM. Investigations of field theories in the presence of a topological $\theta$-term are of great interest in understanding this problem. The topological $\theta$-term is purely imaginary, and these complex action theories are analyzed non-perturbatively in Refs. \cite{Bongiovanni:2014rna, Bongiovanni:2015kia, Hirasawa:2020bnl, Matsumoto:2021zjf}, using the complex Langevin method.  

In Ref. \cite{Pehlevan:2007eq}, a connection between different solutions of the Schwinger-Dyson equations and stationary distributions of the complex Langevin equations was established to study different phases of a QFT. Non-Hermitian but $\mathcal{PT}$-symmetric theories are widely known to have a real and positive spectrum. In Refs. \cite{Bernard:2001wh, Bernard:2004st} equal-time one-point and two-point Green's functions in zero and one dimension were computed to gain insights into a probabilistic interpretation of path integrals in $\mathcal{PT}$-symmetric QFTs. Refs. \cite{Joseph:2019sof, Joseph:2020gdh, Kumar:2022fas} probed the possibility of dynamical SUSY breaking in low-dimensional supersymmetric QFTs with complex actions, including the interesting cases of $\mathcal{PT}$-symmetric superpotentials. 

In Ref. \cite{Basu:2018dtm}, the authors observed Gross-Witten-Wadia (GWW) phase transitions in large-$N$ unitary matrix models using complex Langevin simulations. There have also been studies of spontaneous rotational symmetry breaking in dimensionally reduced super Yang-Mills models with Euclidean signature \cite{Ito:2016efb, Ito:2016hlj, Anagnostopoulos:2017gos}. The authors of Ref. \cite{Ito:2017wun} conducted a comparative study of deformation techniques to circumvent the {\it singular-drift} problem encountered during complex Langevin simulations in the context of matrix models. Ref. \cite{Ito:2016sci} uses numerical simulations based on the Langevin approach to examine the dynamics of spacetime in matrix models. Ref. \cite{Anagnostopoulos:2022dak} provides an extensive review of progress in numerical studies of the IKKT matrix model using complex Langevin and Monte Carlo methods. In Refs. \cite{Anagnostopoulos:2019ptt, Anagnostopoulos:2020xai, Anagnostopoulos:2020ebo, Anagnostopoulos:2020cwo, Kumar:2022giw}, the authors have reported a first-principles study of spontaneous rotational symmetry breaking in Euclidean IKKT matrix model. Ref. \cite{Hatakeyama:2021ake} clarifies the relationship between the Euclidean and Lorentzian versions of the IKKT matrix model. Recent numerical analyses of the Lorentzian IKKT matrix model suggest an expanding $(3+1)$-dimensional universe with exponential behavior at early times and power-law behavior at later times \cite{Aoki:2019tby, Hatakeyama:2019jyw, Nishimura:2019qal, Nishimura:2020blu, Hirasawa:2021xeh, Hatakeyama:2021ake, Hatakeyama:2022ybs}. Since a naive definition of the Lorentzian IKKT matrix model yields spacetime with Euclidean signature, in a promising recent study \cite {Nishimura:2022alt}, the authors have proposed to add a Lorentz invariant mass term. An exponential expansion behavior is observed, consistent with the Lorentzian signature at late times. The authors also observed the expansion of only one of nine spatial directions, corresponding to the $(1+1)$ dimensions of spacetime, which they explained from the perspective of the bosonic action.

Refs. \cite{Aarts:2013fpa, Aarts:2014nxa, Tanizaki:2015rda, Hayata:2015lzj, Fukushima:2015qza, Tsutsui:2015tua, Nishimura:2017eiu, Nishimura:2017vav} provide an excellent side-by-side comparison, and possible unification of complex Langevin and Lefschetz thimble approaches.

	\chapter{Complex Langevin simulations of zero-dimensional supersymmetric quantum field theories} 
\label{chap:chapter1}
\setlength\epigraphwidth{13cm}
\setlength\epigraphrule{0pt}
\epigraph{The chapter is based on the following publication by the author: \\ Anosh Joseph and \textbf{Arpith Kumar},\\ {\it Complex Langevin simulations of zero-dimensional supersymmetric quantum field theories},\\ \href{https://journals.aps.org/prd/abstract/10.1103/PhysRevD.100.074507}{Phys. Rev. D \textbf{100}, 074507 (2019)} {\href{https://arxiv.org/abs/1908.04153}{(arXiv: 1908.04153 [hep-th])}} }{}

QFTs in a spacetime with zero dimensions is the most straightforward starting point to embark on our journey to probe the possibility of spontaneous SUSY breaking. A zero-space dimensional QFT is standard quantum mechanics, often denoted as $d = 0+1$. The $0$ refers to the space dimension (single particle), and the $1$ to the time in which the particle propagates. In this case, we consider zero spacetime dimensions, which is even simpler than the standard quantum mechanics and is better described as a probability distribution of a variable with respect to a non-positive definite weight. This interpretation provides a safe playground for a precise understanding of the evolution and equilibration of field configurations without the extra complications of dimensions. In this chapter, we investigate spontaneous SUSY breaking in zero-dimensional supersymmetric QFTs.

The chapter is organized as follows. In Sec. \ref{sec:bosonic}, we apply complex Langevin dynamics to a class of zero-dimensional bosonic field theories with complex actions. We compute the expectation values of correlation functions and compare them with analytical results. Then, we discuss SUSY breaking in a zero-dimensional model with $\cN = 2$ SUSY and with a general form of the superpotential in Sec. \ref{sec:susy-breaking-mm}. In Sec. \ref{sec:various-sps}, using complex Langevin dynamics, we explore SUSY breaking in these models with real and complex actions for different forms of superpotentials. We also study the correctness criteria of our simulations using the Langevin operator and examine the probability distributions of the magnitude of the drift terms. 

\section{Bosonic models with complex actions}
\label{sec:bosonic}

A class of (Euclidean) scalar quantum field theories that are not symmetric under parity reflection were investigated in Ref. \cite{Bender:1997ps}. The authors considered a two-dimensional Euclidean Lagrangian of the form
\bea
\label{eqn:bos-bender}
{\cal L} = \hf (\partial_\mu \phi)^2 + \hf m^2 \phi^2 + W(\phi),\\
 W(\phi) = - \frac{g}{(2 + \delta)} (i \phi)^{(2 + \delta)},
\eea
for a scalar field $\phi$ with mass $m$. $W(\phi)$ is a $\cal{PT}$-symmetric potential with coupling parameter $g$ and real number $\delta >-2$. Such theories are very interesting from the point of view that they exhibit non-Hermitian Hamiltonian. Even more interesting is that there is numerous evidence that these theories possess energy spectra that are real and bounded below.

In this section, we consider the zero-dimensional version of the above bosonic Lagrangian
\beq
\label{eq:0d-bosonic}
{\cal L} = \hf m^2 \phi^2 + W(\phi)
\eeq 
and for massless theories with $m = 0$, the Euclidean action is nothing but the potential itself
\beq
S =W(\phi) = -\frac{g}{N} (i\phi)^N,
\eeq
where $N = 2 + \delta$. 

The partition function of this zero-dimensional model has the following form
\beq
Z = \frac{1}{2 \pi} \int_{-\infty}^{\infty} d\phi~ e^{-S} = \frac{1}{2 \pi} \int_{-\infty}^{\infty} d\phi~ \exp\left( \left[ \frac{g}{N} (i\phi)^N \right] \right),
\eeq
and the $k$-point correlation functions, $G_k$ can be computed as
\beq
G_k = \langle \phi^k \rangle = \frac{1}{Z} \frac{1}{2 \pi} \int_{-\infty}^{\infty} d\phi ~\phi^k ~ \exp \left( \left[ \frac{g}{N} (i\phi)^N \right] \right).
\eeq

\begin{figure*}[htp]
	
	{\includegraphics[width=.49\textwidth,origin=c,angle=0]{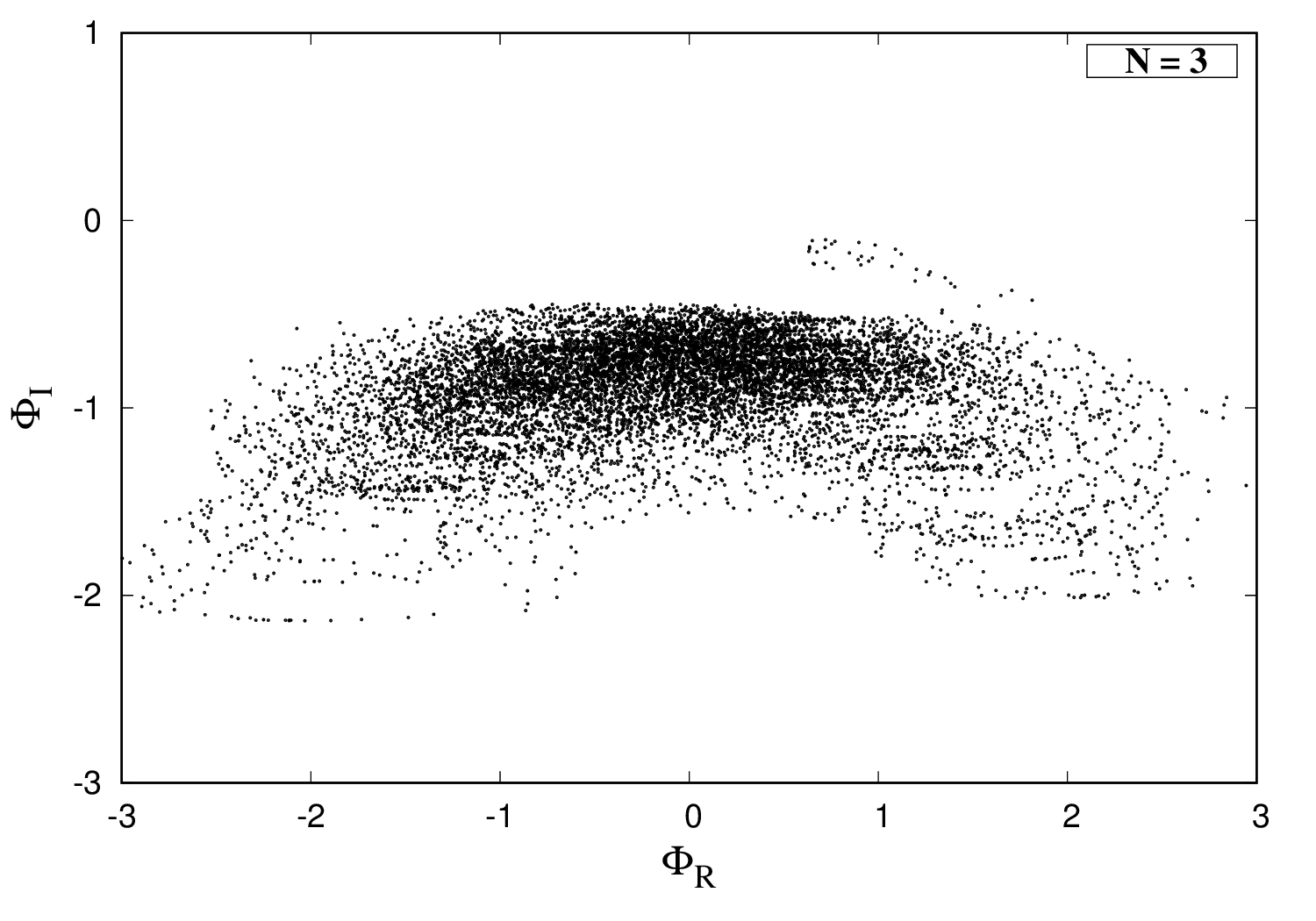}}
	{\includegraphics[width=.49\textwidth,origin=c,angle=0]{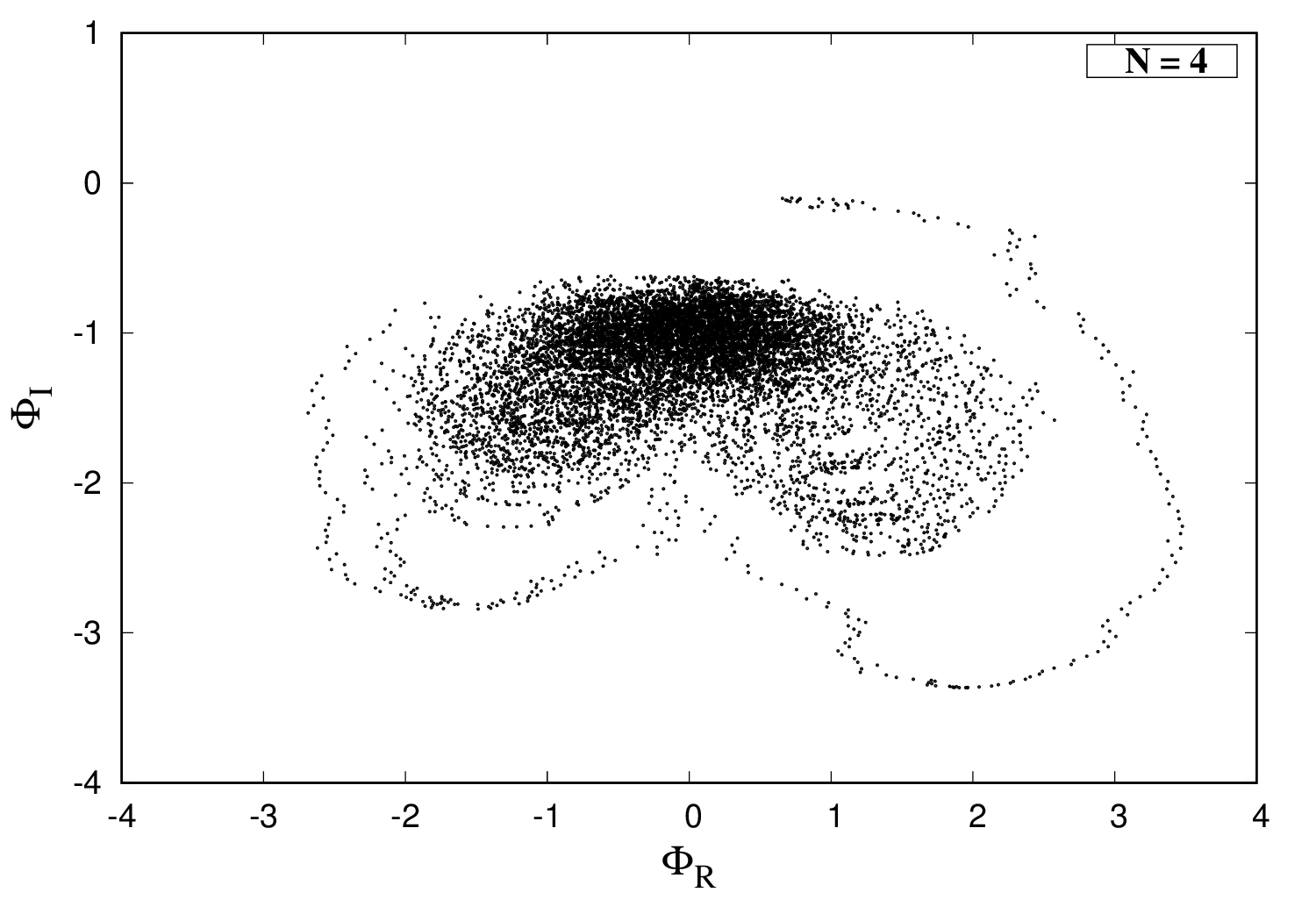}}
	
	\caption[Scatter plot of complexified field configurations on the $\phi_R - \phi_I$ plane for the zero-dimensional $ -\frac{g}{N} \left(i\phi\right)^N$ theory with $g = 0.5$.]{Scatter plot of complexified field configurations on the $\phi_R - \phi_I$ plane for the zero-dimensional $ -\frac{g}{N} \left(i\phi\right)^N$ theory with $g = 0.5$. Black dots represent the trajectories of the fields during complex Langevin evolution. (Left) Case $N = 3$. The field configuration starts at point $(0.5, -0.1)$, and with the aid of stochastic noise, it drifts towards the equilibrium configuration, forming a cloud averaging around $0.0 - i 0.9185$. (Right) Case $N = 4$. The field starts at point $(0.5, -0.1)$, and with the aid of stochastic noise, it drifts towards the equilibrium configuration, forming a cloud averaging around $0.0 -i 1.163$.}
	\label{fig:n3-n4-clouds}
	
\end{figure*}

The one-point correlation function, $G_1$ can be evaluated as \cite{Bender:1999ek}
\bea
G_1 = - \frac{i}{\sqrt{\pi}} \left(\frac{4N}{g}\right)^{1/N}  { \Gamma\left(\frac{1}{N} + \hf\right) \cos \left(\frac{\pi}{N}\right) } ,
\eea
and the two-point correlation function, $G_2$ as
\bea
G_2 = \left(\frac{N}{g}\right)^{2/N} \frac{\Gamma\left(\frac{3}{N}\right)}{{\Gamma\left(\frac{1}{N}\right)}}{ \left[\sin^2\left(\frac{\pi}{N}\right)- 3 \cos^2\left(\frac{\pi}{N}\right)\right]} .
\eea
Similarly, we can compute higher moments of $\phi$. In Table \ref{tab:bosonic}, we compare our results from complex Langevin simulations for $G_1$ and $G_2$ with their corresponding analytical results.

\begin{table*}[t]
	\centering
	{\small \begin{tabular}{|l||	c |c |} 
	\hline
	$$ &   $~~~N=3$   &  $N=4~~~$  \\	
	\hline
	\hline
	$G_1^{\rm exact}$   & $0.0- i0.9185$   	 & $0.0 - i 1.1630$   	\\
	$G_1^{\rm cL}$  & $0.0112(121) - i 0.9183(41) $    & 	$0.0050(58) - i 1.1651(28)$		   \\
	\hline
	$G_2^{\rm exact}$   & $0.0 + i0.0$   	 & $-0.9560 + i 0.0$	    	\\
	$G_2^{\rm cL}$  & $-0.0001(16) - i0.0237(286)$  &	 $-0.9587 (45) -i 0.0122(158)$   \\
	\hline
	\end{tabular}}
	\caption[Simulated values of the correlation functions $G_1$ and $G_2$ obtained from complex Langevin dynamics for zero-dimensional $-\frac{g}{N} (i\phi)^N$ theory for $N = 3, 4$.]{\label{tab:bosonic}Simulated values of the correlation functions $G_1$ and $G_2$ obtained from complex Langevin dynamics for zero-dimensional $-\frac{g}{N} (i\phi)^N$ theory for $N = 3, 4$. The simulations were performed with coupling parameter $g = 0.5$, adaptive Langevin step size $\Delta \tau \leq 0.002$, thermalization steps $N_{\rm therm} = 10^6$, generation steps $N_{\rm gen} = 10^7$, and measurements were taken every $1000$ step. We considered various random initial configurations, and the simulation flows to the same equilibrium fixed point. The table compares these numerically simulated values with the exact results.}
\end{table*}

\begin{figure}[H]
	\centering
	\includegraphics[width=.49\textwidth,origin=c,angle=0]{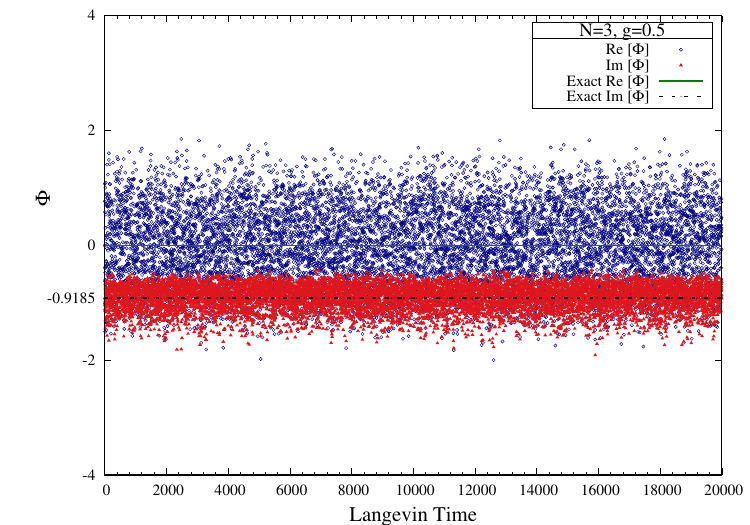}
	\includegraphics[width=.49\textwidth,origin=c,angle=0]{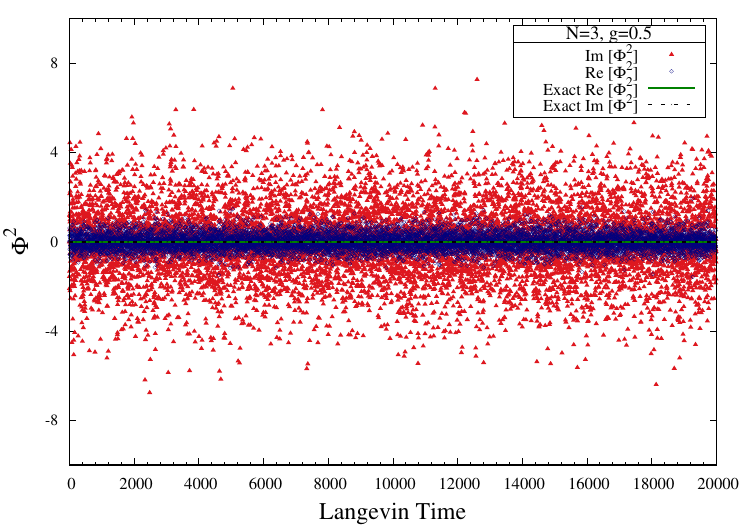}
	\caption[Langevin time history of the one-point (Left) and two-point (Right) correlation functions for the $i \frac{g}{3} \phi^3$ theory at coupling parameter $g = 0.5$.]{Langevin time history of the field variable (one-point correlation function $G_1$) for the $i \frac{g}{3} \phi^3$ theory at coupling parameter $g = 0.5$. Simulations were performed with adaptive Langevin step size $\Delta \tau \leq 0.002$, generation steps $N_{\rm gen} = 10^7$, and measurements were taken every $1000$ step. Solid and dashed lines represent the exact values.}
	\label{fig:n3-history}
\end{figure}

\begin{figure*}[htp]
	
	{\includegraphics[width=.49\textwidth,origin=c,angle=0]{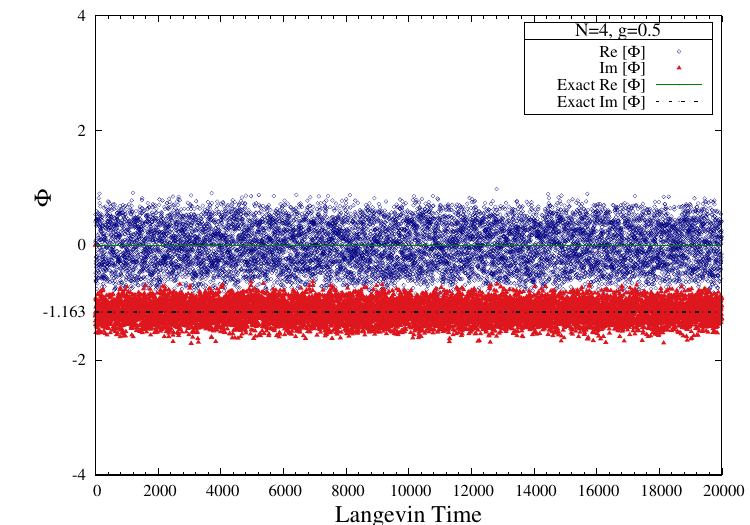}}
	{\includegraphics[width=.49\textwidth,origin=c,angle=0]{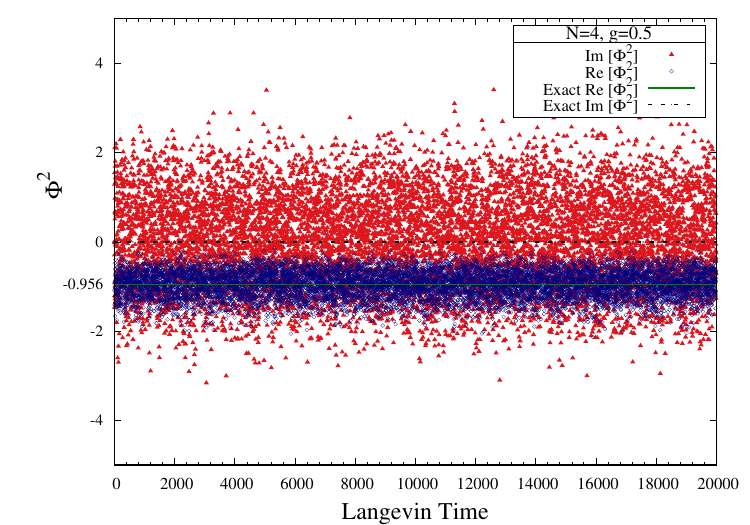}}
	
	\caption[Langevin time history of one-point (Left) and two-point (Right) correlation functions for the $- \frac{g}{4} \phi^4$ theory at fixed coupling constant $g = 0.5$.]{Langevin time history of one-point (Left) and two-point (Right) correlation functions for the $- \frac{g}{4} \phi^4$ theory at fixed coupling constant $g = 0.5$. Simulations were performed with adaptive Langevin step size $\Delta \tau \leq 0.002$, generation steps $N_{\rm gen} = 10^7$, and measurements were taken every $1000$ step. Solid and dashed lines represent the exact values.}
	\label{fig:n4-history}
	
\end{figure*}

In Fig. \ref{fig:n3-n4-clouds}, we show the complexified $\phi$ field configurations on the complex $\phi_R - \phi_I$ plane as it evolves in Langevin time. The Langevin time history of $G_1$ and $G_2$ for the case $N=3$ is shown in Fig. \ref{fig:n3-history}. In Fig. \ref{fig:n4-history}, we show the Langevin time history of $G_1$ and $G_2$ for the case $N=4$.

\section{Supersymmetry breaking in zero-dimensional field theories}
\label{sec:susy-breaking-mm}

One can think of making the Lagrangian in Eq. \eqref{eqn:bos-bender} supersymmetric by adding the right amount of fermions. The two-dimensional supersymmetric Lagrangian can be written as follows
\bea
{\cal L} = \hf \left(\partial_\mu \phi \right)^2 + \hf i \psib \slashed{\partial} \psi + \hf \psib W''(\phi) \psi + \hf \big [ W'(\phi) \big ]^2,\\
W(\phi) = - \frac{g}{(2 + \delta)} (i \phi)^{(2 + \delta)},
\eea
where $\psi,\psib$ are Majorana fermions \cite{Bender:1997ps}. This supersymmetric Lagrangian also breaks parity symmetry. It would be interesting to ask whether the breaking of parity symmetry induces a breaking of SUSY. This question was answered in Ref. \cite{Bender:1997ps}. There, through a perturbative expansion in $\delta$, the authors found that SUSY remains unbroken in this model. We could think of performing non-perturbative investigations on SUSY breaking in this model using the complex Langevin method (Clearly, a non-perturbative investigation based on path integral Monte Carlo fails since the action of this model can be complex, in general.) 

In this section, we consider a zero-dimensional version of the above supersymmetric model. We work with a general form of supersymmetric potential, $W(\phi)$, and the action is given by
\beq
S = \hf {\B}^2 + i \B W' + \bar{\psi} W'' \psi,
\eeq
where $\phi$ is a bosonic field, $\psi$ and $\bar{\psi}$ are fermionic fields, and $\B$ is an auxiliary field. The prime denotes the derivative of the superpotential with respect to $\phi$. SUSY exchanges fermionic fields with bosonic fields in this theory. We can define two independent SUSY charges $\Q$ and $\Qb$ corresponding to an $\cN = 2$ SUSY. This action can be derived from the dimensional reduction of a one-dimensional theory, that is, a supersymmetric quantum mechanics with two supercharges. 

We can see that the above action is invariant under the following SUSY transformations
\bea
	\label{eq:susy-transf-Q}
	\Q \phi  = \psi,&&~~
	\Q \psi = 0, \\
	\Q \bar{\psi} = - i\B,&&~~
	\Q \B = 0,
\eea
and 
\bea
	\label{eq:susy-transf-Qb}
	\Qb \phi = - \bar{\psi},&&~~
	\Qb \bar{\psi} = 0, \\
	\Qb \psi = - i \B,&&~~
	\Qb \B = 0. 
\eea
The supercharges $\Q$ and $\Qb$ satisfy the algebra
\beq
	\{ \Q, \Q \} = 0, ~~
	\{ \Qb, \Qb \} = 0, ~~
	\{ \Q, \Qb \} = 0.
\eeq
We also note that the action can be expressed in $Q$- or $Q \Qb$- exact forms. That is,
\beq
S = \Q \psib \left( \frac{i}{2} \B - W' \right) =  \Q \Qb \left( \hf \psib \psi + W \right),
\eeq
and it is easy to show that the action is invariant under the two SUSY charges
\beq
\Q S = 0,~~\Qb S = 0.
\eeq

The auxiliary field $\B$ has been introduced for off-shell completion of the SUSY algebra and can be integrated out using its equation of motion
\beq
\B = -i W'.
\eeq

The partition function of the model is
\bea
Z & =& \frac{1}{2 \pi} \int d\B d\phi d\psi d\psib \ e^{-S} \nn \\
& =& \frac{1}{2 \pi} \int d\B d\phi d\psi d\psib \exp\left[- \Big( \hf {\B}^2 + i \B W' + \psib W'' \psi \Big) \right],
\eea
and upon completing the square and integrating over the auxiliary field, it becomes
\beq
Z =  \frac{1}{\sqrt{2 \pi}} \int d\phi d\psi d\psib \ \exp \left[ - \left( \hf  {W'}^2 + \psib W'' \psi \right) \right].
\eeq
Finally, after integrating over the fermions, we have
\beq
Z  = - \frac{1}{\sqrt{2 \pi}} \int d\phi \ W'' \ \exp \left[-\hf {W'}^2 \right] .
\eeq
When SUSY is broken, the supersymmetric partition function vanishes. In that case, the expectation values of observables normalized by the partition function could be ill-defined.

The expectation value of the auxiliary field $\B$ is crucial in investigating SUSY breaking. It can be evaluated as
\bea
\langle \B \rangle &=& \frac{1}{Z} \frac{1}{2 \pi}  \int d\B d\phi d\psi d\psib ~ \B ~ e^{-S} \nn  \\
&=& \frac{1}{Z} \frac{i}{\sqrt{2 \pi}} \int d\phi \ W' \ {W}'' \ \exp \left[-\hf {W'}^2 \right] \nn \\
&=& - \frac{1}{Z} \frac{i}{\sqrt{2 \pi}} \int_{-\infty}^{\infty}  d\phi \ \frac{\partial}{\partial \phi}  \ \exp \left[-\hf {W'}^2 \right].
\eea
Thus, when SUSY is broken, the normalized expectation value of $\B$ is indefinite (of the form $0/0$).

In order to overcome this difficulty, we can introduce an external field and eventually take a limit where it goes to zero. We usually introduce some external field to detect the spontaneous breaking of ordinary symmetry so that the ground state degeneracy is lifted to specify a single broken ground state. We take the thermodynamic limit of the theory, and after that, the external field is turned off. The value of the corresponding order parameter then would tell us if spontaneous symmetry breaking happens in the model or not. (Note that to detect the spontaneous magnetization in the Ising model, we use the external field as a magnetic field, and the corresponding order parameter then would be the expectation value of the spin operator.) We will also perform an analogues method to detect SUSY breaking in the system. The introduction of an external field can be achieved by changing the boundary conditions for the fermions to twisted boundary conditions.

\subsection{Theory on a one-site lattice}

Let us consider the above zero-dimensional theory as a dimensional reduction of a one-dimensional theory, which is a supersymmetric quantum mechanics. The action of the one-dimensional theory is integral over a compactified time circle of circumference $\beta$ in Euclidean space. We have the action
\beq
S  =  \int_0^\beta d\tau \left[ \ \hf {\B}^2 + i\B \left(\dot{\phi} + W'\right) + \psib \left( \dot{\psi} + W'' \psi \right) \ \right],
\eeq
where the dot denotes derivative with respect to Euclidean time $\tau \in [0, \beta]$. Note that the $\Qb$ SUSY will not be preserved in the quantum mechanics theory.

Let us discretize the theory on a one-dimensional lattice with $T$ sites, using finite differences for derivatives. We have the lattice action
\bea
S &=& \sum_{n=0}^{T-1} \ \Bigg[\ \hf {\B}^2(n) +  i \B(n) \Big( \phi (n+1) - \phi (n) + W'(\phi (n)) \Big) \nn \\
&&~~~~~~~~~~~~~ + \psib(n) \ \Big( \psi(n+1) - \psi(n)+ W''(\phi (n)) \ \psi(n) \Big) \ \Bigg],
\eea
with $n$ denoting the lattice site. We have rescaled the fields and coupling parameters such that the lattice action is expressed in terms of dimensionless variables. The lattice action preserves one of the supercharges, $\Q$. The $\Qb$ SUSY will not be a symmetry on the lattice when $T\geq 2$.

Let us consider the simplest case of one lattice point, that is, when $T = 1$. The action becomes
\bea
S &=& \Bigg[\ \ \hf {\B}^2(0) +  i \B(0)  \Big( \phi (1) - \phi (0) + W'(\phi (0))  \Big) \nn \\ 
&&~~~~~~~~~~~~~ + \psib(0)  \Big(  \psi(1) - \psi(0)+ W''(\phi (0)) \ \psi(0) \Big) \ \Bigg],
\eea
where $\phi(1)$ and $\psi(1)$ are dependent on the boundary conditions. In the case of periodic boundary conditions, 
\bea
	\phi(1)  =  \phi(0),&&~~
	\psi(1) = \psi(0), \nn \\
	\psib(1) = \psib(0),&&~~
	\B(1) =  \B(0),
\eea
the action reduces to
\beq
S = \hf {\B}^2 + i \B W' + \psib W'' \psi.
\eeq

Thus the action for the zero-dimensional supersymmetric model with $\cN = 2$ SUSY is equivalent to the dimensional reduction of a one-dimensional theory (supersymmetric quantum mechanics) with periodic boundary conditions. 

\subsection{Twisted boundary conditions}

Now, instead of periodic boundary conditions, let us introduce twisted boundary conditions for fermions (analogues to turning on an external field), with the motivation to regularize the indefinite form of the expectation values we encountered earlier\footnote{Twisted boundary conditions were considered in the context of supersymmetric models by Kuroki and Sugino in Refs. \cite{Kuroki:2009yg, Kuroki:2010au}.}. The field configurations are subjected to the following conditions;
\bea
	\phi(1)  =  \phi(0),&&~~
	\psi(1)  = e^{i\alpha} \psi(0), \\
	\psib(1)  = e^{- i\alpha} \psib(0),&&~~
	\B(1)  =  \B(0).
\eea
The action, in this case, has the form
\beq
S_\alpha =  \hf {\B}^2 + i \B W' + \psib \Big( e^{i \alpha} - 1 + W'' \Big) \psi,
\eeq
with SUSY softly broken by the introduction of the twist parameter, $\alpha$, that is
\beq
Q S_\alpha = -i \Qb S_\alpha = \psib \left( e^{i \alpha} - 1 \right) \psi,
\eeq 
and in the limit $\alpha \to 0$ SUSY is recovered.

The partition function is
\bea
\label{eq:Z-twist}
Z_\alpha &=& \frac{1}{2 \pi} \int d\B d\phi d\psi d\psib \ e^{ -S_\alpha } \nn \\
&=& - \frac{1}{\sqrt{2 \pi}} \int d\phi ~ \Big( e^{i \alpha} - 1 + W'' \Big)  \exp \left[ - \hf W'^2 \right].
\eea

The expectation of the auxiliary field, $\B$, observable is given by
\bea
\langle \B \rangle_\alpha &=&  \frac{1}{Z_\alpha} \frac{1}{2 \pi} \int d\B d\phi d\psi d\bar{\psi} \ \B \ e^{ -S_\alpha }  \nn \\
&=&  \frac{1}{Z_\alpha} \frac{i}{\sqrt{2 \pi}} \int d\phi \ W' \Big( e^{i \alpha} - 1 + W'' \Big) \exp \left[ - \hf W'^2 \right].
\eea
It is important to note that the quantity $\langle \B \rangle_\alpha$ is now well defined. Here, the external field $\alpha$ plays the role of a regularization parameter, and it regularizes the indefinite form, $\langle \B\rangle = 0/0$, of the expectation value under periodic boundary conditions and leads to the non-trivial result. Vanishing expectation value of the auxiliary field, $\langle \B \rangle_\alpha$ in the limit $\alpha \to 0$ indicates that SUSY is not broken, while a non-zero value indicates SUSY breaking. Later in Chapter \ref{chap:susy-qm}, when discussing supersymmetric quantum mechanics, we thoroughly explain the indefinite form of auxiliary field and the introduction of twist fields.

The effective action of the model with twisted boundary conditions reads
\beq
S_\alpha^{~\text{eff}} = \hf  W'^2 -  \text{ln} \left[ e^{i \alpha} - 1 + W'' \right],
\eeq
and its gradient, the drift term required for the application of the complex Langevin method in Sec. \ref{sec:various-sps} has the form
\bea
\label{eq:clm-drift}
\frac{\partial S_\alpha^{~\text{eff}}}{\partial \phi} &=& \frac{\partial}{\partial \phi}  \left( \hf  W'^2 - \ln \left[ e^{i \alpha} - 1 + W'' \right] \right) \nn \\
&=& W' W'' - \frac{W'''}{\Big ( e^{i \alpha} - 1 + W''\Big)}.
\eea

\section{Models with various superpotentials}
\label{sec:various-sps}

In this section, we investigate spontaneous SUSY breaking in various zero-dimensional models using the complex Langevin method. Wherever possible, we also compare our numerical results with corresponding analytical results.

\subsection{Double-well potential}

Let us begin with a case where the action is real. We consider the case when the derivative of the superpotential is a double-well potential 
\beq
W' = g \ (\phi^2 + \mu^2),
\eeq
where $g$ and $\mu$ are two parameters in the theory.

When $\mu^2 > 0$, the classical minimum is given by the field configuration $\phi = 0$ with energy
\beq
E_0 = \hf g^2 \mu^4 > 0,
\eeq
implying spontaneous SUSY breaking. The ground state energy can be computed as the expectation value of the bosonic action at the classical minimum
\bea
E_0 \big|_{\phi =0} &=& \langle S_B \rangle = \hf {\B}^2 +i\B W' \nn \\
&=& - \hf (W')^2 + (W')^2 = \hf (W')^2\Big|_{\phi=0} \nn \\
&=& \hf g^2 \mu^4.
\eea
We can also see from SUSY transformations that SUSY is broken in the model, that is
\beq
Q \psib = -g \mu^2,~~ {\rm and}~~\Qb \psi = -g \mu^2.
\eeq

The twisted partition function for the model reads
\bea
Z_\alpha &=&- \frac{1}{ \sqrt{2 \pi} } \int_{-\infty}^\infty d\phi ~ \Big( e^{i \alpha} -1 + W'' \Big) \exp \left[ - \hf (W')^{2} \right] \nn \\
&=&- \frac{1}{\sqrt{2 \pi}} \int_{-\infty}^\infty d\phi ~ \Big( e^{i \alpha} - 1 + 2g   \phi \Big) \exp \left[ - \hf g^2 (\phi^2 + \mu^2)^2 \right] \nn \\
&=& -\frac{\mu \left( e^{i \alpha} - 1 \right) \exp\left[{- \qtr g^2 \mu^4 }\right]}{2 \sqrt{\pi} }   ~{K}_{\qtr} \left( \frac{g^2 \mu^4}{4} \right) ~ \forall ~ \text{Re} \left( g^2 \right) > 0 ~\cap ~\text{Re} \left( g^2 \mu^2 \right) > 0,
\eea
where in the limit $\alpha \rightarrow 0$, we have vanishing partition function, $Z_\alpha \big|_{\alpha = 0} = 0$, implying broken SUSY for $W' = g \ (\phi^2 + \mu^2)$.

The auxiliary field observable expectation values read
\bea
\langle \B \rangle_\alpha &=& -\frac{1}{Z_\alpha} \frac{1}{\sqrt{2 \pi}} \int_{-\infty}^\infty d\phi ~ \left(-iW'\right)~ \Big( e^{i \alpha} -1 + W'' \Big) \exp \left[ - \hf W'^{2} \right] \nn \\
&=& -ig \ \frac{ \int_{-\infty}^\infty d\phi ~(\phi^2 + \mu^2) ~\exp \left[ - \hf g^2 (\phi^2 + \mu^2)^{2} \right] }{ \int_{-\infty}^\infty d\phi ~\exp \left[ - \hf g^2 (\phi^2 + \mu^2)^{2} \right] },
\eea
and once evaluated, becomes
\beq
\langle \B \rangle_\alpha = - \frac{ig \mu^2}{2} \left[1+ \frac{  {K}_{\frac{3}{4}} \left( {g^2 \mu^4}/{4} \right) \Big)}{{K}_{\qtr} \left({g^2 \mu^4}/{4} \right)} \right] ~ \forall ~ \text{Re}\left( g^2 \right) > 0 ~\cap~\text{Re} \left(g^2 \mu^2 \right) > 0. 
\eeq

\begin{figure*}[htp]
	
	{\includegraphics[width=.49\textwidth,origin=c,angle=0]{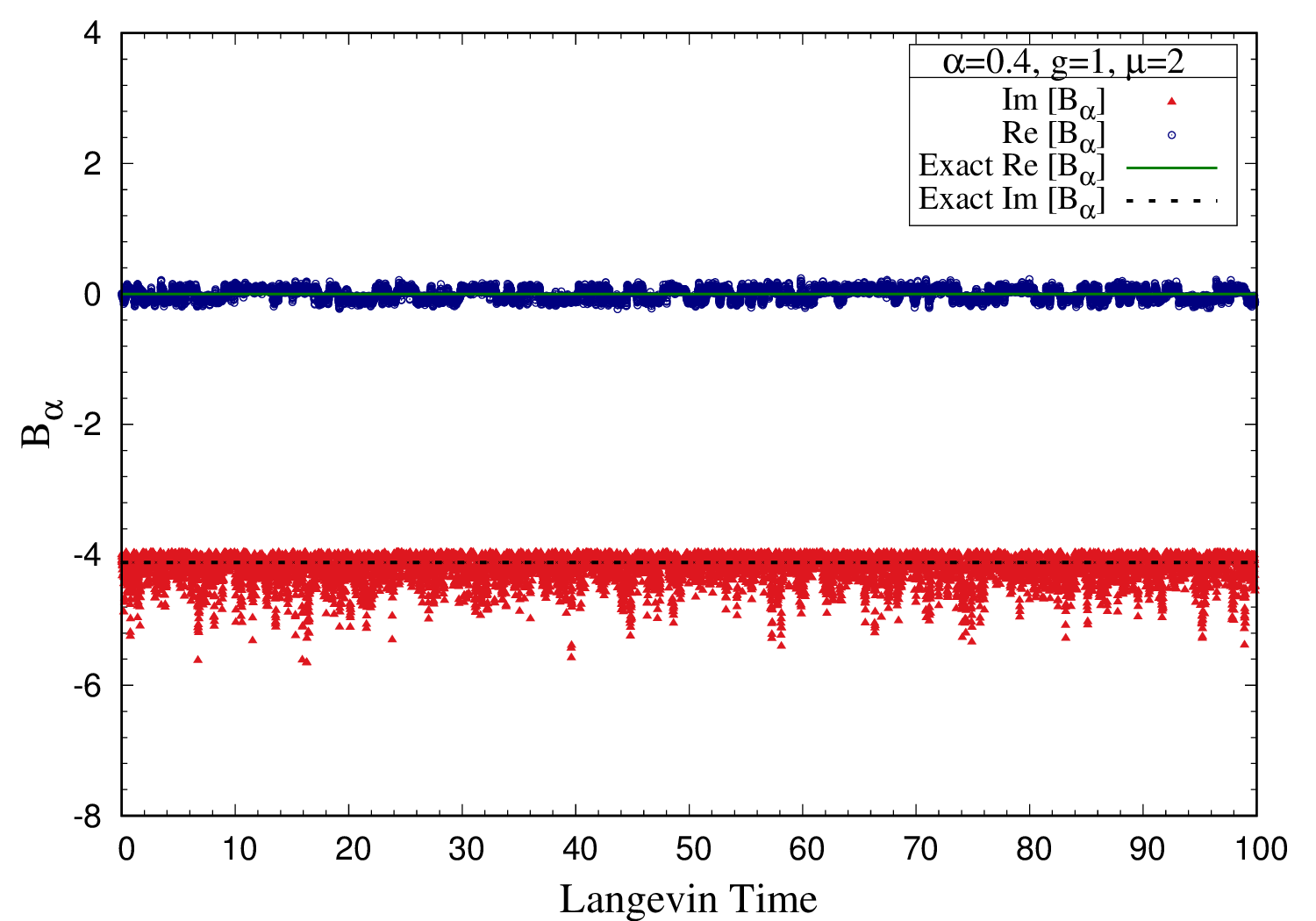}}
	{\includegraphics[width=.49\textwidth,origin=c,angle=0]{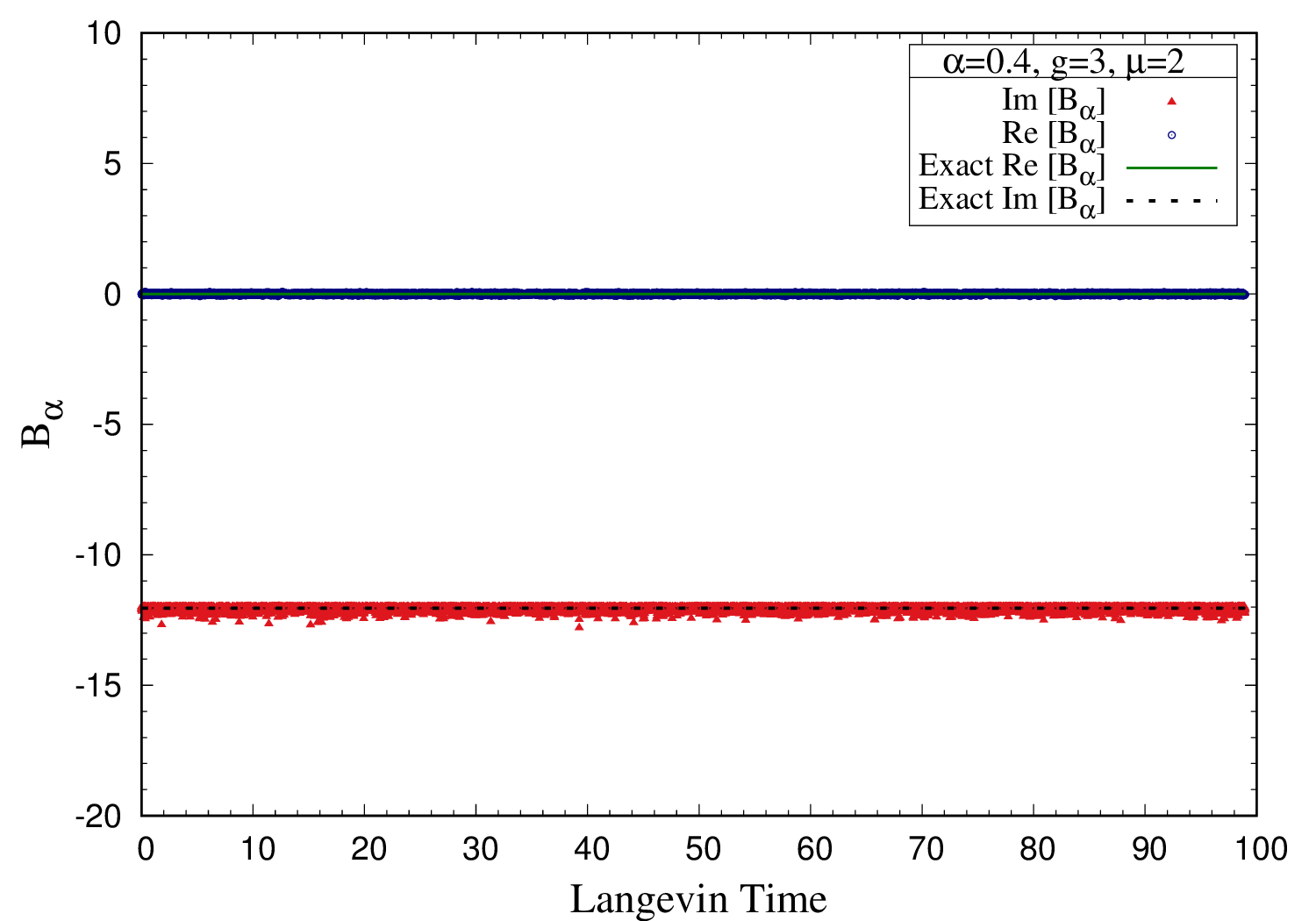}}
	
	\caption[The observable $\B$ against Langevin time for regularization parameter $\alpha = 0.4$. Simulations were performed for superpotential $W' = g \ (\phi^2 + \mu^2)$ with $\mu = 2$.]{The observable $ B$ against Langevin time for regularization parameter $\alpha = 0.4$. Simulations were performed for superpotential $W' = g \ (\phi^2 + \mu^2)$ with $\mu = 2$. In these simulations, we have used adaptive Langevin step size $\Delta \tau \leq 10^{-4}$, generation steps $N_{\rm gen} = 10^6$, and measurements were taken every $100$ step. (Left) Case $g=1$. The exact value is $\langle \B \rangle = 0.0 - i 4.115$, corresponding to a system with broken SUSY.  (Right) Case $g=3$. The exact value is $\langle \B \rangle = 0.0 - i 12.041$ again indicating that SUSY is broken in the model.}
	\label{fig:dw-g1-g3-p0_mu2p0}
	
\end{figure*}

\begin{figure*}[htp]
	
	{\includegraphics[width=.49\textwidth,origin=c,angle=0]{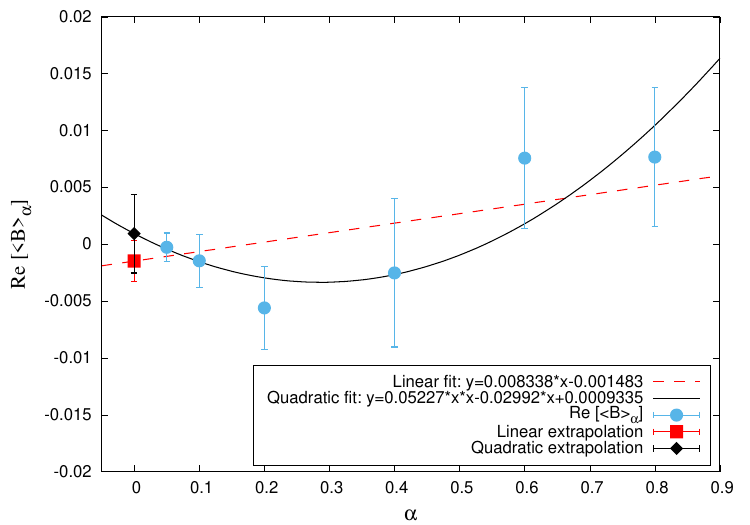}}
	{\includegraphics[width=.49\textwidth,origin=c,angle=0]{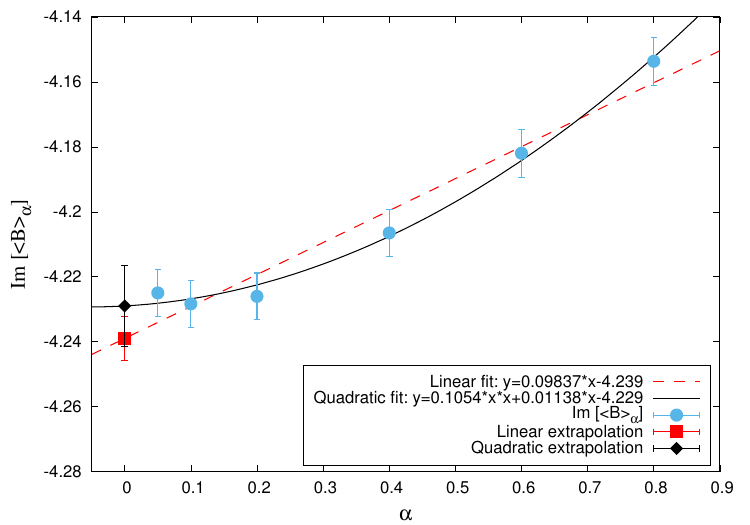}}
	
	\caption[Plot of real (Left) and imaginary (Right) parts of $\langle \B \rangle_\alpha$ against the regularization parameter, $\alpha$ for supersymmetric potential $W' = g \ (\phi^2 + \mu^2)$. Simulations were performed with $g = 1$ and $\mu = 2$.]{Plot of real (Left) and imaginary (Right) parts of $\langle \B \rangle_\alpha$ against the regularization parameter, $\alpha$ for supersymmetric potential $W' = g \ (\phi^2 + \mu^2)$. Simulations were performed with $g = 1$ and $\mu = 2$. We have used adaptive Langevin step size $\Delta \tau \leq 10^{-4}$, thermalization steps $N_{\rm therm} = 10^{4}$, generation steps $N_{\rm gen} = 10^6$, and measurements were taken every $100$ steps. The dashed red lines are the linear fits to $\langle \B \rangle_\alpha$ in $\alpha$, and filled red squares are the linear extrapolation values at $\alpha = 0$.  The solid black lines represent the quadratic fits to $\langle \B \rangle_\alpha$ in $\alpha$, and filled black diamonds are the quadratic extrapolation values at $\alpha = 0$. The $\alpha \to 0$ limit values obtained from these plots are given in Table \ref{tab:sqw_B_mu2p0}.}
	\label{fig:dw_fit_g1p0_mu2p0}
\end{figure*}

\begin{figure*}[htp]
	
	{\includegraphics[width=.49\textwidth,origin=c,angle=0]{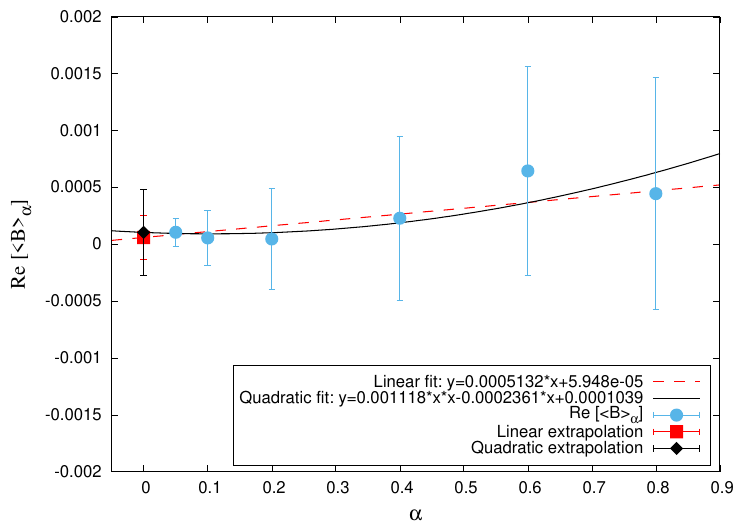}}
	{\includegraphics[width=.49\textwidth,origin=c,angle=0]{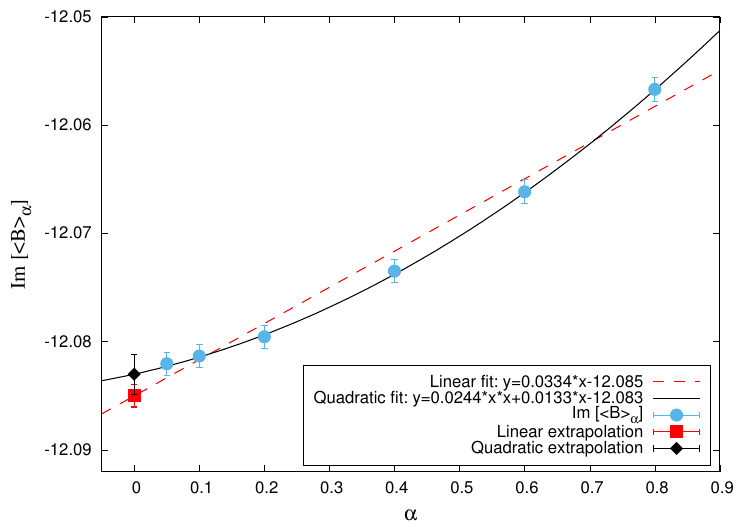}}
	
	\caption[Plot of real (Left) and imaginary (Right) parts of $\langle \B \rangle_\alpha$ against the regularization parameter, $\alpha$ for supersymmetric potential $W' = g \ (\phi^2 + \mu^2)$. Simulations were performed with $g = 3$ and $\mu = 2$.]{Plot of real (Left) and imaginary (Right) parts of $\langle \B \rangle_\alpha$ against the regularization parameter, $\alpha$ for supersymmetric potential $W' = g \ (\phi^2 + \mu^2)$. Simulations were performed with $g = 3$ and $\mu = 2$. We have used adaptive Langevin step size $\Delta \tau \leq 10^{-4}$, thermalization steps $N_{\rm therm} = 10^{4}$, generation steps $N_{\rm gen} = 10^6$, and measurements were taken every $100$ steps. The dashed red lines are the linear fits to $\langle \B\rangle_\alpha$ in $\alpha$, and filled red squares are the linear extrapolation values at $\alpha = 0$. The solid black lines represent the quadratic fits to $\langle \B \rangle_\alpha$ in $\alpha$, and filled black diamonds are the quadratic extrapolation values at $\alpha = 0$. The $\alpha \to 0$ limit values obtained from these plots are given in Table \ref{tab:sqw_B_mu2p0}.}
	\label{fig:dw_fit_g3p0_mu2p0}
	
\end{figure*}

In Fig. \ref{fig:dw-g1-g3-p0_mu2p0}, we show our results from Langevin simulations of this model. We show linear and quadratic extrapolations to $\alpha \to 0$ limit in Figs. \ref{fig:dw_fit_g1p0_mu2p0} and \ref{fig:dw_fit_g3p0_mu2p0}. The results are tabulated in Table \ref{tab:sqw_B_mu2p0}. The simulation results are in good agreement with the analytical predictions and strongly suggest that SUSY is broken for this model. 

\begin{table*}[htp]
	\centering
{\footnotesize	\begin{tabular}{| c | c | c | c  | c |} 
		\hline
		$W'$& $~~\mu~~$  & $~~g~~$ & $~~~~~\alpha~~~~~$ & $~~~~~~\langle \B \rangle |_{\alpha}~~~~~~$ \\ [1.5ex] \hline\hline
		\multirow{10}{*}{$g\Big(\phi^2 + \mu^2\Big)$} &\multirow{10}{*}{$2.0$} & \multirow{5}{*}{$1.0$}
		&  	0.05	 &	 $ -0.0003(12) - i  4.2250 (72)$ \\ \cline{4-5}
		&&		&  	0.1	 	 &	 $ -0.0015(23) - i  4.2283 (72)$ \\ \cline{4-5}
		&&		&  	0.2		 &	 $ -0.0056(37) - i  4.2261 (72)$ \\ \cline{4-5}
		&&	 	&  	0.4		 &	 $ -0.0025(65) - i  4.2065 (72)$  \\ \cline{4-5}
		&&	 	&  	0.6	 	 &	 $ 0.0076(62)  - i  4.1820 (74)$  \\ \cline{4-5}
		&&	 	&  	0.8		 &	 $ 0.0077(61)  - i  4.1537 (74)$   \\ \cline{4-5} 
		&&	&  	$\alpha \rightarrow 0$ &	 $ -0.0003(35) - i 4.2340(123)$  \\ \cline{3-5} 
		&&	\multirow{5}{*}{$3.0$}	
		&  	0.05	 &	 $ 0.0001(1)  - i  12.0820 (11)$ \\ \cline{4-5}
		&&		&  	0.1		 &	 $ 0.0001(2)  - i  12.0813 (11)$ \\ \cline{4-5}
		&&		&  	0.2		 &	 $ 0.0000(4)  - i  12.0796 (11)$	\\ \cline{4-5}
		&&	 	&  	0.4		 &	 $ 0.0002(7)  - i  12.0735 (11) $   \\ \cline{4-5}
		&&	 	&  	0.6		 &	 $ 0.0006(9)  - i  12.0662 (11)$  \\ \cline{4-5}
		&&	 	&  	0.8		 &	 $ 0.0004(10) - i  12.0567 (11) $ \\ \cline{4-5}
		&&	&  	$\alpha \rightarrow 0$ &	 $ 0.0001 (4) - i 12.0840(18)$ \\ \hline
	\end{tabular}}
	\caption[Expectation values $\langle \B \rangle_\alpha$ obtained using complex Langevin simulations for the model with superpotential $W' = g \left( \phi^2 + \mu^2 \right)$.]{\label{tab:sqw_B_mu2p0}Expectation values $\langle \B \rangle_\alpha$ obtained using complex Langevin simulations for the model with superpotential $W' = g \left( \phi^2 + \mu^2 \right)$. In the limit $\alpha \to 0, \langle \B \rangle_\alpha \neq 0 $. Thus SUSY is broken in this model.}
\end{table*}

\begin{figure*}[htp]
	
	{\includegraphics[width=.49\textwidth,origin=c,angle=0]{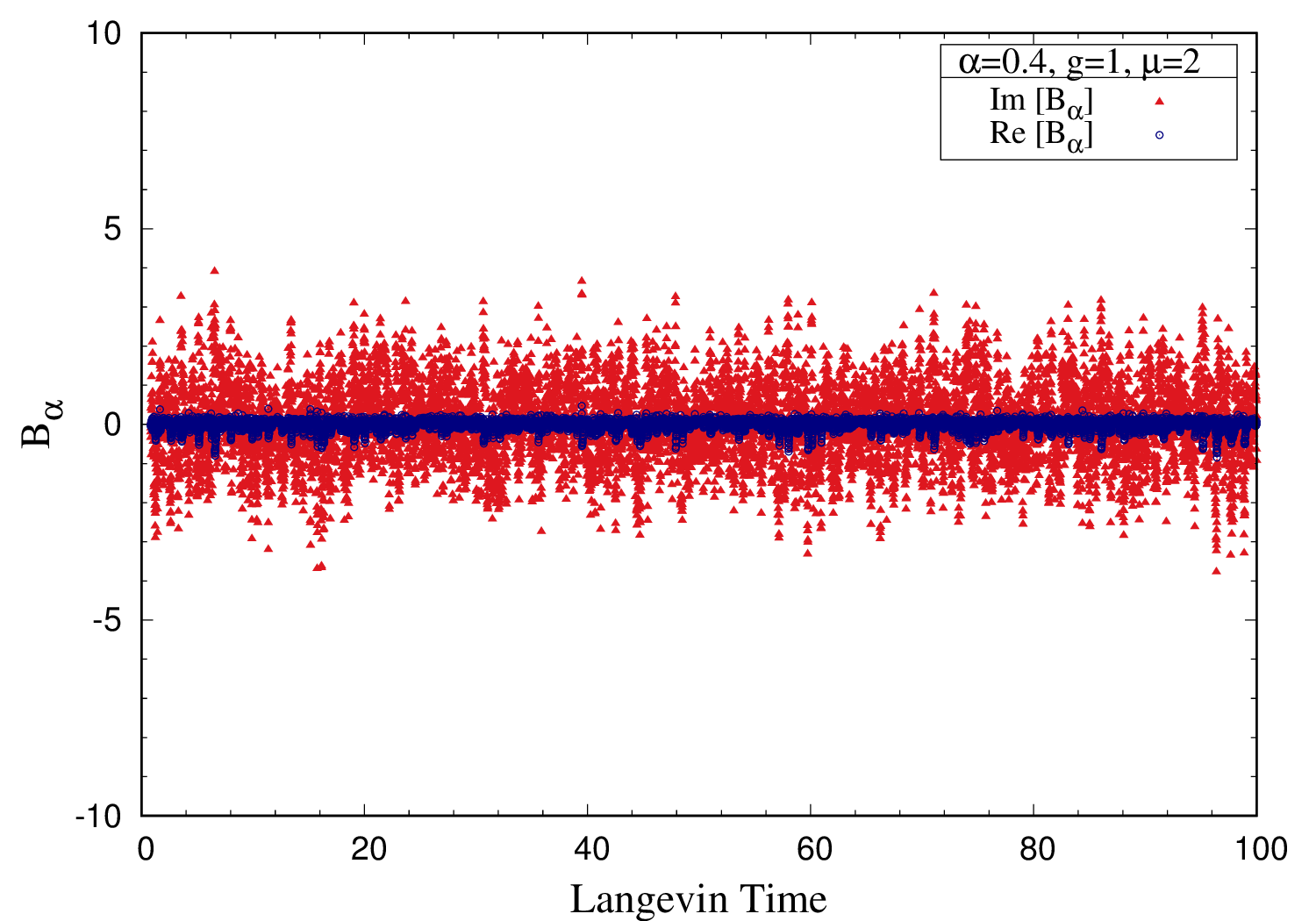}}
	{\includegraphics[width=.49\textwidth,origin=c,angle=0]{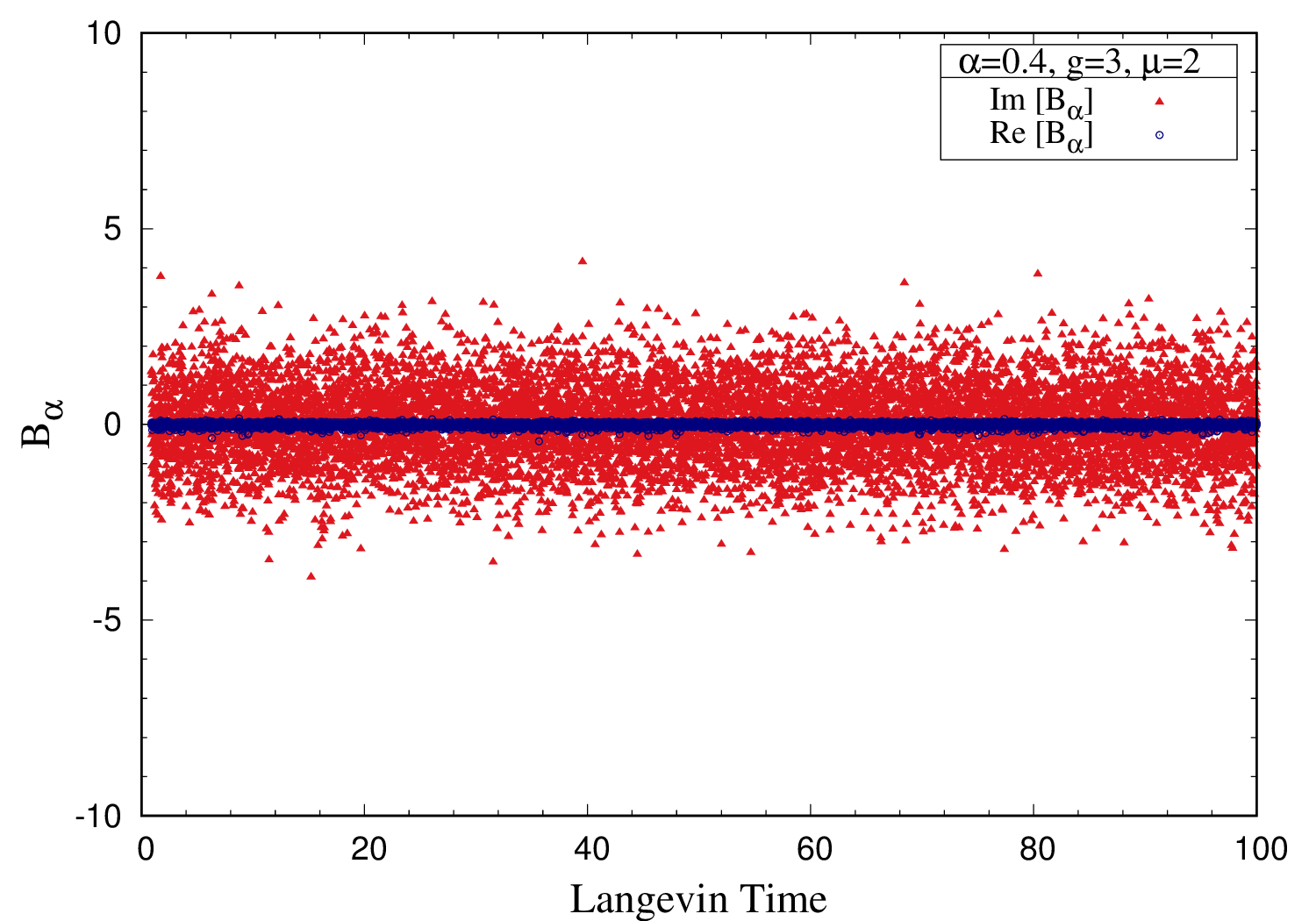}}
	
	\caption[Langevin time history of $\B$ for regularization parameter $\alpha = 0.4$. Simulations were performed for complex superpotential $W' = ig \ (\phi^2 + \mu^2)$ with $\mu = 2$.]{Langevin time history of $\B$ for regularization parameter $\alpha = 0.4$. Simulations were performed for complex superpotential $W' = ig \ (\phi^2 + \mu^2)$ with $\mu = 2$. In these simulations, we have used adaptive Langevin step size $\Delta \tau \leq 10^{-4}$, thermalization steps $N_{\rm therm} = 10^{4}$, generation steps $N_{\rm gen} = 10^6$, and measurements were taken every $100$ step. (Left) $g=1$. (Right) $g=3$.}
	\label{fig:isqw-g1-g3-p0_mu2p0}
	
\end{figure*}

\begin{figure*}[htp]
	
	{\includegraphics[width=.49\textwidth,origin=c,angle=0]{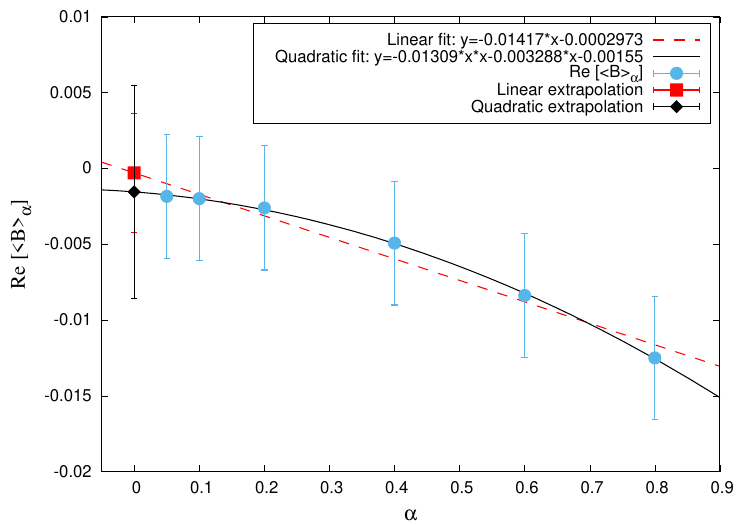}}
	{\includegraphics[width=.49\textwidth,origin=c,angle=0]{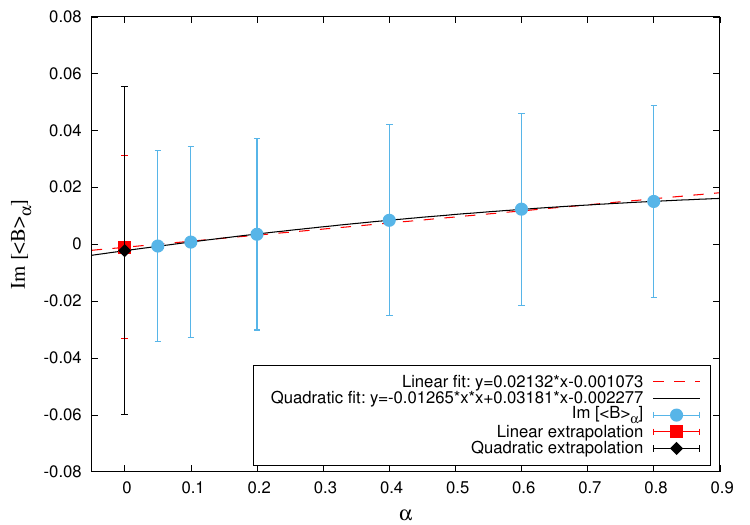}}
	
	\caption[Plot of real (Left) and imaginary (Right) parts of $\langle \B \rangle_\alpha$ against the regularization parameter, $\alpha$ for supersymmetric potential $W' = ig \ (\phi^2 + \mu^2)$. The simulations were performed with parameters $g = 1$ and $\mu = 2$. ]{Plot of real (Left) and imaginary (Right) parts of $\langle \B \rangle_\alpha$ against the regularization parameter, $\alpha$ for supersymmetric potential $W' = ig \ (\phi^2 + \mu^2)$. The simulations were performed with parameters $g = 1$ and $\mu = 2$. We have used adaptive Langevin step size $\Delta \tau \leq 10^{-4}$, thermalization steps $N_{\rm therm} = 10^{4}$, generation steps $N_{\rm gen} = 10^6$, and measurements were taken every $100$ steps. The dashed red lines are the linear fits to $\langle \B \rangle_\alpha$ in $\alpha$, and filled red squares are the linear extrapolation values at $\alpha = 0$. The solid black lines represent the quadratic fits to $\langle \B \rangle_\alpha$ in $\alpha$, and filled black diamonds are the quadratic extrapolation values at $\alpha = 0$. The $\alpha \to 0$ limit values obtained from these plots are given in Table \ref{tab:isqw_mu2p0}.}
	\label{fig:isqw_fit_g1p0_mu2p0}
	
\end{figure*}

\begin{figure*}[htp]
	
	{\includegraphics[width=.49\textwidth,origin=c,angle=0]{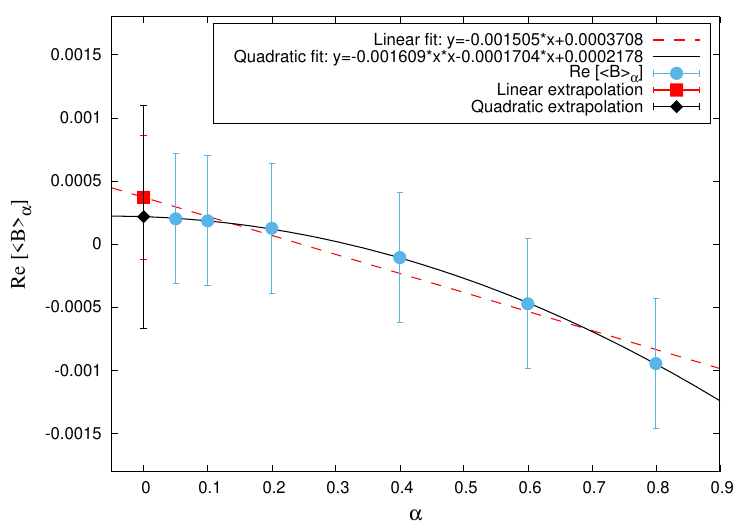}}
	{\includegraphics[width=.49\textwidth,origin=c,angle=0]{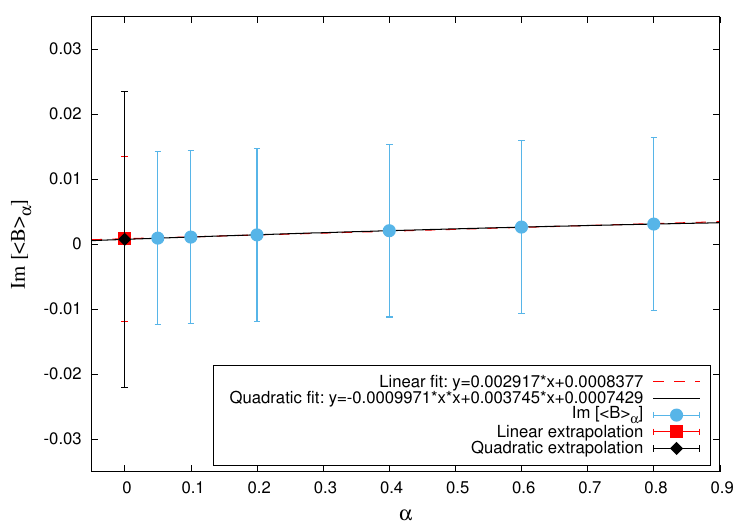}}
	
	\caption[The real (Left) and imaginary (Right) parts of $\langle \B \rangle_\alpha$ against the regularization parameter $\alpha$ for supersymmetric potential $W' = ig \ (\phi^2 + \mu^2)$.]{Real (Left) and imaginary (Right) parts of $\langle \B \rangle_\alpha$ against the regularization parameter $\alpha$ for supersymmetric potential $W' = ig \ (\phi^2 + \mu^2)$. The simulations were performed with $g = 3$ and $\mu = 2$. We have used adaptive Langevin step size $\Delta \tau \leq 10^{-4}$, thermalization steps $N_{\rm therm} = 10^{4}$, generation steps $N_{\rm gen} = 10^6$, and measurements were taken every $100$ steps. The dashed red lines are the linear fits to $\langle \B \rangle_\alpha$ in $\alpha$, and filled red squares are the linear extrapolation values at $\alpha = 0$. The solid black lines represent the quadratic fits to $\langle \B \rangle_\alpha$ in $\alpha$, and filled black diamonds are the quadratic extrapolation values at $\alpha = 0$. The $\alpha \to 0$ limit values obtained from these plots are given in Table \ref{tab:isqw_mu2p0}.}
	\label{fig:isqw_fit_g3p0_mu2p0}
	
\end{figure*}

We also consider the case when the derivative of the superpotential is complex,
\beq
W' = ig \ (\phi^2 + \mu^2),
\eeq
where $g$ and $\mu$ are again two parameters in the theory. We show the Langevin time history of the auxiliary $\B$ field and linear and quadratic extrapolations to $\alpha \to 0$ limit in Figs. \ref{fig:isqw-g1-g3-p0_mu2p0}, \ref{fig:isqw_fit_g1p0_mu2p0} and \ref{fig:isqw_fit_g3p0_mu2p0}, respectively. The results are tabulated in Table \ref{tab:isqw_mu2p0}. We have successfully simulated the complex double-well superpotential using complex Langevin, and our results strongly suggest that SUSY is preserved for this model.

\begin{figure*}[htp]
	
	{\includegraphics[width=.49\textwidth,origin=c,angle=0]{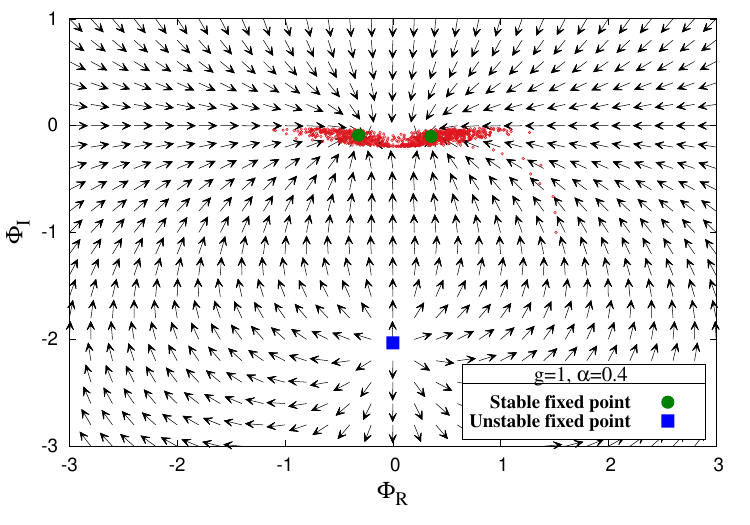}}
	{\includegraphics[width=.49\textwidth,origin=c,angle=0]{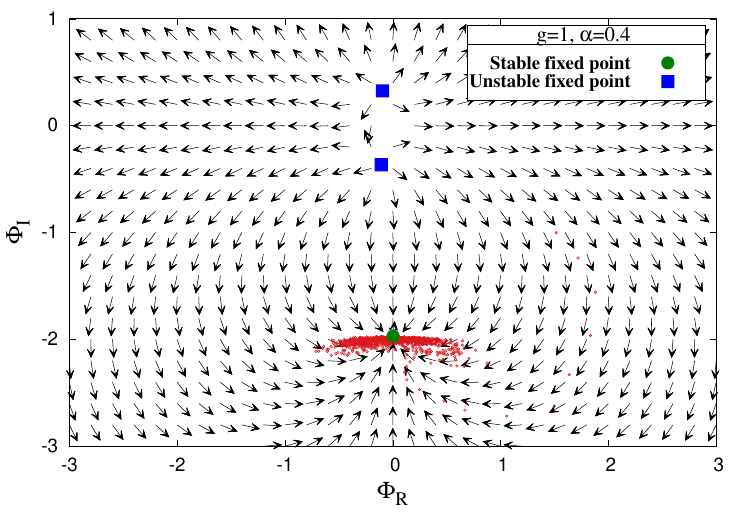}}
	
	\caption[Scatter plot of field configurations (red dots) and classical flow diagram (arrows) on the $\phi_R - \phi_I$ plane. The red dots represent trajectories of the fields during Langevin evolution for superpotential (Left) $W' = g (\phi^2 + \mu^2)$ and (Right) $W' = i g (\phi^2 + \mu^2)$.]{Scatter plot of field configurations (red dots) and classical flow diagram (arrows) on the $\phi_R - \phi_I$ plane. The red dots represent trajectories of the fields during Langevin evolution for superpotential (Left) $W' = g (\phi^2 + \mu^2)$ and (Right) $W' = i g (\phi^2 + \mu^2)$. In these simulations, we have used $g = 1.0, ~\mu = 2.0$ and $\alpha = 0.4$. The first $10^5$ points are plotted with measurements taken every $10^2$ step. In both cases, the field starts at point $(1.5, -1.0)$, and with the aid of a stochastic noise, it drifts towards equilibrium configuration. Filled circles and squares represent the stable and unstable fixed points, respectively.}
	\label{fig:dw_flow}
\end{figure*}

The results mentioned above can be partly motivated by classical dynamics, that is, in the absence of stochastic noise. In Fig. \ref{fig:dw_flow}, we show the classical flow diagrams on the $\phi_R-\phi_I$ plane for the above-discussed double-well models. The arrows indicate the normalized drift term evaluated at the particular field point. In the same figure, we have also shown the scatter plot of complexified field. These plots demonstrate how equilibrium configurations are attained during complex Langevin dynamics.

\begin{table*}[htp]
	\centering
{\footnotesize	\begin{tabular}{| c | c | c | c  | c |} 
		\hline
		$W'$& $~~\mu~~$  & $~~g~~$ & $~~~~~\alpha~~~~~$ & $~~~~~~\langle \B \rangle |_{\alpha}~~~~~~$ \\ [1.5ex] \hline\hline
		\multirow{10}{*}{$ig\Big(\phi^2 + \mu^2\Big)$} &\multirow{10}{*}{$2.0$} & \multirow{5}{*}{$1.0$}
		
		&  	0.05	 &	 $ -0.0018 (41) - i  0.0006 (337)$ \\ \cline{4-5}
		&&		&  	0.1		 &	 $ -0.0020 (41) + i  0.0008 (337)$ \\ \cline{4-5}
		&&		&  	0.2		 &	 $ -0.0026 (41) + i  0.0035 (336)$ \\ \cline{4-5}
		&&	 	&  	0.4		 &	 $ -0.0049 (41) + i  0.0084 (336)$  \\ \cline{4-5}
		&&	 	&  	0.6	 	 &	 $ -0.0084 (40) + i  0.0123 (338)$  \\ \cline{4-5}
		&&	 	&  	0.8		 &	 $ -0.0125 (40) + i  0.0150 (337)$   \\ \cline{4-5} 
		&&	&  	$\alpha \rightarrow 0$ &	 $ -0.0009(70) - i 0.0017(576)$  \\ \cline{3-5} 
		&&	\multirow{5}{*}{$3.0$}	
		&  	0.05	 &	 $ 0.0002(5)   + i  0.0009 (133)$ \\ \cline{4-5}
		&&		&  	0.1		 &	 $ 0.0002(5)   + i  0.0011 (133)$ \\ \cline{4-5}
		&&		&  	0.2		 &	 $ 0.0001(5)   + i  0.0014 (133)$	\\ \cline{4-5}
		&&	 	&  	0.4		 &	 $ -0.0001 (5) + i  0.0021 (133) $   \\ \cline{4-5}
		&&	 	&  	0.6		 &	 $ -0.0005 (5) + i  0.0026 (133)$  \\ \cline{4-5}
		&&	 	&  	0.8		 &	 $ -0.0009 (5) + i  0.0031 (133) $ \\ \cline{4-5}
		&&	&  	$\alpha \rightarrow 0$ &	 $ 0.0003(9) + i 0.0008(227)$ \\ \hline
	\end{tabular}}
	\caption[Expectation values $\langle \B \rangle_\alpha$ obtained using complex Langevin simulations for the model with superpotential $W' = ig\Big(\phi^2 + \mu^2\Big)$.]{Expectation values $\langle \B \rangle_\alpha$ obtained using complex Langevin simulations for the model with superpotential $W' = ig\Big(\phi^2 + \mu^2\Big)$. We see that, in the limit $\alpha \to 0, \langle \B \rangle_\alpha = 0 $. Thus SUSY is preserved in this model.}
	\label{tab:isqw_mu2p0}
\end{table*}

\subsection{General polynomial potential}
\label{subsec:gen-sps-0d}

Let us extend our analyses to the case where the derivative of superpotential, $W'$, is a general polynomial of degree $k$,
\beq
W' = g_k \phi^k + g_{k-1} \phi^{k-1} + \cdots + g_0.
\eeq

The twisted partition function can be written as
\bea
\label{eq:gen-poly-Z}
Z_\alpha &=& -\frac{1}{ \sqrt{2 \pi} } \int_{-\infty}^\infty d\phi ~ \Big( e^{i \alpha} -1 + W'' \Big) \exp \left[ - \hf W'^{2} \right] \nn \\
&=& -\frac{(e^{i \alpha} -1)}{ \sqrt{2 \pi} } \int_{-\infty}^\infty d\phi ~  \exp \left[ - \hf W'^{2} \right]-\frac{1}{ \sqrt{2 \pi} } \int_{-\infty}^\infty d\phi ~  W'' \exp \left[ - \hf W'^{2} \right], 
\eea
where for the second term in the above equation, assuming the coefficients of the polynomial potential to be real, we have
\beq
\frac{1}{\sqrt{2 \pi}} \int_{-\infty}^\infty W'' \exp\left[- \hf W'^2\right] = 
\begin{cases}
	{\rm sgn}(g_k)      & \ k: \text{ odd} \\
	0 & \ k: \text{ even}.
\end{cases}
\eeq
Upon turning off the external field, the first term of Eq. \eqref{eq:gen-poly-Z} vanishes, hence 
\beq
\label{eq:Z-susy-poly-cond}
Z_\alpha \big|_{\alpha \to 0} = 
\begin{cases}
	{-\rm sgn}(g_k)      & \ k: \text{ odd} \\
	0 & \ k: \text{ even}.
\end{cases}
\eeq
The above calculations imply that, for a general polynomial superpotential, $W'$ of the degree even (odd), the SUSY is broken (preserved).

The expectation value of the auxiliary $\B$ field reads
\bea
\label{eq:B_poly}
\langle \B \rangle_\alpha &=& - \frac{1}{ Z_\alpha\sqrt{2 \pi} } \int_{-\infty}^\infty d\phi ~ (-iW')\Big( e^{i \alpha} -1 + W'' \Big) \exp \left[ -\hf W'^{2} \right] \nn \\
&=&\frac{i(e^{i \alpha} -1)}{Z_\alpha \sqrt{2 \pi} } \int_{-\infty}^\infty d\phi ~W'  \exp \left[ - \hf W'^{2} \right] \nn \\
&&~~~~~~~~~~~~~~~+ \frac{i}{Z_\alpha  \sqrt{2 \pi} } \int_{-\infty}^\infty d\phi ~ W'~ W'' \exp \left[ - \hf W'^{2} \right],
\eea
where the second term of Eq. \eqref{eq:B_poly} vanishes for a polynomial superpotential. Since we have twisted partition function in the denominator, this term is not indefinite. Now, we are left with
\beq
\langle \B \rangle_\alpha  = 
\begin{cases}
	{\frac{i{(e^{i \alpha} -1)} \int_{-\infty}^\infty d\phi ~W'  \exp\left[ - \hf W'^{2} \right] }{-{(e^{i \alpha} -1)} \int_{-\infty}^\infty d\phi ~ \exp\left[ - \hf W'^{2} \right]~ - ~ { \sqrt{2 \pi} }~{\rm sgn}(g_k) }}  & k: \text{ odd}\\ \\
	\frac{-i\int_{-\infty}^\infty d\phi ~W'  \exp\left[ - \hf W'^{2} \right]}{\int_{-\infty}^\infty d\phi ~ \exp\left[ - \hf W'^{2} \right]} & k: \text{ even}
\end{cases} 
\eeq
and turning external field off, $\alpha \to 0$, gives
\bea
\label{eq:B-poly-susy-cond}
\langle \B \rangle_\alpha \Big|_{\alpha \to 0} =
\begin{cases}
	~~~~0    & k: \text{ odd} \\
	\frac{-i\int_{-\infty}^\infty d\phi ~W'  \exp\left[ - \hf W'^{2} \right]}{\int_{-\infty}^\infty d\phi ~\exp\left[ - \hf W'^{2} \right]} \neq 0 & k: \text{ even}.
\end{cases}
\eea
The above expressions confirm that SUSY is preserved (broken) for the odd (even) degree of the derivative of a real general polynomial superpotential.

\begin{figure*}[htp]
	
	{\includegraphics[width=.49\textwidth,origin=c,angle=0]{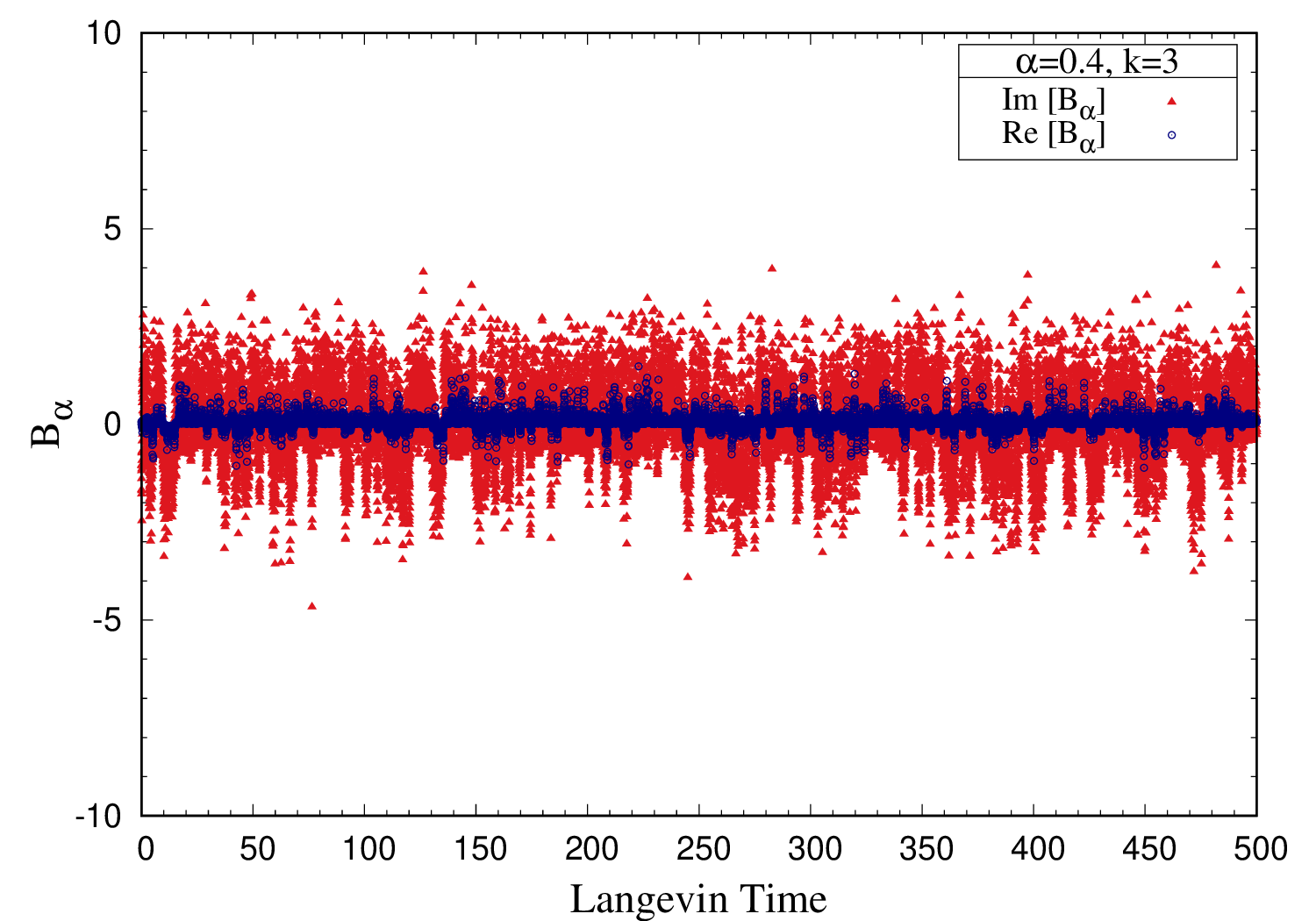}}
	{\includegraphics[width=.49\textwidth,origin=c,angle=0]{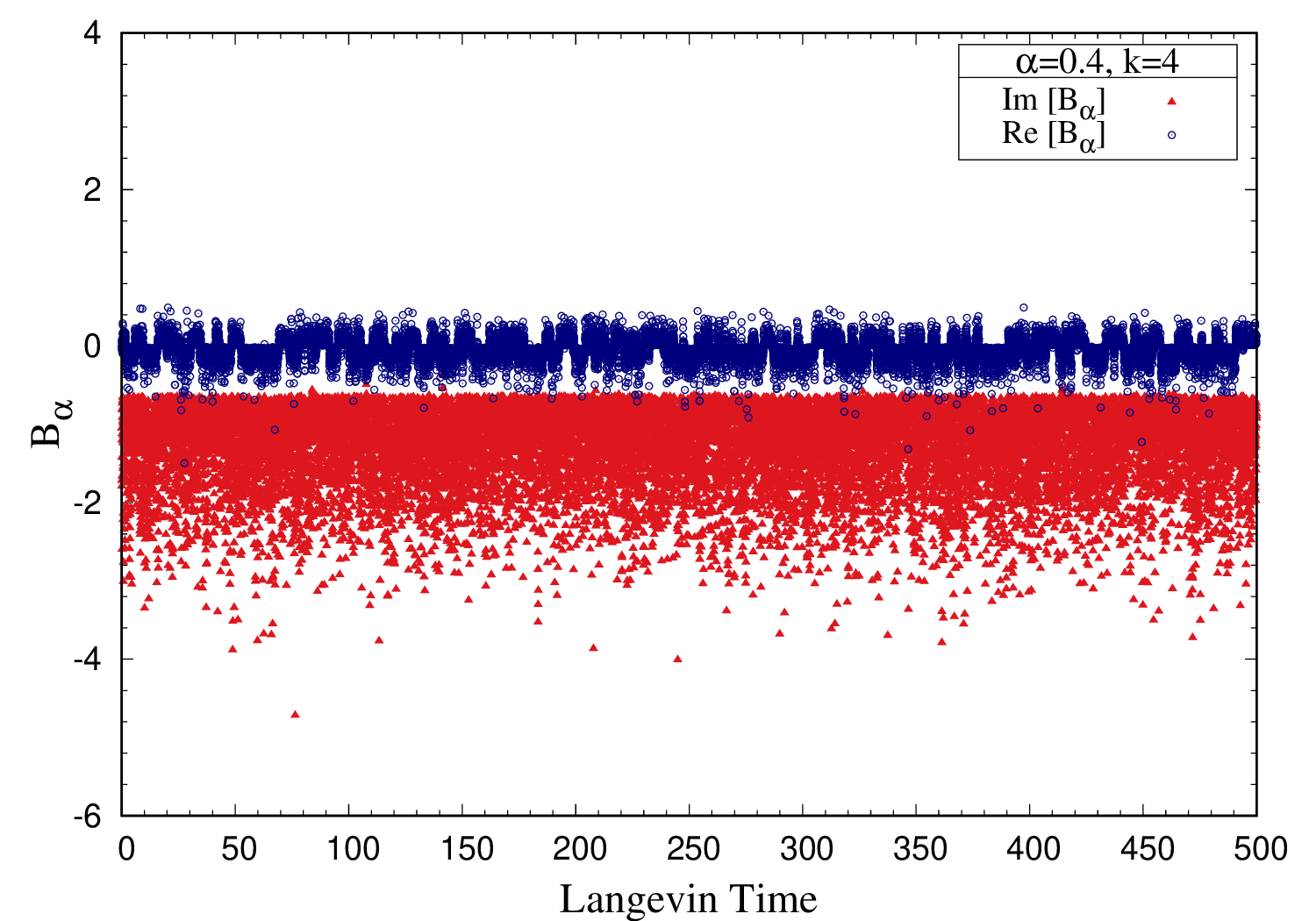}}
	
	\caption[Langevin time history of the field $\B$ for $\alpha = 0.4$. Simulations were performed for superpotential $W'(\phi) = g_k \phi^k + g_{k-1} \phi^{k-1} + \cdots + g_0$ with $g_k = g_{k-1} = \cdots = g_0 = 1$.]{Langevin time history of the field $\B$ for $\alpha = 0.4$. Simulations were performed for superpotential $W'(\phi) = g_k \phi^k + g_{k-1} \phi^{k-1} + \cdots + g_0$ with $g_k = g_{k-1} = \cdots = g_0 = 1$. In these simulations, we have used adaptive Langevin step size $\Delta \tau \leq 5 \times 10^{-5}$, generation steps $N_{\rm gen} = 10^7$, and measurements were taken every $500$ step. (Left) $k = 3$ case. (Right) $k = 4$ case.}
	\label{fig:B-poly-history}
	
\end{figure*}

\begin{figure*}[htp]
	
	{\includegraphics[width=3in]{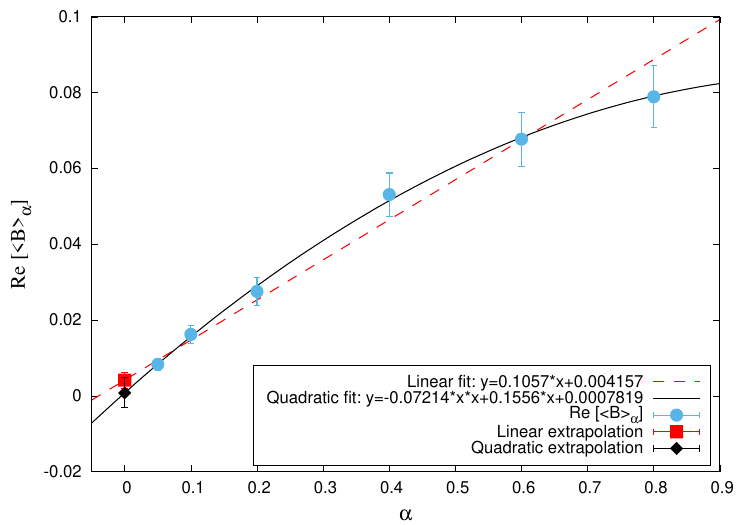}}
	{\includegraphics[width=3in]{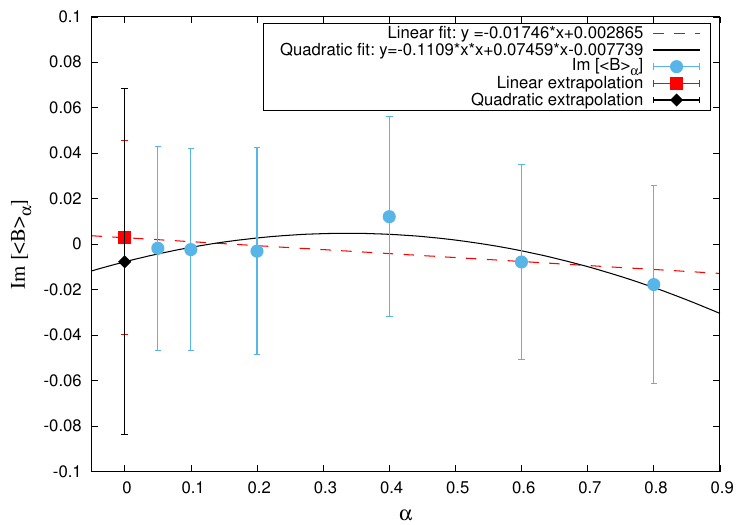}}
	
	{\includegraphics[width=.49\textwidth,origin=c,angle=0]{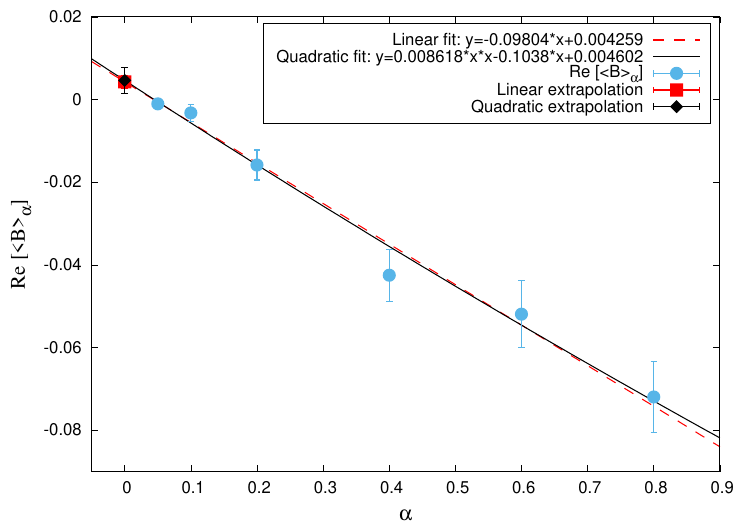}}
	{\includegraphics[width=.49\textwidth,origin=c,angle=0]{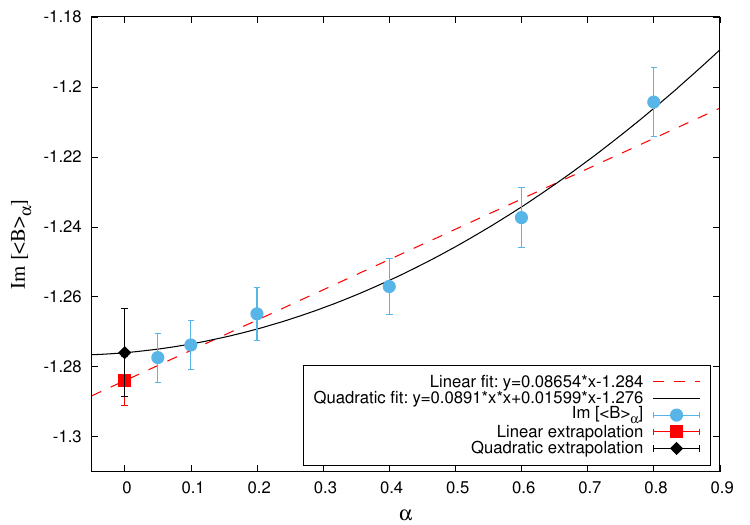}}
	
	\caption[Expectation value, $\langle \B \rangle_\alpha$ against the regularization parameter, $\alpha$ for superpotential $W'(\phi) = g_k \phi^k + g_{k-1} \phi^{k-1} + \cdots + g_0$ with $g_k = g_{k-1} = \cdots = g_0 = 1$.]{Expectation value, $\langle \B \rangle_\alpha$ against the regularization parameter, $\alpha$ for superpotential $W'(\phi) = g_k \phi^k + g_{k-1} \phi^{k-1} + \cdots + g_0$ with $g_k = g_{k-1} = \cdots = g_0 = 1$. (Top-Left) Real part and (Top-Right) imaginary part of $\langle \B \rangle_\alpha$ for $k = 3$. (Bottom-Left) Real part and (Bottom-Right) imaginary part of $\langle \B \rangle_\alpha$ for $k = 4$. The simulations were performed with adaptive Langevin step size $\Delta \tau \leq 5\times10^{-5}$, thermalization steps $N_{\rm therm} = 5 \times 10^4$, generation steps $N_{\rm gen} = 10^7$, and measurements were taken every 500 step. The dashed red lines are the linear fits to $\langle \B \rangle_\alpha$ in $\alpha$, and red dots are the linear extrapolation value at $\alpha = 0$. The solid black lines represent the quadratic fits to $\langle \B \rangle_\alpha$ in $\alpha$, and black dots are the quadratic extrapolation value at $\alpha = 0$. The $\alpha \to 0$ limit values obtained from these plots are given in Table \ref{tab:real-gen-poly}.}
	\label{fig:B_fit_poly}
\end{figure*}

Let us consider polynomial superpotential with real coefficients. In this case, the above argument for SUSY breaking is valid. Later, we will also discuss a specific case of complex polynomial potential. For simplicity we assume  that $g_k = g_{k-1} = \cdots = g_0 = 1$, then for $k = 3, 4$ we have
\beq
W'[k=3] = \phi^3 + \phi^2 + \phi +1,
\eeq
and
\beq
W'[k=4] = \phi^4 + \phi^3 + \phi^2 + \phi +1.
\eeq
We have learned from Eqs. \eqref{eq:Z-susy-poly-cond} and \eqref{eq:B-poly-susy-cond} that SUSY is broken (preserved) for $k = 4 ~ (k = 3)$. In Fig. \ref{fig:B-poly-history}, we show the Langevin time history of $\langle \B \rangle_\alpha$ for the above two polynomial models. We show linear and quadratic extrapolations to $\alpha \to 0$ limit in Fig. \ref{fig:B_fit_poly}. The results are tabulated in Table \ref{tab:real-gen-poly}. The simulation results are in good agreement with the corresponding analytical predictions.

\begin{table*}[htp]
	\centering
{\footnotesize	\begin{tabular}{| c | c  | c | c |} 
		\hline
		$~~~~k~~~~$& $~~~~~\alpha~~~~~$ & $~~~~~~~$ $~~~~~~\langle \B \rangle |_{\alpha}~~~~~~$ $~~~~~~~~~$ & $~~~~$ SUSY $~~~~$ \\ [1.5ex] \hline\hline
		\multirow{6}{*}{$ 3 $}	
		&  	0.05	 &	 $ 0.0083 (15) - i  0.0018 (447)$  & \multirow{6}{*}{Preserved} \\ \cline{2-3}
		&  	0.1		 &	 $ 0.0162 (24) - i  0.0023 (443)$  &\\ \cline{2-3}
		&  	0.2		 &	 $ 0.0275 (37) - i  0.0030 (454)$  &\\ \cline{2-3}
		&  	0.4		 &	 $ 0.0531 (57) + i  0.0121 (440)$  &\\ \cline{2-3}
		&  	0.6	 	 &	 $ 0.0677 (71) - i  0.0078 (428)$  &\\ \cline{2-3}
		&  	0.8		 &	 $ 0.0789 (82) - i  0.0177 (437)$  & \\ \cline{2-3}
		&  	$\alpha \rightarrow 0$ &	 $ 0.0025(40) - i 0.0024(761) $  &\\ \cline{1-4} 
		\multirow{6}{*}{$4 $}	
		&  	0.05	 &	 $ -0.0010 (10) - i  1.2774 (70)$ & \multirow{6}{*}{Broken}\\ \cline{2-3}
		&  	0.1		 &	 $ -0.0032 (20) - i  1.2738 (71)$ &\\ \cline{2-3}
		&  	0.2		 &	 $ -0.0158 (36) - i  1.2649 (76)$ &\\ \cline{2-3}
		&  	0.4		 &	 $ -0.0425 (62) - i  1.2571 (80)$ & \\ \cline{2-3}
		&  	0.6	 	 &	 $ -0.0519 (81) - i  1.2373 (86)$ & \\ \cline{2-3}
		&  	0.8		 &	 $ -0.0719 (85) - i  1.2044 (98)$ &  \\ \cline{2-3}
		&  	$\alpha \rightarrow 0$ &	 $ 0.0044(31) - i 1.2800(126) $  &\\ \hline
	\end{tabular}}
	\caption[Expectation values $\langle \B \rangle_\alpha$ obtained using complex Langevin simulations for the models with superpotentials $W' = g_k \phi^k + g_{k-1} \phi^{k-1} + \cdots + g_0$ with $g_k = g_{k-1} = \cdots = g_0 = 1$ and $k = 3, 4$.]{\label{tab:real-gen-poly}Expectation values $\langle \B \rangle_\alpha$ obtained using complex Langevin simulations for the models with superpotentials $W' = g_k \phi^k + g_{k-1} \phi^{k-1} + \cdots + g_0$ with $g_k = g_{k-1} = \cdots = g_0 = 1$, and $k = 3, 4$.}
\end{table*}

\begin{figure*}[htp]
	
	{\includegraphics[width=.49\textwidth,origin=c,angle=0]{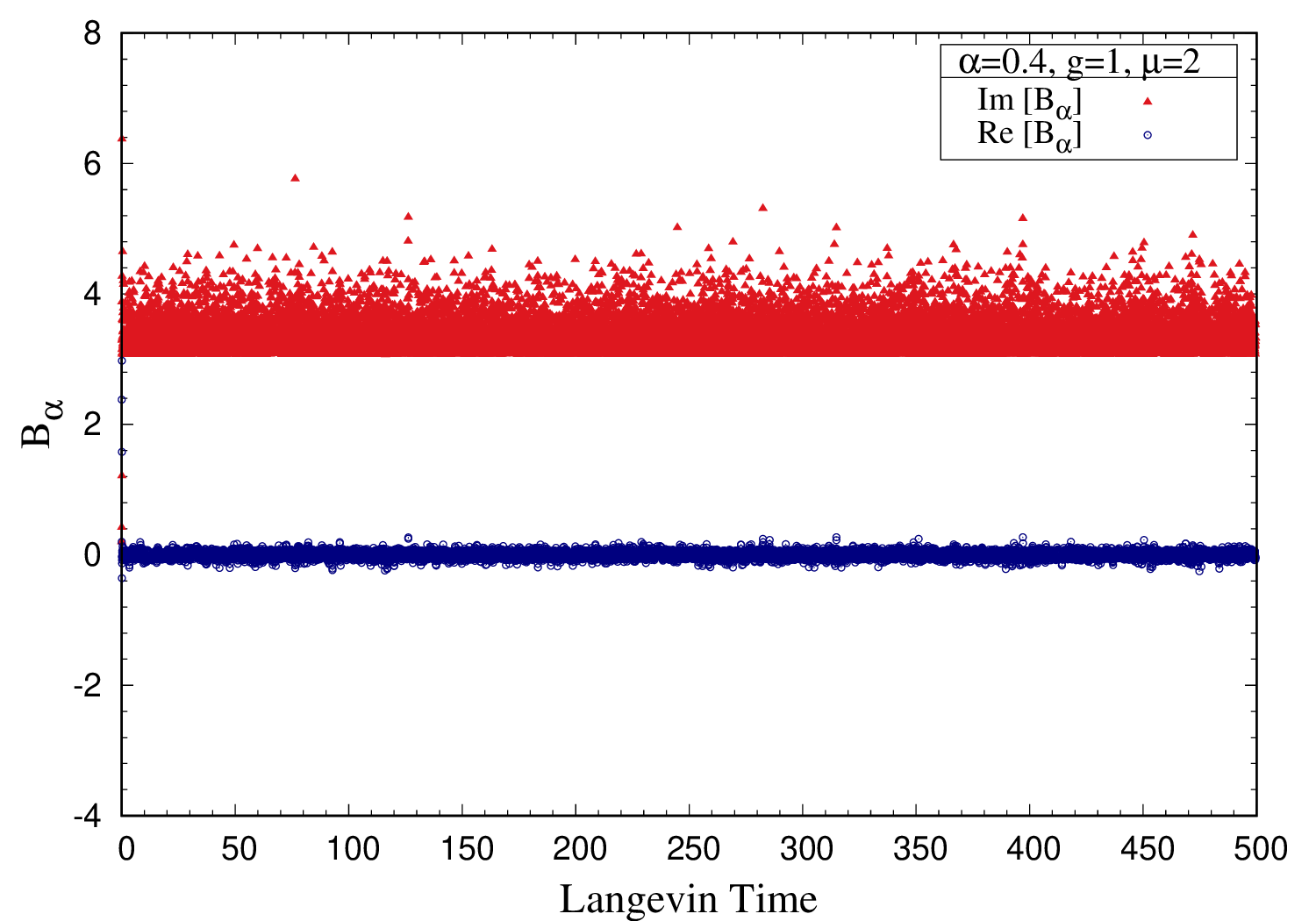}}
	{\includegraphics[width=.49\textwidth,origin=c,angle=0]{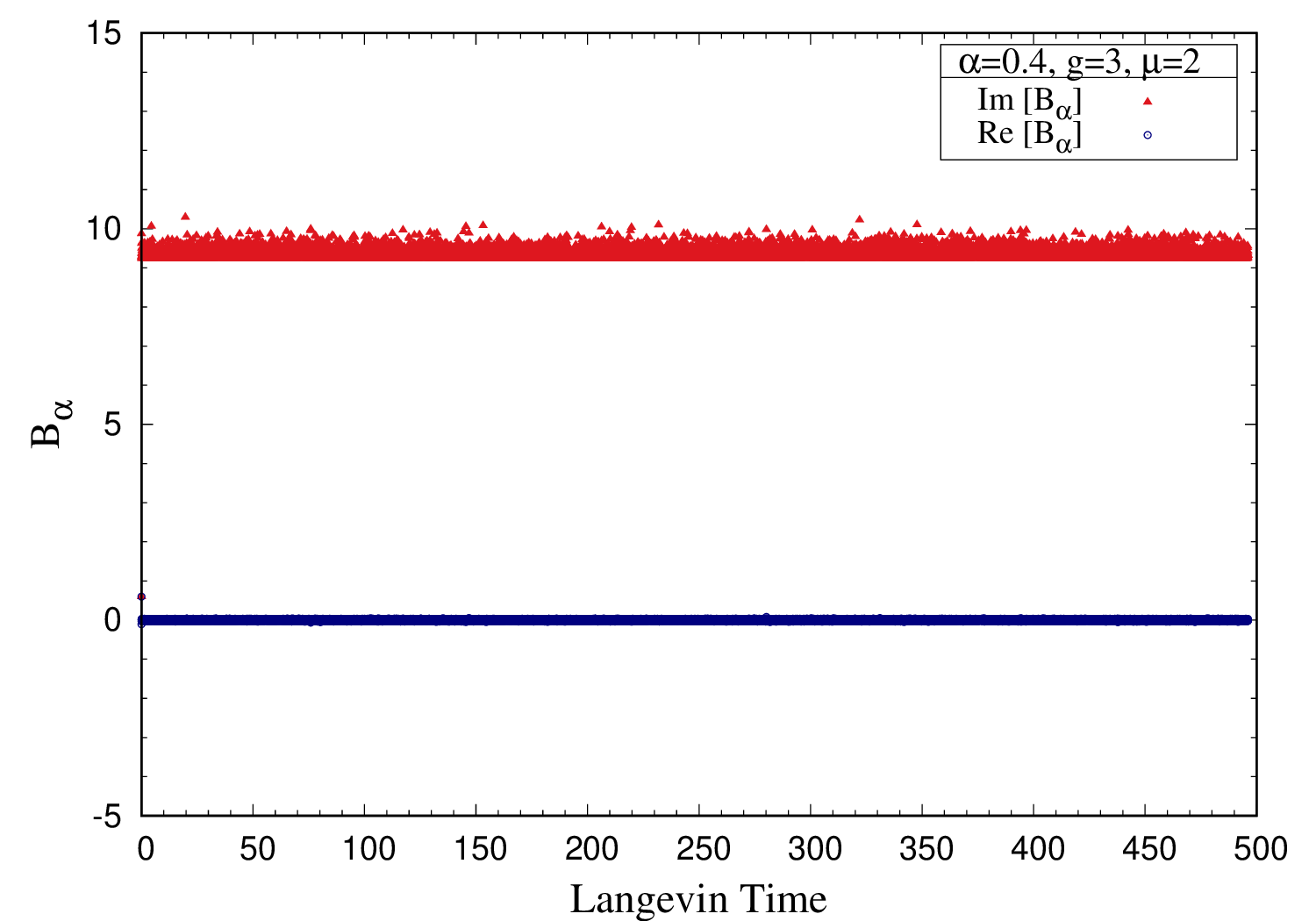}}
	
	\caption[Langevin time history of $\B$ for $\alpha = 0.4$. The simulations were performed for superpotential $W'(\phi) =  ig\phi (\phi^2 +\mu^2) $ with $\mu=2$.]{Langevin time history of $\B$ for $\alpha = 0.4$. The simulations were performed for superpotential $W'(\phi) =  ig\phi (\phi^2 +\mu^2) $ with $\mu = 2$. In these simulations, we have used adaptive Langevin step size $\Delta \tau \leq 5\times10^{-5}$, generation steps $N_{\rm gen} = 10^7$, and measurements were taken every $500$ step. (Left) $g = 1$ case. (Right) $g = 3$ case. }
	\label{fig:sqw_iphi-g1-g3-p0_mu2p0}
	
\end{figure*}

\begin{figure*}[htp]
	
	{\includegraphics[width=3in]{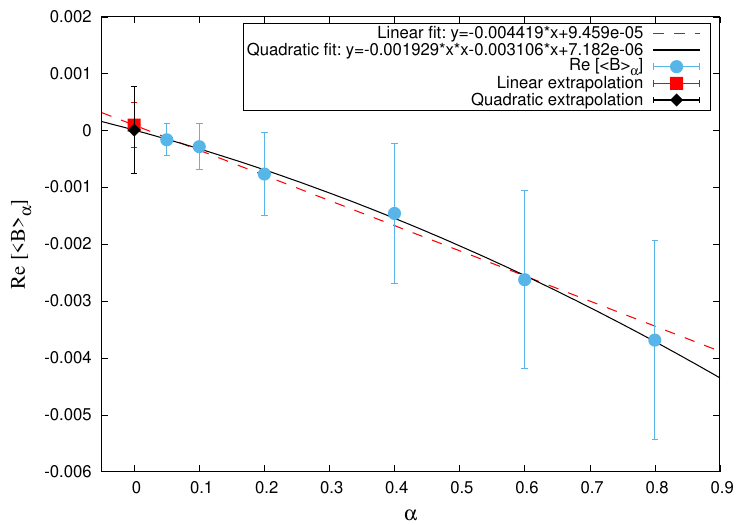}}
	{\includegraphics[width=3in]{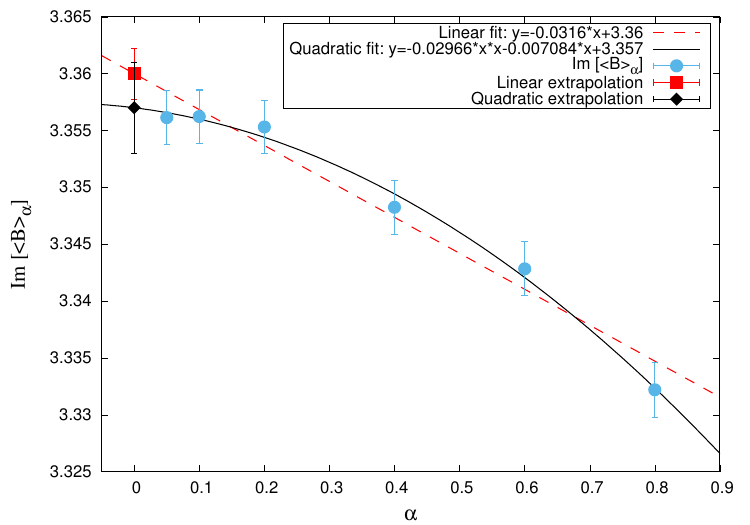}}
	
	\caption[The real (Left) and imaginary (Right) parts of $\langle \B \rangle_\alpha$ against the regularization parameter, $\alpha$ for supersymmetric potential $W' = ig\phi \ (\phi^2 + \mu^2)$.]{Real (Left) and imaginary (Right) parts of $\langle \B \rangle_\alpha$ against the regularization parameter, $\alpha$ for supersymmetric potential $W' = ig\phi \ (\phi^2 + \mu^2)$. Simulations were performed with $g = 1$ and $\mu = 2$. We have used adaptive Langevin step size $\Delta \tau \leq 5\times10^{-5}$, thermalization steps $N_{\rm therm} =  5 \times 10^{4}$, generation steps $N_{\rm gen} = 10^7$, and measurements were taken every $500$ steps. The dashed red lines are the linear fits to $\langle \B \rangle_\alpha$ in $\alpha$, and red dots are the linear extrapolation value at $\alpha = 0$. The solid black lines represent the quadratic fits to $\langle \B \rangle_\alpha$ in $\alpha$, and black dots are the quadratic extrapolation value at $\alpha = 0$. The $\alpha \to 0$ limit values obtained from these plots are given in Table \ref{tab:sqw_iphi_mu2p0}.}
	\label{fig:sqw_iphi_fit_g1p0_mu2p0}
	
\end{figure*}

\begin{figure*}[htp]
	
	{\includegraphics[width=.49\textwidth,origin=c,angle=0]{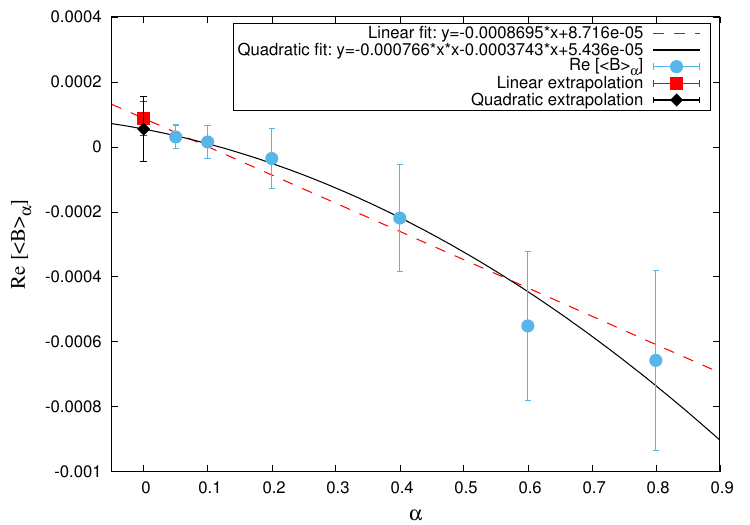}}
	{\includegraphics[width=.49\textwidth,origin=c,angle=0]{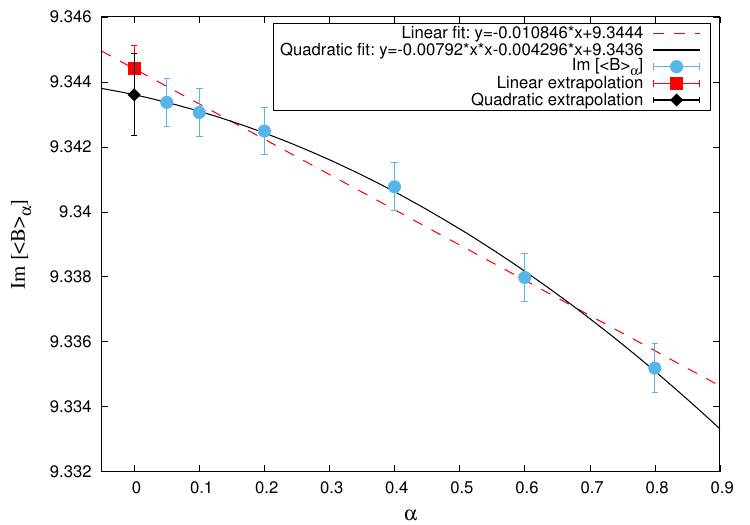}}
	
	\caption[The real (Left) and imaginary (Right) parts of $\langle \B \rangle_\alpha$ against the regularization parameter, $\alpha$ for supersymmetric potential $W' = ig\phi \ (\phi^2 + \mu^2)$.]{Real (Left) and imaginary (Right) parts of $\langle \B \rangle_\alpha$ against the regularization parameter, $\alpha$ for supersymmetric potential $W' = ig\phi \ (\phi^2 + \mu^2)$. Simulations were performed with $g = 3$ and $\mu = 2$. We have used adaptive Langevin step size $\Delta \tau \leq 5\times10^{-5}$, thermalization steps $N_{\rm therm} =  5 \times 10^{4}$, generation steps $N_{\rm gen} = 10^7$, and measurements were taken every $500$ steps. The dashed red lines are the linear fits to $\langle \B \rangle_\alpha$ in $\alpha$, and red dots are the linear extrapolation value at $\alpha = 0$. The solid black lines represent the quadratic fits to $\langle \B \rangle_\alpha$ in $\alpha$, and black dots are the quadratic extrapolation value at $\alpha = 0$. The $\alpha \to 0$ limit values obtained from these plots are given in Table  \ref{tab:sqw_iphi_mu2p0}.}
	\label{fig:sqw_iphi_fit_g3p0_mu2p0}
	
\end{figure*}

Now, let us consider the case with complex polynomial superpotential. We modify the real double-well potential discussed in the previous section as follows,
\beq
W' = ig\phi (\phi^2 +\mu^2),
\eeq
and since it is a complex potential, the argument given in Eqs. \eqref{eq:Z-susy-poly-cond} and \eqref{eq:B-poly-susy-cond} are not valid. We investigate SUSY breaking using complex Langevin dynamics. In Fig. \ref{fig:sqw_iphi-g1-g3-p0_mu2p0}, we show the Langevin time history of the auxiliary $\B$ field for regularization parameter, $\alpha = 0.4$. We show linear and quadratic extrapolations to $\alpha \to 0$ limit in Figs. \ref{fig:sqw_iphi_fit_g1p0_mu2p0} and \ref{fig:sqw_iphi_fit_g3p0_mu2p0}. The results are tabulated in Table \ref{tab:sqw_iphi_mu2p0}. Our simulation results imply that expectation value of the auxiliary field, $\langle \B \rangle_\alpha$ does not vanish in the limit, $\alpha \to 0$. Hence SUSY is broken in this model.

\begin{table*}[htp]
	\centering
{\footnotesize	\begin{tabular}{| c | c  | c | c | c |} 
		\hline
		$W'$& $~~\mu~~$  & $~~g~~$ & $~~~~~\alpha~~~~~$ & $~~~~~~\langle \B \rangle |_{\alpha}~~~~~~$ \\ [1.5ex] \hline\hline
		\multirow{10}{*}{$ig\phi \Big(\phi^2 + \mu^2\Big)$} &\multirow{10}{*}{$2.0$} & \multirow{5}{*}{$1.0$}
		&  	0.05	 &	 $-0.0002(3)  + i 3.3561(23) $ \\ \cline{4-5}
		&&		&  	0.1		 &	 $-0.0003(4)  + i 3.3562(23) $ \\ \cline{4-5}
		&&		&  	0.2		 &	 $-0.0008(7)  + i 3.3553(23) $ \\ \cline{4-5}
		&&	 	&  	0.4		 &	 $-0.0015(12) + i 3.3482(24) $  \\ \cline{4-5}
		&&	 	&  	0.6	 	 &	 $-0.0026(15) + i 3.3428(24) $  \\ \cline{4-5}
		&&	 	&  	0.8		 &	 $-0.0037(17) + i 3.3322(24) $   \\ \cline{4-5} 
		&&	&  	$\alpha \rightarrow 0$ &	 $ 0.0000(8) + i 3.3585(40)$  \\ \cline{3-5} 
		&&	\multirow{5}{*}{$3.0$}	
		&  	0.05	 &	 $ 0.0000(0)  + i 9.3434(7) $ \\ \cline{4-5}
		&&		&  	0.1		 &	 $ 0.0000(0)  + i 9.3430(7) $ \\ \cline{4-5}
		&&		&  	0.2		 &	 $-0.0000(0)  + i 9.3425(7) $ \\ \cline{4-5}
		&&	 	&  	0.4		 &	 $-0.0002(2)  + i 9.3408(7) $  \\ \cline{4-5}
		&&	 	&  	0.6	 	 &	 $-0.0005(2)  + i 9.3380(7) $  \\ \cline{4-5}
		&&	 	&  	0.8		 &	 $-0.0007(3)  + i 9.3352(8) $   \\ \cline{4-5} 
		&&	&  	$\alpha \rightarrow 0$ &	 $ 0.0000(1) + i 9.3440(13)$  \\ \hline
		
	\end{tabular}}
	\caption[Expectation values $\langle \B \rangle_\alpha$ obtained using complex Langevin simulations for the model with superpotential $W' = ig\phi \Big(\phi^2 + \mu^2\Big)$ with $g = 1, 3$ and $\mu = 2$.]{Expectation values $\langle \B \rangle_\alpha$ obtained using complex Langevin simulations for the model with superpotential $W' = ig\phi \Big(\phi^2 + \mu^2\Big)$ with $g = 1, 3$, and $\mu = 2$. We see that SUSY is broken in this model.}
	\label{tab:sqw_iphi_mu2p0}
\end{table*}

\subsection{$\mathcal{PT}$-symmetric models inspired $\delta$-potentials}
\label{subsec7}

Let us consider the superpotential
\beq
W(\phi) = - \frac{g}{(2 + \delta)} (i \phi)^{(2 + \delta)},
\eeq
which is the same as the one we considered earlier for the case of the bosonic models.

The twisted partition function takes the form
\bea
Z_\alpha &=& -\frac{1}{\sqrt{2 \pi}} \int_{-\infty}^\infty d\phi \ \Big( e^{i \alpha} - 1 + W'' \Big) \exp\left[ - \hf W'^2 \right]  \nn \\
&=& -\frac{1}{\sqrt{2 \pi}} \int_{-\infty}^\infty d\phi \ \Big( e^{i \alpha} - 1 + g (1 + \delta) (i \phi)^{\delta} \Big) \exp \left[ \hf g^2 (i \phi)^{2(1+\delta)} \right] .
\eea
The expectation value of the auxiliary field is
\bea
\langle \B \rangle_\alpha &=&- \frac{1}{Z_\alpha} \frac{1}{\sqrt{2 \pi}} \int_{-\infty}^\infty d\phi \ (-iW') \Big( e^{i \alpha} - 1 + W'' \Big) \exp\left[ - \hf W'^2 \right]  \nn \\
&=& \frac{1}{Z_\alpha} \frac{1}{\sqrt{2 \pi}} \int_{-\infty}^\infty d\phi \  g  (i \phi)^{1 + \delta} \Big( e^{i \alpha} - 1 + g (1 + \delta) (i\phi)^\delta \Big) \exp \left[ \hf g^2 (i\phi)^{2(1+\delta)} \right] .~~
\eea
Let us consider various integer cases of $\delta$ and check whether SUSY is broken or preserved in these cases. 

For the case, $\delta = 0$, one can easily perform analytical evaluations. The twisted partition function reads
\bea
Z_\alpha [\delta = 0] &=& -\frac{1}{\sqrt{2 \pi}} \int_{-\infty}^\infty d\phi \ \Big( e^{i \alpha} - 1 + g \Big) \ \exp \left[ - \hf g^2 \phi^2 \right]  \nn \\
&=& -\frac{1}{\sqrt{2 \pi}} \Big( e^{i \alpha} - 1 + g \Big) \sqrt{\frac{2 \pi}{g^2}},
\eea
and turning the external field off, $\alpha \to 0$, we get a non-zero value for the partition function, that is
\beq
Z_{\alpha=0} [\delta = 0] = -\frac{1}{\sqrt{2 \pi}} g \sqrt{\frac{2 \pi}{g^2}} = -1,
\eeq
implying that SUSY is preserved in the system.

Also, we have
\bea
\langle \B \rangle_\alpha   [\delta = 0] &=&  \frac{1}{Z_\alpha} \frac{1}{\sqrt{2 \pi}} \int_{-\infty}^\infty d\phi \ ig \phi \Big( e^{i \alpha} - 1 + g \Big) ~ \exp \left[ -\hf g^2 \phi^{2} \right] \nn \\
&=&-\frac{ ig \int_{-\infty}^\infty d\phi \ \phi \Big( e^{i \alpha} - 1 + g \Big) \exp \left[ - \hf g^2 \phi^2 \right] }{ \Big( e^{i \alpha} - 1 + g \Big) \sqrt{\frac{2 \pi}{g^2}}} \nn \\
&=& -\frac{i g \int_{-\infty}^\infty d\phi \ \phi \ \exp \left[ - \hf g^2 \phi^2 \right]}{ \sqrt{\frac{2 \pi}{g^2}}} = 0
\eea
implying that SUSY is preserved in the theory when $\delta = 0$.

For the case $\delta = 2$, the twisted partition function has the form
\bea
Z_\alpha [\delta = 2] &=&- \frac{1}{\sqrt{2 \pi}} \int_{-\infty}^\infty d\phi \ \Big( e^{i \alpha} - 1 - 3g  \phi^{2} \Big) \exp\left[ -\hf g^2 \phi^{6} \right] \nn \\
&=& -\frac{\Big( e^{i \alpha} - 1\Big)}{\sqrt{2 \pi}} \int_{-\infty}^\infty d\phi \ \exp\left[ -\hf g^2  \phi^{6} \right]  \nn \\
&& ~~~ ~~~ +  \frac{3g}{\sqrt{2 \pi}} \int_{-\infty}^\infty d\phi \ \phi^{2} \exp\left[ -\hf g^2  \phi^{6} \right],
\eea
where turning the external field off, $\alpha \to 0$, we get a non-zero partition function
\bea
Z_{\alpha = 0} [ \delta = 2 ] &=&  \frac{3g}{\sqrt{2 \pi}} \int_{-\infty}^\infty d\phi \ \phi^{2} \exp\left[ -\hf g^2  \phi^{6} \right] \nn \\
&=& 1,
\eea
indicating that SUSY is preserved in the system. The expectation value of the $\B$ field is
\bea
\langle \B \rangle_\alpha [\delta = 2] &=&   - \frac{ig}{Z_\alpha\sqrt{2 \pi}} \int_{-\infty}^\infty d\phi \ \phi^{3} \left( e^{i \alpha} - 1 - 3g \phi^2 \right) \exp\left[ -\hf g^2 \phi^{6} \right] \nn \\ 
&=&0,
\eea
confirming that SUSY is preserved for the case $\delta=2$. One can perform similar calculations for the case $\delta = 4$ and show that SUSY is preserved in the theory.

We simulate the $\delta$-potential using complex Langevin dynamics for $\delta = 1,~2,~3$ and $4$. The drift term coming from the $\delta$-potential is
\bea
\frac{\partial S_\alpha^{~\text{ eff}}}{\partial \phi} &=& \frac{\partial}{\partial \phi}  \left[ \hf  W'^2 - \ln \left( e^{i \alpha} - 1 + W'' \right) \right] \nn \\
&=& W' W'' - \frac{W'''}{\Big ( e^{i \alpha} - 1 + W''  \Big)} \nn \\
&=& - i g^2 (1+ \delta) (i \phi)^{2\delta +1} - \frac{i g \delta (1+\delta)  (i \phi)^{\delta - 1}}{\Big(   e^{i \alpha} - 1 + g (1+\delta) (i \phi)^{\delta}  \Big)}.
\eea

\begin{figure*}[htp]
	
	{\includegraphics[width=.49\textwidth,origin=c,angle=0]{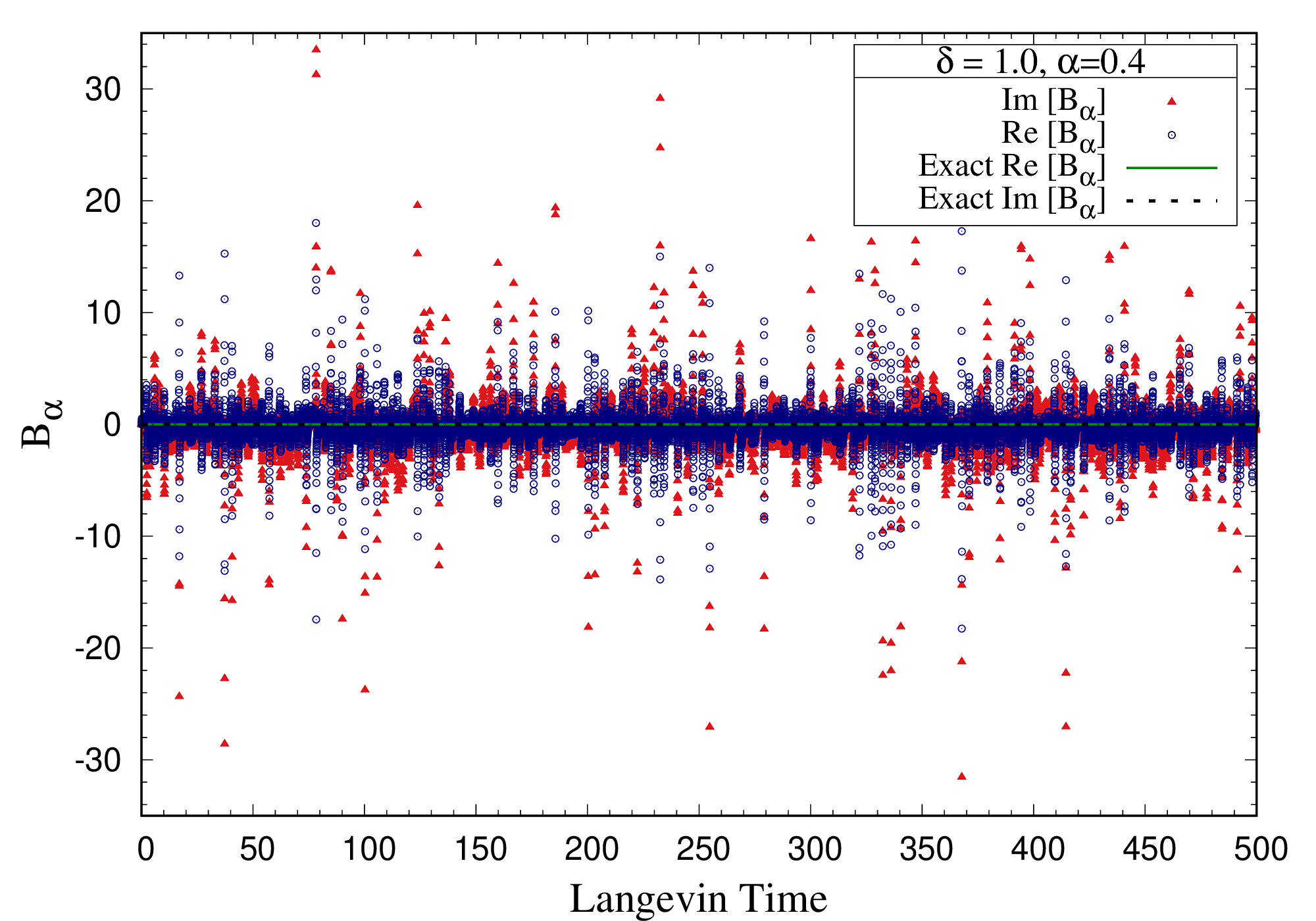}}
	{\includegraphics[width=.49\textwidth,origin=c,angle=0]{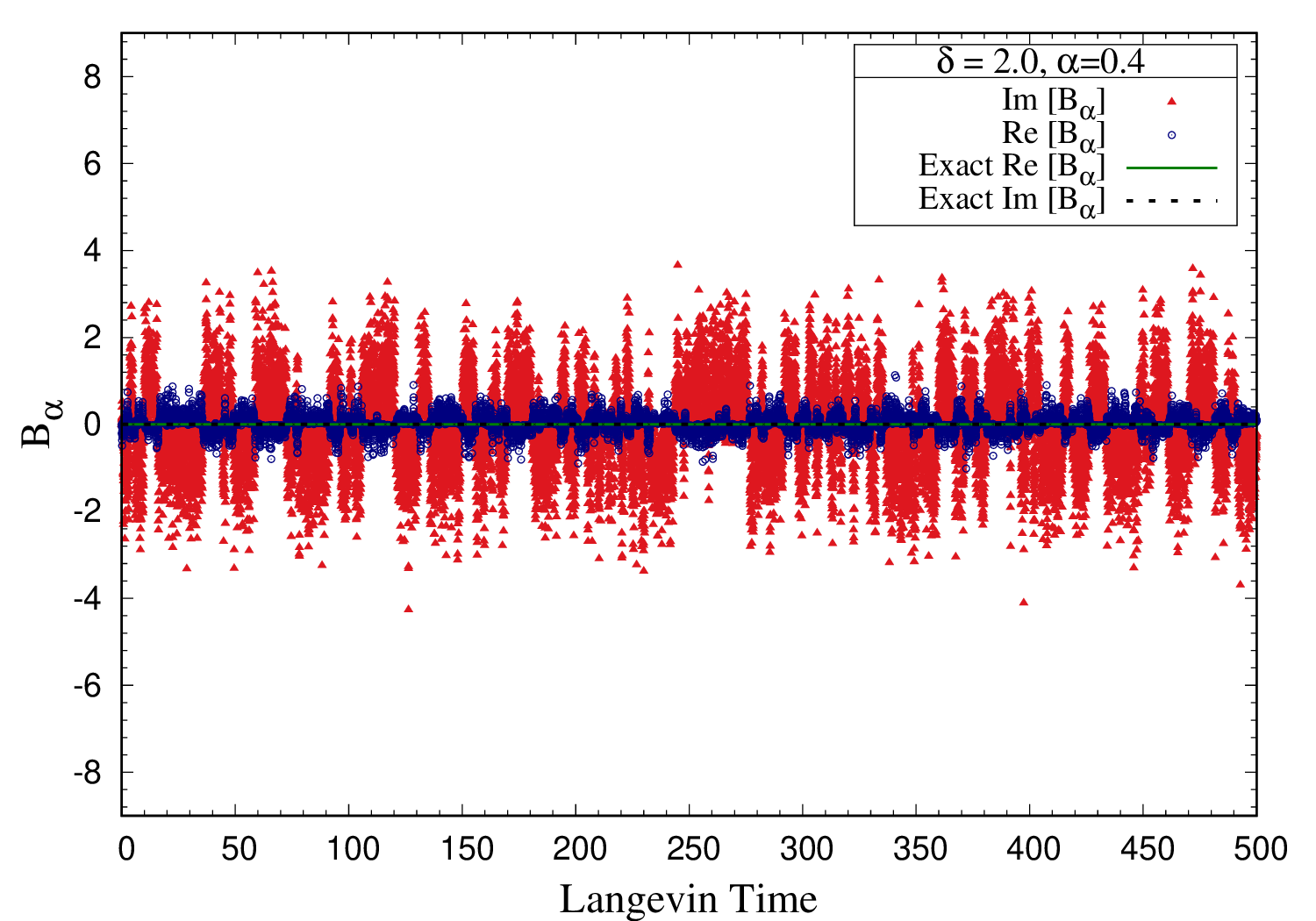}}
	
	{\includegraphics[width=.49\textwidth,origin=c,angle=0]{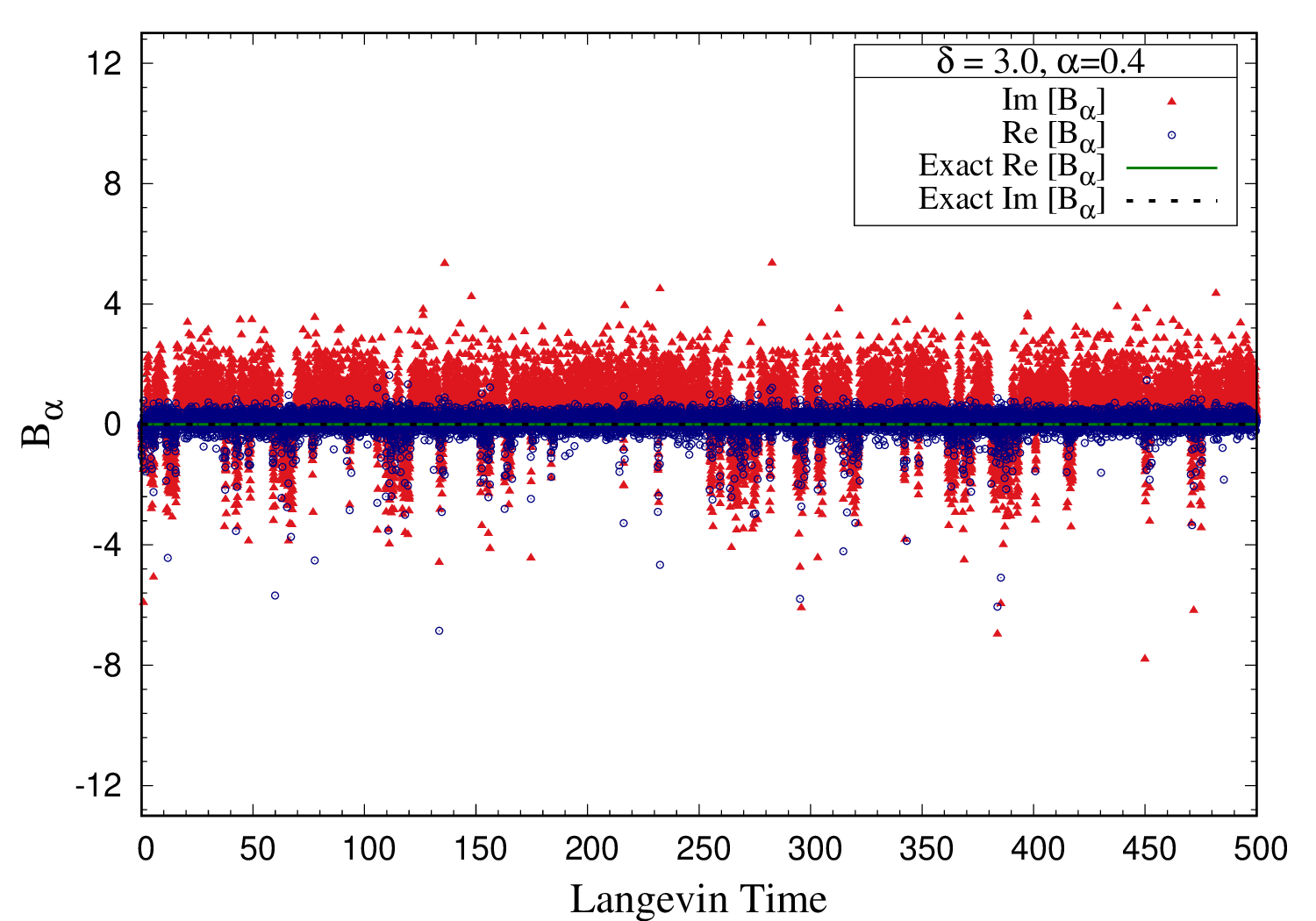}}
	{\includegraphics[width=.49\textwidth,origin=c,angle=0]{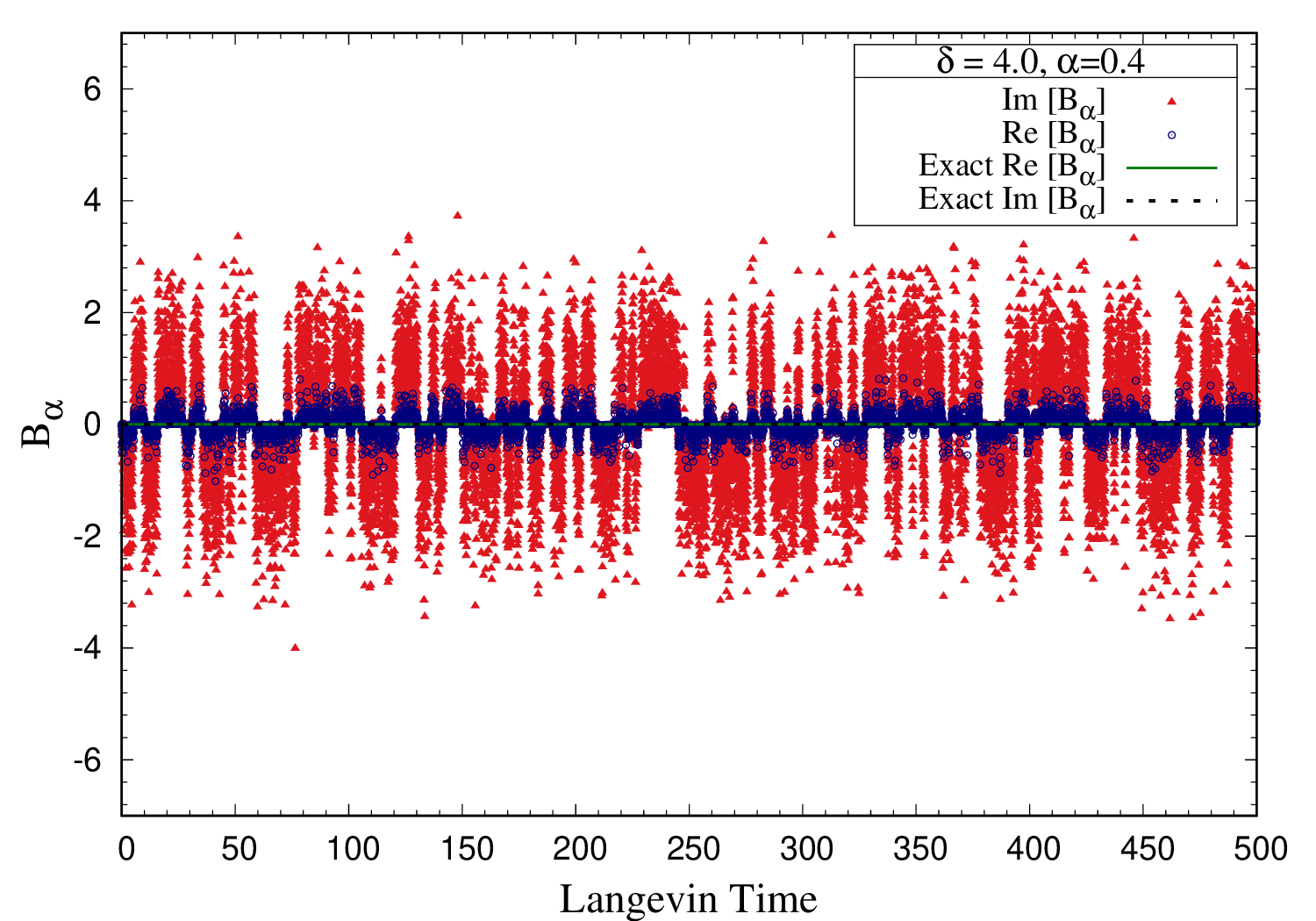}}
	
	\caption[Langevin time history of field $\B$ for $\alpha = 0.4$. The simulations were performed for superpotential $W'(\phi) = -ig (i \phi)^{(1+\delta)}$ with $g = 0.5$.]{Langevin time history of field $\B$ for $\alpha = 0.4$. The simulations were performed for superpotential $W'(\phi) = -ig (i \phi)^{(1+\delta)}$ with $g = 0.5$. In these simulations, we have used adaptive Langevin step size $\Delta \tau \leq 5\times10^{-5}$, generation steps $N_{\rm gen} = 10^7$, and measurements were taken every 500 steps. The plots show $\delta = 1$ case (Top-Left), $\delta = 2$ case (Top-Right), $\delta = 3$ case (Bottom-Left) and $\delta = 4$ case (Bottom-Right).}
	\label{fig:B-delta-history}
	
\end{figure*}

\begin{figure*}[htp]
	
	{\includegraphics[width=.49\textwidth,origin=c,angle=0]{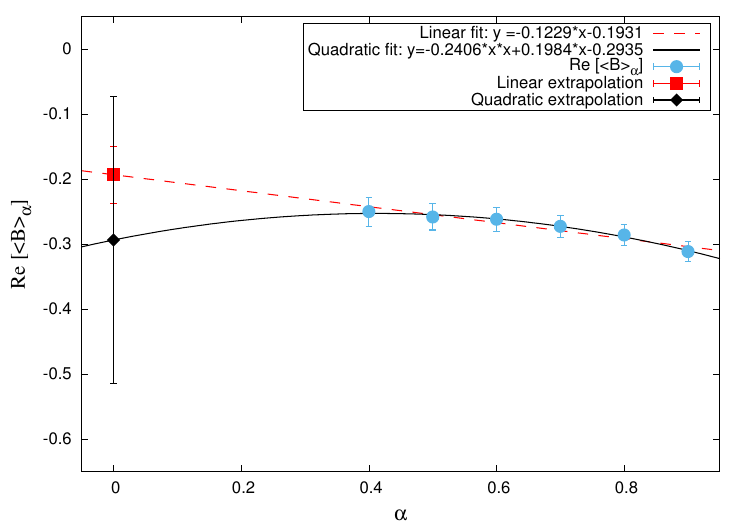}}
	{\includegraphics[width=.49\textwidth,origin=c,angle=0]{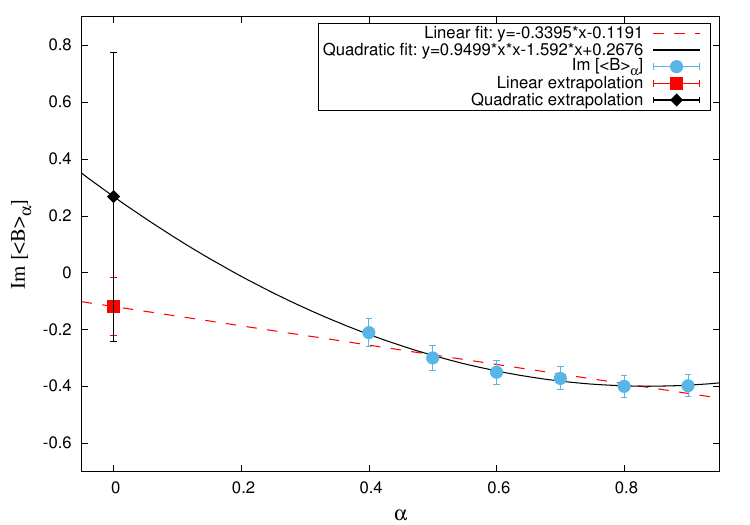}}
	
	{\includegraphics[width=.49\textwidth,origin=c,angle=0]{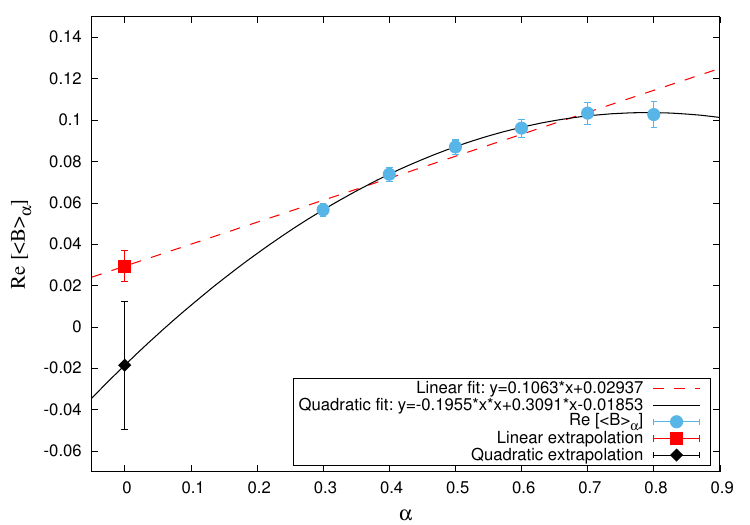}}
	{\includegraphics[width=.49\textwidth,origin=c,angle=0]{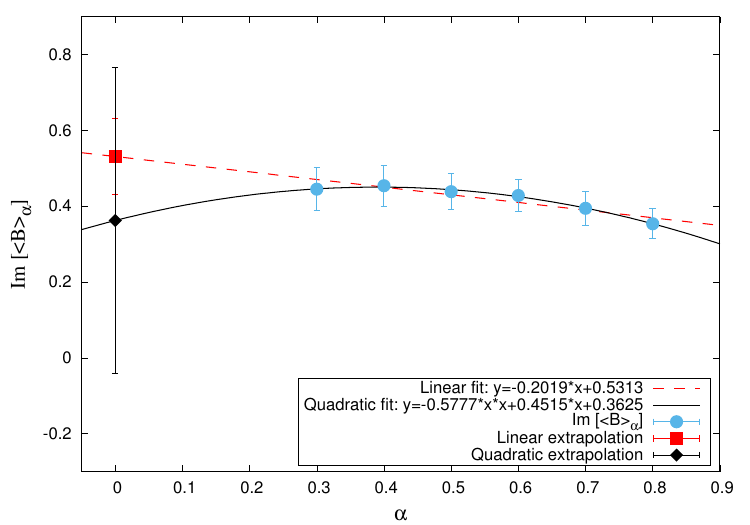}}
	
	\caption[Expectation values of $\B$ against the regularization parameter, $\alpha$ for superpotential $W'(\phi) = -ig (i \phi)^{(1+\delta)}$ with $g=0.5$.]{Expectation values of $\B$ against the regularization parameter, $\alpha$ for superpotential $W'(\phi) = -ig (i \phi)^{(1+\delta)}$ with $g=0.5$. (Top-Left) Real part and (Top-Right) imaginary part of $\langle \B \rangle_\alpha$ for $\delta = 1$. (Bottom-Left) Real part and (Bottom-Right) imaginary part of $\langle \B \rangle_\alpha$ for $\delta = 3$. The simulations were performed with adaptive Langevin step size $\Delta \tau \leq 5 \times 10^{-5}$, thermalization steps $N_{\rm therm} =  5 \times  10^4$, generation steps $N_{\rm gen} = 10^7$, and measurements taken every $500$ steps. The dashed red lines are the linear fits to $\langle \B \rangle_\alpha$ in $\alpha$, and red dots are the linear extrapolation value at $\alpha = 0$.  The solid black lines represent the quadratic fits to $\langle \B \rangle_\alpha$ in $\alpha$, and black dots are the quadratic extrapolation value at $\alpha = 0$. The $\alpha \to 0$ limit values obtained from these plots are given in Table  \ref{tab:delta_1p0_3p0}.}
	\label{fig:delta_fit_1p0_3p0}
\end{figure*}

\begin{figure*}[htp]
	
	{\includegraphics[width=.49\textwidth,origin=c,angle=0]{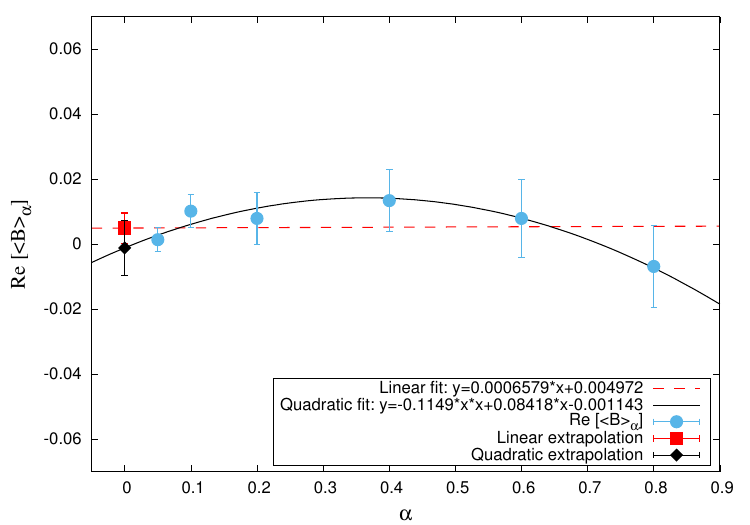}}
	{\includegraphics[width=.49\textwidth,origin=c,angle=0]{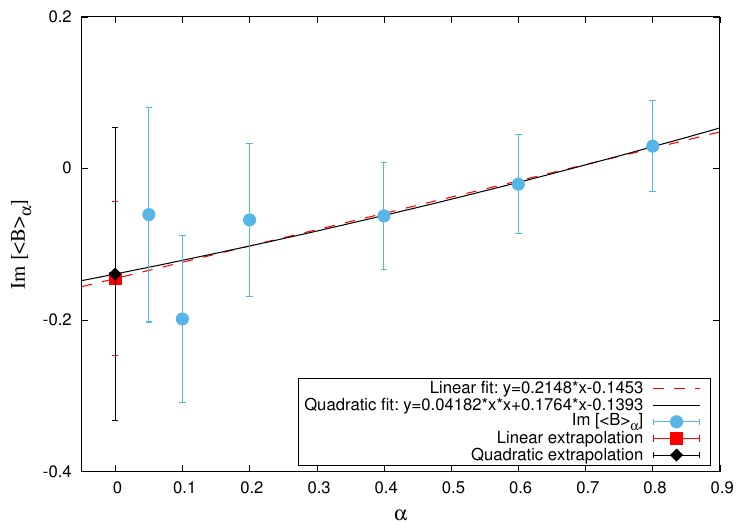}}
	
	{\includegraphics[width=.49\textwidth,origin=c,angle=0]{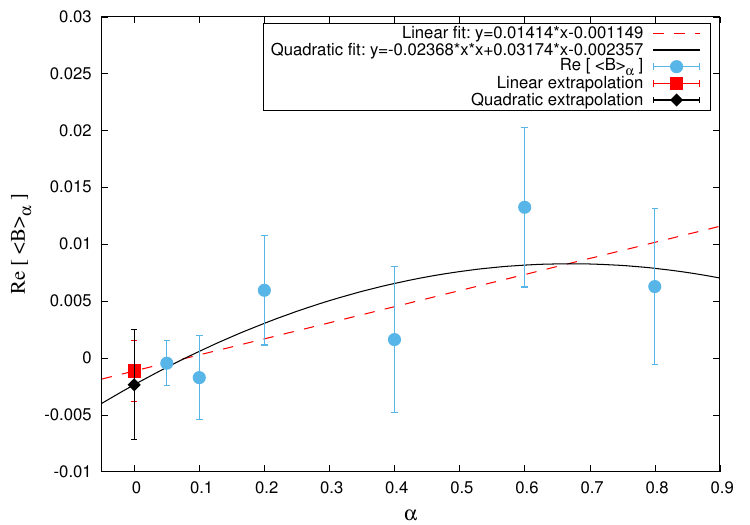}}
	{\includegraphics[width=.49\textwidth,origin=c,angle=0]{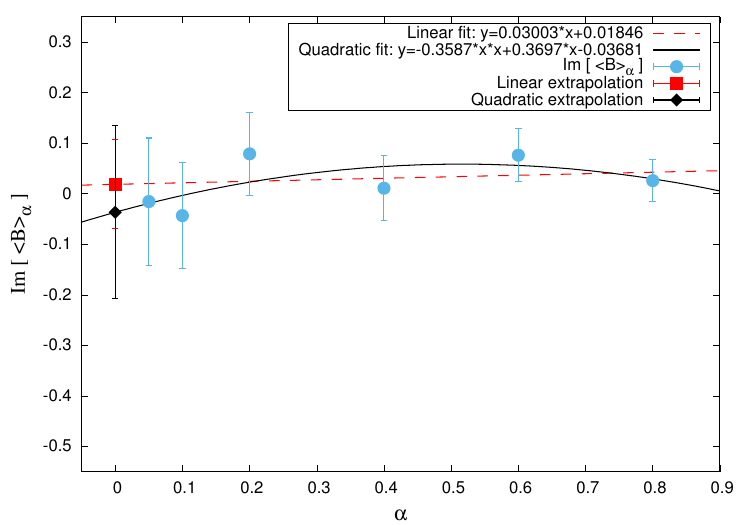}}
	
	\caption[Expectation values of $\B$ against the regularization parameter, $\alpha$ for superpotential $W'(\phi) = -ig (i \phi)^{(1+\delta)}$ with $g=0.5$.]{Expectation values of $\B$ against the regularization parameter, $\alpha$ for superpotential $W'(\phi) = -ig (i \phi)^{(1+\delta)}$ with $g=0.5$. (Top-Left) Real part and (Top-Right) imaginary part of $\langle \B \rangle_\alpha$ for $\delta = 2$. (Bottom-Left) Real part and (Bottom-Right) imaginary part of $\langle \B \rangle_\alpha$ for $\delta = 4$. The simulations were performed with adaptive Langevin step size $\Delta \tau \leq 5 \times 10^{-5}$, thermalization steps $N_{\rm therm} = 5 \times 10^4$, generation steps $N_{\rm gen} = 10^7$, and measurements were taken every $500$ steps. The dashed red lines are the linear fits to $\langle \B \rangle_\alpha$ in $\alpha$, and red dots are the linear extrapolation value at $\alpha = 0$.  The solid black lines represent the quadratic fits to $\langle \B \rangle_\alpha$ in $\alpha$, and black dots are the quadratic extrapolation value at $\alpha = 0$. The $\alpha \to 0$ limit values obtained from these plots are given in Table \ref{tab:delta_2p0_4p0}.}
	\label{fig:delta_fit_2p0_4p0}
	
\end{figure*}

In Fig. \ref{fig:B-delta-history} we show the Langevin time history of the auxiliary $\B$ field for $\delta = 1, 2, 3$ and $4$. We show linear and quadratic extrapolations to $\alpha \to 0$ limit in Fig. \ref{fig:delta_fit_1p0_3p0} for $\delta = 1, 3$ and Fig. \ref{fig:delta_fit_2p0_4p0} for $\delta = 2, 4$, respectively. The results are tabulated in Table \ref{tab:delta_1p0_3p0} and \ref{tab:delta_2p0_4p0}. It is clear from our simulation results that the expectation value of the auxiliary field, $\langle \B \rangle_\alpha$, vanishes in the limit $\alpha \to 0$. Hence we conclude that SUSY is not broken in the model with $\delta$-potential for values of $\delta = 1, 2, 3, 4$.

\begin{table*}[htp]
{\footnotesize \centering
	\begin{tabular}{| c | c | c | c |} 
		\hline
		$ ~~~~~\delta~~~~~$  & $~~~~ \alpha ~~~~$ &  $ \langle \B \rangle |_{\alpha} $  & $~~~~$ SUSY $~~~~$ \\[1.6ex]
		\hline
		\hline
		\multirow{7}{*}{$1.0$}
		&0.4	& $-0.2498 (224)   - i 0.2109 (487) $ & \multirow{7}{*}{Preserved}\\ \cline{2-3}
		&0.5	& $-0.2580 (202)   - i 0.2998 (450) $ & \\ \cline{2-3}
		&0.6	& $-0.2617 (186)   - i 0.3504 (420)$ & \\ \cline{2-3}
		&0.7	& $-0.2726 (172)   - i 0.3719 (403) $ & \\ \cline{2-3}
		&0.8	& $-0.2858 (160)   - i 0.3998 (391) $& \\  \cline{2-3}
		&0.9	& $-0.3113 (149)   - i 0.3978 (391) $ & \\  \cline{2-3}	
		& $\alpha \rightarrow 0$ &	 $ - 0.2433(2213) + i 0.0742(5080)  $  &\\ [1.5ex]
		\hline

		\multirow{7}{*}{$3.0$}
		&0.3	& $ 0.0567 (32) + i 0.4452 (566) $ &  \multirow{7}{*}{Preserved} \\ \cline{2-3}
		&0.4	& $ 0.0738 (32) + i 0.4544 (538) $ & \\ \cline{2-3}
		&0.5	& $ 0.0870 (34) + i 0.4387 (475) $ &   \\  \cline{2-3}
		&0.6	& $ 0.0961 (43) + i 0.4284 (416) $ & \\  \cline{2-3}
		&0.7	& $ 0.1034 (53) + i 0.3946 (441) $ & \\  \cline{2-3}
		&0.8	& $ 0.1027 (64) + i 0.3539 (398) $ & \\  \cline{2-3}	
		& $\alpha \rightarrow 0$ &	 $ 0.0054(311) + i 0.3625(4025)  $ & \\ [1.5ex]
		\hline
	\end{tabular}
	\caption[Expectation values $\langle \B \rangle_\alpha$ obtained using complex Langevin dynamics for the models with superpotential $W'(\phi) = -ig (i \phi)^{(1+\delta)}$ with $g = 0.5$ ad $\delta = 1, 3$, respectively.]{Expectation values $\langle \B \rangle_\alpha$ obtained using complex Langevin dynamics for the models with superpotential $W'(\phi) = -ig (i \phi)^{(1+\delta)}$ with $g = 0.5$ ad $\delta = 1, 3$, respectively.} 
	\label{tab:delta_1p0_3p0}}
\end{table*}

\begin{table*}[htp]
	\centering
{\footnotesize	\begin{tabular}{| c | c | c | c |} 
		\hline
		$ ~~~~~~\delta ~~~~~~$  & $~~~~ \alpha ~~~~$ &  $ ~~~~~~~~~~~~~~~~~~~\langle \B \rangle |_{\alpha}~~~~~~~~~~~~~~~~ $  & $~~~~$ SUSY $~~~~$ \\[1.6ex]
		\hline
		\hline
		\multirow{7}{*}{$2.0$}
		&0.05	& $ 0.0014 (36)  - i 0.0609 (1416) $ & \multirow{7}{*}{Preserved}\\ \cline{2-3}
		&0.1	& $ 0.0102 (50)  - i 0.1986 (1101) $ & \\ \cline{2-3}
		&0.2	& $ 0.0079 (80)  - i 0.0679 (1004) $ & \\ \cline{2-3}
		&0.4	& $ 0.0134 (96)  - i 0.0627 (701) $ & \\ \cline{2-3}
		&0.6	& $ 0.0079 (120) - i 0.0208 (655)$ & \\  \cline{2-3}
		&0.8	& $-0.0068 (126) + i 0.0294 (595) $ & \\  \cline{2-3}	
		& $\alpha \rightarrow 0$ &	 $ 0.0019(84) - i 0.1423(1932)$  &\\ [1.5ex]
		\hline
		
		\multirow{7}{*}{$4.0$}
		&0.05	& $-0.0005 (20) - i 0.0155 (1257) $ & \multirow{7}{*}{Preserved}\\ \cline{2-3}
		&0.1	& $-0.0017 (37) - i 0.0435 (1043) $ & \\ \cline{2-3}
		&0.2	& $ 0.0059 (48) + i 0.0787 (817) $ & \\ \cline{2-3}
		&0.4	& $ 0.0016 (64) + i 0.0108 (648) $ & \\ \cline{2-3}
		&0.6	& $ 0.0132 (70) + i 0.0761 (526)$ & \\  \cline{2-3}
		&0.8	& $ 0.0063 (68) + i 0.0258 (418) $ & \\  \cline{2-3}	
		& $\alpha \rightarrow 0$ &	 $ -0.0018(48) - i 0.0092(1712)$  &\\ [1.5ex]
		\hline
	\end{tabular}}
	\caption[Expectation values $\langle \B \rangle_\alpha$ obtained using complex Langevin dynamics for the models with superpotential $W'(\phi) = -ig (i \phi)^{(1+\delta)}$ with $g = 0.5$ and $\delta = 2, 4$, respectively.]{Expectation values $\langle \B \rangle_\alpha$ obtained using complex Langevin dynamics for the models with superpotential $W'(\phi) = -ig (i \phi)^{(1+\delta)}$ with $g = 0.5$ and $\delta = 2, 4$, respectively.}
	\label{tab:delta_2p0_4p0}
\end{table*}

\section{Reliability of complex Langevin simulations}
\label{app:reliability}

In this section, we would like to justify the simulations used in this work. We look at two of the methods proposed in the recent literature. One is based on the Fokker-Planck equation as a correctness criterion, and the other is based on the probability distribution of the magnitude of the drift term. 

\subsection{Fokker-Planck equation as correctness criterion}
\label{app:FP-correctness-0d}

The holomorphic observables of the theory $\cO[\phi, \tau]$ evolve according to \cite{Aarts:2009uq, Aarts:2011ax, Aarts:2013uza}
\beq
\frac{\partial \cO[\phi, \tau]}{\partial \tau} = \widetilde{L} \cO[\phi, \tau], 
\eeq
where
$\widetilde{L}$ is the Langevin operator
\beq
\widetilde{L} = \left[ \frac{\partial}{\partial \phi}  - \frac{\partial}{\partial \phi} S[\phi] \right] \ \frac{\partial}{\partial \phi}.
\eeq

Once the equilibrium distribution is reached, assuming that it exists, we can remove the $\tau$ dependence from the observables. Then we have
\beq
C_\cO \equiv \langle \widetilde{L} \cO[\phi] \rangle = 0,
\eeq
and this can be used as a criterion for the correctness of the complex Langevin method. This criterion has been investigated in various models in Refs. \cite{Aarts:2009uq, Aarts:2011ax, Aarts:2013uza}.

\begin{figure*}[htp]
	
	{\includegraphics[width=.49\textwidth,origin=c,angle=0]{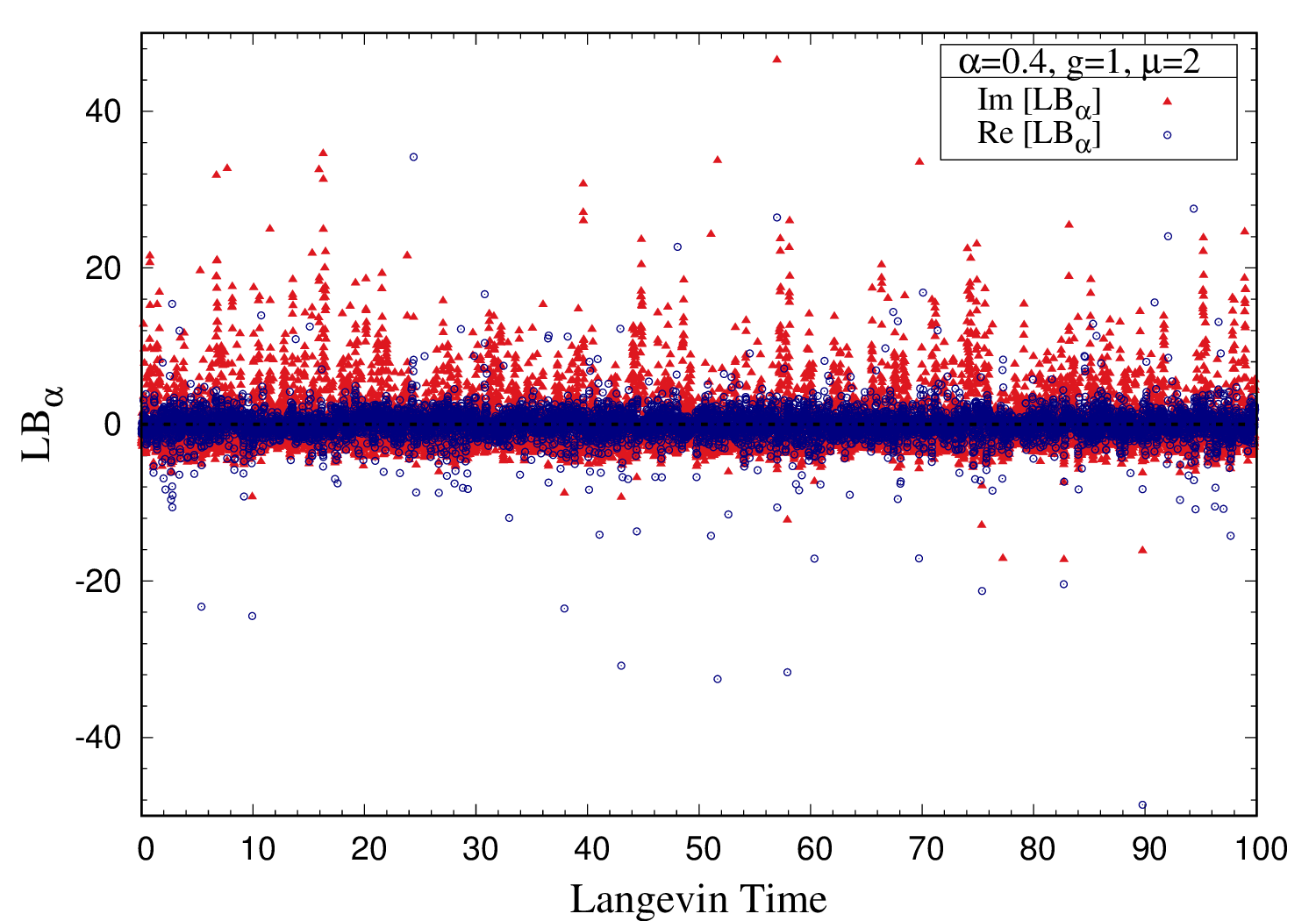}}
	{\includegraphics[width=.49\textwidth,origin=c,angle=0]{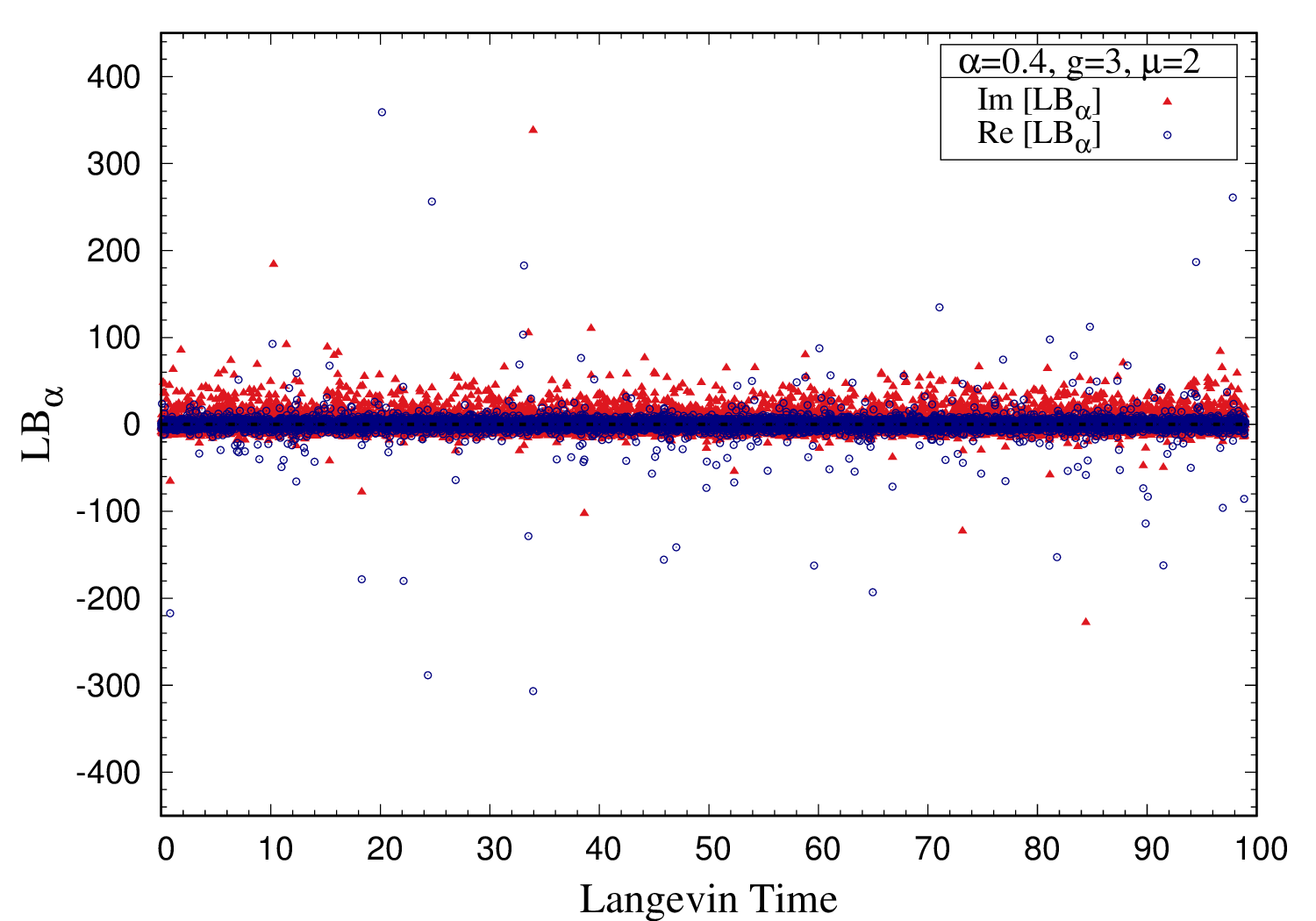}}
	
	\caption[Langevin time history of $\widetilde{L} \B$ for regularization parameter, $\alpha = 0.4$.]{Langevin time history of $\widetilde{L} \B$ for regularization parameter, $\alpha = 0.4$. Simulations were performed for superpotential $W' = g \ (\phi^2 + \mu^2)$ with $\mu = 2$, $g = 1$ (Left) and $g = 3$ (Right). In these simulations, we have used adaptive Langevin step size $\Delta \tau \leq 10^{-4}$, generation steps $N_{\rm gen} = 10^6$, and measurements were taken every $100$ step. The exact value is $\widetilde{L} \B = 0$ at equilibrium distribution.}
	\label{fig:LO_dw}
	
\end{figure*}

\begin{figure*}[htp]
	
	{\includegraphics[width=.49\textwidth,origin=c,angle=0]{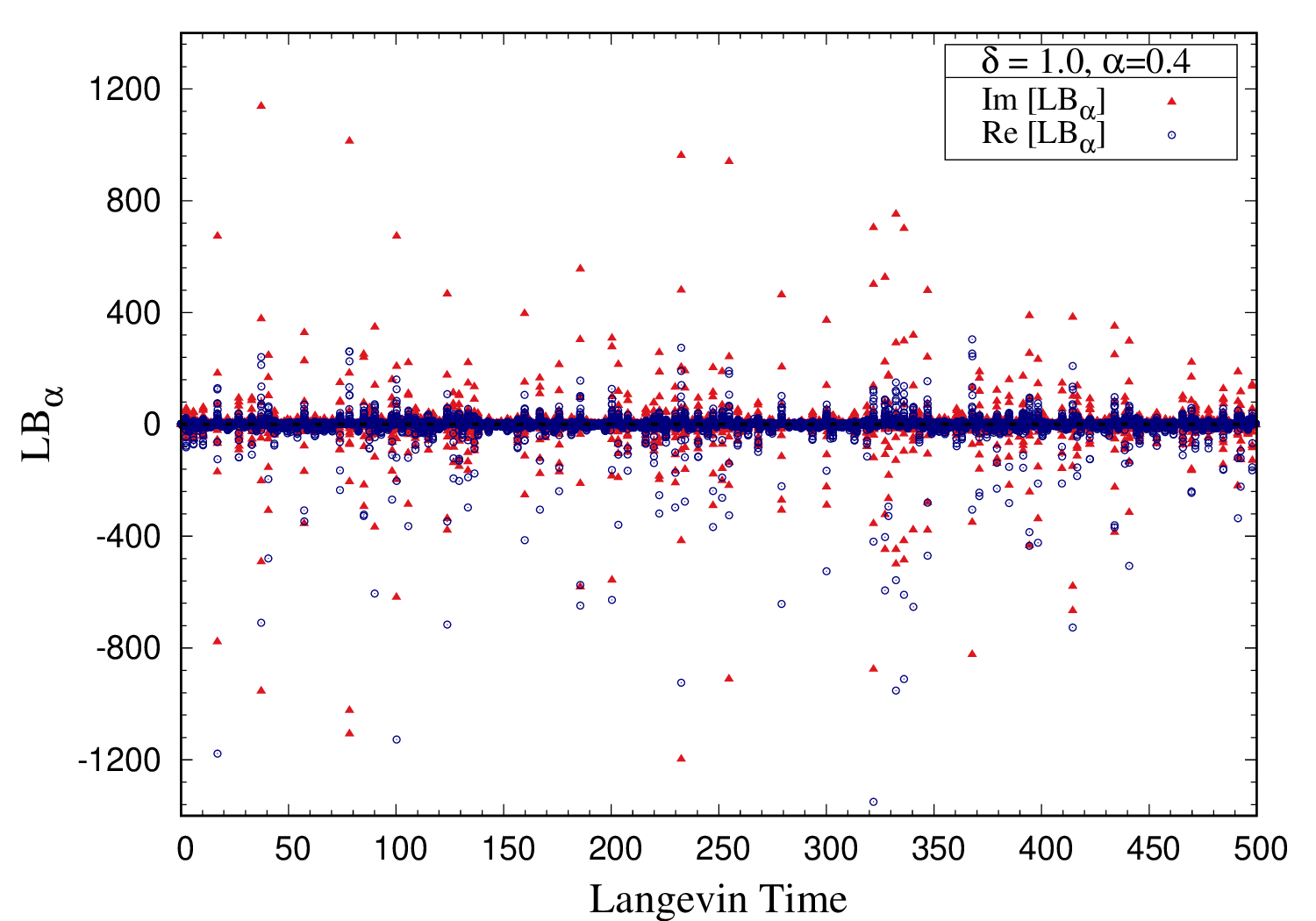}}
	{\includegraphics[width=.49\textwidth,origin=c,angle=0]{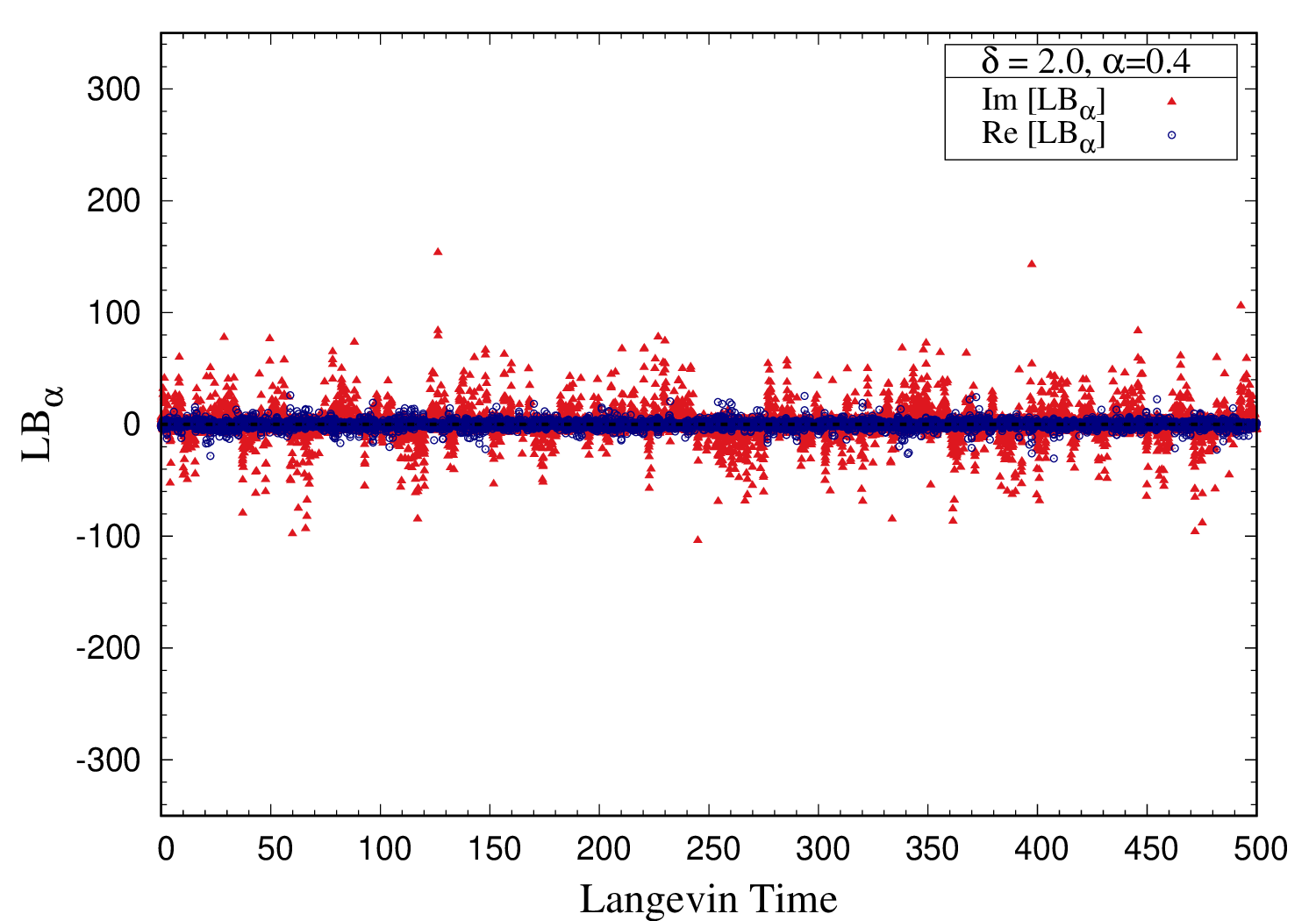}}
	
	{\includegraphics[width=.49\textwidth,origin=c,angle=0]{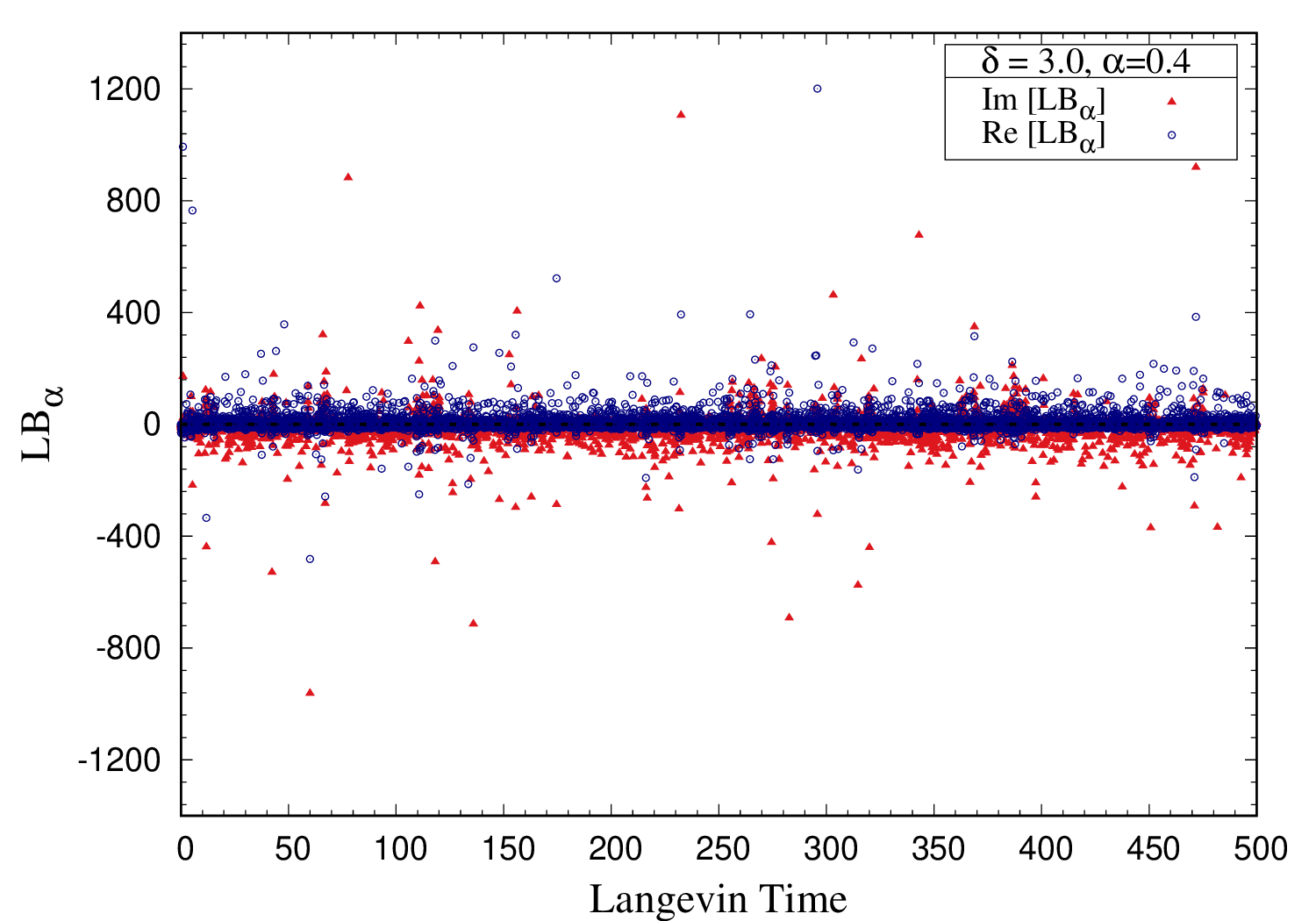}}
	{\includegraphics[width=.49\textwidth,origin=c,angle=0]{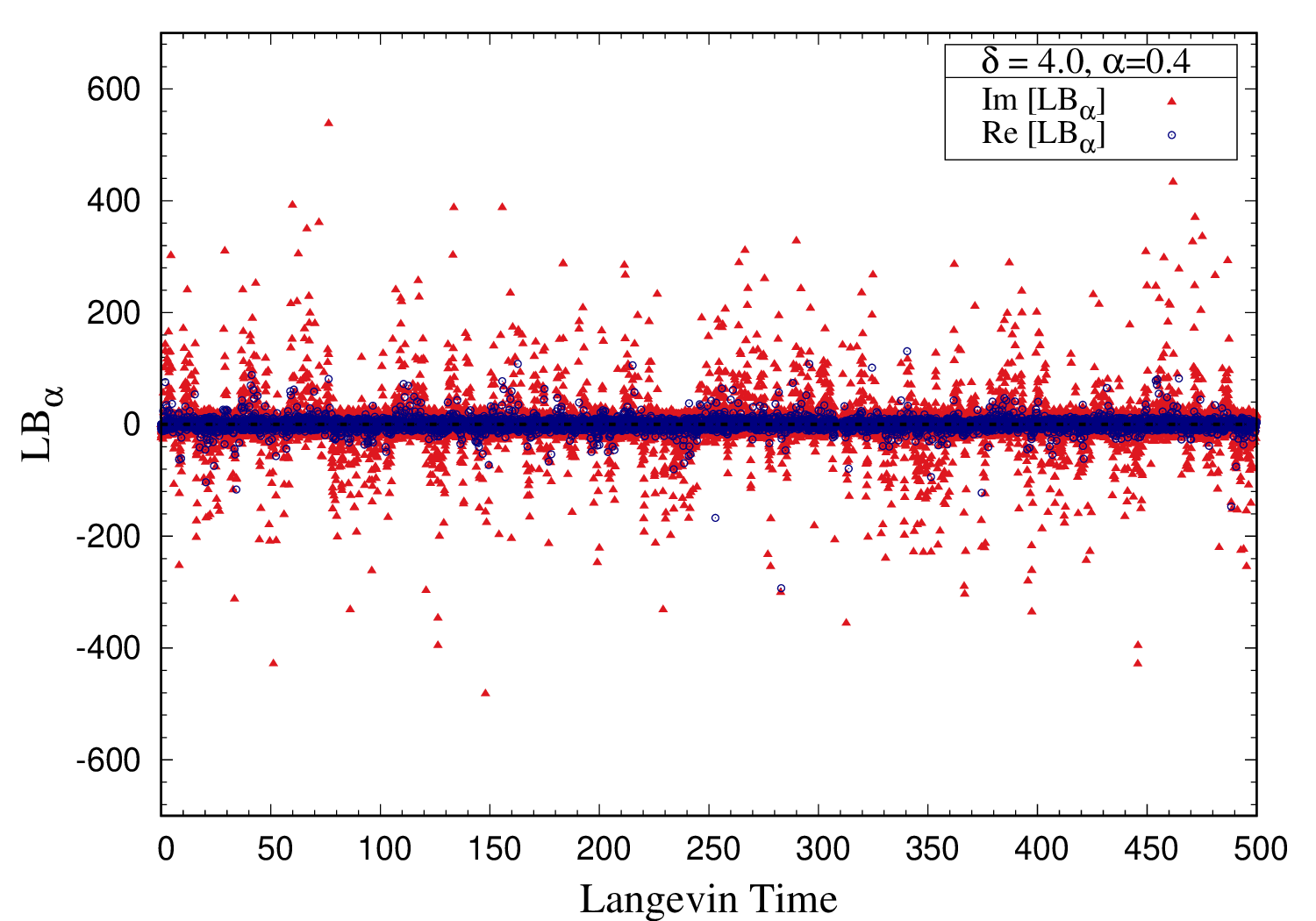}}
	
	\caption[Langevin time history of $\widetilde{L} \B$ for regularization parameter, $\alpha = 0.4$.]{Langevin time history of $\widetilde{L} \B$ for regularization parameter, $\alpha = 0.4$. Simulations were performed for superpotential $W'(\phi) = -ig (i \phi)^{(1+\delta)}$ with $g = 0.5$ for various values of delta: $\delta = 1$ (Top-Left), $\delta = 2$ (Top-Right), $\delta = 3$ (Bottom-Left) and $\delta = 4$ (Bottom-Right). In these simulations, we have used adaptive Langevin step size $\Delta \tau \leq 5\times10^{-5}$, generation steps $N_{\rm gen} = 10^7$, and measurements were taken every $500$ step. The exact value at equilibrium distribution is $\widetilde{L}\B = 0$.}
\label{fig:LO_delta}
	
\end{figure*}

For the observable $\cO$, as the auxiliary $\B$ field, we have
\bea
\widetilde{L} \cO = \widetilde{L} \B =  -i W''' + i W' {W''}^2 - \frac{i W'' W'''}{\Big ( e^{i \alpha} - 1 + W''\Big)}.
\eea
We show the Langevin history of the above-mentioned correctness criterion, $\widetilde{L} \B$, with regularization parameter $\alpha=0.4$ for the superpotentials $W' = g (\phi^2 + \mu^2)$ and $W' = -ig (i\phi)^{1 + \delta}$ in Figs. \ref{fig:LO_dw} and \ref{fig:LO_delta}, respectively. In Table \ref{tab:sqw_LO} we provide the simulated values of $\langle \widetilde{L} \B \rangle_\alpha$ for superpotential $W' = g (\phi^2 + \mu^2)$ with coupling parameter $g = 1,3$ and various values of regularization parameter, $\alpha$. In Tables. \ref{tab:delta_LO_1p0_3p0} and \ref{tab:delta_LO_2p0_4p0}, we tabulate the simulated values of $\langle \widetilde{L} \B \rangle_\alpha$ for superpotential $W'(\phi) = -ig (i \phi)^{(1+\delta)}$ with coupling parameter $g = 0.5$ and various values of regularization parameter $\alpha$.

\begin{table*}[htp]
{\footnotesize \centering
	\begin{tabular}{| c | c | c | c  | c |} 
		\hline 
		$W'$& $~~\mu~~$  & $~~g~~$ & $~~~~~\alpha~~~~~$ & $~~~~~~\langle \widetilde{L}\B \rangle |_{\alpha}~~~~~~$ \\ [1.5ex] \hline\hline
		\multirow{10}{*}{$g\Big(\phi^2 + \mu^2\Big)$} &\multirow{10}{*}{$2.0$} & \multirow{5}{*}{$1.0$}
		
		&  	0.05	 &	 $ -0.0019 (78)   +  i  0.0020 (1379)$ \\ \cline{4-5}
		&&		&  	0.1		 &	 $ -0.0133 (130)  +  i  0.0792 (1388)$ \\ \cline{4-5}
		&&		&  	0.2		 &	 $ -0.0322 (264)  +  i  0.0996 (1368)$ \\ \cline{4-5}
		&&	 	&  	0.4		 &	 $ -0.0090 (420)  +  i  0.0486 (1329)$  \\ \cline{4-5}
		&&	 	&  	0.6	 	 &	 $ -0.0852 (685)  -  i  0.0191 (1444)$  \\ \cline{4-5}
		&&	 	&  	0.8		 &	 $ -0.0252 (539)  +  i  0.0264 (1258)$   \\ \cline{4-5} 
		&&	&  	$\alpha \rightarrow 0$ &	 $ 0.0023 (230) + i 0.0555 (2357) $  \\[1.5ex] \cline{3-5} 
		&&	\multirow{5}{*}{$3.0$}	
		&  	0.05	 &	 $ 0.0257 (250)    - i  0.0304 (1561)$ \\ \cline{4-5}
		&&		&  	0.1		 &	 $-0.0682 (724)    + i  0.0222 (1660)$ \\ \cline{4-5}
		&&		&  	0.2		 &	 $ 0.0678 (966)    - i  0.0088 (1712)$	\\ \cline{4-5}
		&&	 	&  	0.4		 &	 $ 0.1330 (1656)   + i  0.2933 (2790) $   \\ \cline{4-5}
		&&	 	&  	0.6		 &	 $ 0.0816 (2031)   + i  0.4755 (2733)$  \\ \cline{4-5}
		&&	 	&  	0.8		 &	 $-0.2429 (1627)   + i  0.1306 (1682) $ \\ \cline{4-5}
		&&	&  	$\alpha \rightarrow 0$ & $ 0.0098 (778) - i 0.0840 (3020) $ \\[1.5ex] \hline
	\end{tabular}
	\caption[Expectation values $\langle \widetilde{L}\B \rangle_\alpha$ obtained using complex Langevin simulations for the models with superpotential $W' = g\Big(\phi^2 + \mu^2\Big)$.]{\label{tab:sqw_LO}Expectation values $\langle \widetilde{L}\B \rangle_\alpha$ obtained using complex Langevin simulations for the models with superpotential $W' = g \left(\phi^2 + \mu^2\right)$.}}
\end{table*}

\begin{table*}[htp]
	\centering
{\footnotesize	\begin{tabular}{| c | c | c |} 
		\hline
		$~~~~~~\delta~~~~~~$  & $~~~~~~~\alpha~~~~~~~$  & $~~~~~~~~~~~~~~~~~~~~~~~\langle \widetilde{L} \B \rangle |_\alpha~~~~~~~~~~~~~~~~~~~~~$  \tabularnewline
		\hline 
		\hline 
		\multirow{4}{*}{1.0}
		& 0.4  & $-0.6263 (3592)    + i 0.0042 (3062) $  \tabularnewline
		\cline{2-3} 
		& 0.5  & $-0.1442 (2127)    + i 0.0202 (1752)$  \tabularnewline
		\cline{2-3}
		& 0.6  & $-0.0239 (1517)    + i 0.0400 (1375)$  \tabularnewline
		\cline{2-3} 
		& 0.7  & $~~0.0198 (1192)    + i 0.0387 (1171)$  \tabularnewline
		\cline{2-3} 
		& 0.8  & $-0.0107 (1169)    + i 0.0494 (988)$  \tabularnewline
		\cline{2-3} 
		& 0.9  & $-0.0401 (990)     + i 0.0104 (915)$  \tabularnewline
		\cline{2-3} 
		& $\alpha \rightarrow 0$ 
		& $ -1.2716 (2.421) - i 0.1173 (2.122) $  \tabularnewline [1.5ex]
		\hline 
		\multirow{4}{*}{3.0}
		& 0.3  & $ 0.1846 (5176)    + i 0.1366 (3738) $  \tabularnewline
		\cline{2-3} 
		& 0.4  & $-0.3282 (1845)    + i 0.0443 (3164)$  \tabularnewline
		\cline{2-3}
		& 0.5  & $-0.2215 (1856)    + i 0.1869 (2377)$  \tabularnewline
		\cline{2-3} 
		& 0.6  & $-0.2046 (1456)    + i 0.2870 (1969)$  \tabularnewline
		\cline{2-3} 
		& 0.7  & $ 0.0022 (1476)    + i 0.2841 (2076)$  \tabularnewline
		\cline{2-3} 
		& 0.8  & $-0.0483 (1412)    + i 0.1976 (1960)$  \tabularnewline
		\cline{2-3} 
		& $\alpha \rightarrow 0$
		& $-0.3031 (2.181) - i 0.2210 (2.335)  $   \tabularnewline  [1.5ex]
		\hline 
	\end{tabular}}
	\caption[Simulated values of $\widetilde{L} \B_\alpha$ for the models with superpotential $W'(\phi) = -ig (i \phi)^{(1+\delta)}$, with coupling parameter $g = 0.5$ and $\delta = 1, 3$, respectively.]{\label{tab:delta_LO_1p0_3p0}Simulated values of $\widetilde{L} \B_\alpha$ for the models with superpotential $W'(\phi) = -ig (i \phi)^{(1+\delta)}$, with coupling parameter $g = 0.5$ and $\delta = 1, 3$, respectively.}
\end{table*}

\clearpage

\begin{table*}[htp]
	\centering
{\footnotesize	\begin{tabular}{| c | c | c |}
		\hline 
		$~~~~~~\delta~~~~~~$  & $~~~~~~~\alpha~~~~~~~$  & $~~~~~~~~~~~~~~~~~~~~~~~\langle \widetilde{L} \B \rangle |_\alpha~~~~~~~~~~~~~~~~~~~~~$  \tabularnewline
		\hline 
		\hline 
		\multirow{4}{*}{2.0} 
		& 0.05 & $ 0.0036 (49)  - i 0.1572 (1315)$  \tabularnewline
		\cline{2-3}  
		& 0.1 	& $ 0.0082 (94)  - i 0.2145 (1273)$  \tabularnewline
		\cline{2-3}  
		& 0.2  & $ 0.0113 (156) - i 0.1480 (1359)$  \tabularnewline
		\cline{2-3} 
		& 0.4  & $ 0.0066 (246) - i 0.1409 (1300)$  \tabularnewline
		\cline{2-3} 
		& 0.6  & $-0.0014 (312) - i 0.1029 (1280)$  \tabularnewline
		\cline{2-3} 
		&  0.8 & $-0.0023 (348) - i 0.1132 (1245)$  \tabularnewline
		\cline{2-3} 
		& $\alpha \rightarrow 0$ 
		& $  0.0034 (142) - i 0.1906 (2223) $  \tabularnewline
		\hline 
		\multirow{4}{*}{4.0}
		& 0.05 & $-0.0086 (127)  + i 0.3919 (2944)$  \tabularnewline
		\cline{2-3}
		& 0.1	& $-0.0292 (202)  + i 0.3050 (2945)$  \tabularnewline
		\cline{2-3}  
		& 0.2  & $-0.0127 (310)  + i 0.5222 (2910)$  \tabularnewline
		\cline{2-3} 
		& 0.4  & $ 0.0295 (503)  + i 0.4377 (2889)$  \tabularnewline
		\cline{2-3} 
		& 0.6  & $ 0.0497 (595)  + i 0.3674 (2690)$  \tabularnewline
		\cline{2-3} 
		& 0.8  & $-0.0781 (1796) + i 0.1504 (3194)$  \tabularnewline
		\cline{2-3} 
		& $\alpha \rightarrow 0$
		& $  -0.0171 (361) + i 0.3794 (5019) $   \tabularnewline
		\hline 
	\end{tabular}}
	\caption[Simulated values of $\widetilde{L} \B_\alpha$ for the models with superpotential $W'(\phi) = -i g (i \phi)^{(1+\delta)}$, with coupling parameter $g = 0.5$ and $\delta = 2, 4$, respectively.]{Simulated values of $\widetilde{L} \B_\alpha$ for the models with superpotential $W'(\phi) = -i g (i \phi)^{(1+\delta)}$, with coupling parameter $g = 0.5$ and $\delta = 2, 4$, respectively.}
    \label{tab:delta_LO_2p0_4p0}
\end{table*}

\subsection{Decay of the drift terms}
\label{app:drift-decay-0d}

Another method to check the correctness of the complex Langevin dynamics, as proposed in Refs. \cite{Nagata:2016vkn, Nagata:2018net}, is to look at the probability distribution $P(u)$ of the magnitude of the drift term $u$ at large values of the drift. We have the magnitude of the drift term
\beq
u = \left| \frac{\partial S}{\partial \phi} \right|.
\eeq

\begin{figure*}[htp]
	
	{\includegraphics[width=.49\textwidth,origin=c,angle=0]{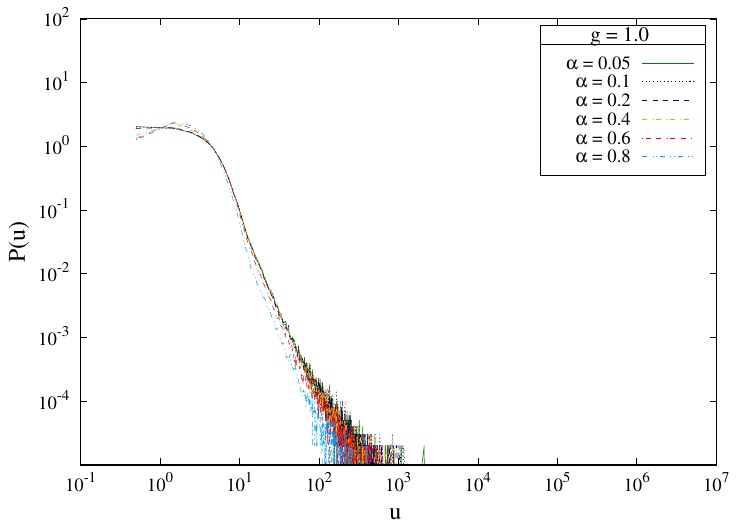}}
	{\includegraphics[width=.49\textwidth,origin=c,angle=0]{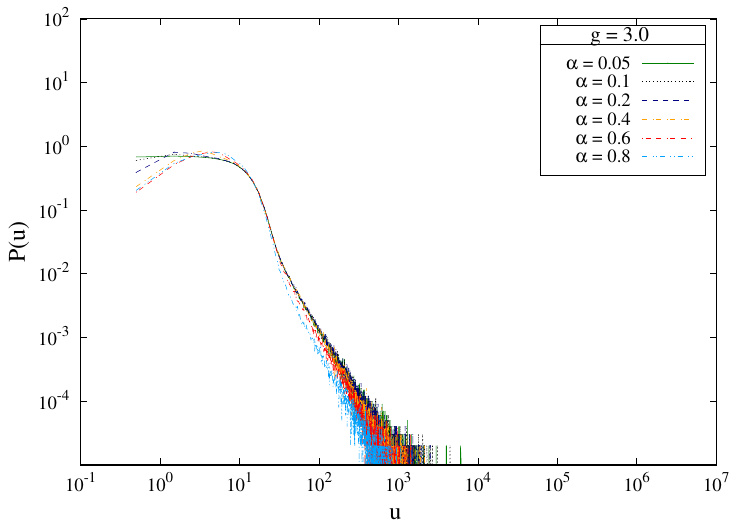}}
	
	\caption[Probability distribution $P(u)$ of the magnitude of the drift term $u$ for the superpotential $W'(\phi) = g (\phi^2 +\mu^2)$ on a log-log plot.]{Probability distribution $P(u)$ of the magnitude of the drift term $u$ for the superpotential $W'(\phi) = g (\phi^2 +\mu^2)$ on a log-log plot. Simulations were performed for $g = 1.0$ (Left) and $g = 3.0$ (Right) with $\mu= 2.0$. We used adaptive Langevin step size $\Delta \tau \leq 10^{-4}$, and generation steps $N_{\rm gen} = 10^6$.}
	\label{fig:drift_dw}
	
\end{figure*}

\begin{figure*}[htp]
	
	{\includegraphics[width=.49\textwidth,origin=c,angle=0]{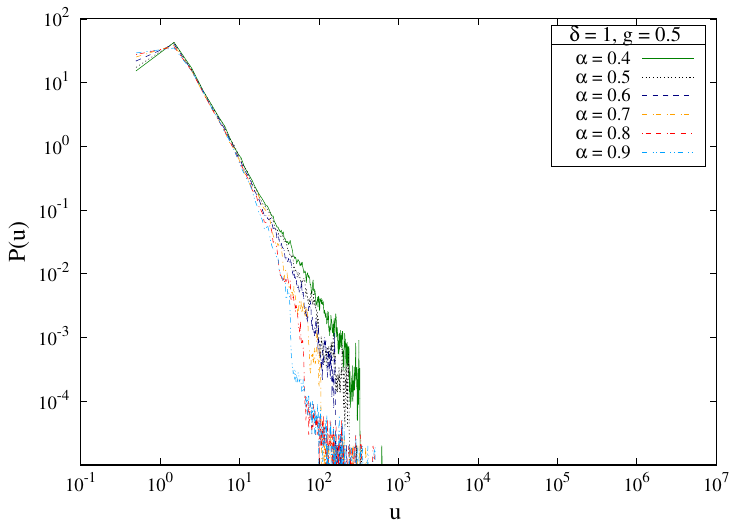}}
	{\includegraphics[width=.49\textwidth,origin=c,angle=0]{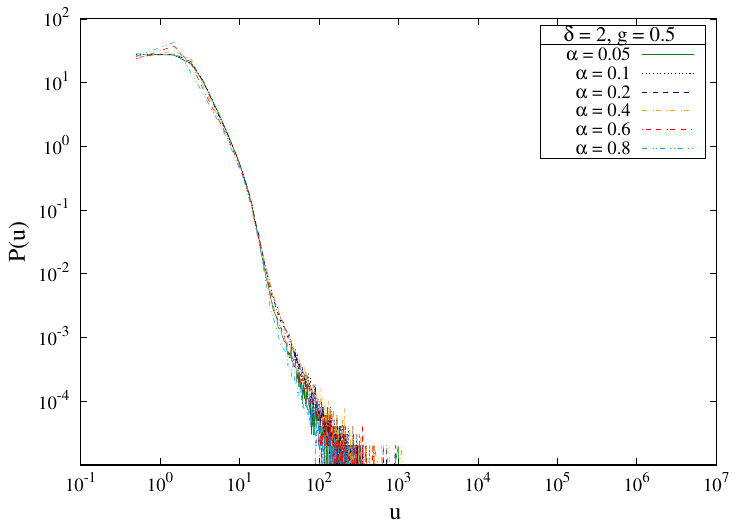}}
	
	{\includegraphics[width=.49\textwidth,origin=c,angle=0]{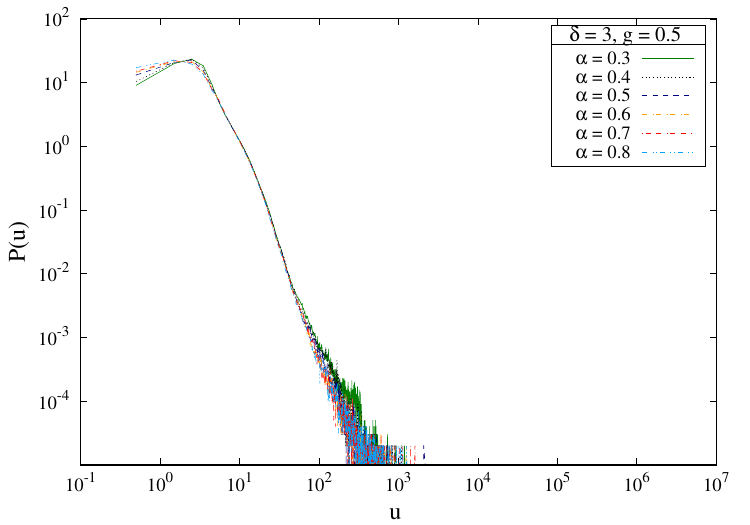}}
	{\includegraphics[width=.49\textwidth,origin=c,angle=0]{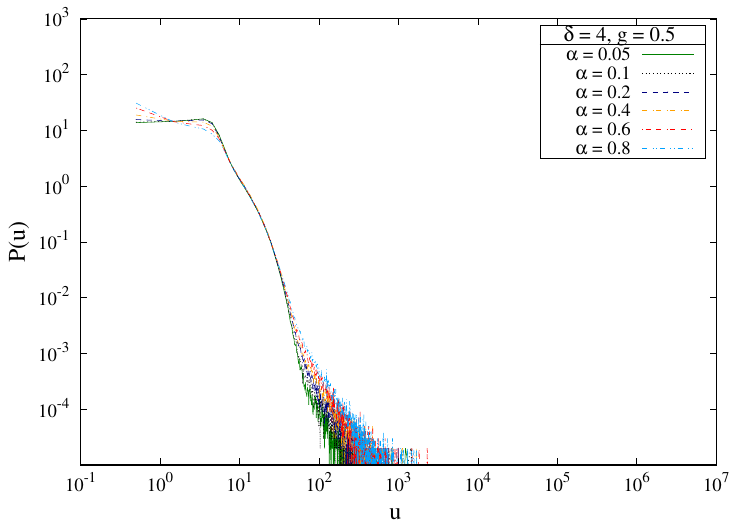}}
	
	\caption[Probability distribution $P(u)$ of the magnitude of the drift term $u$ for the superpotential $W'(\phi) = -ig (i \phi)^{(1+\delta)}$ on a log-log plot.]{Probability distribution $P(u)$ of the magnitude of the drift term $u$ for the superpotential $W'(\phi) = -ig (i \phi)^{(1+\delta)}$ on a log-log plot. Simulations were performed for $\delta = 1$ (Top-Left), $\delta = 2$ (Top-Right), $\delta = 3$ (Bottom-Left) and $\delta = 4$ (Bottom-Right) with coupling constant $g = 0.5$. We used adaptive Langevin step size $\Delta \tau \leq 5 \times 10^{-5}$ and generation steps $N_{\rm gen} = 10^7$.}
	\label{fig:drift_delta}
	
\end{figure*}

In Refs. \cite{Nagata:2016vkn, Nagata:2018net} the authors demonstrated, in a few simple models, that the probability of the drift term should be suppressed exponentially at larger magnitudes in order to guarantee the correctness of the complex Langevin method. However, in the models we investigated in this work, we see that the probability distribution falls off like a power law with $u$, even though we have excellent agreements with corresponding analytical results, wherever applicable. In Fig. \ref{fig:drift_dw} we show the probability distribution $P(u)$ against $u$ for the superpotential $W'(\phi) = g (\phi^2 +\mu^2)$ on a log-log plot. In Fig. \ref{fig:drift_delta}, we show the probability distribution $P(u)$ of the magnitude of drift term $u$ for superpotential $W'(\phi) = -ig (i \phi)^{(1+\delta)}$ on a log-log plot. In both cases, we see that the distribution falls off like a power law for large $u$ values. This needs further investigation, and we will save it for future work.

	\chapter{Complex Langevin dynamics and supersymmetric quantum mechanics} \label{chap:susy-qm}

\setlength\epigraphwidth{11cm}
\setlength\epigraphrule{0pt}
\epigraph{The chapter is based on the following publication by the author: \\ Anosh Joseph and \textbf{Arpith Kumar}, \\{\it Complex Langevin Dynamics and Supersymmetric Quantum Mechanics},\\ \href{https://doi.org/10.1007/JHEP10(2021)186}{J. High Energ. Phys. {\bf 2021}, 186 (2021)} {\href{https://arxiv.org/abs/2011.08107}{(arXiv: 2011.08107 [hep-lat])}} }{}	

A precise understanding of SUSY breaking and the development of theories capable of exhibiting such a phenomenon has proven mathematically daunting. In the early days, physicists started studying SUSY in quantum mechanics models as a testbed for understanding non-perturbative SUSY breaking in field theories. Edward Witten, in 1981, introduced a topological index (now known as the \textit{Witten index}) for studying the dynamical breaking of SUSY. Thereafter, several researchers have examined how non-perturbative mechanisms facilitate SUSY breaking in quantum mechanics. After comprehensive research on various aspects of supersymmetric quantum mechanics, it soon became apparent that SUSY quantum mechanics is not just a mathematical model for testing field theory methods but is, in fact, a fascinating field in its own right.

In this chapter, we investigate spontaneous SUSY breaking in lattice regularized supersymmetric quantum mechanics with two supercharges. The chapter is organized as follows. In Sec. \ref{cont-susy-mechanics}, we briefly introduce supersymmetric quantum mechanics with two supercharges. In Sec. \ref{lattice-susy-theory}, we discuss the lattice regularization of the model. There, we also provide a set of useful observables that can probe SUSY breaking in the system. They include correlation functions and Ward identities. In Sec. \ref{lattice-simulation}, we present the simulation results of various models, including the ${\cal PT}$ symmetric model. Later, we discuss the reliability of simulations using the Langevin operator and observe the fall-off behavior of the probability distributions of the drift term magnitudes. 

\section{Supersymmetric quantum mechanics}
\label{cont-susy-mechanics}

Let us consider the action $S[\phi, \psi, \psib]$ of supersymmetric quantum mechanics with a general superpotential $W(\phi)$. The model is invariant under two supercharges. The degrees of freedom are a scalar field $\phi$ and two fermions $\psi$ and $\psib$. We take the action to be an integral over a compactified time circle of circumference $\beta$ in Euclidean time. It has the form
\bea
\label{cont-action-1d}
S[\phi, \psi, \psib] &=& \int_0^\beta d \tau
\Bigg[ \hf \B(\tau)^2 + i \B \left( \frac{\partial}{\partial{\tau}}{\phi(\tau)} + \frac{\partial }{\partial{\phi}} W(\phi(\tau)) \right)  \nn \\ 
&& \hspace{2cm} + {\psib(\tau)} \left(\frac{\partial}{\partial{\tau}} + \frac{\partial^2 }{\partial{\phi}^2} {W} (\phi(\tau)) \right) \psi(\tau) \Bigg],
\eea
where $\B$ is an auxiliary field and the derivatives with respect to $\tau$, and $\phi$ are denoted by a dot and a prime, respectively.

The action is invariant under the SUSY transformations
\beq
\label{Q-transf}
\Q \phi = \psi,~~~ \Q \psi = 0,~~~ \Q \psib =- i \B,~~~ \Q \B = 0,
\eeq
and
\beq
\label{Qb-transf}
\Qb \phi = - \psib,~~~ \Qb \bar{\psi} = 0,~~~ \Qb \psi = - i \B + 2 \dot{\phi},~~~ \Qb \B = 2 i \dot{\psib}, 
\eeq
where $\Q$ and $\Qb$ are the two supercharges which satisfy the algebra
\beq
\label{eq:algebra}
\{ \Q, \Q \} = 0, \quad \{ \Qb, \Qb \} = 0, \quad \{ \Q, \Qb \} = 2\partial_{\tau}.
\eeq
We also note that the action can be expressed in $\Q$- and $\Q \Qb$-exact forms. That is,
\beq
\label{QQb-cont-exact}
S = \Q  \int_0^\beta d{\tau} ~ \psib \left[ \frac{i}{2} \B - \left(\frac{\partial \phi}{ \partial \tau} + W^{'}(\phi)  \right) \right] = \Q \Qb  \int_0^\beta d\tau \left( \hf \psib \psi + W (\phi) \right).
\eeq

The partition function in path integral formalism is defined by
\beq
\label{eqn:cont-pf}
Z \equiv \int \cD \B \cD \phi \cD \psi  \cD \psib ~e^{-S[\phi, \psi, \psib]},
\eeq
with periodic temporal boundary conditions for all the fields. 

In a system with unbroken SUSY, we can consider a Hamiltonian $H$ corresponding to the Lagrangian in Eq. \eqref{cont-action-1d} having energy levels $E_n$ where $n = 0, 1, 2, \dots$, such that the ground state energy $E_0=0$. Then, the bosonic and fermionic excited states form a SUSY multiplet
\beq
\label{eqn:susy_multiplet_unbroken}
|b_{n+1} \rangle = \frac{1}{\sqrt{2 E_{n+1}}} \Qb |f_n \rangle, ~~~|f_n \rangle = \frac{1}{\sqrt{2 E_{n+1}}} \Q |b_{n+1} \rangle,
\eeq
satisfying the algebra given in Eq. \eqref{eq:algebra}, with $|b_{0} \rangle$ being the ground state of the system. Assuming that the states $|b_n \rangle$ and $|f_n \rangle$ have the fermion number charges $F = 0$ and $F = 1$, respectively, when periodic temporal boundary conditions are imposed for both the bosonic and fermionic fields, $Z$ is equivalent to the Witten index $\Delta_W \left( \beta\right)$ \cite{Witten:1982df}.  It is easy to see that the partition function defined as
\beq
\label{eqn:cont-Witten_index_unbroken}
Z  = \Delta_W \left(\beta \right) = \Tr \left[ (-1)^F e^{-\beta H} \right] 
=  \langle b_{0} | b_{0} \rangle  +\sum_{n = 0}^\infty \left[ \left(\langle b_{n+1} | b_{n+1} \rangle - \langle f_n | f_n \rangle  \right) e^{-\beta E_{n+1}} \right], 
\eeq
does not vanish due to the existence of a normalizable ground state. As a result, the normalized expectation values of observables are well-defined. 

The normalized expectation value of the auxiliary field defined as
\beq
\label{eqn:aux_field_exp_val}
\langle \B \rangle = \frac{1}{\Delta_W \left(\beta \right)} \left[\langle b_{0}|\B | b_{0} \rangle    +  \sum_{n = 0}^\infty \left(\langle b_{n+1}|\B | b_{n+1} \rangle - \langle f_n | \B |f_n \rangle \right) e^{-\beta E_{n+1}} \right],
\eeq
can be used as an order parameter to probe SUSY breaking \cite{Kuroki:2009yg}. (We note that the unpaired state appearing in Eqs. \eqref{eqn:cont-Witten_index_unbroken} and \eqref{eqn:aux_field_exp_val} need not have to be a bosonic state or a unique state.) The auxiliary field was introduced for the off-shell completion of the SUSY algebra. The $\Q$ transformation of $\B$ in Eq. \eqref{Q-transf} and the fact that the ground state is annihilated by the supercharges together imply that in the absence of SUSY breaking, the normalized expectation value of the auxiliary field vanishes. However, in the SUSY broken case, we end up in a not-so-trivial situation.

In a system with SUSY spontaneously broken, the Hamiltonian $H$ corresponding to the Lagrangian in Eq. \eqref{cont-action-1d} has a positive ground state energy ($0 < E_0 < E_1 < E_2 < \dots$), and the SUSY multiplet is defined as
\beq
\label{eqn:susy_multiplet}
|b_n \rangle = \frac{1}{\sqrt{2 E_n}} \Qb |f_n \rangle, ~~~|f_n \rangle = \frac{1}{\sqrt{2 E_n}} \Q |b_n \rangle,
\eeq
satisfying the algebra given in Eq. \eqref{eq:algebra}. Differently from the unbroken SUSY case, when SUSY is broken, the supersymmetric partition function
\beq
\label{eqn:cont-Witten_index}
Z  = \Delta_W = \Tr \left[ (-1)^F e^{-\beta H} \right] = \sum_{n = 0}^\infty \left[ \left(\langle b_n | b_n \rangle - \langle f_n | f_n \rangle  \right) e^{-\beta E_n} \right]
\eeq
vanishes due to the cancellation between bosonic and fermionic states.  As a consequence, the normalized expectation values of observables will be ill-defined. We consider the auxiliary field as an observable, and the normalized expectation value can be computed as
\beq
\label{eqn:cont-B-expectation}
\langle \B \rangle = \frac{1}{\Delta_W \left(\beta \right)} \sum_{n = 0}^\infty \left[ \left(\langle b_n|\B | b_n \rangle - \langle f_n | \B |f_n \rangle\right) e^{-\beta E_n} \right],
\eeq
where the numerator vanishes from $\Q$-supersymmetry $ \left( \langle b_n|\B | b_n \rangle = \langle f_n | \B |f_n \rangle \right)$. Thus, in a system with broken SUSY, the normalized expectation of the auxiliary field admit a $0/0$ indefinite form.

In Ref. \cite{Kuroki:2009yg}, Tsunehide Kuroki and Fumihiko Sugino introduced a regulator that explicitly breaks SUSY and resolves the degeneracy by fixing a single vacuum state in which SUSY is broken. This regulator $\alpha$ (the twist parameter) can be implemented by imposing twisted boundary conditions (TBC) for fermions. That is, 
\beq
\psi (\tau + \beta) = e^{i \alpha} \psi(\tau) \quad {\rm and} \quad \psib(\tau +\beta) = e^{-i \alpha} \psib(\tau).
\eeq
It was shown in Ref. \cite{Kuroki:2009yg} that for a non-zero $\alpha$, the partition function does not vanish, and the normalized expectation value of the auxiliary field is well-defined. In the limit $\alpha \to 0$, PBCs are recovered, and SUSY is restored. Thus $\alpha$ regularizes the indefinite form in Eq. \eqref{eqn:cont-B-expectation}. Now, the vanishing expectation value of the auxiliary field in the limit $\alpha \to 0$ suggests that SUSY is not broken, while a non-zero value suggests that SUSY is broken. We incorporate the twist $\alpha$ when we introduce the lattice regularized theory in Sec. \ref{sec:lattice-theory}. 

It is possible to integrate the auxiliary field using its equation of motion
\beq
\B = - i \left( \frac{\partial \phi}{ \partial \tau} + W^{'}(\phi) \right),
\eeq
to get the on-shell form of the action
\beq
\label{eqn:act-cont}
S = \int_0^\beta d\tau
\left[ \hf  \left({\frac{\partial \phi}{ \partial \tau}} + W^{'}(\phi) \right)^2 + {\psib} \left( \frac{\partial }{ \partial \tau} + W''(\phi) \right) \psi \right].
\eeq

Upon using the Leibniz integral rule and discarding the resultant total derivative term, the action takes the form
\beq
S = \int_0^\beta d\tau
\left[ \hf  {\left(\frac{\partial \phi}{ \partial \tau}\right)}^2 + \hf\left[ W'(\phi) \right]^2 + {\psib} \left( \frac{\partial }{ \partial \tau} + W''(\phi) \right) \psi \right].
\eeq
In the above expression, the total derivative term we omitted was $(\partial \phi / \partial \tau) W'(\phi)$. Note that such an omission is only possible in the continuum theory. When we discretize the theory on a lattice, this term does not vanish, and its presence is crucial to ensure the $\Q$-exact lattice SUSY. Thus, in our lattice analysis, we will use Eq. \eqref{eqn:act-cont} as the continuum target theory.

\section{Lattice regularized models}
\label{lattice-susy-theory}

We discretize the action given in Eq. \eqref{eqn:act-cont} on a one-dimensional lattice. Let us take the lattice to be $\Lambda$, having $T$ number of equally spaced sites with lattice spacing $a$. The integral and continuum derivatives are replaced by a Riemann sum $a  {\Sigma}$ and a lattice difference operator $\nabla$, respectively. The physical extent of the lattice is defined as $\beta \equiv T a$. 

There are several ways to regularize a given theory on a lattice. We will choose the prescription in which the derivatives appearing in the action take the form of a symmetric difference operator
\beq
\nabla^S_{ij} = \hf \left( \nabla^+_{ij} + \nabla^-_{ij} \right),
\eeq
where
\bea
&& \nabla^+_{ij} = \frac{1}{a} (\delta_{i+1, j} - \delta_{i, j}) \quad \longrightarrow \quad \nabla^+_{ij} f_j = \frac{1}{a} ( f_{i+1} - f_i ), {\rm ~ and} \\ 
&& \nabla^-_{ij} = \frac{1}{a} (\delta_{i, j} - \delta_{i-1, j}) \quad \longrightarrow \quad \nabla^-_{ij} f_j = \frac{1}{a} (f_i - f_{i-1}), 
\eea
are the forward and backward difference operators, respectively, and $i, j$ represent lattice sites. However, it is known that the symmetric derivative leads to the so-called fermion doubling problem, and this, in turn, leads to a non-supersymmetric lattice theory. We can use the Wilson discretization prescription to decouple these extra fermionic modes from the system. The difference operator is modified as
\beq
\nabla^W_{ij}(r) = \nabla^S_{ij} - \frac{r a}{2}\square_{ij},
\eeq
where $\square_{ij} = \nabla^+_{ik} \nabla^-_{kj}$ is the usual lattice Laplacian and the Wilson parameter $r \in \left[-1,1\right] / \left\{0 \right\}$ \cite{Baumgartner:2014nka}. For one-dimensional derivatives, it turns out that the standard choice of $r = \pm 1$ yields $\nabla^W_{ij}(\pm 1) = \nabla^{\mp}_{ij}$, thereby suggesting that the doubling problem can be resolved by simply using forward or backward difference operator. The reason is that for any choice of the lattice difference operator, the theories can be made manifestly supersymmetric upon the addition of appropriate improvement terms corresponding to the discretization of continuum surface integrals \cite{Bergner:2007pu}. 

In our analysis, for the standard choice of the Wilson parameter, we follow the symmetric derivative with a Wilson mass matrix suggested in Ref. \cite{Catterall:2000rv}. The lattice regularized action then takes the form
\beq
\label{eqn:lat-reg-action}
\mcS = a \sum_{i = 0}^{T-1} \left[ \hf \left( \sum_{j = 0}^{T-1} \nabla^S_{ij} \phi_j + \Omega'_i  \right)^2 + \psib_i \sum_{j = 0}^{T-1} \left( \nabla^S_{ij} + \Omega''_{ij} \right) \psi_j \right],
\eeq
where the quantity $\Omega'_i$ is defined as 
\beq
\Omega'_i = \sum_{j = 0}^{T-1} K_{ij} \phi_j + W'_i,
\eeq
and its derivative $\Omega''_{ij} $ is $\Omega''_{ij} =  K_{ij} +  W''_{ij} \delta_{ij}$. The Wilson mass matrix $K_{ij}$ has the form $K_{ij} = m \delta_{ij} - \frac{r a}{2} \square_{ij}$.

We can make the variables dimensionless by performing appropriate rescaling. Let us consider the following set of redefinitions for the variables
\beq
\label{eqn:rescale}
\widetilde{\phi} = a^{-1/2} \phi, \quad \widetilde{\nabla}^S = a \nabla^S, \quad \widetilde{\Omega}' = \sqrt{a} \Omega', \quad \widetilde{\Omega}'' = a \Omega''.
\eeq
Under these rescalings the action becomes
\beq
\label{eqn:lat-dimless-action}
\widetilde{\mcS} = \sum_{i = 0}^{T-1} \left[ \hf \left( \sum_{j = 0}^{T-1} \widetilde{\nabla}^S_{ij} \widetilde{\phi}_j + \widetilde{\Omega}'_i \right)^2 + \psib_i \sum_{j = 0}^{T-1} \left( \widetilde{\nabla}^S_{ij} + \widetilde{\Omega}''_{ij}  \right) \psi_j\right].
\eeq

\subsection{Theory on a lattice}
\label{sec:lattice-theory}

For convenience, we will not be using the tilde sign on the dimensionless variables; all variables and fields mentioned from now on are understood to be dimensionless. Physical quantities will be labeled differently.

The SUSY transformations are modified to contain the Wilson mass terms. For a given lattice site $k$ they are given by  
\beq
\label{eqn:lat-Wilson-Q}
\Q \phi_k = \psi_k, \quad \Q \psib_k = - N_k, \quad \Q \psi_k = 0, 
\eeq
and
\beq
\label{eqn:lat-Wilson-Qbar}
\Qb \phi_k = - \psib_k, \quad \Qb \psi_k = \overline{N}_k, \quad \Qb \hspace{0.05cm} \psib_k = 0,
\eeq
where 
\beq
N_k = \nabla^S{\phi}_k + \Omega'_k, \quad {\rm and} \quad \overline{N}_k = \nabla^S{\phi}_k - \Omega'_k.
\eeq
The supercharges satisfy the algebra
\beq
\label{eqn:lat-Q-Qb-algebra}
\{ \Q, \Q \} = 0, \quad \{ \Qb, \Qb \} = 0,\quad{\rm and }\quad \{ \Q, \Qb \} = 2 \nabla^S.
\eeq

The main obstacle that prevents the preservation of exact lattice SUSY is the failure of the Leibniz rule for lattice derivatives. Unlike the continuum action given in Eq. \eqref{eqn:act-cont}, the lattice regularized action
\beq
\label{eqn:lat-action}
\mcS = \sum_{i = 0}^{T-1} \left[\hf \left( \sum_{j = 0}^{T-1} \nabla^S_{ij} \phi_j + \Omega'_i \right)^2 + \psib_i \sum_{j = 0}^{T-1} \left( \nabla^S_{ij} + \Omega''_{ij} \right) \psi_j \right]
\eeq
preserves only the $\Q$ supercharge. The $\Qb$ SUSY is broken for $T \geq 2$. It can also be shown that the action is only $\Q$ invariant. That is, $\Q \mcS = 0 \neq \Qb \mcS$.

The $\Qb$ SUSY is broken for finite lattice size $T$ because it is not possible to define a corresponding $\Qb$ invariant transformation on lattice variables such that the algebra $\{ \Q, \Qb \} = 2 \nabla^S$ still holds \cite{Kanamori:2007yx}. However, the $\Q$-exactness is essential and sufficient to kill any SUSY-breaking counter-terms and thereby suppress lattice artifacts. See Refs. \cite{Catterall:2000rv, Catterall:2003wd, Giedt:2004vb, Kuroki:2009yg} for more discussions on this.

As mentioned in Sec. \ref{cont-susy-mechanics} for the continuum theory, when SUSY is broken, the partition function vanishes. In that case, the expectation values of the observables normalized by the partition function could be ill-defined. To overcome this difficulty in our lattice regularized theory, we will apply periodic boundary conditions for bosons, and twisted boundary conditions for fermions \cite{Kuroki:2009yg, Kuroki:2010au}. 

Introducing the twist, we have $\phi_T = \phi_0, \quad \psi_T = e^{ i \alpha } \psi_0, \quad \psib_T = e^{ - i \alpha } \psib_0$. For the case of dynamically broken SUSY, when $\alpha = 0$, the Witten index vanishes, which dictates that the fermion determinant changes its sign depending on the boson field configurations. This is the sign problem in models with dynamical SUSY breaking.

The partition function given in Eq. \eqref{eqn:cont-pf} takes the following form 
\bea
\label{eqn:lat-pf}
Z_\alpha = \left( \frac{1}{\sqrt{2 \pi}} \right)^T  \int \left( \prod_{k = 0}^{T-1} d\phi_k d\psi_k d\psib_k \right) e^{ - \mcS_\alpha },
\eea
where $\mcS_\alpha$ is the lattice regularized action that respects the twisted boundary conditions. 
We have
\beq
\label{eqn:lat-action-r1}
\mcS_\alpha = \sum_{i = 0}^{T-1} \hf \bigg( \phi_i - \phi_{i-1} + m \phi_i + W'_i \bigg)^2 + \sum_{i = 0}^{T-1} \psib_i \bigg( \psi_i - \psi_{i-1} + \left( m + W''_{ii} \right) \psi_i \bigg).
\eeq
Let us absorb the mass term into the potential $W$, and define a new potential $\Xi$ as
\beq
\label{eqn:lat-anho-pot}
\Xi \equiv \hf m \phi^2 + W.
\eeq
The action with twisted boundary conditions now takes the form
\bea
\label{eqn:lat-action-r1-Xi}
\mcS_\alpha &=& \sum_{i = 0}^{T-1} \hf \bigg( \sum_{j = 0}^{T-1}  \nabla^{-}_{ij} \phi_j + \Xi'_i  \bigg)^2 + \sum_{i = 0}^{T-1} \psib_i  \bigg(\sum_{j = 0}^{T-1}  \nabla^{-}_{ij}  + \Xi^{''}_{ij} \bigg) \psi_j.  
\eea
Also the expressions for $N_i$ and $\overline{N}_i$ become
\bea
N_i = \sum_{j = 0}^{T-1} \nabla^S_{ij} \phi_j + \Omega'_i = \sum_{j = 0}^{T-1} \nabla^{-}_{ij} \phi_j + \Xi'_i, \\
\overline{N}_i =\sum_{j = 0}^{T-1} \nabla^S_{ij} \phi_j - \Omega'_i = \sum_{j = 0}^{T-1} \nabla^+_{ij} \phi_j - \Xi'_i. 
\eea

After integrating out fermions, the fermionic contribution to the partition function given in Eq. \eqref{eqn:lat-pf} has the form
\bea
\label{eqn:z-alpha-F}
Z_\alpha^F &=& \prod_{k = 0}^{T-1} \left(1+ \Xi^{''}_{kk} \right) - e^{i \alpha}.
\eea
This is nothing but the determinant of the twisted Wilson fermion matrix $\mathcal{W}_{\alpha}^F$
\beq
Z_\alpha^F = \det \left[\mathcal{W}_\alpha^F \right].
\eeq
For periodic boundary conditions ($\alpha = 0$ case), this is in agreement with the expression obtained in Ref. \cite{Catterall:2000rv}.  The full partition function takes the form
\beq
\label{eqn:lat-pf-bos}
Z_\alpha = \left( \frac{1}{\sqrt{2\pi}} \right)^T \int \left(\prod_{k = 0}^{T-1} d\phi_k \right) ~ \exp \left[ - \mcS^{\rm ~eff}_\alpha \right],
\eeq
with
\bea
\label{eqn:lat-eff-action}
{\mcS_{\alpha}}^{\text{eff}} &=& {\mcS}^{B} - \ln \left( {\rm det} \left[ \mathcal{W}_\alpha^F \right] \right) \nn \\ 
&=& \sum_{k = 0}^{T-1} \hf \bigg( \phi_k - \phi_{k-1} + \Xi'_k  \bigg)^2 - \ln \left( \prod_{k = 0}^{T-1} \left(1 + \Xi^{''}_{kk} \right) - e^{i \alpha} \right).
\eea
Given an observable $\mathcal{O}$, we can compute its expectation value as
\bea
\label{eqn:lat-exp-obs}
\langle \mathcal{O} \rangle &=& \lim_{\alpha \to 0} \langle \mathcal{O} \rangle_\alpha \nn \\
&=& \lim_{\alpha \to 0} \frac{1}{Z_\alpha} \left( \frac{1}{\sqrt{2\pi}} \right)^T \int \left(\prod_{k = 0}^{T-1} d\phi_k \right) \mathcal{O} ~ \exp \left[{ - \mcS_{\alpha}^{\text{eff}}} \right].
\eea

Note that the gradient of the action has to be computed to update the field configurations in the complex Langevin method. The drift term, given by the negative of the gradient of the action, contains the fermion determinant in the denominator, whose zeroes in the complexified space cause the subtlety for the conditions required for the justification of the complex Langevin method. The dynamical variables may come close to the singularity of the drift term (the singular-drift problem). A recent study highlighted that such a problem is not restricted to logarithmic singularities but is rather generic and may arise where the stochastic process involves a singular-drift term \cite{Nishimura:2015pba}.

\subsection{Correlation functions}
\label{corr-fn}

Using the expression given in Eq. \eqref{eqn:lat-exp-obs} we can compute the correlation functions. The bosonic and fermionic correlation functions are defined as
\beq
G^B_\alpha (k) \equiv \langle \phi_0 \phi_k \rangle_\alpha,
\eeq
and
\beq
G^F_\alpha(k) \equiv \langle \psib_0 \psi_k \rangle_\alpha,
\eeq
respectively, at the site $k$.

The fermionic correlation function can be shown to be 
\bea
\langle \psib_0 \psi_k \rangle_\alpha &=& 
 \frac{1}{Z_\alpha}  \left(  \frac{1}{\sqrt{2 \pi}} \right)^T \int \left(\prod_{t=0}^{T-1} d\phi_t \right) \nn \\
&& \times \underbrace{\left\{\int \left(\prod_{t=0}^{T-1} d\psi_t d\psib_t \right)  \psib_{0} \psi_k \exp \left[-\sum_{t=0}^{T-1} \psib_t \bigg[ \left(1+ \Xi^{''}_{tt}\right) \psi_t  -\psi_{t-1} \bigg] \right] \right\}}_{{\langle \psib_{0} \psi_k \rangle}^{F}}  \nn \\
&& \times \exp\left[ -\sum_{t=0}^{T-1} \hf \bigg( \phi_t -\phi_{t-1} +\Xi'_t  \bigg)^2  \right] \\
&=&\frac{1}{Z_\alpha} \left( \frac{1}{\sqrt{2\pi}} \right)^T \int \left( \prod_{i = 0}^{T-1} d\phi_i \right) \underbrace{ \left( -\frac{{ \langle \psib_0 \psi_k \rangle^F} }{ \det \left[ \mathcal{W}_\alpha^F \right] } \right) }_{\left[ \psib_0 \psi_k \right]^L_\alpha} ~ \exp \left[ -{ \mcS_\alpha }^{\text{eff}} \right]
\eea
where we can integrate out the fermions and compute the fermionic part explicitly as
\beq
\langle \psib_{0} \psi_k \rangle^{F}
= -\left(\prod_{t=k+1}^{T-1} \left[1+ \Xi^{''}_{tt}\right]  \right). 
\eeq
Then, upon comparison with Eq. \eqref{eqn:lat-exp-obs}, we define $\left[ \psib_0 \psi_k \right]^L_\alpha$ as our Langevin observable corresponding to the fermionic correlator $\langle \psib_0 \psi_k \rangle_\alpha$. That is,
\beq
{\left[\psib_0 \psi_k \right]^L_\alpha} = \left(-  \frac{ \prod_{i = k + 1}^{T-1} \left[1 + \Xi^{''}_{ii} \right] }{ \prod_{i = 0}^{T-1} \left[ 1 + \Xi^{''}_{ii} \right] - e^{i \alpha}} \right).
\eeq

Now, for the bosonic correlation function, the computation is rather straightforward. The Langevin observable is the bosonic correlation function itself. For the $k$-th lattice site, ${ \left[ \phi_0 \phi_k \right]^L} = \phi_0 \phi_k $, such that
\bea
\langle \phi_0 \phi_k \rangle_\alpha = \frac{1}{ Z_\alpha } \left( \frac{1}{ \sqrt{2 \pi} } \right)^T \int \left( \prod_{i = 0}^{T-1} d\phi_i \right) \phi_0 \phi_k ~ \exp \left[ { - \mcS_\alpha }^{ \text{eff}} \right]. 
\eea

\subsection{Ward identities}
\label{ward-id}

Another set of observables that would help us in the investigations on SUSY breaking is the Ward identities. For the supersymmetric variation of the fields, Eqs. \eqref{eqn:lat-Wilson-Q} and \eqref{eqn:lat-Wilson-Qbar}, the invariance of the lattice action guides us to a set of Ward identities that connect the bosonic and fermionic correlators. The partition function in Eq. \eqref{eqn:lat-pf}, upon addition of the source terms  ($J, \chi, \overline{\chi}$), becomes
\bea
\label{eqn:lat-pf-source}
Z_\alpha \left( J, \chi, \overline{\chi} \right) &=& \left( \frac{1}{ \sqrt{2 \pi} } \right)^T \int \left( \prod_{k = 0}^{T-1} d\phi_k d\psi_k d\psib_k \right) \nn \\ 
&&\hspace{2cm} \times \exp \left[{ - \mcS_\alpha + \sum_{k = 0}^{T-1} \left( J_k \phi_k + \chi_k \psib_k + \overline{\chi}_k \psi_k \right) } \right].
\eea

It is easy to see that the variation of the partition function under the $\Q$-transformations vanishes upon turning off the external sources. That is, $\Q Z_\alpha \left( J, \chi, \overline{\chi} \right) = 0$. In fact, the variation of any derivative of the partition function with respect to these external source terms also vanishes (upon turning off the sources). Taking the derivative of the partition function with respect to the source terms $J_j$ and $\chi_i $, we get the following set of non-trivial supersymmetric Ward identities,
\beq
\langle \psib_i \psi_j \rangle + \langle N_i \phi_j \rangle = 0.
\eeq

We will consider
\beq
\label{eqn:lat-ward-iden}
\mathscr{W}_1 : 
\begin{array}{l}
	\langle \psib_0 \psi_k \rangle  +  \langle  N_0 \phi_k \rangle = 0
\end{array}
\eeq
to probe the SUSY breaking.

\section{Complex Langevin simulations}
\label{lattice-simulation}

Let us look at the relevant observables we used in the simulations. One crucial observable is the expectation value of the auxiliary field
\bea
\B_\alpha = -i \left( \nabla^S_{ij} \phi_j + \Omega'_i \right)  = -i \left( \nabla^{-}_{ij} \phi_j + \Xi'_i \right).
\eea

Studies have shown that the auxiliary field expectation value can be used as an order parameter to reliably predict dynamical SUSY breaking \cite{Kuroki:2009yg, Kuroki:2010au, Joseph:2019sof}. The non-vanishing (vanishing) nature of the auxiliary field indicates that SUSY is broken (preserved) in the system. That is,
\begin{equation}
	\langle \mathcal{B} \rangle =\lim_{\alpha \to 0} \langle \mathcal{B}_{\alpha} \rangle  
	\begin{cases}
		\neq 0 & \text{SUSY broken} \\
		=0 & \text{SUSY preserved}. 
	\end{cases}
\end{equation}
However, this vanishing nature could be accidental for some models. In \cite{Kuroki:2009yg}, higher powers of $\B$ were considered to confirm SUSY breaking for zero-dimensional models. Thus, we also analyze other significant observables to confirm SUSY-breaking predictions.

We next consider the bosonic action $\mcS^B_\alpha$. It has been studied that for exact lattice SUSY, the expectation value of the bosonic action is independent of the interaction couplings \cite{Catterall:2001fr}. Thus, the bosonic action expectation value simply counts the number of degrees of freedom on the lattice \cite{Catterall:2001fr, Catterall:2009it}. That is, $\langle \mathcal{S}^B \rangle = \hf N_{\rm d.o.f}$. In supersymmetric quantum mechanics, it is expected that $\langle \mcS \rangle = T$, and $\langle \mcS^B \rangle = T/2$,  where $T$ is the number of sites. Thus we have
\beq
\langle \mcS^B \rangle =\lim_{\alpha \to 0} \langle \mathcal{S}^B_\alpha \rangle
\begin{cases}
	\neq {T}/{2} & \text{SUSY broken} \\
	=T/2 & \text{SUSY preserved}. 
\end{cases}
\eeq

The third indicator is the equality of the fermionic and bosonic mass gaps. The mass gaps can be extracted either by a $\cosh \big[ m a( t - \frac{T}{2}) \big]$ fit for the $t$-th lattice site, or a simple exponential fit over say, the first or last $T/4$ time slices of the respective correlation functions \cite{Catterall:2001fr, Giedt:2004vb}.

The last set of observables involves the Ward identity. They can be used to confirm the exact lattice SUSY successfully. See Refs. \cite{Catterall:2000rv, Catterall:2001fr, Catterall:2003wd, Kadoh:2018ele}. We expect that $\mathscr{W}_1$, given in Eq. \eqref{eqn:lat-ward-iden}, to hold (not to hold) for theories with SUSY preserved (broken). That is,
\beq
\lim_{\alpha \to 0} \mathscr{W}_1:
\begin{cases}
	-\langle  \psib_0 \psi_k \rangle_\alpha  \neq \langle  N_0  \phi_k \rangle_\alpha & \text{SUSY broken} \\
	-\langle  \psib_0 \psi_k \rangle_\alpha = \langle  N_0  \phi_k \rangle_\alpha  &  \text{SUSY preserved}. 
\end{cases}
\eeq

Only in the limit $\alpha \to 0$ can we comment, using the above set of observables, if the system possesses exact lattice SUSY. Since the partition function is a well-defined quantity for models with SUSY preserved, as expected, we were able to compute the normalized expectation values of observables, and hence perform numerical investigations for $\alpha = 0$ (PBC) case. The issue in working without the twist field arises only in models where SUSY is spontaneously broken, since the partition function vanishes, and normalized expectation values of the observables are ill-defined. Hence in Sec. \ref{subsec:gen-sps}, we perform complex Langevin simulations for various values of the twist parameter to verify the consistency of our results for the $\alpha = 0$ case. 

\subsection{Supersymmetric anharmonic oscillator}
\label{subsec:susy-anho}

The model we consider in this section, the supersymmetric anharmonic oscillator, has a real action. Our goal is to put the simulation code to the test by comparing our results with those given in Ref. \cite{Catterall:2000rv}.

The model has the potential 
\beq
\label{eqn:sim-anho}
\Xi(\phi) = \hf m \phi^2 + \qtr g \phi^4.
\eeq
This model has been investigated in great detail with the help of the Hybrid Monte Carlo (HMC) algorithm in the lattice and non-lattice formalisms, respectively, in Refs. \cite{Catterall:2000rv} and \cite{Hanada:2007ti}. It was concluded that SUSY is preserved in this model for a finite value of the coupling. In the case with $\alpha = 0$, the action is real, and the evolution of field configurations in the system is governed by real Langevin dynamics. 

\begin{figure}[tbp]
	\begin{center}
		{
			
			\includegraphics[width=.7\textwidth,origin=c,angle=0]{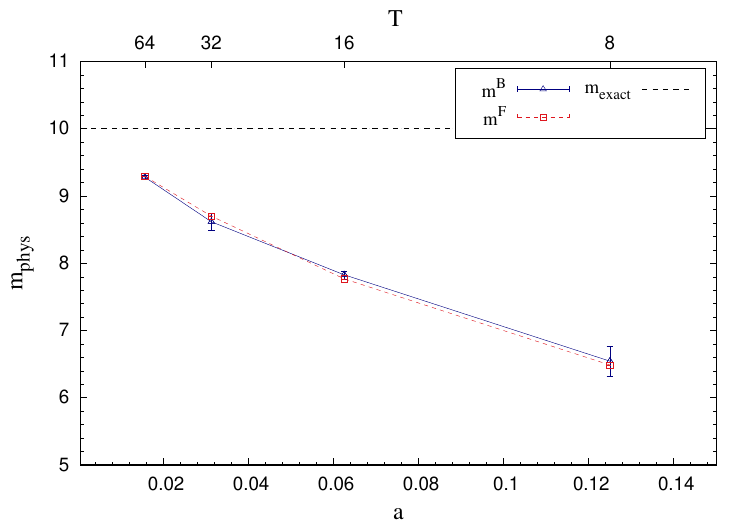}
			
			\caption{Bosonic and fermionic mass gaps for SUSY harmonic oscillator with physical parameters $m_{\rm phys} = 10$ and $g_{\rm phys} = 0$ versus lattice spacing ($a$) and lattice size ($T$).}
			\label{fig:lat-susy-ho-m10g0-massgaps}	
		}
	\end{center}
\end{figure}

First, we simulate SUSY harmonic oscillator for physical parameters $m_{\rm phys} = 10$ and $g_{\rm phys} = 0$, and $\alpha = 0$. Simulations were performed for different lattice spacings keeping the (physical) circle size $\beta = 1$. Figure \ref{fig:lat-susy-ho-m10g0-massgaps} shows bosonic (blue triangle) and fermionic (red square) physical mass gaps versus lattice spacing ($a$) and lattice size ($T$). The Black dashed line shows the continuum value of SUSY harmonic oscillator mass gaps for the physical parameters $m_{\rm phys} = 10$, $g_{\rm phys} = 0$, that is, $m_{\rm exact} = 10$. We see that boson and fermion masses are degenerate within statistical errors, and furthermore, as the lattice spacing $a \to 0$, the common mass gap approaches the correct continuum value. The simulations confirm that the free action has an exact SUSY at finite lattice spacing, which is responsible for the degenerate mass gaps. 

Now we simulate the SUSY anharmonic oscillator for physical parameters $m_{\rm phys} = 10$ and $g_{\rm phys} = 100$ and $\alpha = 0$. Simulations were performed for different lattice spacings keeping the (physical) circle size $\beta = 1$. In Table \ref{tab:lat-susy-anho-m10g100}, we provide the bosonic and fermionic mass gaps. Here also, we have $m^B_{\text{phys}} \approx m^F_{\text{phys}}$ indicating that SUSY is preserved in the model. Table \ref{tab:lat-susy-anho-m10g100-action-B} contains the expectation values of the auxiliary field $\B_{\alpha}$ and bosonic action $\mcS^B_{\alpha}$. The mean expectation value $\langle \B \rangle $ vanishes in the simulations and thus indicates exact lattice SUSY. This Table also contains the expectation value of the bosonic action $\mcS^B_{\alpha}$. For this model, we observe $\langle \mcS^B \rangle = \hf T$, and it was independent of physical parameters $g_{\rm phys}$ and $m_{\rm phys}$, which again suggests SUSY is preserved in this model. 

\begin{figure}[tbp]
	\centering
	\includegraphics[width=.75\textwidth,origin=c,angle=0]{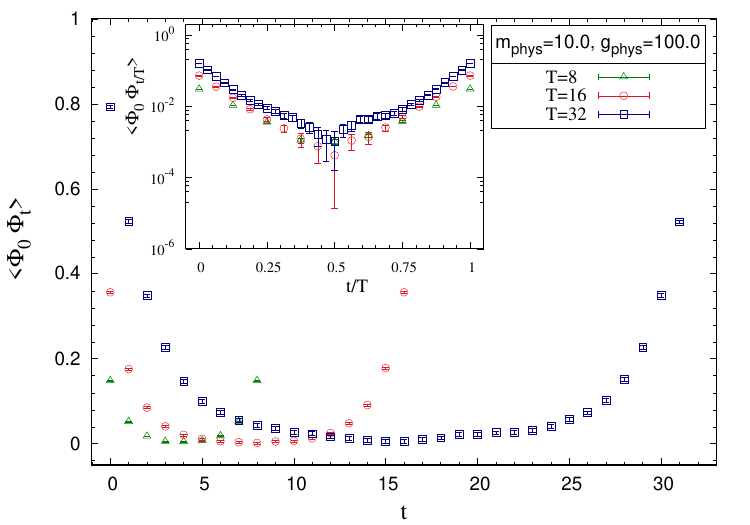}\vsphf
	
	\includegraphics[width=.75\textwidth,origin=c,angle=0]{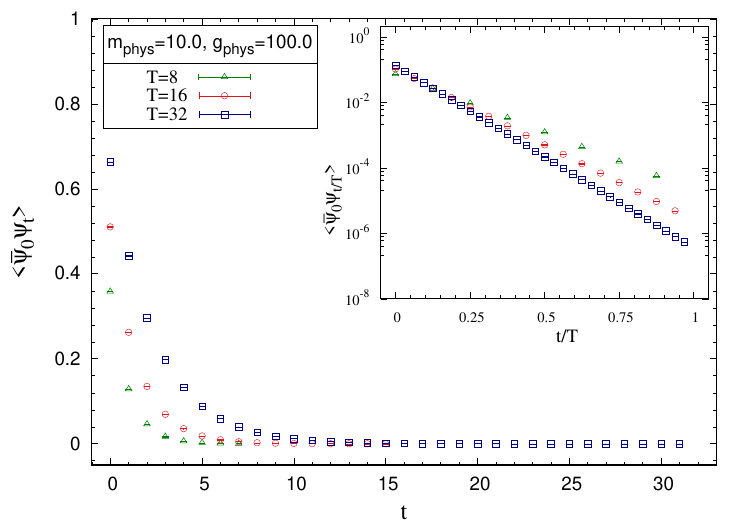}
	
	\caption{Bosonic (top) and fermionic (bottom) correlation functions for SUSY anharmonic oscillator. The parameters used are $m_{\rm phys} = 10$ and $g_{\rm phys} = 100$.}
	\label{fig:lat-susy-anho-m10g100-corr}	
\end{figure}

\begin{figure*}[tbp]
	\centering
	\includegraphics[width=.75\textwidth,origin=c,angle=0]{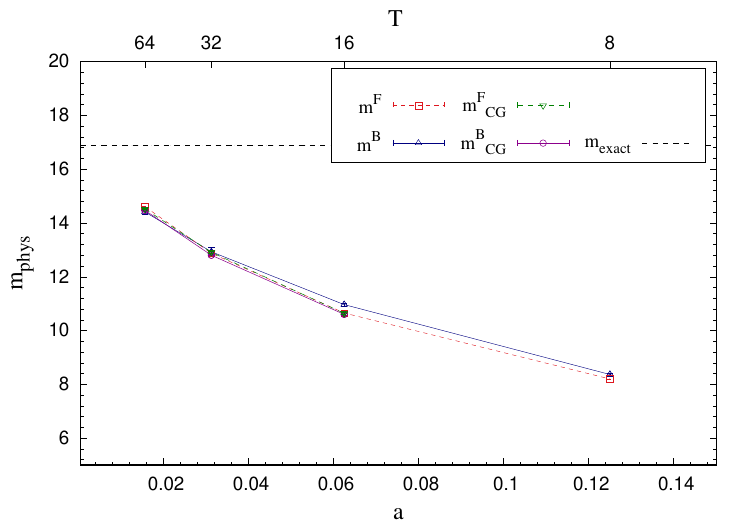}
	
	\caption[Bosonic and fermionic mass gaps for SUSY anharmonic oscillator. The parameters used are $m_{\rm phys} = 10$ and $g_{\rm phys} = 100$.]{Bosonic and fermionic mass gaps for SUSY anharmonic oscillator. The parameters used are $m_{\rm phys} = 10$ and $g_{\rm phys} = 100$. Here $m^{B}_{\rm CG}$ and $m^{F}_{\rm CG}$, respectively represent bosonic and fermionic mass-gap results from Ref. \cite{Catterall:2000rv}.}
	\label{fig:lat-susy-anho-m10g100-massgaps}	
\end{figure*}

\begin{figure*}[tbp]
	\centering
	\includegraphics[width=.48\textwidth,origin=c,angle=0]{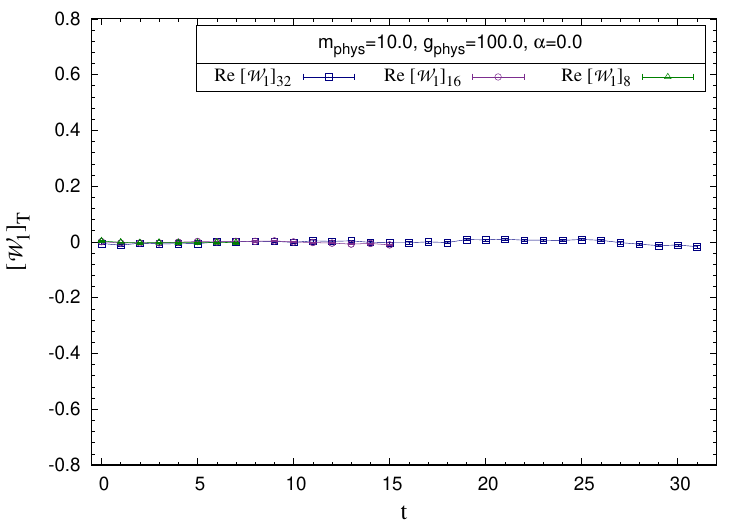}
	\includegraphics[width=.48\textwidth,origin=c,angle=0]{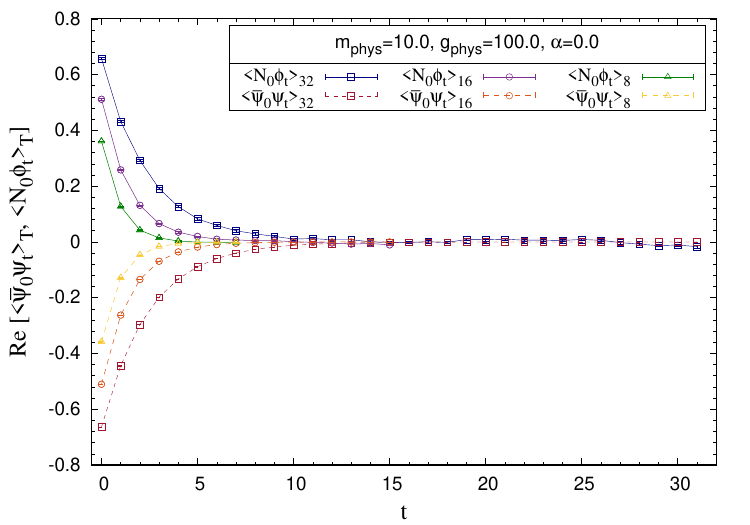}
	
	\caption[Real part of the Ward identity (left) and real part of the bosonic and fermionic contributions to the Ward identity (right), versus lattice site $t$ for lattices with $T$.]{Real part of the Ward identity (left) and real part of the bosonic and fermionic contributions to the Ward identity (right), versus lattice site $t$ for lattices with $T$. Simulations were performed for SUSY anharmonic oscillator with physical parameters $m_{\rm phys} = 10$ and $g_{\rm phys} = 100$ for lattice sizes $T = 8, 16$, and $32$.}
	\label{fig:lat-susy-anho-m10g100-ward}	
\end{figure*}

Figure \ref{fig:lat-susy-anho-m10g100-corr} shows the bosonic (top) and fermionic (bottom) correlation functions (used to compute the respective mass gaps) versus lattice site ($t$) for various lattice size ($T$) for the SUSY anharmonic oscillator. In Fig. \ref{fig:lat-susy-anho-m10g100-massgaps}, we show the bosonic and fermionic physical mass gaps versus lattice spacing ($a$) and lattice size ($T$). Here we also compare our results with those obtained by Simon Catterall and Eric Gregory \cite{Catterall:2000rv}, ($m^{B}_{\rm CG}$, $m^{F}_{\rm CG}$), and find excellent agreement. The Black dashed line shows the continuum value of SUSY anharmonic oscillator mass gaps for the physical parameters $m_{\rm phys} = 10$, $g_{\rm phys} = 100$ that is $m_{\rm exact}$ = 16.865 \cite{Bergner:2007pu}. We see that boson and fermion masses are degenerate within statistical errors, and furthermore, as the lattice spacing $a \to 0$, the common mass gap approaches the correct continuum value. The exact lattice SUSY and approaching correct continuum limit results are nontrivial, and this becomes apparent when compared to the results of a naive discretization of the continuum action. See Ref. \cite{Catterall:2000rv} for a detailed comparison. In naive discretization, the extracted mass gaps differ widely at finite lattice spacing. As we reduce lattice spacing, they diverge and do not approach the correct continuum limit implying that the quantum continuum limit is not supersymmetric.  In Fig. \ref{fig:lat-susy-anho-m10g100-ward}, we plot the real part of Ward identity $\mathscr{W}_1$  (left), and its bosonic and fermionic contributions (right), given in Eq. \eqref{eqn:lat-ward-iden}, versus the lattice site $t$ for lattices with $T$ values. We observe that the respective bosonic and fermionic contributions cancel each other out within statistical errors, and hence $\mathscr{W}_1$ is satisfied. Our results confirm that the SUSY anharmonic oscillator has an exact SUSY, which is responsible for the degenerate mass gaps. 

\begin{table}[tbp]
	\begin{center}
		{\small	
			\begin{tabular}{|l	r|	l	r	|l	r|} 
				\hline
				$T$ &  $a = T^{-1}$  &  $m^B$   &  $m^B_{\text{phys}} = a^{-1} m^B$&  $m^F$   &  $m^F_{\text{phys}} = a^{-1} m^F$  \\	
				\hline
				\hline
				8   & 0.125   	 & $1.0457(65)$	  &$8.3656(520)$	&   $1.0247(4)$	  &$8.1976(32)$    	\\
				16  & 0.0625   	 & $0.6852(45)$  &  $10.9632(720)$	& $0.6657(1)$	  &$10.6512(16)$	   \\
				32  & 0.03125  	 & $0.4040(54)$  & $12.9280(1664)$	 & $0.4023(2)$	  &$12.8736(64)$	\\
				64  & 0.015625 	 & $0.2252(13)$  &  $14.4128(832)$	& $0.2282(3)$	  &$14.6048(192)$	\\
				\hline
		\end{tabular}}
		\caption{Bosonic and fermionic mass gaps for SUSY anharmonic oscillator. The parameters used are $m_{\rm phys} = 10$ and $g_{\rm phys} = 100$.}
		\label{tab:lat-susy-anho-m10g100}
	\end{center}
\end{table}

\begin{table}[tbp]
	\begin{center}
		{\small
			\begin{tabular}{|c |l	r|	c|	l	r|} 
				\hline
				$\Xi'(\phi)$	&$T$ &  $a = T^{-1}$  &  \makecell{ $\alpha$ } &  $~\langle \B_{\alpha} \rangle$   &  $~\langle \mathcal{S}^B_{\alpha} \rangle$ \\   
				\hline
				\hline
				$m \phi + g \phi^3$
				&$8$ &  $0.1250$  & $0.00$ &  $0.0(0) -i0.0008(38)$ & $4.0672(67) + i 0.0(0)$ \\
				&$16$ &  $0.0625$  & $0.00$ &  $0.0(0) +i0.0003(68)$ & $8.0698(95) + i 0.0(0)$ \\
				&$32$ &  $0.03125$  & $0.00$ &  $0.0(0) -i0.0038(131)$ & $16.1589(147) + i 0.0(0)$ \\
				&$64$ &  $0.015625$  & $0.00$ &  $0.0(0) -i0.0162(245)$ & $32.2293(252) + i 0.0(0)$ \\
				\hline
		\end{tabular}}
		\caption{Expectation value of the auxiliary field $\B_\alpha$ and the bosonic action $\mcS^B_\alpha$ for SUSY anharmonic oscillator. The parameters used are $m_{\rm phys} = 10$ and $g_{\rm phys} = 100$.}
		\label{tab:lat-susy-anho-m10g100-action-B}
	\end{center}
\end{table}

\subsection{General polynomial potential}
\label{subsec:gen-sps}

As a check of our code, we consider the model with a degree-$k$ polynomial potential with real coefficients
\beq
{\Xi{'}}^{(k)} = g_k \phi^k + g_{k-1} \phi^{k-1} + \cdots + g_0.
\eeq
In these systems, it is well known that SUSY is spontaneously broken if the count of zeroes of the potential is even, that is, when ${\Xi}^{(k)}(-\infty)$ and ${\Xi}^{(k)}(+\infty)$ have opposite signs \cite{Witten:1981nf}. For simplicity, we assume the form $g_k = g$, $g_{k-1} = \cdots = g_2 = 0$, $g_1 = m$ and $g_0 = g \mu^2$. Then, for degree $k = 4$ and $5$, we have
\bea
\label{eqn:lat-gen-pot-k4}
{\Xi{'}}^{(4)} &=& g \phi^4 + m \phi + g \mu^2, \\
\label{eqn:lat-gen-pot-k5}
{\Xi{'}}^{(5)} &=& g \phi^5 + m \phi + g \mu^2.
\eea

Using the complex Langevin method, we confirm the analytical prediction that SUSY is broken for even-($k = 4$) and preserved for odd-($k = 5$) degree real-polynomial potentials.

For even-degree potential, we encounter {\it singular-drift problem} when $\alpha = 0$. This shows up in the simulations as the complex Langevin correctness criteria involving the probability of absolute drift not being satisfied. For $\alpha = 0$, we obtained a power law decay instead of an exponential decay. Therefore we have performed simulations for various non-zero $\alpha$. Our simulations suggest that above a particular $\alpha$ value, the correctness criteria are satisfied, and the probability of absolute drift falls off exponentially. 

\begin{figure*}[tbp]
	\centering
	\includegraphics[width=.48\textwidth,origin=c,angle=0]{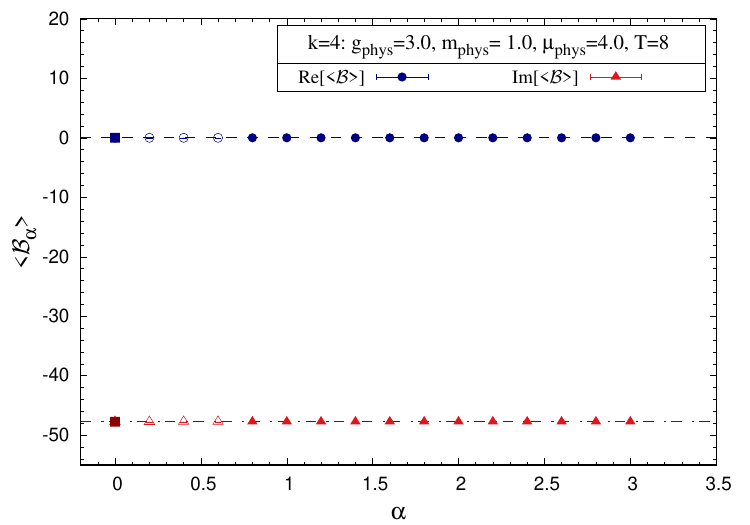}  \includegraphics[width=.48\textwidth,origin=c,angle=0]{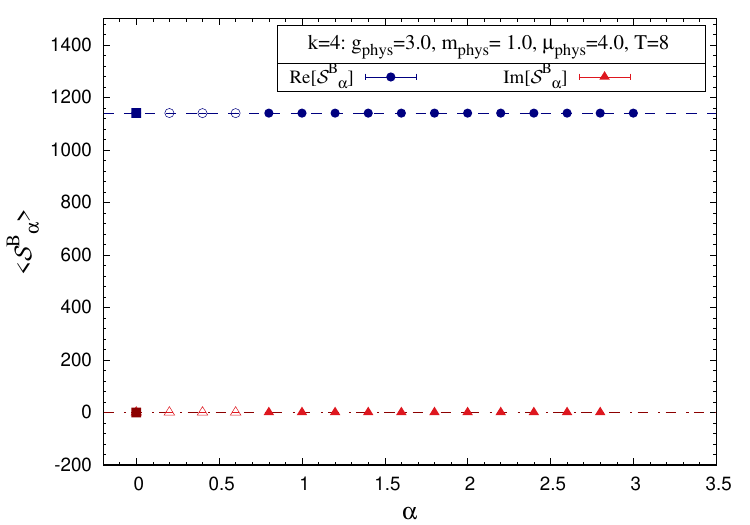}
	
	\caption[Expectation values of auxiliary field $\B_\alpha$ (left) and the bosonic action $\mathcal{S}^B_\alpha$ (right) for various $\alpha$ values on a lattice with $T= 8$. The simulations are for the model with even-degree real polynomial potential ${\Xi{'}}^{(4)}$ given in Eq. \eqref{eqn:lat-gen-pot-k4}.]{Expectation values of the auxiliary field $\B_\alpha$ (left) and the bosonic action $\mathcal{S}^B_\alpha$ (right) for various $\alpha$ values on a lattice with $T= 8$. The simulations are for the model with even-degree real polynomial potential ${\Xi{'}}^{(4)}$ given in Eq. \eqref{eqn:lat-gen-pot-k4}. The parameters used are $g_{\rm phys} = 3$, $m_{\rm phys} = 1$, and $\mu_{\rm phys} = 4$. 
	}
	\label{fig:lat-susy-k4-b-sb}	
\end{figure*}

In Fig. \ref{fig:lat-susy-k4-b-sb} we show the expectation value of the auxiliary field $\langle \B_\alpha \rangle$ (left) and the bosonic action $\langle \mcS^{B}_\alpha \rangle$ (right) on a $T = 8$ lattice for various $\alpha$ values. We consider the parameter space where complex Langevin simulations are justified and take the limit $\alpha \to 0$. The filled data points (red triangles for the imaginary part and blue circles for the real part) represent expectation values of observables for parameter space where complex Langevin can be trusted, while for unfilled data points, complex Langevin correctness criteria are not satisfied. The lines represent a linear fit of observables for parameter space where complex Langevin simulations are justified, and solid squares represent values of respective observables in the limit $\alpha \to 0$. These results indicate that $\langle \B  \rangle$ does not vanish. The expectation value of the bosonic action $\langle \mcS^B \rangle \neq \hf T$ and is found to be dependent on the physical parameters. These results indicate that SUSY is broken for the even-degree real-polynomial superpotential. 

\begin{figure*}[tbp]
	\centering
	
	\includegraphics[width=.48\textwidth,origin=c,angle=0]{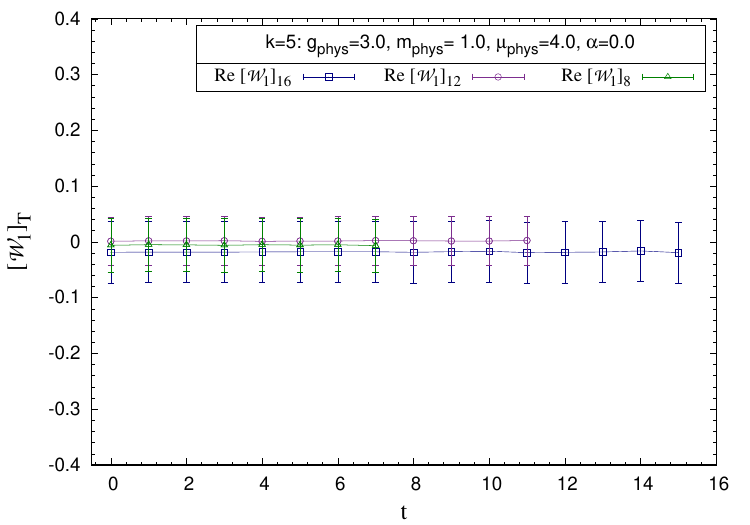}
	\includegraphics[width=.48\textwidth,origin=c,angle=0]{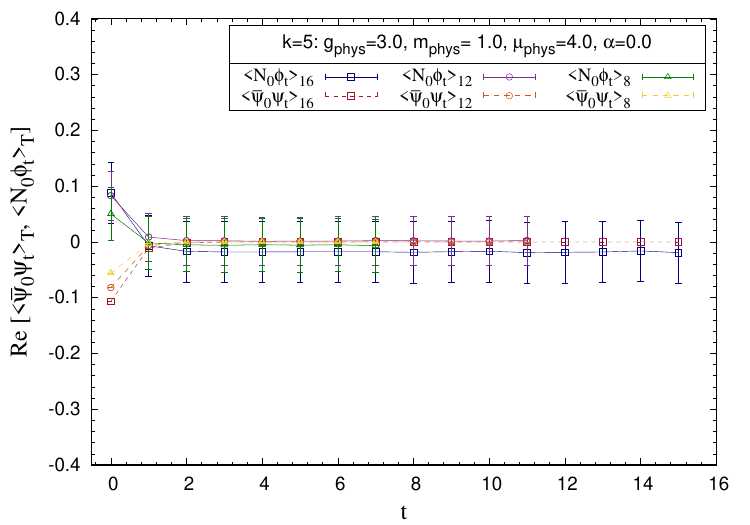}
	
	\caption[Real part of the Ward identity (left) and real part of bosonic and fermionic contributions to the Ward identity (right) for the model with odd-degree real polynomial potential.]{Real part of the Ward identity (left) and real part of bosonic and fermionic contributions to the Ward identity (right) for the model with odd-degree real polynomial potential. The parameters used are $g_{\rm phys} = 3$, $m_{\rm phys} = 1$, and $\mu_{\rm phys} = 4.0$ on lattices with $T = 8, 12$, and $16$.}
	\label{fig:k5ward}	
\end{figure*}

For the model with odd-degree potential, we observe that for $\alpha = 0$ the complex Langevin correctness criteria are satisfied. In Table \ref{tab:k5bSb} we provide the simulation results for physical parameter $g_{\rm phys} = 3,~m_{\rm phys} = 1~ {\rm and}~ \mu_{\rm phys} = 4.0$, on lattices with $T = 8, 12,16$, and $\alpha = 0$. Our simulations gave a vanishing value for the auxiliary field expectation value $\langle \B \rangle$. We also observed that the expectation value of bosonic action is $\langle \mcS^B \rangle = \hf T$ within errors. It is also independent of the coupling $g$. These results indicate that SUSY is preserved for these models. In Fig. \ref{fig:k5ward}, we plot the Ward identities for this model. In these plots, on the left, we show the complete Ward identity (real part), and on the right, we present the respective bosonic and fermionic contributions to the Ward identity. For this model, the bosonic and fermionic contributions cancel each other out within statistical errors, and the Ward identity is thus satisfied. These results indicate unbroken SUSY in this model.

\begin{table}[tbp]
	\begin{center}
		{\small
			\begin{tabular}{|c| l r| c c|} 
				\hline
				$\Xi'(\phi)$&$T$ &  $a = T^{-1}$ &    $~\langle \B_{\alpha} \rangle$   &  $~\langle \mcS^B_{\alpha} \rangle$ \\   
				\hline
				\hline
				&${8}$	&${0.25}$	&
				$0.0000(0) 	+ i0.0042(101)$ & $4.0214(202)+ i0.0000(0)$ \\	
				${\Xi{'}}^{(5)}$
				&${12}$ &${0.125}$	
				& $0.0000(0) 	+ i0.0044(74)$ & $5.9942(176)+ i0.0000(0)$ \\ 	
				&${16}$	&${0.0833}$ &
				$0.0000(0) 	- i0.0026(86)$ & $8.0398(210)+ i0.0000(0)$\\
				\hline
			\end{tabular}
		}
		\caption{Expectation value of the auxiliary field $\B_{\alpha}$ and the bosonic action $\mcS^B_{\alpha}$ for the odd-degree superpotential ${\Xi{'}}^{(5)}$. The parameters used are $g_{\rm phys} = 3$, $m_{\rm phys} = 1$, and $\mu_{\rm phys} = 4.0$.}
		\label{tab:k5bSb}
	\end{center}
\end{table}

\subsection{$\mathcal{PT}$-symmetric models}
\label{subsec:pt-sps}

In this section, we simulate an interesting class of complex actions that exhibit, in addition to SUSY, $\mathcal{PT}$-symmetry. 

Quantum mechanics and quantum field theory are conventionally formulated using Hermitian Hamiltonians and Lagrangians, respectively. In recent years there has been an increasing interest in extensions to non-Hermitian quantum theories \cite{Bender:2002vv}, particularly those with $\mathcal{PT}$-symmetry \cite{Bender:1998ke, Bender:2005tb}, which have real spectra. Such theories have found applications in many areas, such as optonics \cite{PhysRevLett.105.013903, Longhi_2017} and phase transitions \cite{Ashida_2017, Matsumoto_2020}. Recently, it has been shown that it is possible to carry over the familiar concepts from Hermitian quantum field theory, such as the spontaneous symmetry breaking and the Higgs mechanism in gauge theories, to $\mathcal{PT}$-symmetric non-Hermitian theories \cite{PhysRevD.98.045001, PhysRevD.99.045006, FRING2020114834}. In Ref. \cite{Alexandre:2020wki}, the authors constructed $\mathcal{PT}$-symmetric $\cN = 1$ supersymmetric quantum field theories in $3+1$ dimensions. There, they found that even though the construction of the models is explicitly supersymmetric, they offer a novel non-Hermitian mechanism for soft SUSY breaking.

Imposing $\mathcal{PT}$-symmetric boundary conditions on the functional-integral representation of the four-dimensional $- \lambda \phi^4$ theory can give a spectrum that is bounded below \cite{Bender:1999ek}. Such an interaction leads to a quantum field theory that is perturbatively renormalizable and asymptotically free, with a real and bounded spectrum. These properties suggest that a $- \lambda \phi^4$ quantum field theory might be useful in describing the Higgs sector of the Standard Model. 

We hope that our investigations will serve as a starting point for exploring the non-perturbative structure of these types of theories in higher dimensions.

The models we consider here have the following potential 
\beq
\label{eqn:lat-pt-symm-pot}
\Xi(\phi) = - \frac{g}{(2 + \delta)} \left(i \phi \right)^{\left(2 + \delta \right)},
\eeq
with $\Xi^{'}(\phi) = - i g ~ (i \phi)^{(1 + \delta)}$ and $\delta$ is a continuous parameter. The supersymmetric Lagrangian for this $\mathcal{PT}$-symmetric theory breaks parity symmetry, and it would be interesting to ask whether the breaking of parity symmetry induces a breaking of SUSY. This question was answered for the case of a two-dimensional model in Ref. \cite{Bender:1997ps}. There, through a perturbative expansion in $\delta$, the authors found that SUSY remains unbroken in this model. We investigate the absence or presence of non-perturbative SUSY breaking in the one-dimensional cousins of these models using the complex Langevin method. Clearly, such an investigation based on path integral Monte Carlo fails since the action of this model can be complex, in general\footnote{In Ref. \cite{Dhindsa:2020ovr} it was shown that $\mathcal{PT}$ symmetry is preserved in supersymmetric quantum mechanics models with $\delta = 0, 2, 4$ using Monte Carlo simulations. See Refs. \cite{Kadoh:2015zza, Kadoh:2018ivg, Kadoh:2018ele, Kadoh:2019bir} for other related work on supersymmetric quantum mechanics on the lattice.}.

\subsection*{Even $\delta$ case}

We note that when $\delta = 0$, the model becomes the supersymmetric harmonic oscillator discussed in Sec. \ref{subsec:susy-anho}. 

In Table \ref{tab:d2-4bSb} we provide the simulation results for $\delta = 2,~4$. Simulations were performed for physical parameter $g_{\rm phys} = 0.5$, on lattices with $T = 4, 8, 12$, and $\alpha = 0$. We noticed that the auxiliary field expectation value $\langle \B \rangle$ vanishes. We also observe that the expectation value of bosonic action is $\langle \mcS^B \rangle = \hf T$ within errors. It is also independent of the coupling $g$. These results indicate that SUSY is preserved in these models.

\begin{table}[tbp]
	\begin{center}
		{\small
			\begin{tabular}{|c| l r| c c|} 
				\hline
				$\Xi'(\phi)$&$T$ &  $a = T^{-1}$ &    $~\langle \B_{\alpha} \rangle$   &  $~\langle \mcS^B_{\alpha} \rangle$ \\   
				\hline
				\hline
				&${4}$	&${0.25}$	&
				$0.0000(0) 	+ i0.0005(282)$ & $2.0130(102)+ i0.0000(0)$ \\	
				$\delta = 2$
				&${8}$ &${0.125}$	
				& $0.0000(0) 	+ i0.0128(750)$ & $4.0326(157)+ i0.0000(0)$ \\ 	
				&${12}$	&${0.0833}$ &
				$0.0000(0) 	- i0.0071(263)$ & $6.0354(58)+ i0.0000(0)$\\
				\hline
				&${4}$	&${0.25}$	&
				$0.0000(0) 	+ i0.0167(679)$ & $1.9975(47)+ i0.0000(0)$ \\		
				$\delta = 4$
				&${8}$ &${0.125}$	
				& $0.0000(0) 	+ i0.0142(567)$ & $4.0058(54)+ i0.0000(0)$ \\	
				&${12}$	&${0.0833}$ &
				$0.0000(0) 	- i0.0309(1022)$ & $6.0018(74)+ i0.0000(0)$ \\ 
				\hline
			\end{tabular}
		}
		\caption{Expectation values of the auxiliary field $\B_\alpha$ and the bosonic action $\mcS^B_{\alpha}$ for the $\mathcal{PT}$-symmetric models with $\delta = 2, 4$.}
		\label{tab:d2-4bSb}
	\end{center}
\end{table}

\begin{figure*}[tbp]
	\centering	
	\includegraphics[width=.6\textwidth,origin=c,angle=0]{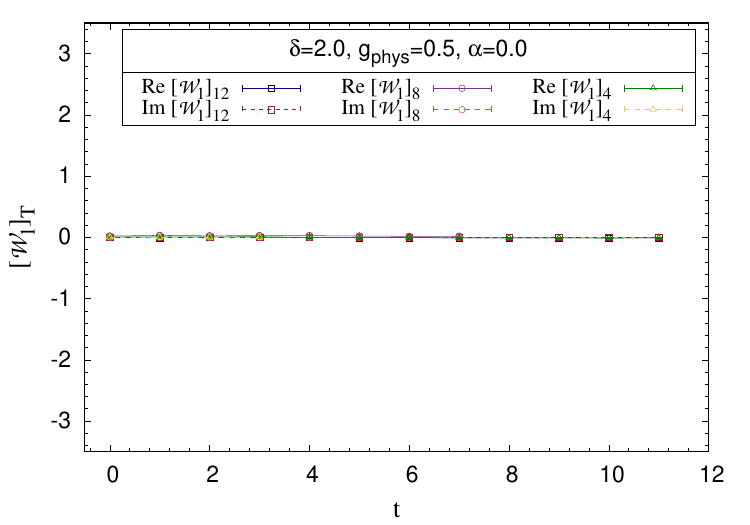}\\
	\includegraphics[width=.6\textwidth,origin=c,angle=0]{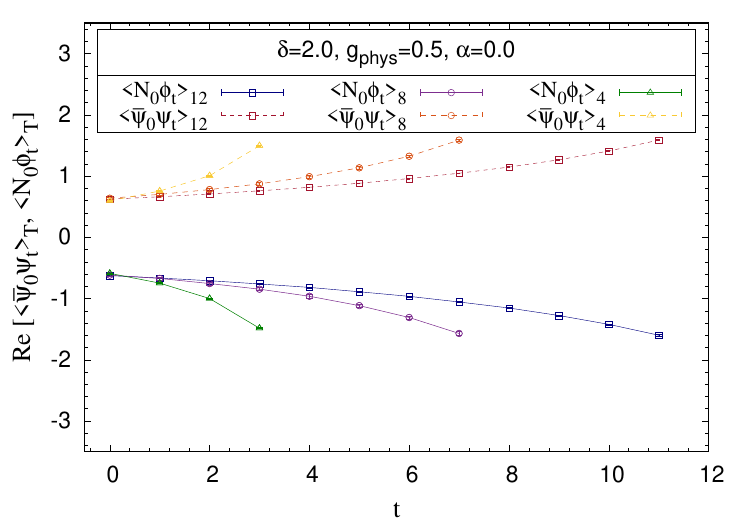}
	\includegraphics[width=.6\textwidth,origin=c,angle=0]{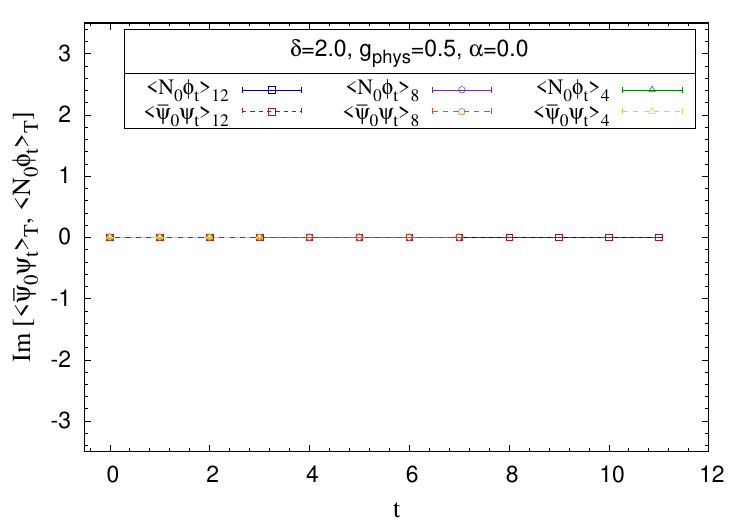}
	
	\caption {The  $\mathcal{PT}$-symmetric model with $\delta = 2$. The full Ward identity (top) and real (middle) and imaginary (bottom) parts of bosonic and fermionic contributions to Ward identity, for lattices with $T = 4, 8$, and $12$.}
	\label{fig:lat-susy-d2-T8-B-Sb}	
\end{figure*}

\begin{figure*}[tbp]
	\centering	
	\includegraphics[width=.6\textwidth,origin=c,angle=0]{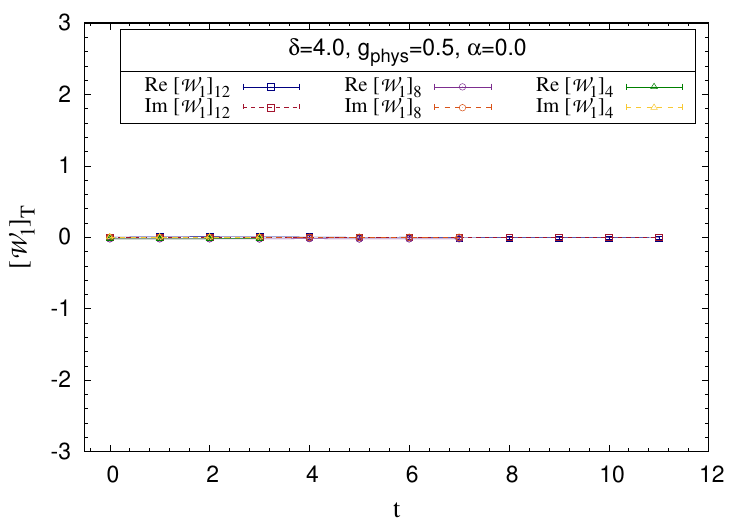}\\
	\includegraphics[width=.6\textwidth,origin=c,angle=0]{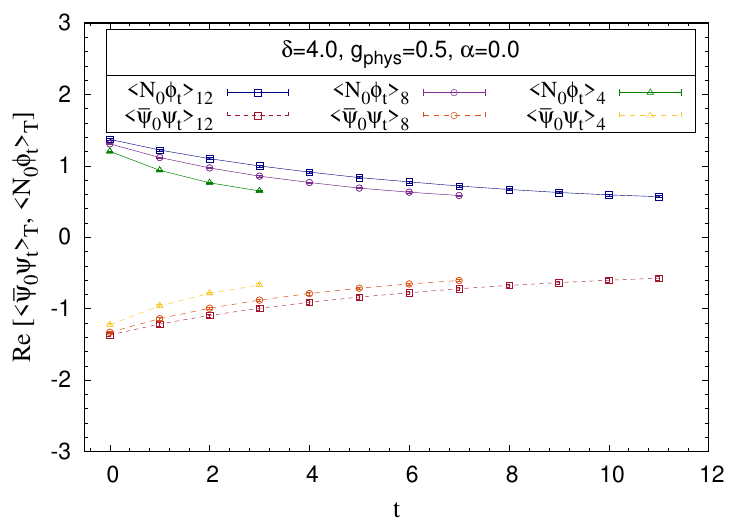}\\
	\includegraphics[width=.6\textwidth,origin=c,angle=0]{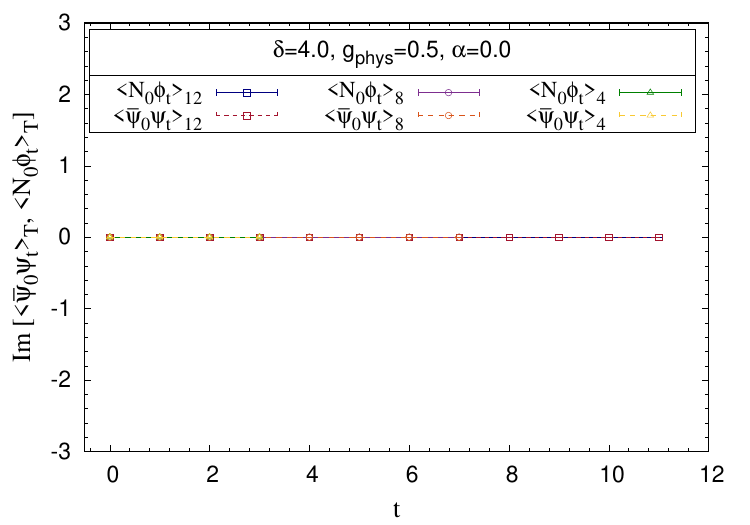}
	
	\caption {The $\mathcal{PT}$-symmetric model with $\delta = 4$. The full Ward identity (top) and real (middle) and imaginary (bottom) parts of bosonic and fermionic contributions to Ward identity, for lattices with $T = 4, 8$, and $12$.}
	\label{fig:lat-susy-d4-T8-B-Sb}	
\end{figure*}

In Figs. \ref{fig:lat-susy-d2-T8-B-Sb} and \ref{fig:lat-susy-d4-T8-B-Sb} we show the Ward identities for $\delta = 2,~4$, respectively, on a $T = 8$ lattice. The full Ward identity $\mathscr{W}_1$ is shown on the left panel, and the real and imaginary parts of the bosonic and fermionic contributions to the Ward identity are shown in the middle and right panels. Our simulations show that the bosonic and fermionic contributions cancel each other out within statistical errors, and hence the Ward identities are satisfied. All these results clearly suggest that SUSY is preserved in models with $\mathcal{PT}$-symmetry inspired $\delta$-even potentials. 

\subsection*{Odd $\delta$ case}

In these models, we had to introduce a deformation parameter to handle the singular-drift problem. The results are extracted in the vanishing limit of this parameter.

\begin{figure*}[tbp]
	\centering
	\includegraphics[width=.48\textwidth,origin=c,angle=0]{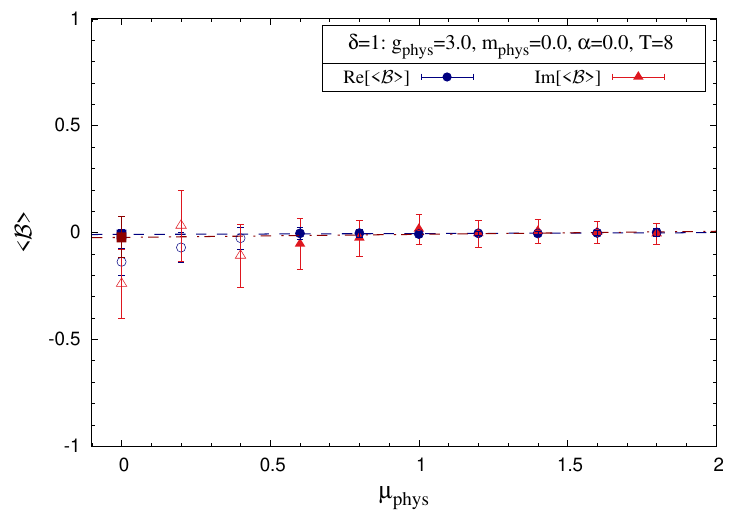} \includegraphics[width=.48\textwidth,origin=c,angle=0]{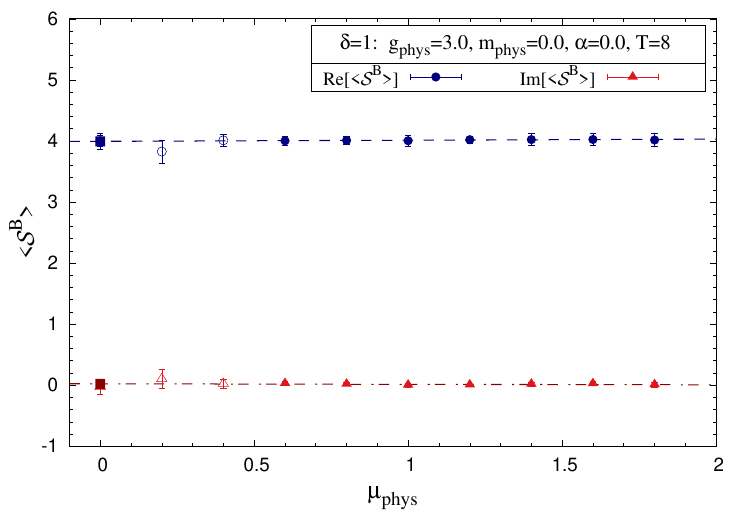}
	
	\caption{The $\mathcal{PT}$-symmetric model with $\delta = 1$ model. Expectation values of $\B_\alpha$ (left) and $\mathcal{S}^B_\alpha$ (right) for various $\mu_{\rm phys}$ and $\lim \mu_{\rm phys} \to 0$ on a $T = 8$ lattice. Parameters used in the simulations are $g_{\rm phys} = 3$, $m_{\rm phys} = 0$, and $\alpha = 0$. }
	\label{fig:d1bSb}	
\end{figure*}

The simulations were carried out for various non-zero $\mu$, and then the $\delta = 1$ model is recovered by taking ${\mu_{\rm phys} \to 0}$ limit. Our simulations suggest that when $\mu_{\rm phys}$ is above a particular value, the correctness criteria are satisfied, and the probability of absolute drift falls off exponentially. We take into account the $\mu_{\rm phys}$ parameter space where complex Langevin simulations are justified and consider the limit $\mu_{\rm phys} \to 0$. In Fig. \ref{fig:d1bSb}, we show the expectation values $\langle \B \rangle$ (left) and $\langle \mcS^{B} \rangle$ (right) for a lattice with $T = 8$ and for various $\mu_{\rm phys}$ values. The filled data points (red triangles for the imaginary part and blue circles for the real part) represent expectation values of observables for the parameter space where complex Langevin can be trusted, while for unfilled data points where the complex Langevin correctness criteria are not satisfied. Lines represent the linear fit of observables for parameter space where complex Langevin simulations are justified, and the solid squares represent values of respective observables in the $\lim \mu_{\rm phys} \to 0$. 

These simulation results indicate that $\langle \B  \rangle$ vanishes in the limit $\mu_{\rm phys} \to 0$. Also, the expectation value of the bosonic action $\langle \mcS^B \rangle = \hf T$ in this limit. It is also found to be independent of the physical parameters. Thus we conclude that SUSY is preserved in the $\delta = 1$ model. 

Now for the $\delta = 3$ model, inspired by the idea mentioned in Ref. \cite{Ito:2016efb} (which was successfully applied in Refs. \cite{Anagnostopoulos:2017gos, Anagnostopoulos:2020xai}) to handle the singular-drift problem, we introduce a fermionic deformation term in the action. The fermionic action then becomes
\beq
\label{eqn:df}
\mcS^F = \sum_{i = 0}^{T-1} \psib_i  \bigg(\sum_{j = 0}^{T-1}  \nabla^{-}_{ij} + d_f + \Xi^{''}_{ij} \bigg) \psi_j,
\eeq
where $d_f$ is the deformation parameter. The values of $d_f$ are chosen such that complex Langevin correctness criteria are satisfied. The $\delta = 3$ model is recovered as ${d_f \to 0}$. Our simulations suggest that above a particular $d_f$ value, the correctness criteria are satisfied, and the probability of absolute drift falls off exponentially. 

\begin{figure*}[tbp]
	\centering	
	\includegraphics[width=.48\textwidth,origin=c,angle=0]{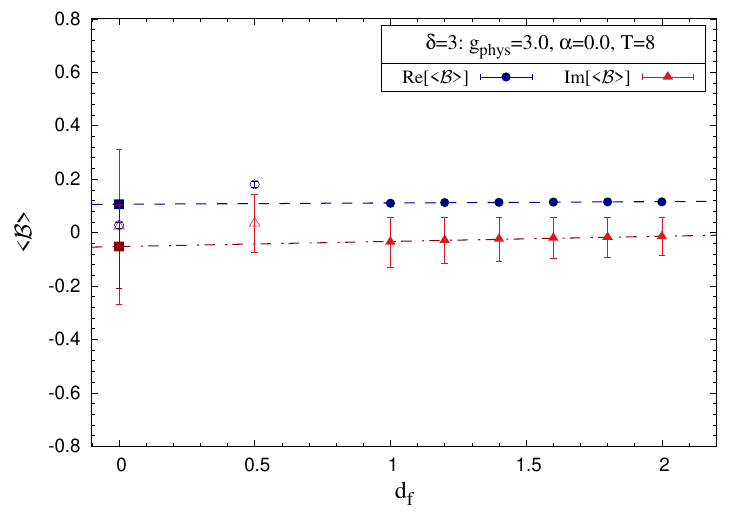}	
	\includegraphics[width=.48\textwidth,origin=c,angle=0]{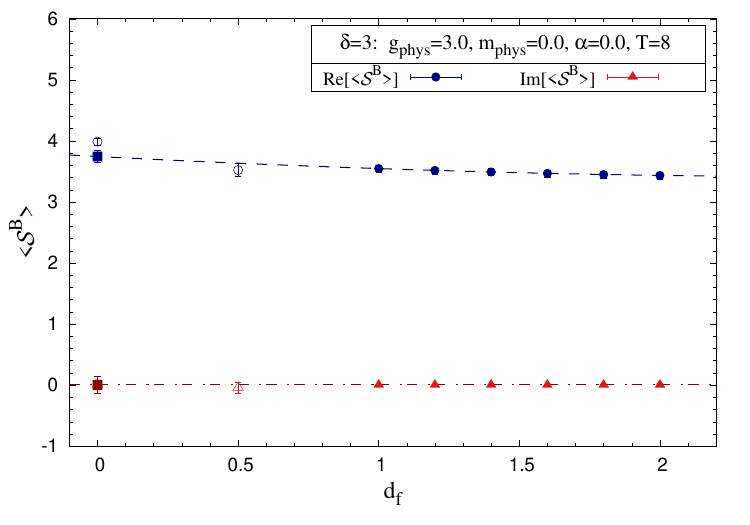}
	
	\caption[The $\mathcal{PT}$-symmetric model with $\delta = 3$ model. Expectation values of $\B_\alpha$ (left) and $\mathcal{S}^B_\alpha$ (right) against mass deformation $d_f$ parameter on a $T = 8$ lattice. ]{The $\delta = 3$ model. Expectation values of $\B_\alpha$ (left) and $\mathcal{S}^B_\alpha$ (right) against mass deformation $d_f$ parameter on a $T = 8$ lattice. The parameters used are $g_{\rm phys} = 3$, $m_{\rm phys} = 0$, and $\alpha = 0.0$. The dashed curves represent extrapolations to the $d_f \to 0$ limit.}
	\label{fig:d3bSb}	
\end{figure*}

In Fig. \ref{fig:d3bSb}, we show $\langle \B \rangle$ (left) and $\langle \mcS^{B} \rangle$ (right) on a $T = 8$ lattice for various values of $d_f$. The filled data points (red triangles for the imaginary part and blue circles for the real part) represent the expectation values of the observables for the parameter space where complex Langevin can be trusted, while the corresponding unfilled data points represent the simulation data that do not satisfy the correctness criteria. The dashed curves represent a linear fit for $\langle \B \rangle$ data and a quadratic fit for $\langle \mcS^{B} \rangle$ data. The solid squares represent the values of respective observables in the $\lim d_f \to 0$ limit. The simulation results suggest that $\langle \B  \rangle$ vanishes in the $\mu_{\rm phys} \to 0$ limit. Also, $\langle \mcS^B \rangle = \hf T$ within error bars in this limit. It is also found to be independent of the physical parameters of the model. These results indicate that SUSY is preserved in the $\delta = 3$ model. 

\section{Reliability of simulations}
\label{subsec:reliability}

In order to monitor the reliability of simulations, we use the two recently proposed methods: one tracks the vanishing nature of the Fokker-Planck operator \cite{Aarts:2009uq, Aarts:2011ax, Aarts:2013uza} and the other monitors the decay of the probability distribution of the drift term magnitude \cite{Nagata:2016vkn, Nagata:2018net}. 

\subsection{Langevin operator on observables}
\label{app:FP-correctness}

The observables of the theory $\cO_i[\phi, \theta]$ at $i$-th site evolve in the following way 
\beq
\frac{\partial \cO_i[\phi, \theta]}{\partial \theta} = \widetilde{L}_i \cO_i[\phi, \theta], 
\eeq
where
$\widetilde{L}_i$ is the Langevin operator for the $i$-th site. It is defined as
\beq
\widetilde{L}_i = \left(\frac{\partial}{\partial \phi_i}  - \frac{\partial \mcS^{\rm eff}[\phi]}{\partial \phi_i}  \right) \ \frac{\partial}{\partial \phi_i}.
\eeq

Once the equilibrium distribution is reached, we can remove the $\theta$ dependence from the observables. Then we have $C_{\cO_i} \equiv \langle \widetilde{L}_i  \cO_i [\phi] \rangle = 0$, and this can be used as a criterion for the correctness of the simulations.

If we take $\B_i$ at the $i$-th site as the observable then we have
\beq
\widetilde{L}_i \B_i =  -i \Xi^{'''}_{iii} + i \Xi^{''}_{ii} \frac{\partial \mcS^{\rm eff}}{\partial \phi_i}.
\eeq
Note that $\widetilde{L} \B$ respects translational symmetry on the lattice, and hence we can monitor the value obtained by averaging over all lattice sites. 

In Table \ref{tab:LB-anho}, we provide the expectation values of $\widetilde{L} \B$ for the simulations of the supersymmetric anharmonic oscillator model discussed in Sec. \ref{subsec:susy-anho}. We see that this observable is zero within error bars, and thus we can trust the simulations. 

\begin{table}[tbp]
	\begin{center}
		{\small
			\begin{tabular}{|c| l r |c|} 
				\hline
				$\Xi'(\phi)$&		$T$ &  $a = T^{-1}$   &  $~\langle \widetilde{L}\B_{\alpha}\rangle $  \\
				\hline
				\hline
				$ m \phi + g  \phi^3$
				&{8}
				&{0.125}& $0.0000(0) 	+ i0.0268(910)$ \\
				&{16}
				&{0.0625}&$0.0000(0) 	- i0.0379(450)$ \\
				&{32}
				&{0.03125}&$0.0000(0) 	+ i0.0131(232)$ \\
				&{64}
				&{0.015625}&$0.0000(0) 	+ i0.0056(132)$ \\ 
				\hline
			\end{tabular}
		}	
		\caption[Expectation value of $\widetilde{L} \B_\alpha$ for supersymmetric anharmonic oscillator with parameters $m_{\rm phys} = 10.0$ and $g_{\rm phys} = 100.0$.]{Expectation value of $\widetilde{L} \B_\alpha$ for supersymmetric anharmonic oscillator with parameters $m_{\rm phys} = 10.0$ and $g_{\rm phys} = 100.0$. Simulations were performed for different lattice spacings with $\beta = 1$ and $\alpha = 0$.}
		\label{tab:LB-anho}
	\end{center}
\end{table}

\begin{figure}[tbp]
	\centering
	\includegraphics[width=.48\textwidth,origin=c,angle=0]{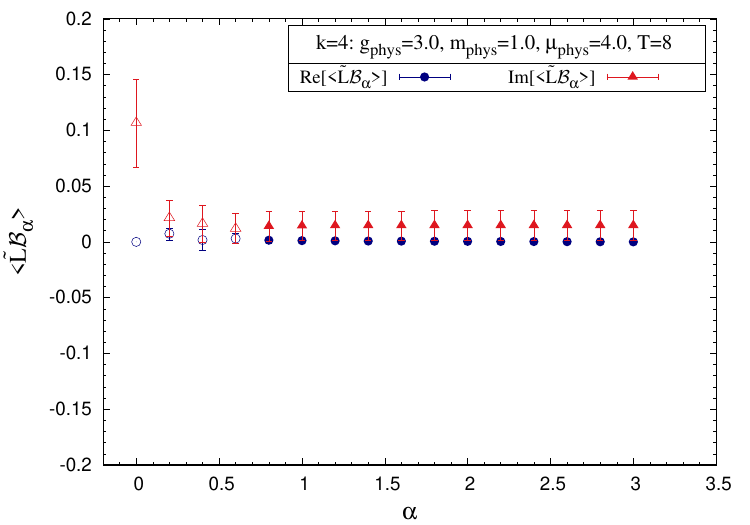}
	
	\caption[Expectation values of $\widetilde{L} \B_\alpha$ for various $\alpha$ values, on a $T = 8$ lattice, for the model with even-degree ($k = 4$) polynomial potential, given in Eq. \eqref{eqn:lat-gen-pot-k4}.]{Expectation values of $\widetilde{L} \B_\alpha$ for various $\alpha$ values, on a $T = 8$ lattice, for the model with even-degree ($k = 4$) polynomial potential, given in Eq. \eqref{eqn:lat-gen-pot-k4}. The parameters used are $g_{\rm phys} = 3$, $m_{\rm phys} = 1$, and $\mu_{\rm phys} = 4.0$.}
	\label{fig:lbk4}
\end{figure}

In Fig. \ref{fig:lbk4}, we show the expectation values $\widetilde{L} \B$ for various twist parameter values for the model with the case of real even-degree ($k = 4$) polynomial potential (discussed in Sec. \ref{subsec:gen-sps}). The filled data points (red triangles for the imaginary part and blue circles for the real part) represent expectation values of $\widetilde{L} \B$ for the parameter space where complex Langevin can be trusted, when the second criterion, decay-of-the-drift-term, to be discussed in the next section, is applied. The unfilled data points represent the expectation values of $\widetilde{L} \B$ in the parameter space where the decay-of-the-drift-term criterion is not satisfied.

In Table \ref{tab:lbk5}, we provide the expectation values of $\widetilde{L} \B$ for the simulations of real odd-degree ($k = 5$) polynomial potentials discussed in Sec. \ref{subsec:gen-sps}. We see that this observable is zero within error bars, and thus we can trust the simulations.

\begin{table}[tbp]
	\begin{center}
		{\small
			\begin{tabular}{|c| l r |c|}  
				\hline
				$\Xi'(\phi)$&		$T$ &  $a = T^{-1}$  &  $~\langle \widetilde{L}\B_{\alpha}\rangle $ \\     	
				\hline
				\hline
				&{8}
				&{0.125}& $0.0000(0) 	- i0.0010(729)$ \\	
				${\Xi{'}}^{(5)}$
				&{12}
				&{0.0833}& $0.0000(0) 	+ i0.0127(508)$ \\	
				&{16}
				&{0.0625}& $0.0000(0) 	- i0.0301(396)$ \\ 	
				\hline
			\end{tabular}
		}
		\caption {Expectation values of $\widetilde{L} \B_\alpha$ for the model with odd-degree ($k = 5$) real-polynomial potential given in Eq. \eqref{eqn:lat-gen-pot-k5}. The parameters used are $g_{\rm phys} = 3$, $m_{\rm phys} = 1$, and $\mu_{\rm phys} = 4.0$.}
		\label{tab:lbk5}
	\end{center}
\end{table}

\begin{figure}[tbp]
	\centering
	\includegraphics[width=.48\textwidth,origin=c,angle=0]{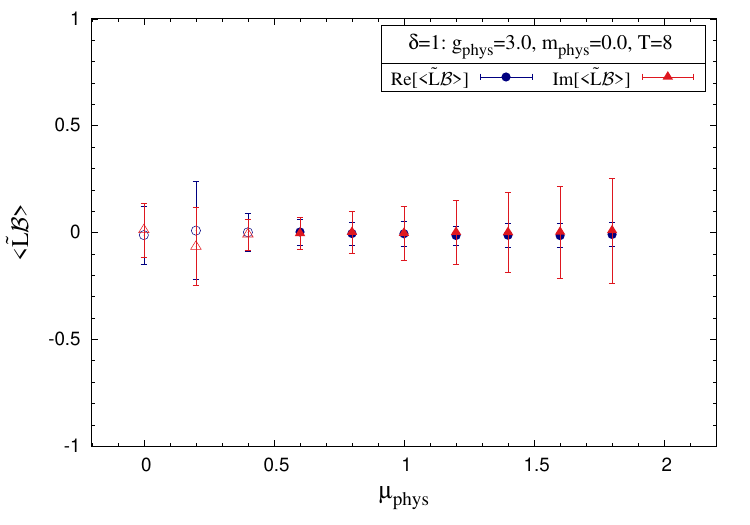} $~$ \includegraphics[width=.48\textwidth,origin=c,angle=0]{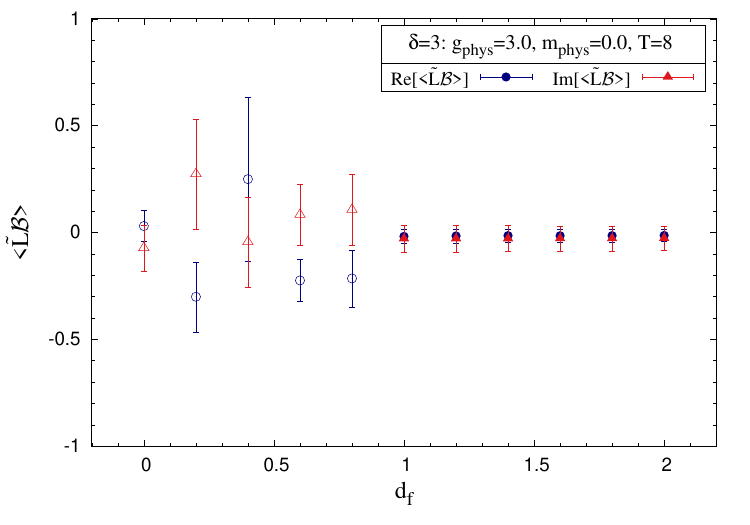}	
	
	\caption{Expectation values of $\widetilde{L} \B_\alpha$ for the $\mathcal{PT}$ symmetric model on a $T = 8$ lattice. (Left) Case $\delta = 1$. (Right) Case $\delta = 3$. The parameters used are $g_{\rm phys} = 3$, $m_{\rm phys} = 0$, and $\alpha = 0.0$.}
	\label{fig:lbd1_fig:lbd3}
\end{figure}

In Table \ref{tab:lat-susy-pt-symm-d1-4-LB} we show the expectation values of $\widetilde{L} \B$ for $\mathcal{PT}$ symmetric models with $\delta = 2$ and 4. 
\begin{table}[tbp]
	\begin{center}
		{\small
			\begin{tabular}{|c| l r |c|}  
				\hline
				$\Xi'(\phi)$&$T$ &  $a=T^{-1}$   &  $\langle \widetilde{L}\B_{\alpha} \rangle$  \\   
				\hline
				\hline
				&${4}$&${0.25}$  & $-0.0000(0) - i0.0104(72)$ \\
				$\delta={2}$&${8}$&${0.125}$ & $-0.0000(0) + i0.0006(59)$ \\
				&${12}$&${0.0833}$  & $-0.0000(0) - i0.0104(72)$ \\
				\hline
				&${4}$&${0.25}$ & $0.0000(0) + i0.0403(244)$ \\
				$\delta={4}$&${8}$&${0.125}$  & $0.0000(0) + i0.0027(91)$ \\
				&${12}$&${0.0833}$  & $0.0000(0) - i0.0098(64)$ \\ 
				\hline
			\end{tabular}
		}
		\caption{Expectation value of $\widetilde{L}\B_\alpha$ for the $\mathcal{PT}$-symmetric potentials given in Eq. \eqref{eqn:lat-pt-symm-pot} with $\delta = 2$ and 4. The simulation parameters used are $\beta = 1$, $g_{\rm phys} = 0.5$, and $\alpha = 0$.}
		\label{tab:lat-susy-pt-symm-d1-4-LB}
	\end{center}
\end{table}
In Fig. \ref{fig:lbd1_fig:lbd3} (left), we show the expectation values for various $\mu_{\rm phys}$ values of $\widetilde{L} \B$ for the simulations of $\delta = 1$ model. The filled (unfilled) data points represent the simulations that do (do not) respect the decay-of-the-drift-term criterion. In Fig. \ref{fig:lbd1_fig:lbd3}, (right) we show $\widetilde{L} \B$ for various deformation parameter $d_f$ for the $\delta = 3$ model. Again, the filled (unfilled) data points represent the simulations that do (do not) respect the decay-of-the-drift-term criterion. 

\subsection{Decay of the drift terms}
\label{app:drift-decay}

The decay-of-the-drift-term criterion was proposed in Refs. \cite{Nagata:2016vkn, Nagata:2018net}. There, the authors demonstrated, in a few simple models, that the probability of the drift term should be suppressed exponentially or faster at larger magnitudes to guarantee the correctness of the complex Langevin method.

The magnitude of the mean drift is defined as
\beq
u \equiv \sqrt{\frac{1}{T} \sum_{i = 0}^{T-1} \left| \frac{\partial \mcS^{\text{eff}}}{\partial \phi_i}\right|^2}. 
\eeq

In our work, to avoid the singular-drift problem, we introduced appropriate deformation parameters in the theory. The final results are obtained after extrapolating to the vanishing limits of deformation parameters.

\begin{figure}[tbp]
	\centering
	\includegraphics[width=.48\textwidth,origin=c,angle=0]{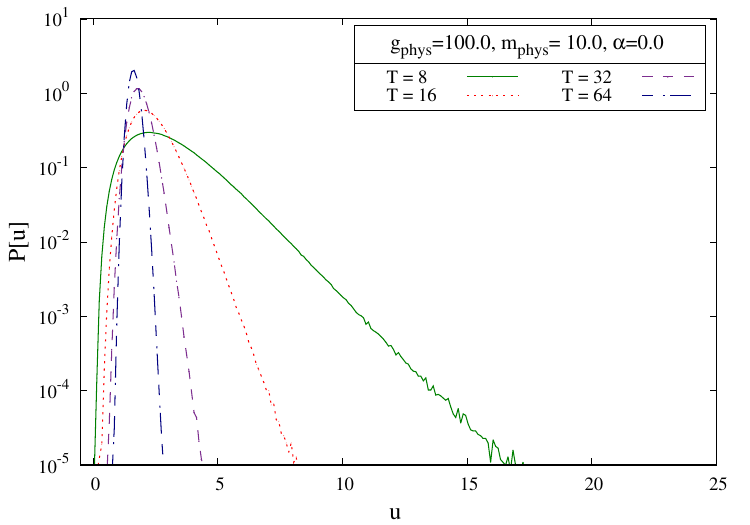}	$~$
	\includegraphics[width=.48\textwidth,origin=c,angle=0]{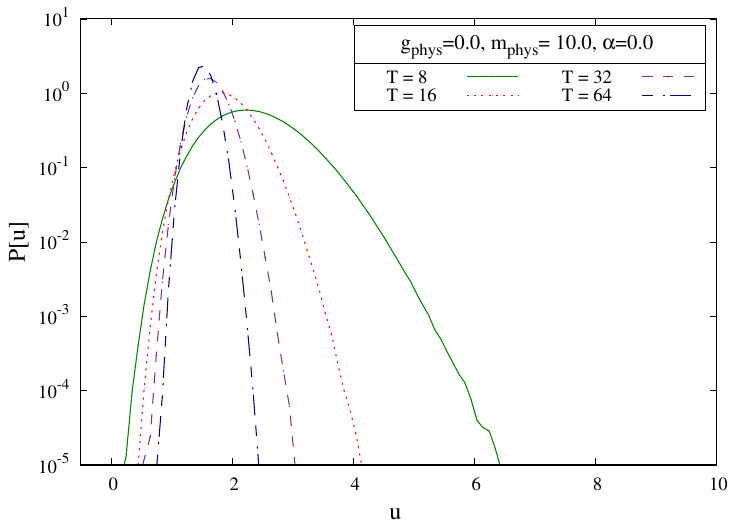}
	
	\caption{Decay of the drift terms. (Left) Supersymmetric anharmonic oscillator with the parameters $m_{\rm phys} = 10.0$ and $g_{\rm phys} = 100.0$. (Right) supersymmetric harmonic oscillator with parameters $m_{\rm phys} = 10.0$ and $g_{\rm phys} = 0.0$.}
	\label{fig:lat-anho-drift}
	
\end{figure}

In Fig. \ref{fig:lat-anho-drift}, we show the probability distribution $P(u)$ against $u$ for the simulations of supersymmetric anharmonic (left) and harmonic (right) potentials discussed in Sec. \ref{subsec:susy-anho}. We see that the drift terms decay exponentially or faster in these models, and thus the simulations can be trusted.

\begin{figure}[htbp]
	\centering
	\includegraphics[width=.48\textwidth,origin=c,angle=0]{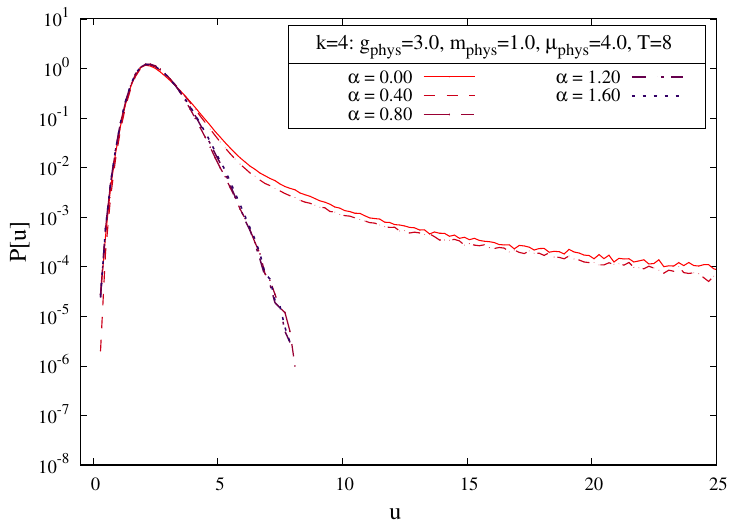}		$~$ \includegraphics[width=.48\textwidth,origin=c,angle=0]{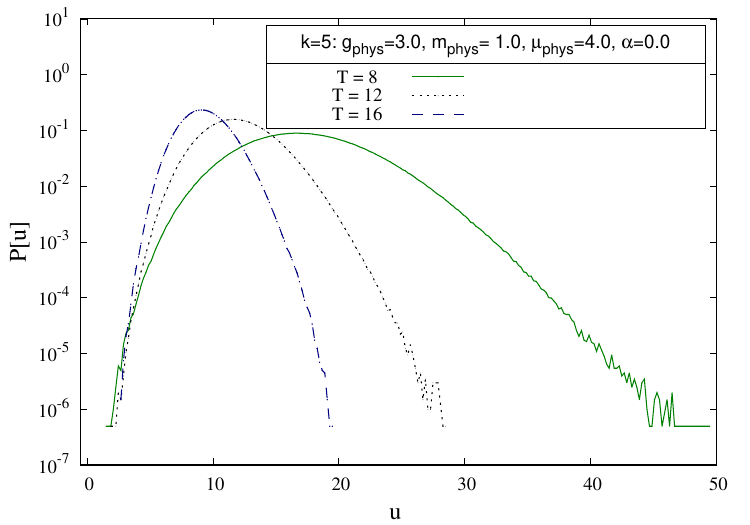}
	\caption[(Left) Decay of the drift term for various $\alpha$ values for the model with even-degree ($k = 4$) polynomial potential on a $T = 8$ lattice. (Right) The decay of the drift terms for the model with odd-degree ($k = 5$) real polynomial potential on $T = 8,12,16$ lattices.]{(Left) Decay of the drift term for various $\alpha$ values for the model with even-degree ($k = 4$) polynomial potential on a $T = 8$ lattice. The parameters used are $g_{\rm phys} = 3$, $m_{\rm phys} = 1$, and $\mu_{\rm phys} = 4.0$. (Right) The decay of the drift terms for the model with odd-degree ($k = 5$) real polynomial potential on $T = 8,12,16$ lattices. The parameters used are $g_{\rm phys} = 3$, $m_{\rm phys} = 1$, and $\mu_{\rm phys} = 4.0$.}
	\label{fig:pdSk4_fig:pdSk5}
	
\end{figure}

In Fig. \ref{fig:pdSk4_fig:pdSk5} (left), we show the decay of drift terms for various $\alpha$ values, for the model with even-degree ($k = 4$) real polynomial potential. We see that the drift terms decay exponentially or faster when $\alpha \geq 0.8$, and the simulations can be trusted in this parameter regime. In Fig. \ref{fig:pdSk4_fig:pdSk5} (right), we show the decay of drift terms for various $\alpha$ values, for the model with even-degree ($k = 5$) real polynomial potential. We see that the drift terms decay exponentially or faster when $\alpha = 0$, and the simulations can be trusted in this parameter regime.

\begin{figure}[htbp]
	\centering
	\includegraphics[width=.48\textwidth,origin=c,angle=0]{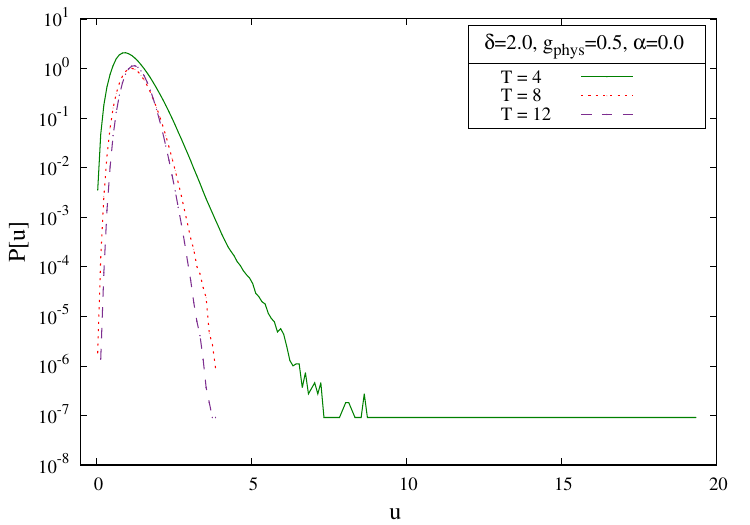}	$~$ \includegraphics[width=.48\textwidth,origin=c,angle=0]{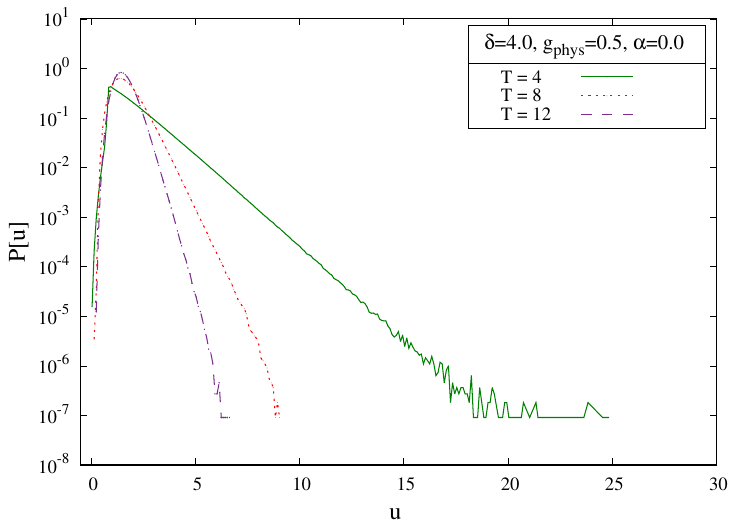}	
	\caption{Decay of the drift terms for the $\mathcal{PT}$ symmetric models with $\delta =  2$ (left) and $\delta =  4$ (right). The parameters used are $g_{\rm phys} = 0.5$ and $\alpha = 0$.}
	\label{fig:pdSd2-4}
\end{figure}

In Fig. \ref{fig:pdSd2-4} we show the drift term decay for the $\mathcal{PT}$ symmetric models with $\delta = 2$ (left) and $\delta = 4$ (right). The drift terms decay exponentially or faster when $\alpha = 0$. Thus the simulations can be trusted in this parameter regime.

\begin{figure}[t]
	\centering	
	\includegraphics[width=.48\textwidth,origin=c,angle=0]{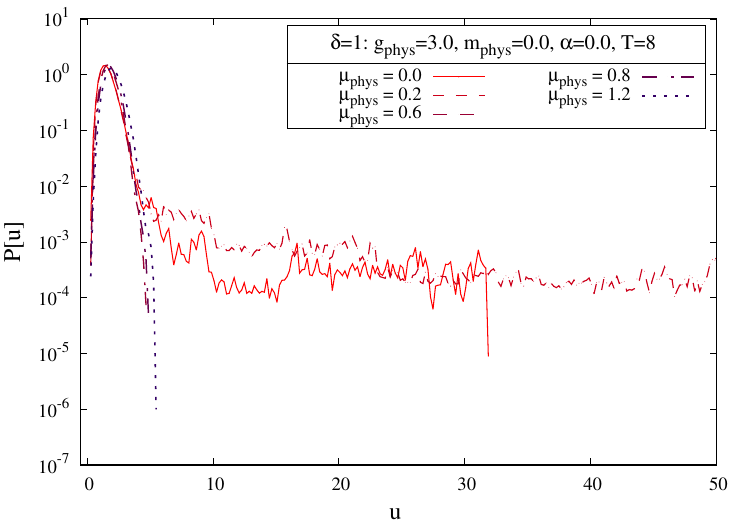} 	$~$ \includegraphics[width=.48\textwidth,origin=c,angle=0]{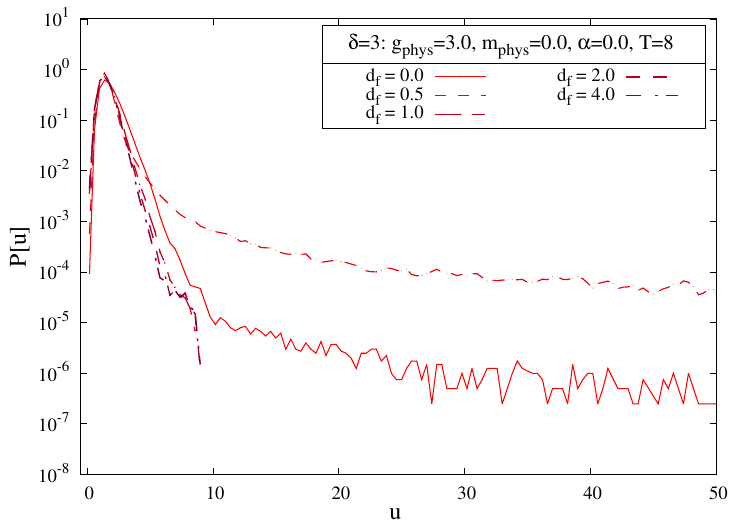}
	\caption[Decay of the drift terms for the $\mathcal{PT}$ symmetric models with odd $\delta$ on $T=8$ lattice. (Left) The $\delta = 1$ case. (Right) The $\delta = 3$ case.]{The decay of the drift terms for the $\mathcal{PT}$ symmetric models with odd $\delta$ on $T=8$ lattice. The parameters used were $g_{\rm phys} = 3$, $m_{\rm phys} = 0$, and $\alpha = 0.0$. (Left)  The $\delta = 1$ case. Simulations were performed for various $\mu_{\rm phys}$ values. (Right) The $\delta = 3$ case. Simulations were performed for various values of the fermionic mass deformation parameter $d_f$.}
	\label{fig:pdSd1_fig:pdSd3}
\end{figure}

In Fig. \ref{fig:pdSd1_fig:pdSd3} (left), we show the drift term decay for the $\mathcal{PT}$ symmetric model with $\delta = 1$, for various $\mu_{\rm phys}$ values. The drift terms decay exponentially or faster when $\mu_{\rm phys} \geq 0.6$. Thus the simulations can be trusted in this parameter regime. In Fig. \ref{fig:pdSd1_fig:pdSd3} (right), we show the decay of the drift terms for various values of the fermionic mass deformation parameter $d_f$, for the $\mathcal{PT}$ symmetric model with $\delta = 3$. Here we see that the drift terms decay exponentially or faster when $d_f > 1.0$, and thus the simulations can be trusted in this parameter regime.

\vfill
	\chapter{Complex Langevin analysis of two-dimensional quantum field theories} \label{chap:susy-qft-2d}

\setlength\epigraphwidth{10.5cm}
\setlength\epigraphrule{0pt}
\epigraph{The chapter is based on the following publication by the author: \\ \textbf{Arpith Kumar} and Anosh Joseph,\\ {\it Complex Langevin simulations for $\mathcal{PT}$-symmetric models},\\ \href{https://doi.org/10.22323/1.396.0124
}{  PoS LATTICE2021 \textbf{124} (2022) }{\href{https://doi.org/10.48550/arXiv.2201.12001}{(arXiv: 2201.12001 [hep-lat])}} }{}

Quantum field theories such as QCD with quark chemical potential, Chern-Simons gauge theories, chiral gauge theories, and field theories with topological terms can suffer from the infamous sign problem. The list also includes an interesting class of non-Hermitian and self-interacting quantum field theories that exhibit $\mathcal{PT}$-invariance. Although formulated using complex actions, these theories can possess real and bounded below energy spectra. In Refs. \cite{Bender:1997ps, Bender:1998gh} Bender and Milton considered a new class of $\mathcal{PT}$-invariant (Euclidean) quantum field theories with interactions of the form $\lambda( i \phi )^{(2 + \delta)}$. These theories are physically admissible; that is, they possess a real and bounded below energy spectra. But for these interactions, parity in itself is manifestly broken. 
\vsphf

In this chapter, we study quantum field theories in two dimensions with the help of the complex Langevin method. We lattice regularized a minimal supersymmetric model, namely $\mathcal{N} = 1$ Wess-Zumino model in two dimensions. We discuss the case when the superpotential is a double-well potential. We also discuss our ongoing simulations of the $\mathcal{PT}$-symmetric superpotentials. Before moving on to supersymmetric models in Sec. \ref{sec:susy_models}, in Sec. \ref{sec:Two-dimensional_scalar_field_theories} we discuss two-dimensional scalar field theories with $\phi^4$ and $\mathcal{PT}$-symmetric potentials.

\section{Two-dimensional scalar field theories}
\label{sec:Two-dimensional_scalar_field_theories}

Consider the Lagrangian of a two-dimensional Euclidean scalar field theory 
\beq
\mathcal{L}_E = \hf \partial_\mu \phi  \partial_\mu \phi +\hf m^2 \phi^2 + W(\phi),
\eeq
where $\phi$ is a dimensionless scalar, $m$ is the mass parameter, and $W(\phi)$ is the interaction potential. The Euclidean action is 
\beq
S_E = \int d^2 x~\mathcal{L}_E.
\eeq 

To simulate the model using the complex Langevin method, we first discretize the model on a two-dimensional toroidal lattice. The temporal and spatial extents, $\beta_t$ and $\beta_x$, respectively, can be expressed as $\beta_t = \beta_x = La$ with $L$ denoting the number of lattice sites in each direction and $a$ denoting the lattice spacing. We have $\int d^2 x \longrightarrow a^2 \sum_x$. The periodicity of the lattice enables us to write
\beq
\left(\partial_\mu \phi \right)^2 = - \phi \partial_\mu^2 \phi = -\frac{1}{a^2} \left[ \phi_{x}\phi_{x+\mu} + \phi_{x}\phi_{x-\mu} -2{\phi_{x}}^2\right],
\eeq
where $\phi_{x\pm \mu}$ represents the field at the neighboring site in $\pm \mu$-th direction. 

Using the complex Langevin method, we can study these models for various interaction potentials, including the $\mathcal{PT}$-invariant potentials. Complex Langevin update for field configurations at a lattice site $x$, for Langevin time $\theta$, with step-size $\epsilon$ is given by
\beq
\phi_{x, \theta + \epsilon} = \phi_{x, \theta} + \epsilon v_{x, \theta} + \eta_{x, \theta} \sqrt{\epsilon}, 
\eeq
where the drift term is obtained as $v_{x, \theta} = - { \partial S_E }/{ \partial \phi_{x, \theta} }$ and $\eta_{x, \theta}$ is a real Gaussian noise.

\subsection{Model with $\phi^4$ potential}

\begin{figure}
	\centering
	\includegraphics[width=.6\textwidth]{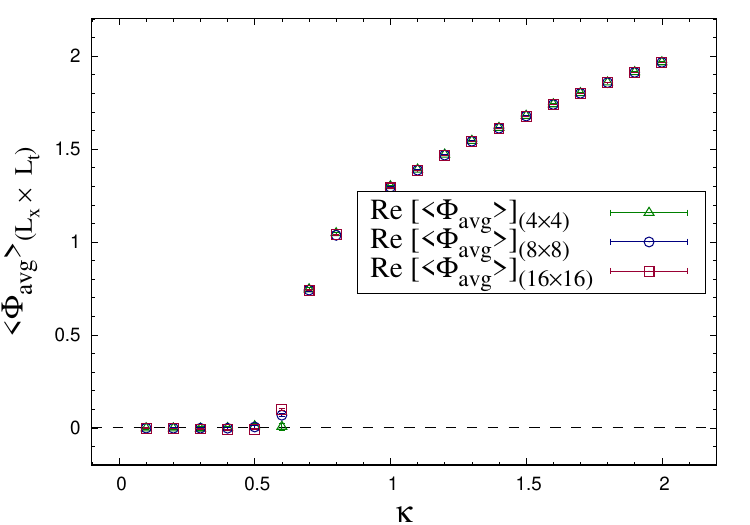}
	\includegraphics[width=.6\textwidth]{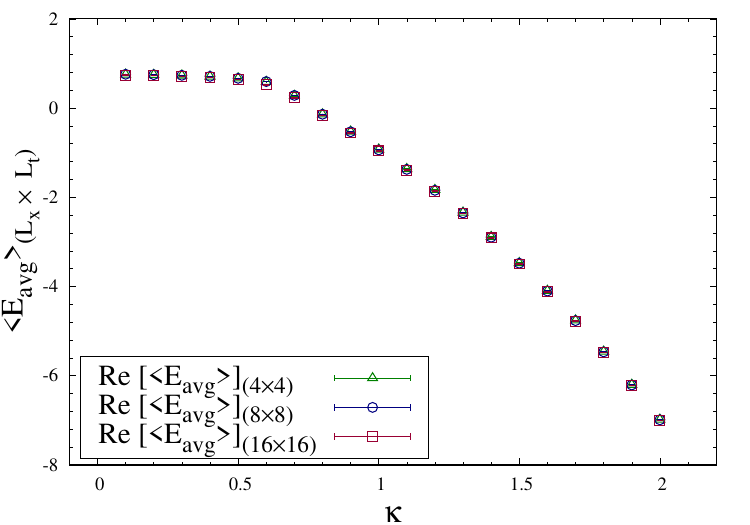}
	\includegraphics[width=.6\textwidth]{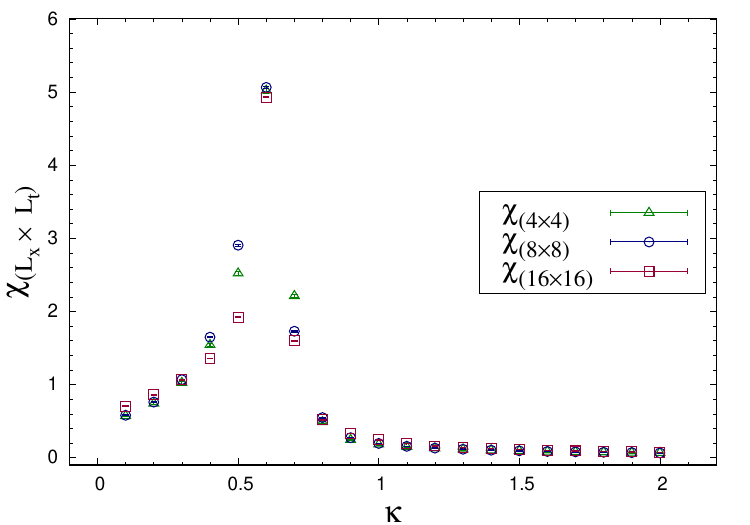}
	\caption{Model with $\phi^4$ potential. Expectation values of the order parameter $\Phi$ (top panel), energy $E$ (middle panel), and susceptibility $\chi$ (bottom panel) against $\kappa$ for different lattice extents and fixed $\tilde{\lambda} = 0.5$.}
	\label{fig:phi4}
\end{figure}

Consider the potential $W(\phi) = \lambda \phi^4$. Classically, the model is invariant under the discrete $\mathbb{Z}_2$ symmetry, that is, $\phi \to -\phi$. However, in quantum theory, this symmetry may be broken dynamically. The expectation value of the scalar field, $\langle \phi \rangle$ can be regarded as an order parameter. If $\langle \phi \rangle = 0$, the theory is in a symmetric phase; otherwise, it is in a symmetry-broken phase. There exist comprehensive studies of the $\phi^4$ theory on the lattice \cite{De:2005ny, Schaich:2009jk, Wozar:2011gu, Sakai:2018xkx}. We will utilize this model as a testbed for our Langevin analysis. We employ a lattice parameterization with dimensionless lattice parameters $m_0^2 = m^2 a^2$ and $\lambda_0 = \lambda a^2$, and in addition, we introduce a new set of parameters $\kappa$ and $\tilde{\lambda}$ \cite{De:2005ny}, 
\beq
m_0^2 \to \frac{1-2\tilde{\lambda}}{\kappa} -4, ~~ \lambda_0 \to 6\frac{\tilde{\lambda}}{\kappa^2}, ~~ {\rm and} ~~ \phi \to \sqrt{2\kappa} \Phi.
\eeq
The above parameterization leads to the lattice action 
\beq
S = - 2 \kappa \sum_x \sum_\mu \Phi_x \Phi_{x + \mu} + \sum_x \Phi_x^2 + \tilde{\lambda} \sum_x \left( \Phi_x^2 - 1 \right)^2.
\eeq
In our simulations of the model, we monitor the following observables as $\kappa$ is varied: the average of the field $\Phi$ as an order parameter, energy $E$, and susceptibility $\chi$. The simulation results are shown in Fig. \ref{fig:phi4} for different lattice extents and a fixed $\tilde{\lambda} = 0.5$. The results indicate that the model possesses a phase transition around $\kappa = 0.6$ and $ \langle  \Phi_{\rm avg} \rangle  \neq 0$ for $\kappa \ge 0.6$ implying $\mathbb{Z}_2$ broken phase.

\subsection{Model with $\mathcal{PT}$-invariant potential}

Next we move onto $\mathcal{PT}$-invariant scalar field theory with the potential $W(\phi) = - \lambda (i\phi)^{(2 + \delta)}$, where the coupling $\lambda$ has $m^2$ dimension and $\delta$ is a real parameter. It is fascinating to note that these models possess a real and bounded below spectra for $\delta > 0$ with a non-zero mass parameter. The positivity of the spectrum can be understood from a theoretical point of view. 

As an example, we consider the theory for $\delta = 1$. The Lagrangian is
\beq
\mathcal{L}_E = \hf \left(\partial_\mu \phi\right)^2 +\hf m^2 \phi^2 + i \lambda \phi^3.
\eeq
For a conventional real $\lambda \phi^3$ theory, in the weak-coupling expansion, Green's functions can be expressed as a formal power series in $\lambda^2$. This power series, although real, does not alternate in sign, and hence, is not Borel summable. The non-summability of the perturbation series reflects the fact that the spectrum is not bounded below. Upon replacing the coupling $\lambda$ $\to$ $i\lambda$, the theory becomes $\mathcal{PT}$-symmetric. The power series remains real, and also it alternates signs. As a consequence, the perturbation series becomes summable, suggesting that the underlying theory possesses a real positive spectrum \cite{Bender:1997ps, Bender:1998gh, Milton:2003av}.

The action for such $\mathcal{PT}$-symmetric theories is complex in general. Path integral Monte Carlo requires the action to be real, and hence a non-perturbative lattice study of these theories is hindered due to a sign problem or \textit{complex phase problem}. We use the complex Langevin method to overcome this difficulty. For the $\delta = 1$ model, the lattice action can be expressed as
\beq
S = -\sum_x \sum_\mu \phi_x \phi_{x + \mu} + \left( 2 + \frac{m_0^2}{2} \right) \sum_x \phi_x^2 + i \lambda_0 \sum_x \phi_x^3,
\eeq
where $m_0$ and $\lambda_0$ are dimensionless mass and coupling parameters, respectively.

\begin{figure}
	\centering
	\includegraphics[width=.49\textwidth]{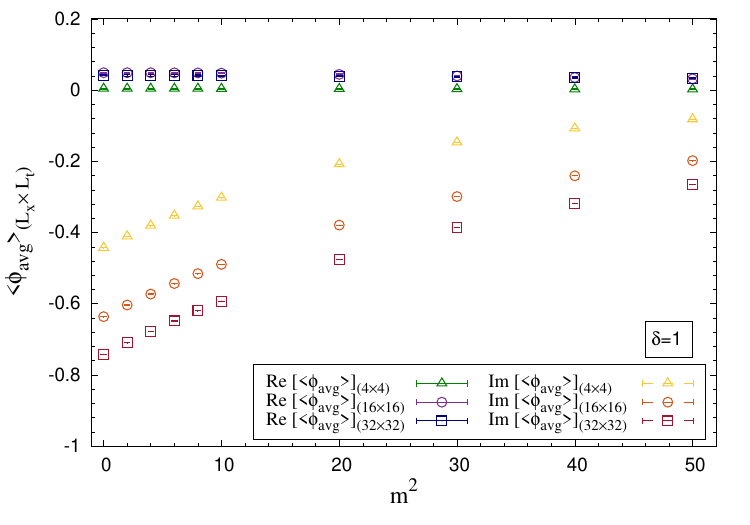}
	\includegraphics[width=.49\textwidth]{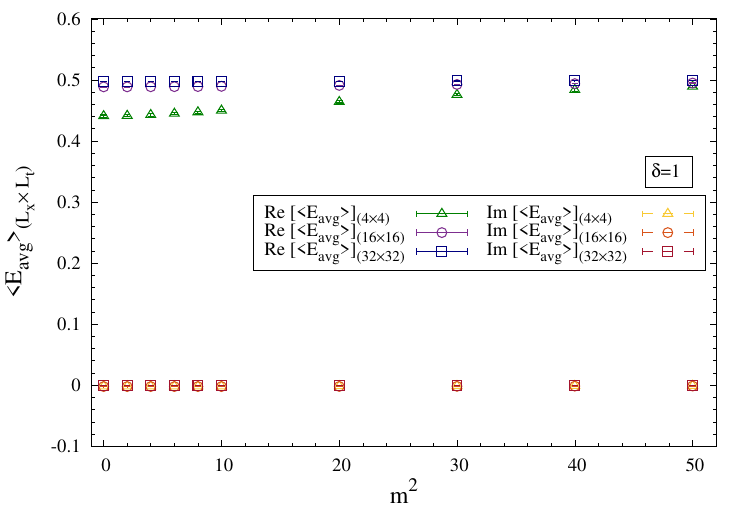}
	
	\includegraphics[width=.49\textwidth]{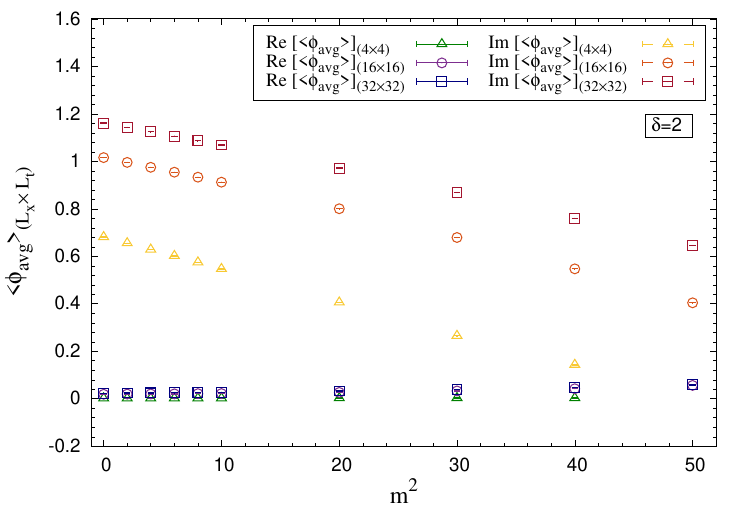}
	\includegraphics[width=.49\textwidth]{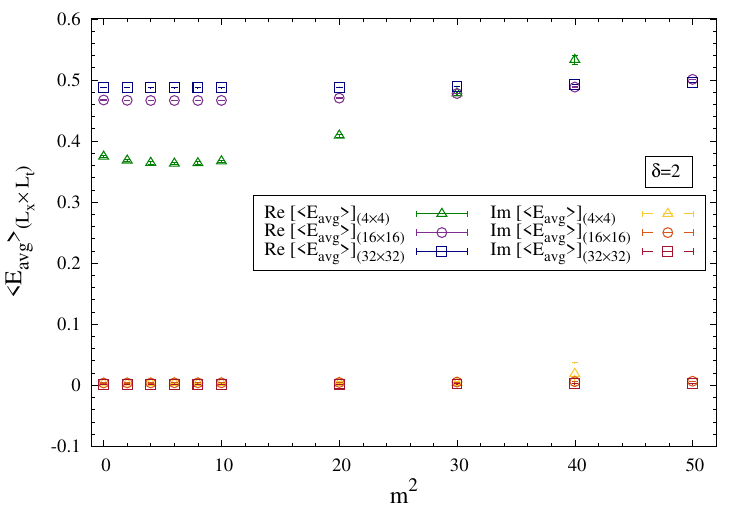}
	
	\caption{Bosonic $\mathcal{PT}$-symmetric model with $\delta = 1$ (top) and $\delta = 2$ (bottom) potential. The expectation values of the real and imaginary parts of the order parameter $\phi$ (left panel) and the energy $E$ (right panel) against physical mass parameter $m^2$ for different lattice extents and fixed physical coupling $\lambda = 10.0$.}
	\label{fig:pt1-2}
\end{figure}

In Fig. \ref{fig:pt1-2}, we show our simulation results for the bosonic $\mathcal{PT}$-symmetric theory with $\delta = 1$ (top) and $\delta = 2$ (bottom) potential. On the left panel, the expectation values of the real and imaginary parts of the average field $\phi$ (order parameter) against physical mass $m^2$ for different lattice extents and fixed physical coupling $\lambda = 10.0$ is shown. On the right panel, we show the ground state energy $E$ against $m^2$ for different lattice extents and fixed physical coupling $\lambda = 10.0$. These preliminary results suggest $ \langle  \phi_{\rm avg} \rangle  \neq 0$, that is, parity is manifestly broken for $\delta=1,~2$. The expectation value of energy is real and positive, Re [$\langle {\rm E_{avg}} \rangle] > 0$ and Im $[\langle {\rm E_{avg}} \rangle] = 0$, indicating a real bounded below spectra for this class of interactions. Our simulation results are in accordance with the analytical predictions \cite{Bender:1998gh}.

\section{Two-dimensional $\mathcal{N} = 1$ Wess-Zumino model}
\label{sec:susy_models}

In this section, we study a supersymmetric version of the model discussed in the previous section. (The zero- and one-dimensional cousins of this model were studied recently in Refs. \cite{Joseph:2020gdh, Joseph:2019sof}.) We add fermions to the Lagrangian and consider the simplest two-dimensional supersymmetric quantum field theory, the $\mathcal{N} = 1$ Wess-Zumino model. The theory involves a minimalistic set of fields, that is, a scalar field $\phi$ and a two-component Majorana spinor $\psi$. The on-shell model in Euclidean spacetime has the action 
\beq
S_E = \int d^2 x ~\hf \left[ \left(\partial_\mu \phi\right)^2 + \bar{\psi} \mathcal{M} \psi + W^2\left( \phi \right) \right],
\eeq
where $\mathcal{M} = \gamma^\mu \partial_\mu + W'\left(\phi \right)$ is referred to as the fermion matrix and the potential $W(\phi)$ is actually the derivative of the superpotential. The action is invariant under a single SUSY given by the transformations
\beq
\delta \phi = \bar{\epsilon} \psi, ~~\delta \psi = \big[ \gamma^\mu \partial_\mu \phi- W\left(\phi \right)  \big] \epsilon, ~~ \delta \bar{\psi} = 0.
\eeq
The Majorana spinor satisfies the relation, $\bar{\psi} = \psi^T \mathcal{C}$, where $\mathcal{C}$ is the `charge conjugation' operator in Euclidean space. It is given as
\beq
C =	\begin{pmatrix}
	0 & -1 \\
	1 & 0 
\end{pmatrix}.
\eeq 
It is crucial to note that this theory does not have a $\mathcal{Q}$-exact formulation or a Nicolai map. The model is not obtained by dimensional reduction, unlike its $\mathcal{N} = 2$ supersymmetric version. Another interesting property is that for periodic boundary conditions for fermions, dynamical breaking of SUSY is possible; that is, the vanishing of the Witten index, $\Delta = 0$, can happen. 

For a non-perturbative analysis of the model, we place the theory on a symmetric toroidal lattice discussed in the previous section. We consider a particular lattice formulation of the model introduced by Maarten Golterman and Donald Petcher \cite{Golterman:1988ta}. After integrating the fermions, the lattice representation of the Euclidean continuum action has the following bosonic and fermionic components
\bea
\label{eqn:n1wz-Slat}
S &=& S_b + S_f,~~ {\rm with} \\ 
S_b &=& \hf \left( - \phi_r \Box^2_{rr'} \phi_{r'} + W^2_r   \right), \\
S_f &=& \ln \left[{\rm Pf}\mathcal{M}\right] = - \hf {\rm tr} \left[ \ln \mathcal{M} \right],
\eea
where $r, r'$ are the lattice vectors, the fermion matrix is 
\beq
\mathcal{M} \equiv \mathcal{M}^{\alpha \beta}_{r r'} = \gamma^{\mu}_{\alpha \beta} \mathcal{D}^{\mu}_{r r' } + \delta_{\alpha \beta} W'_{rr'}, 
\eeq
and ${\rm Pf} \mathcal{M}$ is the Pfaffian of the fermion matrix. We use symmetric difference operators defined as follows
\bea
\mathcal{D}^\mu_{rr'} &=& \hf \left[\delta_{r + e_\mu , r' } - \delta_{r - e_\mu, r' } \right], \\
\Box^n_{rr'} &=& \hf \sum_\mu \left[\delta_{r + ne_\mu, r' } + \delta_{r - ne_\mu, r' } - 2 \delta_{rr' } \right].
\eea  
Since the action Eq. \eqref{eqn:n1wz-Slat} can be complex in general, we apply the complex Langevin method to study the theory for various superpotentials. We use the Euler discretized Langevin equation for the $s$-th lattice vector at Langevin time $\theta$. The drift term is defined as
\beq
v_{s, \theta} = - \frac{\partial S}{\partial \phi_{s, \theta}} = \Box^2_{s r'} \phi_{r', \theta} - W_{r'} W'_{r' s} + \left( \frac{\partial \mathcal{M}}{\partial \phi_s} \right)^{\alpha \beta}_{rr'} \left( \mathcal{M}^{-1} \right)^{\beta \alpha}_{r'r}.
\eeq

In order to test the reliability of complex Langevin simulations, we check the correctness criteria \cite{Nagata:2016vkn} based on the decay of the distribution $P(u)$ of the absolute value $u$ of the drift term. We have at a particular Langevin time $\theta$, the drift-term magnitude
\beq
u \equiv u_\theta = \sqrt{ \frac{1}{L^2}  \sum_s \Big | v_{s,\theta} \Big |^2}.
\eeq
We can trust the simulations if the distribution $P(u)$ of $u$ falls off exponentially or faster.

\subsection{Double-well superpotential}

We begin with considering a quadratic interaction potential or a double-well superpotential \cite{Catterall:2003ae, Baumgartner:2011jw, Wozar:2011gu} of the form
\beq
W(\phi) = \lambda \phi^2 -\frac{m^2}{4\lambda}, ~~~ \lambda \neq 0. 
\eeq 
The theory has two classical vacua at $\phi = \pm m/2\lambda$. In the lattice theory, we consider dimensionless couplings $\lambda_0$ and $m_0$, related to their continuum counterparts through $\lambda_0 = \lambda a$ and $m_0 = ma$. The potential and its derivative take the following form
\bea
W_r &=& \lambda_0 \phi_r^2 - \frac{m_0^2}{4\lambda_0} -\hf \Box^1_{rr'} \phi_r,\\
W'_{r r'} \equiv \frac{\partial P_r}{\partial \phi_{r'}} &=& 2 \lambda_0 \phi_r \delta_{rr'} - \Box^1_{r r'},
\eea
where $\Box^1_{r r'}$ is the Wilson mass operator, which vanishes in the continuum limit but eliminates the fermion doubling problem at a finite lattice spacing. Due to the introduction of the Wilson term, the lattice action is no longer invariant under parity, implying that the two vacuum states are not equivalent. It is expected that field configurations would reside in the vicinity of one of the classical vacua.	In the large values of ${m_0}^2/\lambda_0$, the $Z_2$ symmetry is spontaneously broken (in infinite volume), and $\phi$ settles down to a definite ground state.

\begin{figure}
	\centering
	\includegraphics[width=.6\textwidth,origin=c,angle=0]{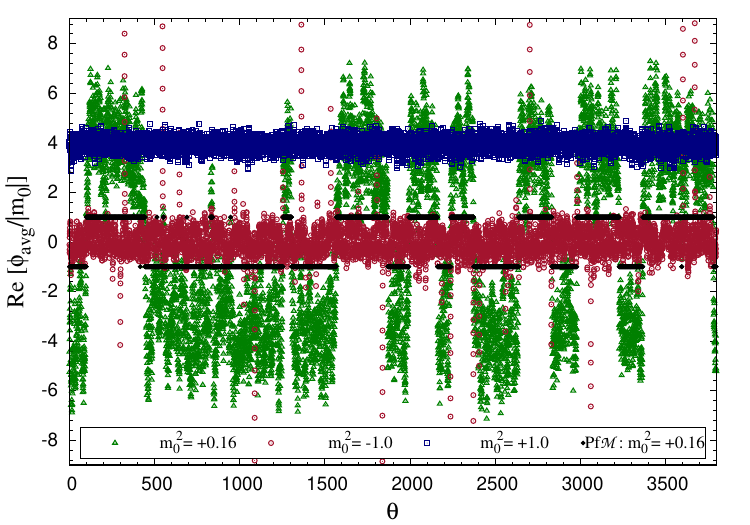} \\
	
	\includegraphics[width=.49\textwidth,origin=c,angle=0]{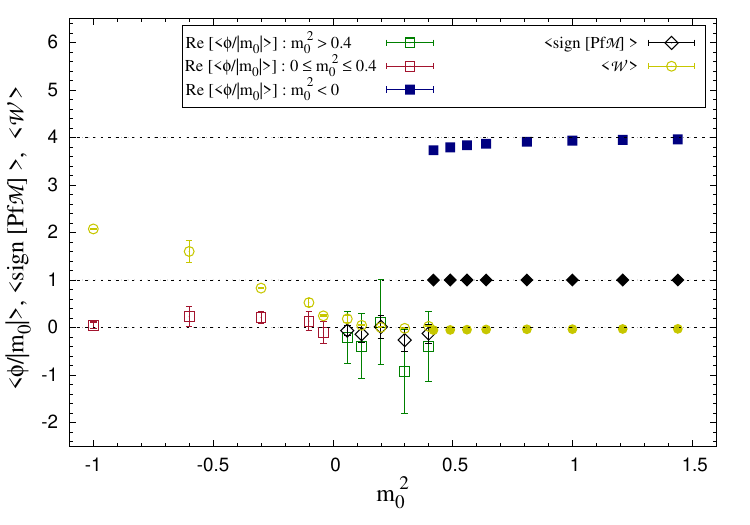}	
	\includegraphics[width=.49\textwidth,origin=c,angle=0]{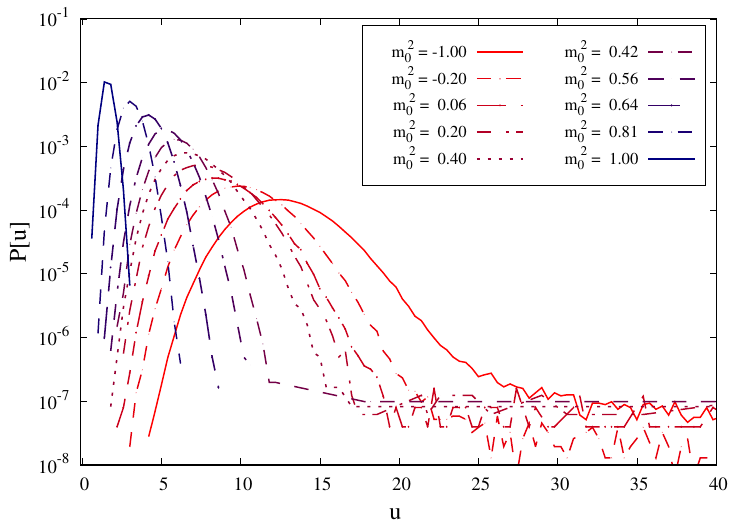}	
	
        \caption{$\mathcal{N} = 1$ Wess-Zumino model with a double-well superpotential. (Top) Langevin time histories of $\phi$ for various lattice mass $m_0^2$. The sign of the Pfaffian is also plotted for $m_0^2 = 0.16$. (Bottom-Left) Field $\langle \phi \rangle$, the sign of the Pfaffian $\langle {\rm sign} \left[ {\rm Pf} \mathcal{M}\right] \rangle$, and the Ward identity $\langle \mathcal{W} \rangle$ against lattice mass $m_0^2$. (Bottom-Right) Decay of the absolute drift for various lattice mass $m_0^2$. The plots are for a fixed lattice extent $L = 4$ and lattice coupling $\lambda_0 = 0.125$.}
	\label{fig:wz_quad}
\end{figure}

In Fig. \ref{fig:wz_quad}, we show our simulation results. On the top panel, we have the scalar field $\phi$ (order parameter) against the Langevin time $\theta$ for various $m_0^2$ values, depicting the different phases of the theory. We have fixed the lattice extent to $L= 4$, and lattice coupling to $\lambda_0 = 0.125$. For $m_0^2 = +1$, the field configurations (blue squares) are confined, along with small fluctuations, to one of the classical vacua, $\phi = {m_0}/2{\lambda_0}$, implying that the theory is in an $\mathbb{Z}_2$ broken phase. At $m_0^2 = +0.16$, we observe the tunneling behavior, the field configurations (green triangles) undergo large fluctuations, and they oscillate in between the two classical vacua at $\phi = \pm m_0/2\lambda_0$. The change in the sign of the Pfaffian (black diamonds) clearly illustrates this behavior. For $m_0^2 = -1$, the field configurations (red circles) suffer from small fluctuations around a single vacuum state, respecting the $\mathbb{Z}_2 $ symmetry. 

On the bottom left panel, we have the real part of the field $\langle \phi \rangle$ (order parameter), the sign of the Pfaffian $\langle {\rm sign}~{\rm Pf} \mathcal{M} \rangle$, and the simplest Ward identity $\langle \mathcal{W} \rangle$ against the lattice mass $m_0^2$ for a fixed lattice coupling $\lambda_0 = 0.125$. In the infinite-volume continuum theory, for large values of $m^2/\lambda$, the scalar field chooses a single unique ground state indicating a broken $\mathbb{Z}_2$ symmetry and unbroken SUSY in the model. We see that for mass larger than some critical value, $m_0^2 \ge m_{0, c}^2$, the scalar field (blue squares) selects the ground state $+ m_0/2\lambda_0$ and the sign of the Pfaffian (black diamonds) approaches $+1$. As $m_0^2$ is decreased, tunneling effects to the other vacuum state are observed, and the expectation value of the scalar field (green squares) vanishes $\langle \phi \rangle \sim 0$. This effect is a direct consequence of the Pfaffian flipping sign, reflected in $\langle {\rm sign}~{\rm Pf} \mathcal{M} \rangle \sim 0$. These results hint towards the restoration of $\mathbb{Z}_2$ symmetry and dynamical SUSY breaking. The above argument is supported by the Ward identity (yellow circles). As $m_0^2$ is decreased, we observe that the Ward identity no longer vanishes, that is, $\langle \mathcal{W} \rangle \neq 0$, indicating a transition from unbroken to broken SUSY phase. For $m_0^2 < 0$, we notice a $\mathbb{Z}_2$ symmetric phase with scalar field (red squares) $\langle \phi \rangle \sim 0$, and broken SUSY with $\langle \mathcal{W} \rangle \neq 0$. We show the decay of the absolute drift on the bottom right panel for our simulations. We observe exponential or faster decay for $m_0^2 > 0.42$ (illustrated by filled data points in the bottom left panel) and a power-law behavior for $m_0^2 \le 0.42$ (illustrated by unfilled data points in the bottom left panel). This could be pertaining to the singular-drift problem, and we are looking further into it.
\vspace{-2mm}
\subsection{Model with $\mathcal{PT}$-symmetric  potential}
Our main goal is to cross-check the results obtained by Bender and Milton in Ref. \cite{Bender:1997ps}. There they have looked at a two-dimensional supersymmetric model with four supercharges with the superpotential $W(\phi) = - i \lambda (i \phi)^{(1+\delta)}$. Parity symmetry is broken in this model. The authors tried to answer the question of whether the breaking of parity induces the breaking of SUSY with the help of a perturbative expansion in parameter $\delta$. They found, through the second order in $\delta$, that SUSY remained unbroken in the model, and suggested that SUSY could remain intact to all orders in powers of $\delta$. We plan to verify these results with the help of complex Langevin simulations of the model. 

	\chapter{Complex Langevin study of spontaneous symmetry breaking in the IKKT matrix model} \label{chap:mmodel-0d}

\setlength\epigraphwidth{13.2cm}
\setlength\epigraphrule{0pt}
\epigraph{The chapter is based on the following publication by the author: \\  \textbf{Arpith Kumar}, Anosh Joseph, and Piyush Kumar, \\{\it Complex Langevin study of spontaneous symmetry breaking in IKKT matrix model},\\ \href{https://doi.org/10.22323/1.430.0213}{PoS LATTICE2022 \textbf{213} (2023)}, accepted for publication  {\href{https://doi.org/10.48550/arXiv.2209.10494}{(arXiv: 2209.10494 [hep-lat])}} 	 }{}

Non-perturbative studies of ten-dimensional superstring theories are essential to understand the emergence of spacetime. In particular, the dynamical compactification of six extra dimensions is critical for such theories to be phenomenologically admissible. Matrix models are standard tools to investigate the non-perturbative aspects of superstrings. The IKKT (type IIB) matrix model was proposed in 1996 as a constructive definition of the ten-dimensional type IIB superstring theory \cite{Ishibashi:1996xs}. The action is a matrix regularization of the type IIB superstring action in the Schild gauge \cite{Green:1983wt}. The zero-volume limit of the ten-dimensional $\mathcal{N} = 1$ super Yang-Mills with SU($N$) gauge group formally yields the IKKT matrix model. The equivalence between the IKKT matrix model and type IIB superstring holds in the large-$N$ limit. The ten-dimensional extended $\mathcal{N} = 2$ SUSY ensures that gravity is included. The $N \times N$ bosonic matrices are analogous to the gravitational degrees of freedom, where the eigenvalues of the matrices denote the spacetime points. In this model, spacetime does not exist a priori but is dynamically generated from the matrix degrees of freedom. In the large-$N$ limit, a smooth spacetime manifold is expected to emerge from the eigenvalues. The compactification of the extra dimensions suggests that the distribution of eigenvalues should collapse to a lower-dimensional manifold. When this occurs in the Euclidean signature, the ten-dimensional rotational symmetry of the model must be spontaneously broken. 

In this work, we investigate the possibility of spontaneous symmetry breaking (SSB) of SO$(10)$ symmetry in the Euclidean version of the IKKT matrix model. The model has a severe sign problem; the Pfaffian obtained after integrating out fermions is inherently complex. The phase of the Pfaffian plays a critical role in determining the correct vacuum of the model. Unfortunately, Monte Carlo methods are unreliable for studying complex action matrix models. In recent years, the complex Langevin method \cite{Klauder:1983nn, Parisi:1984cs} has emerged as a successful candidate for tackling models with the sign problem. While applying the complex Langevin method to the Euclidean IKKT matrix model, we encounter problems that hamper the reliability of the simulations. The singular-drift problem is one of them. To avoid this problem, we suggest introducing SUSY-preserving deformations to the IKKT model. Complex Langevin simulations are performed to probe the nature of spontaneous rotational symmetry breaking at finite values of the mass deformation parameters, and then zero-mass extrapolation is taken to recover the original IKKT matrix model.

In this chapter, we present our preliminary results from the complex Langevin analysis of the IKKT matrix model in Euclidean signature. In Sec. \ref{sec:ikkt-model}, we briefly discuss the mathematical formalism of the model and the associated sign problem. Sec. \ref{sec:clm-ikkt} explains the problems associated with the complex Langevin study of the model. We introduce SUSY-preserving deformations in Sec. \ref{sec:susy-def} and discuss the simulation results. 

\section{Review of Euclidean IKKT matrix model}
\label{sec:ikkt-model}
The Euclidean IKKT matrix model, obtained by Wick rotation of the Lorentzian version, has a finite well-defined partition function \cite{Krauth:1998xh, Austing:2001pk},
\bea
\label{eqn:ikkt-action}
Z = \int dX d\psi e^{-S_{\rm IKKT}} \quad {\rm where~~} S_{\rm IKKT} = S_\text{b} + S_\text{f}, \\
S_\text{b} = -\qtr N ~\text{tr} \left([X_\mu, X_\nu]^2 \right), \\ 
S_\text{f} = -\hf N ~\text{tr} \left(\psi_\alpha (\mathcal{C} \Gamma^\mu)_{\alpha\beta}[X_\mu,\psi_\beta] \right).
\eea
The $N \times N$ traceless Hermitian matrices, $X_\mu (\mu = 1, 2, 3 \dots,~ 10)$ and $\psi_\alpha (\alpha = 1, 2, 3 \dots,~ 16)$ transform as vectors and Majorana-Weyl spinors under SO(10) transformations, respectively. We consider the following Weyl projected representation of Gamma matrices $\Gamma^\mu$ in ten dimensions \cite{Ambjorn:2000dx}, \\
\begin{minipage}{.25\linewidth}
	\vspace{-5mm}
	\begin{flushleft}
		\bea
			~~~~~~~~\Gamma^1 &=& i \sigma^2 \otimes \sigma^2 \otimes \sigma^2 \otimes \sigma^2, \nn \\
			~~~~~~~~\Gamma^3 &=& i \sigma^2 \otimes \sigma^2 \otimes \bf{1} \otimes \sigma^3, \nn \\
			~~~~~~~~\Gamma^5 &=& i \sigma^2 \otimes \sigma^3 \otimes \sigma^2 \otimes \bf{1}, \nn \\
			~~~~~~~~\Gamma^7 &=& i \sigma^2 \otimes \bf{1} \otimes \sigma^3 \otimes \sigma^2, \nn \\
			~~~~~~~~\Gamma^9 &=& i \sigma^3 \otimes \bf{1} \otimes \bf{1} \otimes \bf{1}, \nn
		\eea
		\end{flushleft}
\end{minipage}
\hfill
\begin{minipage}{.65\linewidth}
	\vspace{-5mm}
	\begin{flushright}
		\bea
			\Gamma^2 &=& i \sigma^2 \otimes \sigma^2 \otimes \bf{1} \otimes \sigma^1, \nn \\
			\Gamma^4 &=& i \sigma^2 \otimes \sigma^1 \otimes \sigma^2 \otimes \bf{1}, \nn \\
			\Gamma^6 &=& i \sigma^2 \otimes \bf{1} \otimes \sigma^1 \otimes \sigma^2, \nn \\
			\Gamma^8 &=& i \sigma^1 \otimes \bf{1} \otimes \bf{1} \otimes \bf{1}, \nn \\
			\Gamma^{10} &=&  \bf{1} \otimes \bf{1} \otimes \bf{1} \otimes \bf{1},
		\eea
		\end{flushright} 
\end{minipage}
\vspace*{4mm}
\newline
where $\sigma^{i} (i = 1, 2, 3)$ are the Pauli matrices and $\bf{1}$ is the $2 \times 2$ identity matrix. In the above representation, the charge conjugation matrix $\mathcal{C}$, satisfying $\mathcal{C} \Gamma^{\mu} \mathcal{C}^{\dagger} = \left(\Gamma^{\mu}\right)^{T}$ and $\mathcal{C}^{T} = \mathcal{C}$,  becomes an identity matrix.  The action is invariant under the SU($N$) gauge symmetry, the extended $\mathcal{N} = 2$ SUSY, and the SO(10) symmetry. 

The partition function, after integrating out the fermions, reads
\beq
Z = \int dX ~ {\rm Pf} \mathcal{M}~ e^{-S_{\rm b}}= \int dX ~ e^{-S_{\rm eff}}; ~~ S_{\rm eff} = S_\text{b} - \ln  {\rm Pf} \mathcal{M},  
\eeq
where the fermionic operator, $\mathcal{M}$ is a $16(N^2 -1) \times 16(N^2 -1)$ anti-symmetric matrix. In order to get the explicit form of $\mathcal{M}$, we expand $X_\mu$ and $\psi_\alpha$ in terms of the $N^2-1$ generators, $\{t^a\}$ of SU($N$) as follows,
\beq
X_\mu = \sum_{a=1}^{N^2-1} X_\mu^a t^a ~~{\rm and}~~ \psi_\alpha = \sum_{b=1}^{N^2-1} \psi_\alpha^b t^b,
\eeq
where $X_\mu^a$ and $\psi_\alpha^b$ are real and Grassmann numbers, respectively. The traceless, Hermitian generators are normalized as ${\rm tr} \left( t^a t^b \right) = \delta^{ab}$.
Using the properties of SU($N$) structure constants, we have
\beq
\mathcal{M}_{\alpha a, \beta b} =  \frac{N}{2} \Gamma_{\alpha \beta}^\mu ~{\rm tr} \Big ( X_\mu  \left[ t^a,t^b \right] \Big ).
\eeq
The interpretation of eigenvalues with spacetime points allows us to define the `radial extent’ of spacetime in each direction as follows,
\beq
\label{eqn:extent}
\langle \lambda_\mu \rangle = \Big \langle  \frac{1}{N} {\rm tr} \left( X_\mu^2 \right) \Big  \rangle.
\eeq
We consider $\lambda_\mu$ as an order parameter for investigating SSB. In the large-$N$ limit, if the extents are not equivalent, i.e., if they grow along some directions, $d < 10$, and shrink along others, we say that the SO(10) symmetry spontaneously breaks to SO($d$). The bosonic IKKT model was studied using Monte Carlo and $1/D$ expansion, and no SSB was observed \cite{Hotta:1998en}. Later, phase-quenched Monte Carlo studies were performed, and again no SSB was evident \cite{Ambjorn:2000dx, Anagnostopoulos:2012ib}. These studies implicate that the complex phase of the Pfaffian plays a crucial role in SSB. The phase fluctuates wildly, suggesting that the sign problem is severe; hence phase-quenched approximations are inexact. There exist only a few methods that are capable of incorporating the complex phase and tackling the associated sign problem. The complex Langevin method is one such promising approach.

\section{Applying complex Langevin to the IKKT model}
\label{sec:clm-ikkt}
In this section, we discuss the application of complex Langevin to the Euclidean IKKT model. 

The complex Langevin update of bosonic matrices $X_\mu$ at fictitious Langevin time $\theta$ reads
\bea
\frac{d(X_{\mu})_{ij}}{d\theta} = -\frac{\partial S_{\rm eff}}{\partial(X_{\mu})_{ji}} + (\eta_{\mu})_{ij}(\theta),
\eea
where $\eta_{\mu}(\theta)$ is a Hermitian Gaussian noise obeying the probability distribution 
\beq
P[\eta_{\mu}(\theta)] \sim \exp \left(\qtr \int {\rm tr} \left( \eta^2_{\mu}(\theta) \right) \right),
\eeq 
and the gradient of the action can be computed using the following expressions;  
\bea
\label{eqn:grad-act}
{\partial {S}_{\rm eff}}/{\partial (X_\mu)_{ji}} &=& {\partial {S}_\text{b}}/{\partial (X_\mu)_{ji}} - \hf {\partial \left( {\rm tr} \left[
		\ln \mathcal{M} \right] \right)} /{\partial (X_\mu)_{ji}}, \\
\label{eqn:grad-bos-act}
{\partial {S}_\text{b}}/{\partial (X_\mu)_{ji}} &=& -N \left( {\Big [} X_\nu,\left[X_\mu, X_\nu\right] {\Big ]} \right)_{ij}, \\
\label{eqn:grad-fer-act}
- \hf {\partial \left( {\rm tr} \left[
		\ln \mathcal{M} \right] \right)} /{\partial (X_\mu)_{ji}}  &=& -\hf  {\rm tr} \left[ \mathcal{M}^{-1} \frac{\partial \mathcal{M}}{\partial (X_\mu)_{ji}} \right], \\
\frac{\partial \mathcal{M}}{\partial (X_\mu)_{ji}} &=& \frac{N}{2} \Gamma_{\alpha \beta}^{\mu} \Big (  \left[ t^a,t^b \right] \Big )_{ij}.
\eea
The complex Langevin method sometimes yields wrong results due to incorrect convergence. Fortunately, there exist certain correctness criteria \cite{Seiler:2012wz, Nagata:2016vkn}, which can validate the simulation results. The more recent one is based on the localized distribution of the probability of complex field configurations. The distribution of the magnitude of drift 
\beq
u = \sqrt{\frac{1}{10N^2} \sum_{\mu=1}^{10} \sum_{i,j=1}^{N} \left| \frac{\partial S_{\rm eff}}{\partial(X_{\mu})_{ji}}\right|^2},
\eeq
should be suppressed exponentially or faster to ascertain the reliability of simulations. 
\begin{figure}[htbp]
	\centering
	\includegraphics[width=.49\textwidth,origin=c,angle=0]{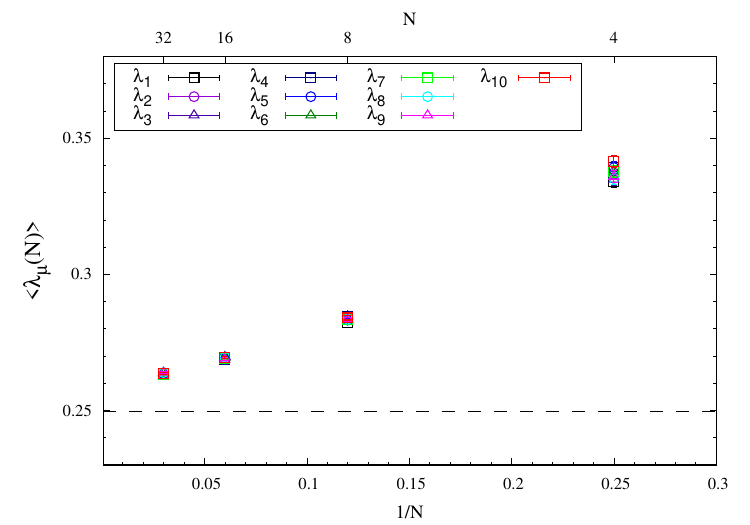}
	\includegraphics[width=.49\textwidth,origin=c,angle=0]{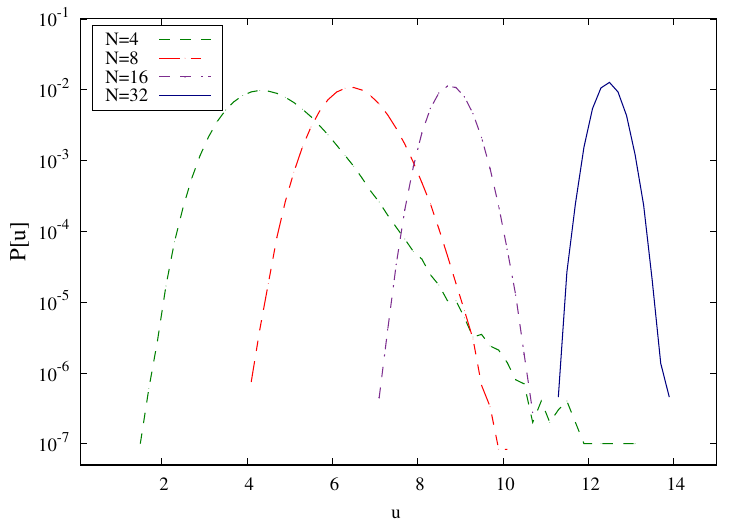}
	\caption{Bosonic IKKT model. (Left) The expectation value of order parameter $\lambda_{\mu}$ and (Right) the corresponding probability of magnitude of drift for various $N$.} 
	\label{fig:bos}
\end{figure}

In Fig. \ref{fig:bos}, we present the complex Langevin simulation results for the bosonic IKKT model. From the plot on the left panel, we infer that the SO(10) symmetry is intact even for finite-$N$ and approaches the analytical value in the large-$N$ limit \cite{Hotta:1998en}. The probability of drift plotted on the right panel falls off exponentially or faster, which implies that the simulation results are reliable. While applying the complex Langevin method to the Euclidean IKKT model, we encountered two major problems, namely the {\it excursion problem} and the {\it singular-drift problem}, that violate the above-mentioned correctness criterion. In the upcoming subsections, we briefly discuss these problems and also ways to circumvent them.

\subsection{Excursion problem and gauge cooling}
\label{subsec:excursion-problem}
The inherently complex nature of the Pfaffian advocates excursion of the bosonic matrices $X_\mu$ into anti-Hermitian directions, enlarging the group space from  SU($N$) to ${\rm SL}(N, \mathbb{C})$. We encounter the excursion problem when $X_\mu$ wanders too far away from SU($N$). A proposed solution to this problem is gauge cooling \cite{Seiler:2012wz}.
 
We define the `Hermiticity norm' \cite{Nagata:2016vkn}
\beq
\mathcal{N}_{\rm H} = -\frac{1}{10N} \sum_{\mu}{\rm tr} \left( \left[ X_\mu - X_\mu^\dagger \right]^2 \right),
\eeq
to track deviation of $X_{\mu}$ from Hermitian configurations. The matrix fields $X_\mu$ are invariant under the enlarged gauge symmetry,
\bea
X_\mu \rightarrow g X_\mu g^{-1},~ g \in {\rm SL}(N,\mathbb{C}),  \\
g = {\rm e}^{-\alpha \delta{\mathcal{N}}_{\rm H}}, ~~ \delta{\mathcal{N}}_{\rm H} = \frac{1}{N} \sum_\mu\left[ X_\mu, X_\mu^\dagger \right], ~~\alpha \in \mathbb{R}^+,  
\eea
while $\mathcal{N}_{\rm H}$ is not invariant. We utilize this property and successively apply the above gauge transformation at each Langevin step until $\mathcal{N}_{\rm H}$ is minimized. The gauge cooling procedure has been proven to respect complex Langevin correctness criteria \cite{Nagata:2016vkn}. In our simulations, we observe that after applying gauge cooling, $\mathcal{N}_{\rm H}$ is well under control.  

\subsection{Singular-drift problem and mass deformations}
\label{subsec:singular-drift-problem}
The gradient of the effective fermionic action contains the inverse of the fermion operator, $\mathcal{M}^{-1}$. The singular-drift problem arises when the eigenvalues of $\mathcal{M}$ accumulate densely near zero. One way to avoid this problem is to shift the eigenvalues of the fermion operator away from the origin. This shift can be introduced by adding fermion bilinear mass deformation terms to the action \cite{Ito:2016efb}. In general, the deformations have the following form
\beq
\Delta S = \frac{N}{2}\epsilon m_\mu  {\rm tr}\left( X_\mu ^2 \right) +  \frac{N}{2}  {\rm tr} \Big ( \psi_\alpha (\mathcal{CA})_{\alpha\beta} \psi_\beta \Big ),
\eeq
where $m_\mu$ is the mass vector and $\mathcal{A}$ is a complex $16 \times 16$ anti-symmetric matrix. Majorana-Weyl spinors severely limit the allowed ranks of gamma matrices in ten dimensions. This implies that only bilinears of rank three and seven tensor (equivalent due to the duality relations) survive \cite{Wetterich:1982eh}, that is, $\mathcal{A} = i m_{\rm f} \epsilon_{\mu\nu\sigma} \Gamma_{\mu} \Gamma_{\nu}^\dagger \Gamma_\sigma$ with totally anti-symmetric $\epsilon_{\mu \nu \sigma}$ 3-form. Here $\epsilon$ and $m_{\rm f}$ are the deformation parameters. Apart from explicitly breaking the SO(10) symmetry, such deformations induce SUSY breaking. The extended $\mathcal{N} = 2$ SUSY is crucial for the model to include gravity. Similar deformations were considered in a recent study \cite{Anagnostopoulos:2020xai}, where the authors concluded that the SO(10) symmetry was spontaneously broken down to SO(3) (consistent with Gaussian expansion method results \cite{Nishimura:2011xy}). Studying SSB with this deformation requires three-step extrapolations, $N \rightarrow \infty$, $\epsilon \rightarrow 0$, $m_{\rm f} \rightarrow 0$, which introduce systematic errors. In this work, we suggest SUSY-preserving deformations that reduce the number of extrapolations to just two.

\section{Supersymmetry-preserving mass deformations}
\label{sec:susy-def}
We introduce SUSY-preserving deformations \cite{Metsaev:2001bj, Blau:2001ne, Bonelli:2002mb, Cvetic:2002si,  Austing:2003kd} which includes a Myers term, to the original IKKT model $(S_{\rm IKKT})$. We obtain the following deformed model 
\bea
\label{eqn:susy-deformed-ikkt-action}
S &=& S_{\rm IKKT} +  S_{\Omega},\nn \\
S_{\rm  \Omega} &=& N~ {\rm tr} \left( M^{\mu \nu} X_{\mu} X_{\nu} +i N^{\mu \nu \gamma} X_\mu \left[X_{\nu}, X_{\gamma}  \right]  +\frac{i}{8} \psib N_3 \psi \right),
\eea
where $N_3 = \Gamma^{\mu \nu \gamma} N_{\mu \nu \gamma}$, with $N^{\mu \nu \gamma}$ denoting a totally anti-symmetric tensor, and $M_{\mu\nu}$ is the mass matrix. The deformed model is a zero-dimensional analog of the quantum mechanical BMN matrix model \cite{Berenstein:2002jq} in the pp-wave background and preserves half of the SUSY. The action is invariant under the following SUSY transformations
\bea
\label{eqn:susy-deformed-ikkt-transformations}
\delta X^{\mu} &=& -\hf \overline{\varepsilon} \Gamma^{\mu} \psi,\\
\delta \psi &=& \qtr \left[ X^{\mu}, X^{\nu} \right] \Gamma_{\mu \nu} \varepsilon - \frac{i}{16} X^{\mu} \left( \Gamma_{\mu} N_3 + 2N_3 \Gamma_{\mu} \right)\varepsilon,
\eea 
provided a mass/flux constraint, 
\beq
\left[ N_3(\Gamma^{\mu} N_3 +2 N_3 \Gamma^{\mu}) + 4^3 M^{\mu \nu} \Gamma_{\nu} \right]\varepsilon = 0,
\eeq
is satisfied. A straightforward solution to this constraint is to consider 
\bea 
\label{eqn:susy-deformed-ikkt-solution}
N_{3} &=& -\Omega\Gamma^{8}{\Gamma^{9}}^{\dagger}\Gamma^{10}, \\ ~N^{\mu \nu \gamma} &=&  \frac{\Omega}{3!} \sum_{\mu,\nu,\gamma=8}^{10}{\epsilon^{\mu \nu \gamma}}, \\ M &=&  \frac{\Omega^2}{4^3} \left( \mathbb{I}_7 \oplus 3\mathbb{I}_3  \right), 
\eea
which explicitly breaks the ten dimensional rotational symmetry, SO(10) to SO(7) $\times$ SO(3). One can obtain the original IKKT matrix model and study the spontaneous breaking of rotational symmetry by extrapolating $\Omega \to 0$ in the large-$N$ limit. 

The gradient of the supersymmetric deformed model, in addition to the original IKKT model in Eq. \eqref{eqn:grad-act}, has the following bosonic deformation contributions 
\bea
\label{eqn:bos-mass-matrix}
{-N M ^{\mu \nu} \partial \left( X_{\mu} X_{\nu}\right) }/{\partial X_{\sigma}} &= & -2N M ^{\sigma \nu}X_{\nu}^{T} \\ &=& {\frac{2\Omega^2 N}{4^3} } ~  
\begin{cases}
	X_{\sigma}^{T} \rm{, ~} \sigma = 1\text{-}7\\
	3X_{\sigma}^{T} \rm{, ~} \sigma = 8,9,10,
\end{cases}
\\
{iN \partial \left(  N^{\mu \nu \gamma} X_\mu \left[X_{\nu}, X_{\gamma}  \right] \right) }/{\partial X_{\sigma}} &= & i{\Omega}N \sum_{\nu\gamma=8}^{10} \left(   \epsilon^{\sigma\nu\gamma} X_{\nu} X_{\gamma}   \right) \\ 
&=& { i{\Omega}N }
\begin{cases}
	\left[ X_{9},X_{10} \right]^{T} \rm{, ~} \sigma = 8\\
	\left[ X_{10},X_{8} \right]^{T} \rm{, ~} \sigma = 9\\
	\left[ X_{8},X_{9} \right]^{T} \rm{, ~} \sigma = 10.
\end{cases}
\eea
The fermion bilinear deformation modifies the fermion operator in the following manner;
\bea
\mathcal{M}_{\alpha a \beta b}  \to \tilde{\mathcal{M}}_{\alpha a \beta b} &=& \frac{N}{2}  \Gamma_{\alpha \beta}^\mu{\rm tr} \left( X_\mu \left[t^a,t^b \right] \right) -\frac{i\Omega N}{8} \left( \Gamma^{8} {\Gamma^{9} }^{\dagger}\Gamma^{10} \right)_{\alpha\beta} \delta_{ab}.   
\eea
\begin{figure}[htbp]
	\centering
	\includegraphics[width=.6\textwidth,origin=c,angle=0]{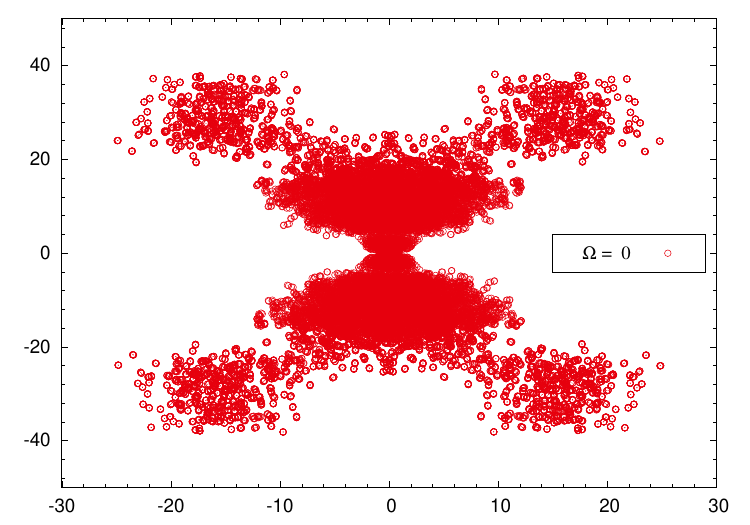}
	\includegraphics[width=.6\textwidth,origin=c,angle=0]{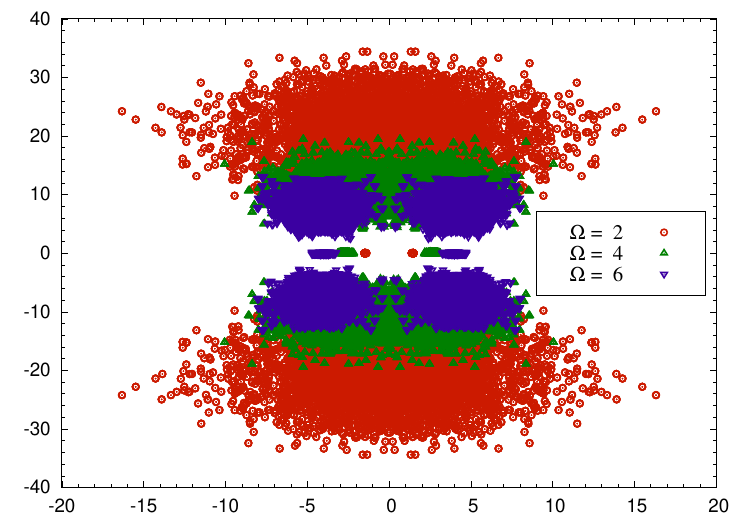}
	\includegraphics[width=.6\textwidth,origin=c,angle=0]{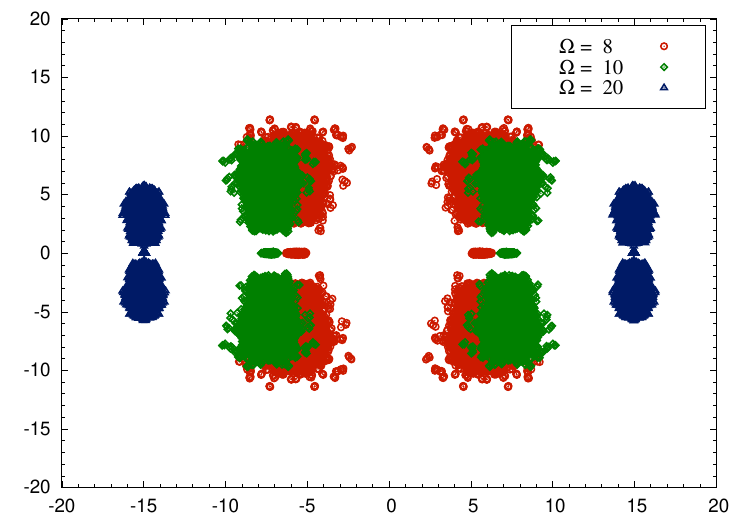}
	\caption{IKKT model with SUSY-preserving mass deformations. Scatter plot of real versus imaginary part of eigenvalues of the fermion operator $\mathcal{M}$. The plots are for various mass deformation parameters $\Omega$ and fixed $N=6$.} 
	\label{fig-eigen-nc6.png}
\end{figure}
In Fig. \ref{fig-eigen-nc6.png}, we plot the eigenvalue distribution of the fermion operator $\mathcal{M}$ for the SUSY-preserving mass deformed IKKT model. The singular-drift problem is apparent for mass deformation parameter $\Omega = 0$, that is, the original IKKT model. As we increase $\Omega$, the trend suggests that the eigenvalue distribution shifts further and further away from the origin. These results strongly indicate that the SUSY-preserving mass deformations indeed evade the singular-drift problem. 

\subsection{Bosonic IKKT deformed model with Myers term}
\label{subsec:bos-ikkt-myers}
We append the bosonic Gaussian mass deformation terms along with a Myers term to the bosonic IKKT matrix model. The deformed model action reads $ S_{\rm b} = S_{\rm bIKKT} + S_{\rm G} +S_{\rm Myers},$ 
where
\beq
S_{\rm G} = \frac{\Omega^2 N}{4^3}~ {\rm tr} \left( \sum_{i=1}^{7} X_{i}^2 + 3 \sum_{a=8}^{10}  X_{a}^2 \right),
\eeq
and
\beq
 S_{\rm Myers} = \frac{i\Omega N}{3!}~ {\rm tr} \left( \sum_{a,b,c=8}^{10} X_{a} \left[X_{b}, X_{c}  \right] \right).
\eeq
We perform complex Langevin simulations for various mass deformation parameters $\Omega$ and investigate whether the ten-dimensional rotational symmetry is intact in the limit $\Omega \to 0$. We notice that the order parameter $\lambda_\mu(\Omega) $ has an inverse order dependence on mass deformation parameter $\Omega$. As a consequence, $\lambda_\mu (\Omega)$ blows up in the limit $\Omega \to 0$. To resolve this issue, we consider the normalized extent values defined as
\beq
\label{eqn:normalized_extent}
\langle \rho_{\mu}(\Omega) \rangle  = \Big \langle  \frac{\lambda_\mu (\Omega) \rangle}{\sum_{\mu}\lambda_\mu (\Omega) } \Big \rangle.
\eeq
The normalized extents cancel a significant part of the dependency on the deformation parameter. In the case of broken SO(10) symmetry, the normalized extents $\rho_\mu$ will not be equal in all directions. 
\begin{figure}[htbp]
	\centering
	\includegraphics[width=.49\textwidth,origin=c,angle=0]{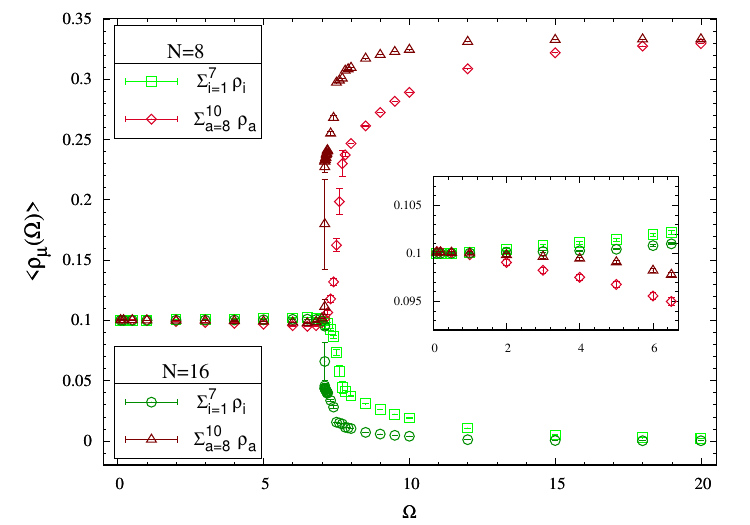}
	\includegraphics[width=.49\textwidth,origin=c,angle=0]{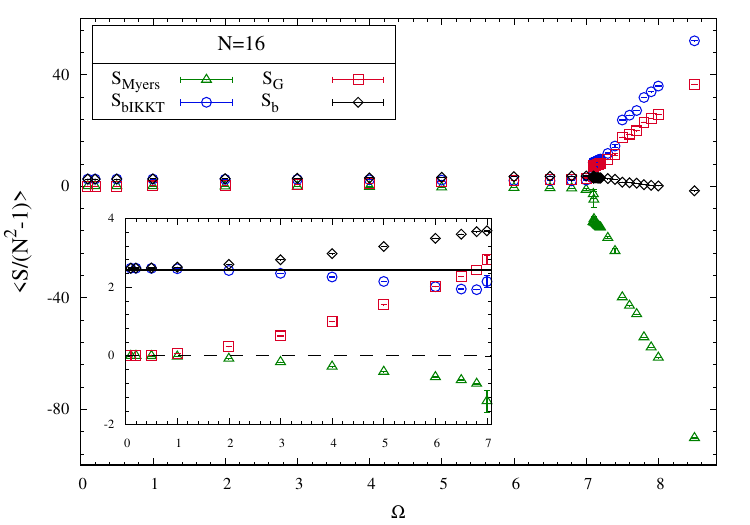}
	\caption{Bosonic IKKT deformed model with Myers term. (Top) The averaged extents, $\frac{1}{7}\sum_{i=1}^{7} \rho_{i}(\Omega)$ and $\frac{1}{3}\sum_{a=8}^{10} \rho_{a}(\Omega)$ versus mass deformation parameter $\Omega$ for $N=8, 16$. (Bottom) The bosonic action terms versus mass deformation parameter $\Omega$ for $N=16$. } 
	\label{fig:bos-myers}
\end{figure}

In this model, we observe an explicit symmetry breaking of SO(10) $\to$ SO(7) $\times$ SO(3) for large enough $\Omega$ values, and thus, we have considered the averaged extents, that is, $\frac{1}{7}\sum_{i=1}^{7} \rho_{i}(\Omega)$ and $\frac{1}{3}\sum_{a=8}^{10} \rho_{a}(\Omega)$ as the order parameters. The averaged extents are shown on the left panel of Fig. \ref{fig:bos-myers}. In the limit $\Omega \to 0$, the two averaged extents converge, and the SO(10) symmetry of the original bosonic IKKT model is restored. These results demonstrate that the bosonic mass deformation and the Myers term do not play any role in the SSB of SO(10) symmetry. We also notice a first-order phase transition around $\Omega \sim 7.1$ for $N = 16$. We believe this is a consequence of the change in the saddle point configurations due to the Myers term. On the right panel of Fig. \ref{fig:bos-myers}, we see that the dominant nature of the Myers term is apparent after $\Omega \sim 7.1$. The inset plot shows that in the limit $\Omega \to 0$, the contributions from the Gaussian deformation and the Myers term vanish, and we obtain the bosonic IKKT model.   

\subsection{IKKT model with SUSY-preserving mass deformations}
\label{subsec:ikkt-susy-def}
In this section, we report our results from complex Langevin simulations of the IKKT model with SUSY-preserving mass deformations. In these simulations, the expectation values of extents are rendered complex because of the complexity of the action. We apply the gauge cooling algorithm discussed in Sec. \ref{subsec:excursion-problem} to stay as close as possible to the Hermitian directions. In the context of extents (or normalized extents), we only refer to the real part and note that the imaginary part is negligible.

\begin{figure}[htbp]
	\centering
	\includegraphics[width=.49\textwidth,origin=c,angle=0]{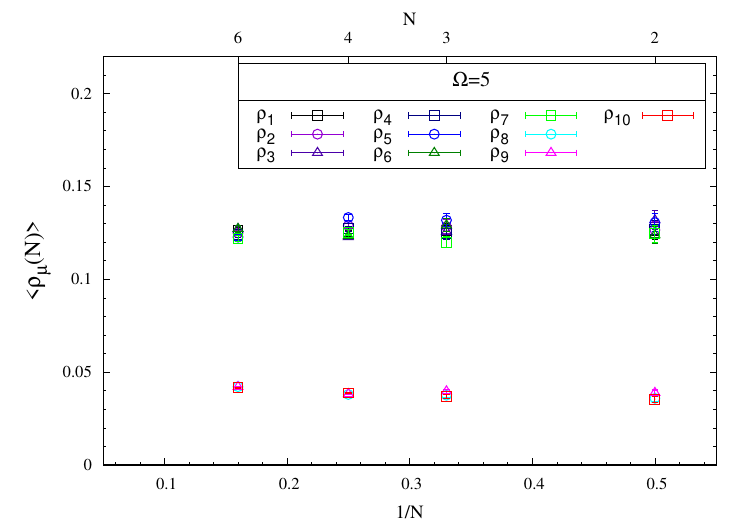}
	\includegraphics[width=.49\textwidth,origin=c,angle=0]{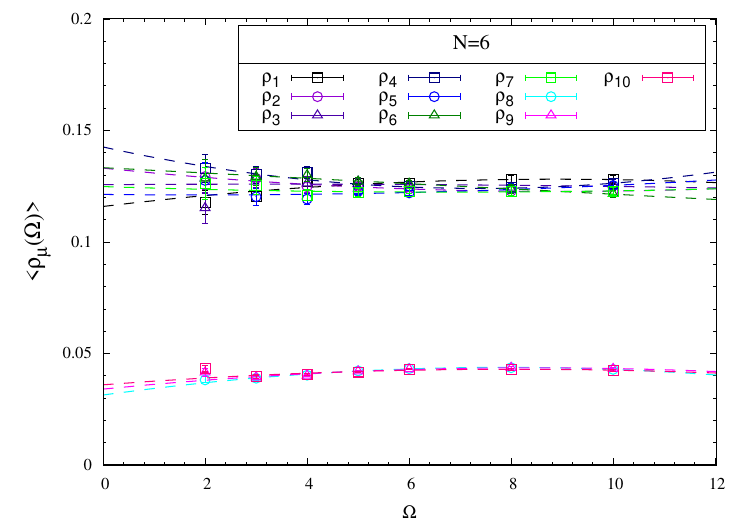}
	\caption{IKKT model with SUSY-preserving mass deformations. (Left) The normalized extents (order parameter) $\rho_{\mu}$ versus $N$ for fixed $\Omega=5$. (Right) The normalized extents $\rho_{\mu}$ versus $\Omega$ for fixed $N=6$. } 
	\label{fig:rho-N-Omega}
\end{figure}

On the left panel of Fig. \ref{fig:rho-N-Omega}, we plot the normalized extents $\rho_\mu$ for fixed $\Omega = 5$ and various matrices size $N$. For a large enough $\Omega$ value, as expected, we observe an explicit SO(7) $\times$ SO(3) symmetry breaking. Our finite-$N$ results suggest that the extents $\rho_\mu$ are almost independent of $N$, but we require large-$N$ computations to comment on the exact behavior concretely. 

The estimation of $\mathcal{M}^{-1}$ has a computational time complexity of O($N^6$) and is the bottleneck of the algorithm. In this study, we consider $N = 6$ as the large-$N$ limit and take the mass deformation parameter $\Omega \to 0$ limit on the right panel of Fig. \ref{fig:rho-N-Omega}. We observe that the complex Langevin simulations are unreliable for $\Omega < 2$. In the limit $\Omega \to 0$, we recover the original IKKT matrix model, and even for $N = 6$, the spontaneous breaking of SO(10) $\to$ SO(7) $\times$ SO(3) symmetry is apparent. Interestingly, we notice that the SO(7) symmetry appears to further break down into smaller subgroups as $\Omega \to 0$, indicating a SO($d$) symmetric vacuum with $d < 7$. To investigate the exact nature of the symmetric vacuum of the IKKT matrix model, we need to consider large-$N$ extrapolations. 

	\fancyhead[RO]{\small Chapter \thechapter \ \ \ \textit{\leftmark}  \ \ \ \   \ \textbf{\thepage}}
	\chapter{Conclusions and Future Prospects}

In this chapter, we discuss the summary of our studies, along with some future prospects.

	\subsection*{Complex Langevin simulations of zero-dimensional supersymmetric quantum \\ field theories}
	
	We implemented complex Langevin dynamics with stochastic quantization to investigate SUSY breaking in a class of zero-dimensional $\cN = 2$ supersymmetric models with real and complex actions. We looked at double-well superpotential, general polynomial superpotential, and $\mathcal{PT}$-symmetric models inspired $\delta$-potentials. In some cases, we were able to cross-check the presence or absence of SUSY breaking with the existing analytical results. Our simulations strongly suggest that SUSY is preserved for $\mathcal{PT}$-symmetric models inspired $\delta$-potentials. We have also investigated the reliability of complex Langevin simulations by monitoring the Fokker-Planck equation as a correctness criterion and also by looking at the probability distributions of the magnitude of the drift terms.
	\vsphf
	
	It would be interesting to study complex Langevin dynamics in the above models, generalized to non-Abelian cases, for example, with SU$(N)$ symmetry. SUSY may be restored in the large-$N$ limit of these models. It would also be interesting to explore spontaneous SUSY breaking when $\delta$ in the superpotential is a continuous parameter. 
	
	\subsection*{Complex Langevin dynamics and supersymmetric quantum mechanics}
	
	With the help of the complex Langevin method, we investigated dynamical SUSY breaking in quantum mechanics models with real and complex actions. When periodic boundary conditions are used in quantum mechanics with broken SUSY, the expectation values of observables are ill-defined. We resolved this problem by using twisted boundary conditions \cite{Kuroki:2009yg} in our non-perturbative lattice simulations. We find that for the case of real actions, SUSY is preserved in models with harmonic and anharmonic oscillator potentials, and with odd-powered polynomial potential. SUSY is dynamically broken for the case of even-powered polynomial potential. For the case of the supersymmetric anharmonic oscillator, our simulations reproduced the earlier results obtained through Hamiltonian Monte Carlo \cite{Catterall:2000rv}.
	\vsphf
	
	We then moved on to simulate an interesting class of supersymmetric models with complex actions that are ${\cal PT}$-symmetric. Our simulations suggested that SUSY is preserved in these models. We noticed that during the complex Langevin simulations, some of these models suffered from the singular-drift problem. In order to overcome this difficulty, we introduced appropriate deformation parameters in the models, such that the criteria of the correctness of complex Langevin simulations are respected. The target theories are recovered by taking the limits in which the deformation parameters go to zero. The reliability of the simulations was checked by studying the Langevin operator on observables and the exponential fall-off of the drift terms.  Our conclusion is that the complex Langevin method can be used reliably to probe non-perturbative SUSY breaking in various quantum mechanics models with real and complex actions. 
	\vsphf
	
	It would be interesting to extend our investigations to models in higher dimensions, especially quantum field theory systems in four dimensions, such as QCD with finite temperature and baryon/quark chemical potentials. Another long-term hope would be to apply these methods to ${\cal PT}$-symmetric supersymmetric quantum field theories in higher dimensions. We hope that these studies may find applications in fundamental physics. 
	
	Formulating supersymmetric field theories on the lattice is difficult due to naive discretization, which breaks SUSY, and the Leibniz rule's failure. An exciting prospect is to adopt the elegant non-lattice approach, based on the Fourier decomposition of fields, to study supersymmetric quantum mechanical models. The goal will be to perform complex Langevin analysis of various models with complex potentials, including the ones exhibiting $\mathcal{PT}$-symmetry.

	\subsection*{Complex Langevin analysis of two-dimensional quantum field theories}
	
	We investigated two-dimensional scalar field theories with various interactions, including the interesting cases of $\mathcal{PT}$-invariant potentials. We laid out the lattice construction of the models and then studied the bosonic versions with $\phi^4$ and $\mathcal{PT}$-symmetric potentials. After that, we looked at a model with minimal SUSY, the two-dimensional $\cN = 1$ Wess-Zumino model. Our simulations for the model with double-well superpotential suggest that SUSY is preserved in this model when the mass parameter $m_0^2$ is greater than some critical value.   
	\vsphf
	
	Another motivation for considering the two-dimensional $\cN = 1$ Wess-Zumino model is to cross-check the results obtained by Carl Bender and Kimball Milton in Ref. \cite{Bender:1997ps}. There, they examined a two-dimensional supersymmetric model with four supercharges and a parity-breaking superpotential of the form $W(\phi) = - i \lambda (i \phi)^{(1+\delta)}$. There, the authors tried to answer whether parity breaking would induce the breaking of SUSY with the help of an expansion in parameter $\delta$. They proved that SUSY remained unbroken through a second-order expansion in $\delta$, suggesting that SUSY could remain intact across all orders. An immediate prospect is to verify these results with the help of complex Langevin simulations. 
	
	\subsection*{Complex Langevin study of spontaneous symmetry breaking in IKKT matrix model}
	
	We conducted a first-principles analysis of the Euclidean IKKT matrix model using the complex Langevin method. Our main objective was to inspect the spontaneous breaking of ten-dimensional rotational symmetry. As a preliminary step, we performed simulations of the bosonic IKKT model and found no evidence of spontaneous symmetry breaking. This result is consistent with earlier research that employed Monte Carlo and $1/D$ expansions. Our next step was to incorporate fermions into the bosonic model and perform simulations for the IKKT model. During these simulations, we encountered the singular-drift problem, which we resolved with the help of SUSY-preserving mass deformations. Based on the results of our analysis, mass deformations with a Myers term respect SUSY and can successfully resolve the singular-drift problem. To understand the nature of the contribution from the Myers term, we performed simulations for the bosonic IKKT model with Gaussian deformations and the Myers term. Here, we verified that the presence of the Myers term does not play any role in the spontaneous SO$(10)$ symmetry breaking.
	\vsphf
	
	Finally, we studied the spontaneous symmetry breaking in the Euclidean IKKT matrix model in the vanishing deformation parameter limit. Our analysis indicates that the Pfaffian phase triggers the spontaneous breaking of ten-dimensional rotational symmetry. Analysis for $N = 6$ indicated the SO$(7)$ symmetric vacuum realization. In addition, we observed hints of smaller subgroups of SO$(d)$ symmetric vacua with $d < 7$. A more comprehensive and robust large-$N$ investigation is in progress to identify the exact nature of the vacuum. 
 
 We are considering various efficient techniques to compute the $\mathcal{M}^{-1}$ operator. An alternative is to approximate the gradient of fermionic contributions to the effective action, that is ${\rm tr} \left[\frac{\partial \mathcal{M}}{\partial {(X_\mu)}_{ji}} \mathcal{M}^{-1} \right]$, with the help of {\it stochastic estimation} (also known as {\it noisy estimator method}). The direct estimation of $\mathcal{M}^{-1}$ has a computational time complexity of O($N^6$), while the stochastic estimation of the fermion gradient using an explicit representation of $\mathcal{M}$ requires only O($N^4$) arithmetic operations. First, we construct $16 (N^2 -1)$-dimensional stochastic vectors $\eta_a$ obeying the following property
	\beq
		\langle \eta_a^* \eta_b \rangle_{\eta} = \delta_{ab}, 
	\eeq
	where we consider
	\beq
		\eta_a = \frac{\chi_a + i\xi_a}{\sqrt{2}}~~{\rm and}~~ \eta_a^{*} = \frac{\chi_a - i\xi_a}{\sqrt{2}},  
	\eeq
	with $\chi_a$ and $\xi_a$ adhere to the distribution ${\rm N}(0,1)$, that is $\langle \chi_a^2 \rangle = \langle \xi_a^2 \rangle =1$. Then, we approximate the trace in the fermion gradient in the following manner
	\bea
	\label{eqn:trace-approx-noisy-estimator}
		{\rm tr} \left[\frac{\partial \mathcal{M}}{\partial {(X_\mu)}_{ji}} \mathcal{M}^{-1} \right] 
		&=& \sum_{a,b=1}^{16 (N^2 -1)} {\left[\frac{\partial \mathcal{M}}{\partial {(X_\mu)}_{ji}} \mathcal{M}^{-1} \right]}_{ab} \delta_{ab}  \\
		&=& \sum_{a,b=1}^{16 (N^2 -1)} {\left[\frac{\partial \mathcal{M}}{\partial {(X_\mu)}_{ji}} \mathcal{M}^{-1} \right]}_{ab} \Big \langle \eta_a^* \eta_b  \Big \rangle_{\eta} \\	
		&=& \sum_{a,b=1}^{16 (N^2 -1)} \Big \langle \eta_a^*{\left[\frac{\partial \mathcal{M}}{\partial {(X_\mu)}_{ji}} \mathcal{M}^{-1} \right]}_{ab}  \eta_b \Big \rangle_{\eta} \\
		&=& \Big \langle \eta^{*} { \frac{\partial \mathcal{M}}{\partial {(X_\mu)}_{ji}}  } \mathcal{M}^{-1}  \eta \Big \rangle_{\eta} 
		\label{eqn:noisy-estimator-avg}\\
		&\simeq& \eta^* \frac{\partial \mathcal{M}}{\partial {(X_\mu)}_{ji}} \underbrace{\mathcal{M}^{-1} \eta}_{\zeta},   
		\label{eqn:noisy-estimator-approx}
	\eea
	where the average ${\langle \cdot  \rangle}_{\eta}$ is with respect to the stochastic vectors $\eta$ and the quantity $\zeta$ is estimated by solving the linear equation 	
	\beq
	\left( \mathcal{M}^{\dagger} \mathcal{M} \right) \zeta = \mathcal{M}^{\dagger}\eta  
	\eeq
    using the conjugate gradient method. Most importantly, the approximation in Eq. \eqref{eqn:noisy-estimator-approx} implies that only a single noisy vector is taken into account rather than an average. Such an approximation is allowed in complex Langevin dynamics since the associated Fokker-Planck equation is unchanged, and it does not yield any systematic errors \cite{Batrouni:1985jn}. Moreover, performing stochastic estimation using the linear transformation property of the fermion matrix $\mathcal{M}$, that is,
    \beq
    \Psi_{\alpha} \to (\mathcal{M}\Psi)_{\alpha} \equiv (\Gamma_{\mu})_{\alpha \beta} \left[ X_{\mu}, \Psi_{\beta} \right],
    \eeq
    operating on the linear space of $N \times N$ traceless complex matrices $\Psi_\alpha$, can reduce the computational cost down to O($N^3$)  \cite{Anagnostopoulos:2017gos}. We hope to report on the results of ongoing simulations soon.

        \clearpage\mbox{}\clearpage

	\chapter*{List of Publications}
\begin{enumerate}
	\item Anosh Joseph and \textbf{Arpith Kumar}, {\it Complex Langevin simulations of zero-dimensional supersymmetric quantum field theories}, \href{https://journals.aps.org/prd/abstract/10.1103/PhysRevD.100.074507}{Phys. Rev. D \textbf{100}, 074507 (2019)} {\href{https://arxiv.org/abs/1908.04153}{(arXiv: 1908.04153 [hep-th])}} 

	\item Anosh Joseph and \textbf{Arpith Kumar}, {\it Complex Langevin Dynamics and Supersymmetric Quantum Mechanics}, \href{https://doi.org/10.1007/JHEP10(2021)186}{J. High Energ. Phys. {\bf 2021}, 186 (2021)} {\href{https://arxiv.org/abs/2011.08107}{(arXiv: 2011.08107 [hep-lat])}} 	
	
	\item \textbf{Arpith Kumar} and Anosh Joseph, {\it Complex Langevin simulations for $\mathcal{PT}$-symmetric models}, \href{https://doi.org/10.22323/1.396.0124
	}{  PoS LATTICE2021 \textbf{124} (2022) }{\href{https://doi.org/10.48550/arXiv.2201.12001}{(arXiv: 2201.12001 [hep-lat])}}
	
	\item \textbf{Arpith Kumar}, Anosh Joseph, and Piyush Kumar, {\it Complex Langevin study of spontaneous symmetry breaking in IKKT matrix model}, \href{https://doi.org/10.22323/1.430.0213}{PoS LATTICE2022 \textbf{213} (2023)}, accepted for publication  {\href{https://doi.org/10.48550/arXiv.2209.10494}{(arXiv: 2209.10494 [hep-lat])}} 	  	
\end{enumerate}
	\thispagestyle{empty}
	\clearpage
	\thispagestyle{empty}
	\mbox{}\clearpage
	\thispagestyle{empty}
	
	\bibliography{thesis.bib}
	\bibliographystyle{JHEP.bst}
	
	
\end{document}